\documentclass[fleqn,usenatbib]{mnras}
\usepackage{newtxtext,newtxmath}
\usepackage{graphicx}
\usepackage{xspace}
\usepackage[T1]{fontenc}
\usepackage{ae,aecompl}
\usepackage{tabularx}
\usepackage{array}
\usepackage[figuresleft]{rotating}
\usepackage{capt-of}

\newlength{\GSBDLandscapeTableWidth}
\newcommand{\GSBDtablecaption}[2]{%
  \refstepcounter{table}%
  \label{#1}%
  \noindent\textbf{Table~\thetable.}\enspace #2\par\vspace{12pt}%
}

\newlength{\GSBDLandscapeXShift}
\usepackage{amsmath}
\usepackage{pdflscape}
\usepackage{rotating}
\usepackage{longtable}
\usepackage{booktabs}

\newcommand{\orcid}[1]{\href{https://orcid.org/#1}{\textsuperscript{\includegraphics[width=8pt]{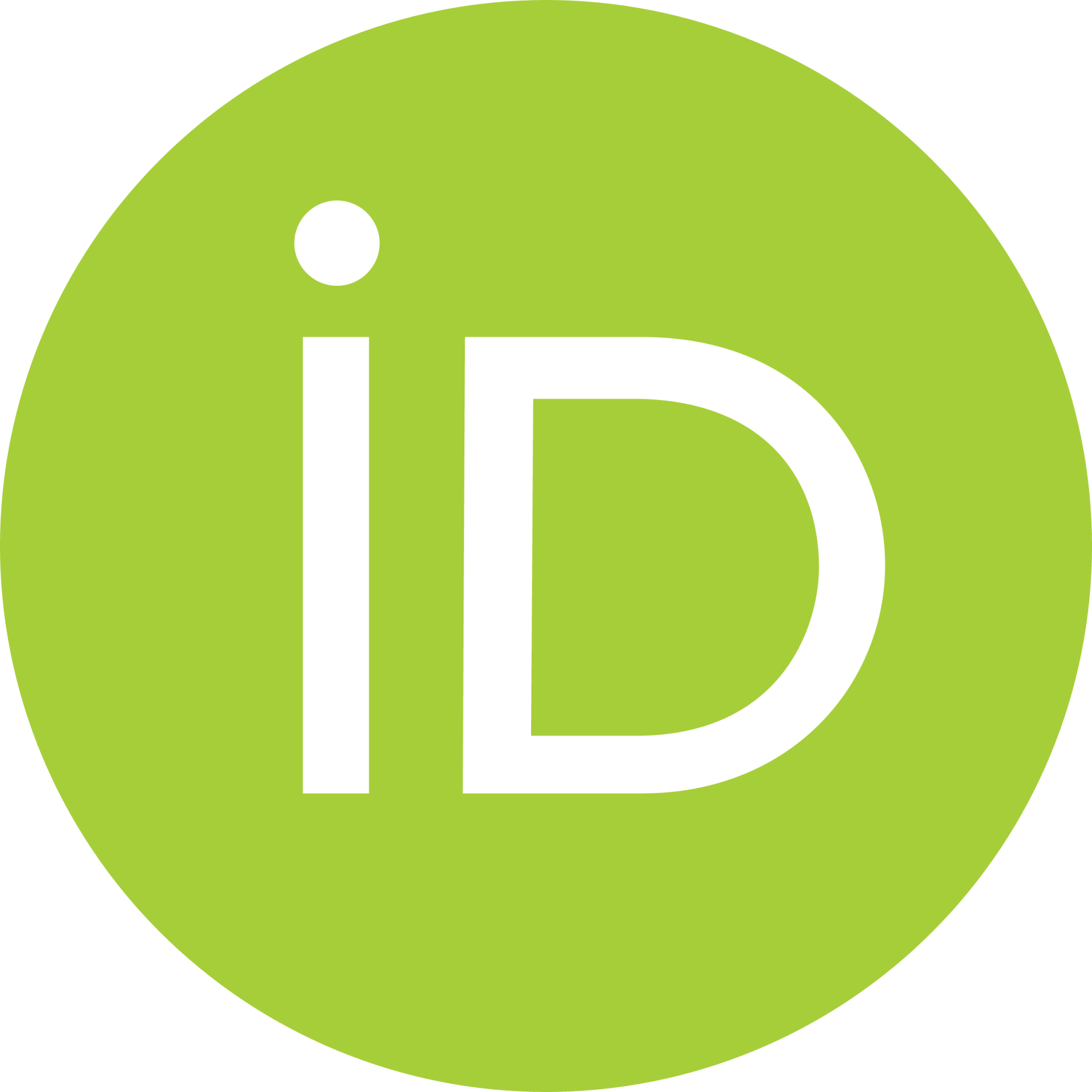}}}}

\title[Parallax of an extremely cold Y dwarf]
{The cold substellar frontier in \textit{JWST} Extragalactic surveys: a robust parallax measurement yields the most distant known ultra-cold Y dwarf}
\author[F.-Y. F. Liu et al.]
{Feng-Yuan Frey Liu$^{1}$\orcid{0000-0003-1386-3676}$^{1}$\thanks{E-mail: Fengyuan.Liu@ed.ac.uk}, Derek J. McLeod$^{1}$\orcid{0000-0003-4368-3326}, Xueqing Chen$^{1}$\orcid{0009-0005-9339-2369}, Ben J. Sutlieff$^{1,2}$\orcid{0000-0002-9962-132X}, Ho-Hin Leung$^{1}$\orcid{0000-0003-0486-5178}, \newauthor Cl\'emence Fontanive$^{1,2}$\orcid{0000-0002-2428-9932}, James S. Dunlop$^{1}$\orcid{0000-0002-1404-5950}, Ross J. McLure$^{1}$\orcid{0009-0005-9742-2318}, Trent J. Dupuy$^{1,2}$\orcid{0000-0001-9823-1445}, Adam C. Carnall$^{1}$\orcid{0000-0002-1482-5818}, \newauthor Thomas M. Stanton$^{1}$\orcid{0000-0002-0827-9769}, Beth A. Biller$^{1,2}$\orcid{0000-0003-4614-7035}, Isaac J. B. Holst$^{1}$\orcid{0009-0006-3003-9688}, Callum T. Donnan$^{3}$\orcid{0000-0002-7622-0208}, Ryan Begley$^{4}$\orcid{0000-0003-0629-8074}
\\
$^{1}$Institute for Astronomy, University of Edinburgh, Royal Observatory, Blackford Hill, Edinburgh, EH9 3HJ, UK\\
$^{2}$Centre for Exoplanet Science, University of Edinburgh, Edinburgh, EH9 3HJ, UK\\
$^{3}$ NSF's National Optical-Infrared Astronomy Research Laboratory, 950 N. Cherry Avenue, Tucson, AZ 85719, USA\\
$^{4}$ Armagh Observatory and Planetarium, College Hill, Armagh, BT61 9DG, N. Ireland, UK\\}

\date{Accepted XXX. Received YYY; in original form ZZZ}

\pubyear{\the\year{}}

\begin{document}
\label{firstpage}
\pagerange{\pageref{firstpage}--\pageref{lastpage}}
\maketitle

\begin{abstract}
We present an analysis of JADES-GS-BD-5, an extremely cold Y dwarf identified in the JADES/GOODS-South field. Using multi-epoch \textit{JWST}/NIRCam imaging spanning a $\simeq$ 2-yr baseline, we calibrate the \textit{JWST} astrometry and obtain the first robust trigonometric parallax measurement for a brown dwarf discovered in a \textit{JWST} extragalactic deep field. We measure an absolute proper motion of $\mu_{\rm tot}=49.8^{+0.3}_{-0.3}$~mas~yr$^{-1}$ and a significant parallax of $\varpi=15.7^{+1.4}_{-1.3}$~mas, corresponding to a geometric distance of $d=64^{+6}_{-5}$~pc, $\sim3\times$ further than that of the previously known most distant Y dwarfs with parallax-measured distances. We hence determine a tangential velocity of $v_{\rm tan}=15.0^{+1.4}_{-1.3}$~km~s$^{-1}$, consistent with a Galactic thin-disk object. We find no evidence for binary acceleration, significant photometric variability, or a plausible widely-separated stellar or substellar host. The $J$-band spectral morphology of JADES-GS-BD-5 is consistent with existing Y-dwarf templates, while its extremely faint luminosity and overall near-infrared spectral profile confirm that it is one of the coldest dwarfs discovered to date. We therefore adopt a classification of $>$Y1, consistent with a NIRCam colour-based proxy of {\it c}Y1.5. The robust parallax-determined distance of JADES-GS-BD-5 allows us to determine that its bolometric luminosity is $\log(L_{\rm bol}/L_{\odot})=-7.08^{+0.09}_{-0.09}$. Evolutionary modelling places JADES-GS-BD-5 in the planetary-mass regime and makes it a rare benchmark for the study of cold giant-planet atmospheres. Current atmospheric models fail to reproduce its SED in detail, pointing to the importance of disequilibrium chemistry and vertical mixing. Water deficiency may also contribute, but the evidence for this is currently unclear. Finally, we also report additional proper motion measurements for eight of the other nine known JADES Y-dwarf candidates, highlighting \textit{JWST} extragalactic surveys as a powerful route to building a statistically useful, parallax-calibrated sample of Y dwarfs.
\end{abstract}


\section{Introduction}

Trigonometric parallax remains the only model-independent route to derive the distances of individual brown dwarfs and isolated planetary-mass objects, and such distance measurements are of crucial importance for inferring their physical properties. As such objects never reach a stable hydrogen-burning main sequence and instead cool continuously, their physical properties at any time are governed by a combination of mass, age, and composition \citep{Burrows2001, Bowler2016, Dupuy2017, Chabrier2023}. Measurements of their distances allow us to convert their observed apparent magnitudes into absolute magnitudes and ultimately luminosities. This provides an anchor for empirical spectral type--magnitude and colour--magnitude relations, which can then be used to deduce the physical properties of objects without parallaxes \citep[e.g.,][]{Dupuy2012, Dupuy2013, Faherty2012, Best2021, Fontanive2023, Fontanive2025, Fontanive2026}. Substellar objects with well-constrained properties supply the valuable empirical benchmarks against which ultracool atmosphere and evolutionary models are tested \citep[e.g.][]{Phillips2020, Marley2021, LacyBurrows2023, Beiler2024, Mukherjee2024, Leggett2025, Mang2026, Radcliffe2026}, and volume-complete samples further allow us to place crucial constraints on the substellar luminosity function, potentially providing further insight into the low-mass cut-off of the stellar initial mass function (IMF) \citep[e.g.][]{Kirkpatrick2012, Kirkpatrick2021, Best2024, hainline2026b}. Accurate distances also allow measurements of proper motion to be converted into tangential velocities, and hence can provide valuable kinematic information \citep[e.g.][]{Faherty2009, Burningham2013, Fontanive2025}.

These measurements have, however, so far been confined to the immediate solar neighbourhood for the coldest brown dwarfs, i.e., late-T and Y dwarfs. They are intrinsically fainter and predominantly radiate into the thermal near and mid-infrared (IR). Y dwarfs, with $T_{\rm eff}\lesssim500$\,K and $\log(L_{\rm bol}/{\rm L_\odot})\lesssim-6$ \citep[e.g.][]{Dupuy2013, Leggett2021, Fontanive2026}, emit the bulk of their flux at 4--5\,$\mu$m, where the thermal background renders ground-based astrometry difficult and where the \textit{WISE} survey that discovered them is complete only to ${\sim}10$--$15$\,pc \citep{Cushing2011, Kirkpatrick2012, Kirkpatrick2019}. Measurements of Y dwarf parallaxes have consequently required dedicated space-based campaigns with \textit{Spitzer} and \textit{HST} \citep[e.g.][]{Marsh2013,Dupuy2013,Beichman2014,Martin2018,Kirkpatrick2019,Kirkpatrick2021,Fontanive2021,Fontanive2025,Marocco2026}. The result of their faintness and the consequent difficulty to measure parallaxes is that only ${\sim}50$ Y dwarfs are known to date, all within 20\,pc of the Sun, and even fewer have a measured parallax.

Within the Y-class, the census of known objects is even less established towards later subtypes. Brown dwarfs of spectral types later than Y1 are known only in small numbers, essentially all within ${\sim}15$\,pc \citep{Kirkpatrick2021,Leggett2026}, and the cold extreme of the sequence almost solely relies on a single object, WISE~J085510.83$-$071442.5 at 2.28\,pc \citep[WISE~0855 for short;][]{Luhman2014, Skemer2016, Kirkpatrick2019}. The absolute magnitude, luminosity, and effective temperature scales for these subtypes are consequently limited, calibrated by only a handful of parallax measurements \citep{Beiler2024,Fontanive2025}. This impacts our interpretation of their atmospheres, which have effective temperatures overlapping those of the Solar system giant planets and the coldest directly imaged exoplanets \citep[e.g.,][]{Matthews2024, BardalezGagliuffi2025, Lagrange2025, Crotts2025}, and are thus expected to have atmospheres with complex molecular absorption features, disequilibrium chemistry, variability, and even water clouds in the lowest-temperature cases \citep[e.g.][]{Morley2014, Miles2020, Phillips2020, Mang2022, LacyBurrows2023, Lee2025, Tan2025, Fontanive2026, Leggett2026}. Expanding the number of parallax-measured extremely cold brown dwarfs, particularly those comparable to WISE~0855, is therefore a key goal of substellar science.
For Galactic studies, these objects trace the low-mass terminus of the star-formation process, where the shape of the mass function encodes the physics of fragmentation and the minimum mass for opacity-limited collapse \citep{Kirkpatrick2019, Kirkpatrick2024}. 
A parallax-calibrated sample beyond the local volume would also provide not only a larger census, but also a direct probe of a different Galactic environment \citep{Burgasser2024,hainline2026b}.

Deep extragalactic imaging with \textit{JWST} has opened an unanticipated route to these objects. Surveys such as JADES \citep{Eisenstein2023}, CEERS \citep{Finkelstein2023}, PRIMER (\textcolor{black}{Dunlop et al., in prep.}), COSMOS-Web \citep{Casey2023}, and UNCOVER \citep{Bezanson2024} reach of order nJy 1-$\sigma$ depths in 1--5\,$\mu$m bands in which cold brown-dwarf spectral energy distributions (SEDs) peak, and their fields are revisited repeatedly across cycles (e.g., follow-up of PRIMER with MINERVA; \citealt{Muzzin2025}). At the high Galactic latitudes probed by these surveys, many cold substellar objects have now been detected: photometric searches have produced dozens of T- and Y-type candidates at implied distances of ${\sim}0.1$--4\,kpc \citep{Nonino2023, Wang2023, Hainline2024a, hainline2026b, A_Chen2025}, a growing subset of which have been confirmed spectroscopically, several revealing subsolar metallicities consistent with thick-disc or halo membership \citep{Langeroodi2023, Burgasser2024, Hainline2024b,Tu2025a,Tu2025b,Li2026,Morrissey2026}. Indeed, cold dwarfs are effective interlopers in extragalactic samples. Their red near-IR colours mimic those of high-redshift galaxies and dust-reddened active galactic nuclei (AGN), a degeneracy familiar from \textit{HST} deep-field work \citep{Pirzkal2005,Dunlop2013,Aganze2022} and now recurring in \textit{JWST} samples. Photometrically selected candidate ``little red dots'' (LRDs), massive (stellar mass $\rm M_{*}>10^{10.5}\,M_{\odot}$) $z>4$ galaxies and ultra-high-redshift ($z>20$) galaxies are all classes of object that have recently proven to be substellar objects \citep[e.g.,][]{Langeroodi2023,Hainline2024b, Tu2025a}. An outstanding case is that of Capotauro, a redshift $z\sim32$ candidate \citep{Gandolfi2026b} in the \textit{JWST} Cosmic Evolution Early Release Science Survey (CEERS) field, which has been recently confirmed to be a cold brown dwarf through astrometric proper motion detection by \citet{Liu2026}. From an extragalactic perspective, the aforementioned deficiency of a large, parallax-calibrated late-T and Y-dwarf templates for SED fitting therefore compounds the misinterpretation.

The multi-epoch cadence of these fields has already yielded some of the most precise astrometry achieved for cold substellar objects. Proper motions have been measured for a growing number of deep-field brown dwarfs, from \textit{HST}--\textit{JWST} and \textit{JWST}--\textit{JWST} baselines, with implied transverse velocities of tens to a couple of hundred km\,s$^{-1}$ \citep{Hainline2024a,Hainline2024b,hainline2026b,Liu2026}. A proper motion alone, however, leaves the distance, and therefore the luminosity and tangential velocity, resting on model-dependent spectrophotometric estimates, which for cold brown dwarfs are uncertain at the tens-of-per-cent level and degenerate with metallicity and gravity \citep{Hainline2024a,Beiler2024, Mukherjee2024, Leggett2025}. Parallax measurements for objects at such distances require milliarcsecond-level relative astrometry with adequate sampling of the parallactic ellipse, which mosaic-level catalogue positions and sparse annual revisits have not so far delivered. To date, no parallax has been reported for any brown dwarf discovered in a \textit{JWST} deep extragalactic field. 

In this paper, we report such a measurement: the first robust parallax detection for the ultracool, distant brown dwarf JADES-GS-BD-5 in the \textit{JWST} Advanced Deep Extragalactic Survey (JADES) Great Observatories Origins Deep Survey (GOODS)-South field. JADES-GS-BD-5, hereafter GSBD5, was initially reported by \citet{Hainline2024a} at ICRS J2000 coordinates RA = 03:32:20.170 and Dec = $-$27:50:21.65, and was identified photometrically as a cold brown-dwarf candidate with an overall motion of $50\pm20~$mas~yr$^{-1}$. It was indeed the coldest dwarf found in the JADES fields \citep{hainline2026b}, and recent spectroscopic follow-up classified it as a Y0--Y1 dwarf \citep{Hainline2026c}. Here, we use the latest available \textit{JWST} imaging and spectroscopic data to measure its astrometric motion and constrain its physical properties. We further analyse a broader JADES Y-dwarf sample, and demonstrate the current frontier of the ultracool Galactic substellar population in deep \textit{JWST} extragalactic survey fields. 

This paper is structured as follows. In Section~\ref{sec:data} we describe the NIRCam, MIRI and NIRSpec data and their reduction. We present the astrometric measurement in Section~\ref{sec:astrometry}, including the relative and absolute reference-frame calibrations and the joint proper-motion and parallax fit. The photometric measurement of GSBD5 with both NIRCam and MIRI is then presented in Section~\ref{sec:photometry}. In Section~\ref{sec:specphoto_analysis}, we investigate the spectral type of GSBD5, estimate its physical properties with evolutionary models, and explore atmospheric models to reproduce its spectrophotometric SED. In Section~\ref{sec:discussion}, we discuss the evidence of GSBD5 being an isolated planetary mass object and an ultracool giant-planet analogue, the possible physical scenarios needed to explain its unique SED, and a sample of nine other Y dwarf candidates discovered in JADES in the context of using \textit{JWST} extragalactic survey fields to survey this coldest dwarf population. We summarise our conclusions in Section~\ref{section:conclusion}. All magnitudes are quoted in the AB system \citep{Oke1974, Oke1983} unless stated otherwise.

\section{Data}\label{sec:data}

\subsection{Imaging data}\label{subsec:img}

Here we briefly describe the imaging data targeting the sample of brown dwarf candidates in the GOODS-South and GOODS-North areas, including our main target, GSBD5, and other similar objects discussed in later sections. We use our own reductions for all NIRCam imaging data sets, using the software \textsc{pencil}, which is a customised version of \textit{JWST} pipeline version 1.18.0 (Magee et al., in prep). We use the Calibration
Reference Data Syste (CRDS) context \textit{jwst\_1414.pmap}. In Table \ref{tab:programmes}, we provide a summary of the different imaging programs and observation dates for each source covered by our study. Pixels are approximately $0.031''\times0.031''$ in NIRCam short-wavelength (SW) bands, i.e., F070W to F210M; $0.063''\times0.063''$ in NIRCam long-wavelength (LW) bands, i.e., F250M to F444W; and $0.060''\times0.060''$ in MIRI bands.

\subsubsection{NIRCam and MIRI imaging for GSBD5}
There have been seven observing programs with the \textit{JWST} \citep{Gardner2006, Gardner2023} Near Infrared Camera \citep[NIRCam,][]{Rieke2003, Rieke2023a}, spanning a $\simeq2$ year window that have targeted the region within which GSBD5 resides. We group them as five epochs. 

In Epoch 1, as part of the JADES survey \citep{rieke2023,Eisenstein2023} GTO~1180 (PI Eisenstein), ultra-deep imaging in F090W, F115W, F150W, F200W, F277W, F356W, F410M and F444W was taken on 7th - 8th October, 2022 UT. Shortly after, between 20th - 24th October 2022, deeper imaging in these same bands, plus F335M, was taken as part of GTO~1210 (PI Luetzgendorf), forming Epoch 2. Approximately one month later, GSBD5 was imaged in F182M, F210M, and F444W as part of FRESCO (GO~1895; PI Oesch; \citealt{oesch2023}), and F200W+F444W with GO~2198 \citep{Barrufet2025}, both taken on 18th November 2022. We consider both of these programs to be Epoch 3.

Eleven months later, on 14th October 2023, GSBD5 was imaged in F115W, F150W, F200W, F277W, F356W and F444W as part of the pure parallel program PANORAMIC (GO~2514; PIs Williams, Oesch; \citealt{williams2025}) as Epoch 4, along with ultra-deep data in medium bands F162M, F182M, F210M, F250M, F300M and F335M taken around the same time from GO~3215 (PI Eisenstein), i.e. the JADES Origins Field \citep{eisenstein2025}. Finally, a further year later, imaging was taken in F070W, F150W, F200W and F444W as part of GTO~4540 (PI Eisenstein) on 19th October 2024, forming Epoch 5. The programs and observations dates are summarised in Table~\ref{tab:programmes}, and the epoch coverage of each band is tabulated in Table~\ref{tab:photometry}.

We also used Mid-Infrared Instrument \citep[MIRI,][]{Glasse2015, Wright2023} imaging data, from SMILES (GTO~1207; PI Rieke; \citealt{Rieke2024}). These data were taken in December 2022 in F560W, F770W, F1000W, F1280W, F1500W, F1800W, F2100W and F2550W. We used the DR1 mosaic products, the data reduction process for which is detailed in \citet{Alberts2024}.

\begin{table}
\centering
\caption{A list of the various NIRCam imaging programs targeting each of the brown dwarfs featured in this study, along with the date of observation. They are all the Y dwarf candidates identified in JADES DR5 fields by \citet{hainline2026b}.}
\label{tab:programmes}
\setlength{\tabcolsep}{5pt}
\begin{tabularx}{\columnwidth}{@{}l l >{\raggedright\arraybackslash}X@{}}
\hline
Source ID & PID & Observation Dates\\
\hline
JADES-GS-BD-4 & 1180 & Oct 2022\\
       & 1286 & Oct 2023\\
       & 4540 & Oct 2024\\
\hline
JADES-GS-BD-5 & 1180 & Oct 2022\\
       & 1210 & Oct 2022\\
       & 1895 & Nov 2022\\
       & 2198 & Nov 2022\\
       & 2514 & Oct 2023\\
       & 3215 & Oct 2023\\
       & 4540 & Oct 2024\\
\hline
JADES-GS-BD-8 & 1180 & Oct 2022, Oct 2023, Nov 2023, Jan 2024\\
       & 1895 & Nov 2022\\
       & 6541 & Nov 2023\\
       & 6630 & Oct 2024\\
       & 8882 & Oct 2025, Dec 2025\\
\hline
JADES-GS-BD-18 & 2514 & Oct 2023\\
\hline
JADES-GS-BD-21 & 1180 & Oct 2022, Oct 2023\\
        & 1286 & Oct 2023\\
        & 2514 & Oct 2023\\
\hline
JADES-GN-BD-5 & 1181 & Feb 2023\\
       & 1895 & Feb 2023\\
       & 2674 & Mar 2023\\
       & 3577 & Feb 2024\\
       & 7404 & Jan 2026, Mar 2026, Apr 2026\\
\hline
JADES-GN-BD-8 & 1181 & May 2023\\
       & 1264 & Dec 2023\\
       & 5398 & Feb 2025, May 2025\\
       & 7404 & Jan 2026\\
\hline
JADES-GN-BD-9 & 1181 & Feb 2023\\
        & 2674 & Mar 2023\\
        & 6434 & Jan 2025\\
\hline
JADES-GN-BD-11 & 1181 & Feb 2023\\
        & 2514 & Mar 2023\\
        & 6434 & Jan 2025, Feb 2025\\
\hline
JADES-GN-BD-16 & 1181 & Apr 2023\\
        & 2514 & Mar 2023\\
        & 2674 & Mar 2023\\
\hline
\end{tabularx}
\end{table}

\subsubsection{NIRCam imaging for other sources}

In addition to GSBD5, we analyse the other nine Y dwarf candidates reported in JADES \citep{hainline2026b} in the later section of this paper. For the candidates located in the GOODS-South region, as well as the aforementioned JADES GTO~1180, FRESCO and PANORAMIC datasets, we include imaging from GTO~1286 (PI Luetzgendorf) and DD 6541 (PI Egami), as well as the calibration programs, CAL/NIRCAM~6630 (PI Boyer) and CAL/NIRCAM~8882 (PI Boyer).

For the brown dwarf candidates in GOODS North, we include JADES imaging from GTO~1181 (PI Eisenstein; see \citealt{Eisenstein2023}) and GTO~1264 (PI Colina Robledo). GOODS North also includes pointings from FRESCO (GO~1895), CONGRESS (GO~3577; PI Egami), GO~2674 (PI Arrabal Haro), TWINKLE (GO~7404; PIs Naidu, Matthee, Chisholm), and the pure-parallel imaging data sets PANORAMIC, Sapphires (GO~6434; PI Egami) and POPPIES (GO~5398; PI Kartaltepe).

\subsection{Spectroscopic data}\label{subsec:spec}
GSBD5 was observed with the {\it JWST} Near Infrared Spectrograph \citep[NIRSpec;][]{Jakobsen2022, Boker2023} using the Micro-Shutter Assembly \citep[MSA;][]{Ferruit2022} and the low resolution $R\simeq30$--330 PRISM setting as part of the OASIS program (GO~5997; PI Looser). It received 27.7 hours of total exposure time from 2025-01-06 00:37:02.566 UTC (MJD 60681.02572414352) to 2025-01-07 06:57:08.550 UTC (MJD 60682.2896822921). The observations were split between two visits on the two dates, and 36 exposures were obtained in three dithered configurations. The MSA target coordinates were (03:32:20.165, $-$27:50:21.75), i.e., $\sim0.1\arcsec$ southward and $\sim0.05\arcsec$ westward from the coordinates observed in October 2022.

To perform spectroscopic data reduction, we use version 1.20.2 of the pipeline \citep{Bushouse2025} and CRDS\_CTX=\texttt{jwst\_1481.pmap} version of the {\it JWST} CRDS files. First, we acquired the level 1 ``uncal'' data products from the Mikulski Archive for Space Telescopes (MAST). We then processed these products with the level 1 pipeline (\texttt{calwebb\_detector1}) following the methods for MSA slit spectroscopy in \cite{Leung2026a}, where we used the default configurations for all steps except we turned on the \texttt{clean\_flicker\_noise} step to remove the effects of 1/f noise from the exposures. For this \texttt{clean\_flicker\_noise} step, we set \texttt{fit\_method} to \texttt{median}, \texttt{mask\_science\_region} to \texttt{True}, \texttt{background\_method} to \texttt{None} and \texttt{n\_sigma} to 2. Next, we ran the level 2 pipeline (\texttt{calwebb\_spec2}) using the default configurations. We then masked any pixels in the resulting level 2b ``cal'' files that had at least one flag among a list of selected data quality flags.\footnote{Bits: 0, 1, 3, 4, 6, 7, 10, 11, 12, 13, 16, 17, 18, 19, 20, 21, 24, 26, 27, 29 and 30 (see \url{https://jwst-pipeline.readthedocs.io/en/latest/jwst/references_general/references_general.html\#data-quality-flags})} Finally, we run the level 3 pipeline (\texttt{calwebb\_spec3}) using the default configurations to obtain a combined, rectified 2D spectrum, and extract the 1D spectrum via optimal extraction \citep{Horne1986}. 

\section{Astrometric measurement}\label{sec:astrometry}

\subsection{Reference-frame calibrations}
\label{subsec:ref_frame_calib}

Reference calibration is essential for improving \textit{JWST}/NIRCam astrometry beyond the standard pipeline-calibrated precision and robustly measuring the motion of GSBD5. We first aligned the individual \textit{JWST} exposures to establish a consistent relative reference frame, and subsequently tied this frame to the International Celestial Reference System \citep[ICRS;][]{Arias1995}. Both steps introduce correlated uncertainties that must be propagated into the motion of GSBD5. All astrometric measurements were performed in a projected common tangent plane, $(\xi,\eta)$, whose coordinate system was initially inherited from the \textit{JWST} pipeline and subsequently calibrated as described below. Its positive directions point eastwards and northwards, respectively.

\subsubsection{Relative calibration}
\label{subsubsec:rela_calib}

\begin{figure*}
    \centering
    \includegraphics[width=1.5\columnwidth]{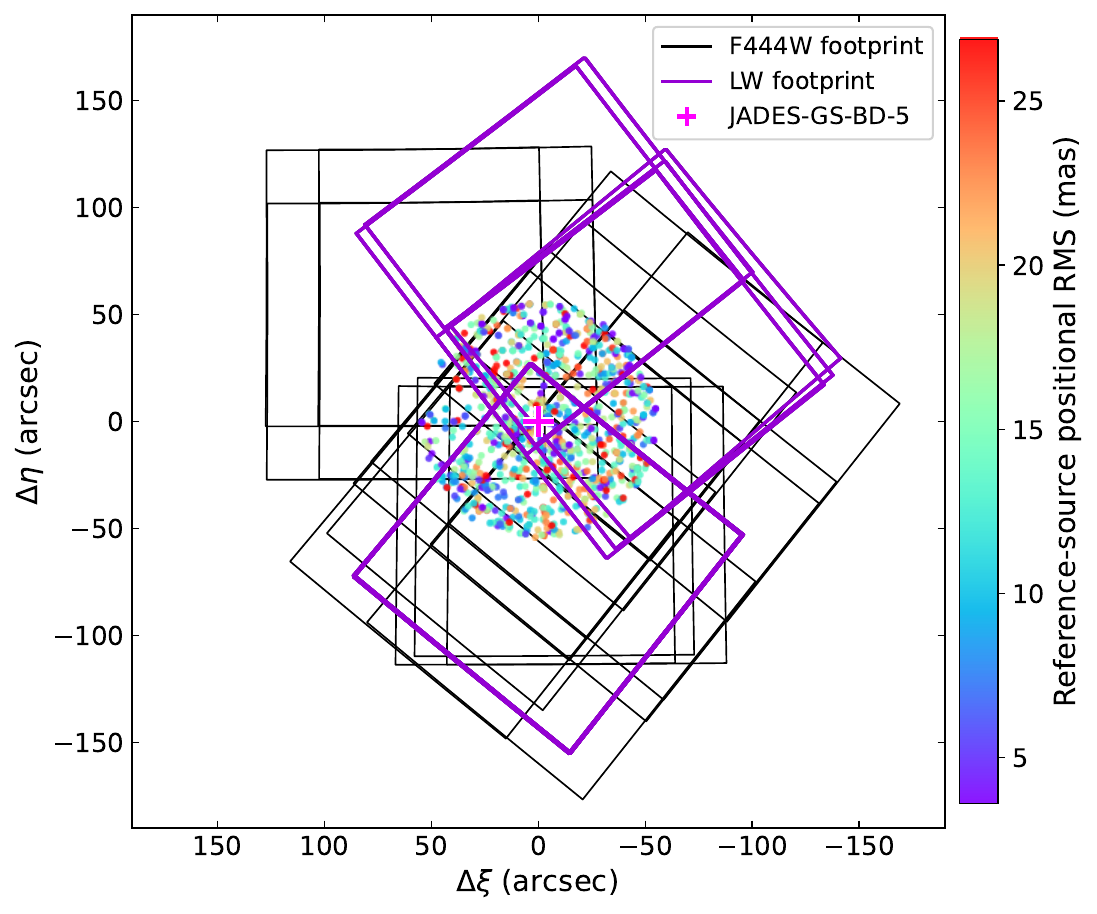}
    \caption{
    Per-exposure footprints and astrometric reference sources. Thin black outlines show the WCS footprints of the 50 F444W exposures, while thin purple frames show the 130 exposures obtained in other three detection LW bands, F277W, F356W, and F410M. The footprints in these three additional long-wavelength filters are almost entirely contained within the combined F444W footprint. Coloured points mark the 850 sources retained in the F444W reference-frame solution. Their colours indicate the 2-D rms positional residuals after fitting the relative astrometric network, providing an empirical measure of each source's cross-exposure positional stability. The pink cross marks the position of JADES-GS-BD-5.
    }
    \label{fig:reference_frame}
\end{figure*}

The relative calibration largely follows \citet{Liu2026}, but the present data require two modifications to the workflow. First, the images of GSBD5 were obtained by several \textit{JWST} programs with markedly different footprints (Figure~\ref{fig:reference_frame}), instead of by surveys designed to repeat a common footprint (as was the case for Capotauro; \citealt{Liu2026}). A single set of reference sources is therefore not present in every exposure. Secondly, GSBD5 is detected independently in several LW filters, permitting a direct cross-band test of the inferred motion. F444W provides the highest target S/N, the largest number of exposures, and coverage of all five epochs, and is therefore used for the adopted astrometric measurement. The remaining detection LW filters F277W, F300M, F356W and F410M, each with final $\mathrm{S/N>10}$ (see Section~\ref{sec:photometry}), are retained as independent consistency tests.

Fifty F444W exposures, or frames, were obtained over five epochs, and GSBD5 was significantly detected in every exposure. Reference candidates were detected independently with Source Extraction and Photometry \citep[\textsc{sep};][]{Barbary2018}, the \textsc{Python} implementation of \textsc{Source Extractor} \citep[][]{Bertin1996}, with centroids measured by Gaussian windows. We retained detections within $55\arcsec$ of GSBD5 with $0.2''$-radius aperture $\mathrm{S/N}\geq8$, $2.1\leq\mathrm{FWHM}\leq12$ detector pixels, axis ratio $q\geq0.25$, and a clear separation from the target. Anomalous pixels were excluded. At most the 500 highest-S/N candidates were retained in each exposure. These deliberately broad cuts admit stars, compact galaxies, and some extended but core-dominated galaxies, whilst suppressing highly elongated blends and diffuse structures.

The preliminary pipeline-produced World Coordinate System (WCS) was used to initialise the cross-exposure associations. We adopted a reference-source matching radius of $0.06\arcsec$ and required each source to occur in at least two exposures. This produced 858 initial F444W reference candidates. Seven sources were rejected by visible, large offsets between exposures, which include five \textit{Gaia} stars (see Section~\ref{subsubsec:abso_calib}) and two galaxies with noise-affected centroids, leaving 851 reference sources and 11,412 retained per-exposure detections. One remaining \textit{Gaia} star was removed to further reduce motion-related uncertainty, leaving 850 sources, for which there were 11,371 detections. These partially overlapping detections form a connected network: although they are not all commonly present in every exposure, the overlapping sources connect all five epochs. For the other three GSBD5-detected LW bands, 1408 reference sources contributed $\sim43,000$ detections across 85 exposures, with at least 47 reference detections in every exposure. The distribution of these reference sources is illustrated in Figure~\ref{fig:reference_frame}.

For each exposure, we fitted an affine correction among reference sources,
\begin{equation}
\begin{pmatrix}
\Delta\xi_{\rm relative}\\
\Delta\eta_{\rm relative}
\end{pmatrix}
=
\begin{pmatrix}
a_{0} & a_{1} & a_{2}\\
b_{0} & b_{1} & b_{2}
\end{pmatrix}
\begin{pmatrix}
1\\
\xi\\
\eta
\end{pmatrix}.
\end{equation}
which accounts for translation, rotation, differential scale, and shear. This allowed us to correct the frames within the simultaneous cross-frame reference-source network.

Reference sources were inverse-variance weighted during the correction fit, with the positional uncertainties reported by \textsc{sep}. However, the nominal centroid uncertainties can become unrealistically small for bright sources and do not include residual PSF mismatch, morphology, or detector-related errors. We therefore imposed an empirical lower bound on the positional uncertainty of each reference source, $\sigma_{\rm used}=\max(\sigma_{\rm SEP},3.7~{\rm mas})$. This value was determined from the frame-to-frame scatter of the reference sources measured relative to their mean positions after the first iteration of relative calibration with only $\sigma_{\rm SEP}$. The scatter has a median of 5.15~mas and a 16\% lower bound of 4.77~mas, corresponding to a 4-$\sigma$ equivalent lower-tail value of $\sim3.7$~mas. This choice makes uncertainties below the floor statistically negligible among the $\sim11,000$ detections. Such a conservative prescription caps exceptionally small values and prevents nominally highest-S/N sources dominating the reference sample. We found our reference-source positional uncertainty in good agreement with the approximately 5-mas bright-source centroid scatter relative to \textit{Gaia} reported by JADES DR5 \citep{Johnson2026}.

Neighbouring reference sources can share background, PSF, and residual distortion errors, so the $\sim11,000$ detections cannot be treated as independent. We found weak evidence for residual spatial coherence on a scale of approximately $9.6\arcsec$, and adopted a conservative $20\arcsec$ block size to grid the reference field and group the reference sources in each block. Specifically, to construct these blocks, we drew an independent two-dimensional grid zero point uniformly over $[0,20\arcsec)^2$, assigned the 850 sources to the resulting occupied blocks, and resampled the occupied blocks with replacement. The continuously randomised division produced 30--36 occupied blocks over the reference-source-covered region and prevents the uncertainty estimate from depending on one arbitrary placement of the block boundaries. These blocks were used in later bootstrapping as basic units.

Evaluating the affine transformation of each exposure at the position of GSBD5 by bootstrapping realisations, we derived that the relative corrections to a common zero point for the five epochs are $(\Delta\xi_{\rm relative},\, \Delta\eta_{\rm relative}) = (-10.29,+17.49)$, $(+5.37,-17.76)$, $(+10.91,+0.84)$, $(+1.73,+0.23)$, and $(-7.73,+0.80)$~mas, respectively. Their uncertainties after the correction are $(\delta\xi_{\rm relative},\, \delta\eta_{\rm relative}) = (0.38, 0.51)$, $(0.44, 0.34)$, $(0.83, 0.64)$, $(0.58, 0.42)$, and $(0.47, 0.45)$~mas. Therefore, the relative calibration substantially improved between-frame target position scattering by roughly an order of magnitude from the native \textit{JWST}-pipeline alignment. The relative corrections remain correlated between epochs and thus enter the motion measurement. It is propagated to the final measurements in a similar block-level bootstrapping approach, as detailed in Section~\ref{subsec:motion}.

\subsubsection{Absolute calibration}
\label{subsubsec:abso_calib}

The relative solution aligns the exposures and corrects their imperfect sky-to-detector projections, but it remains tied to the initial \textit{JWST}-pipeline WCS and, effectively, the weighted mean apparent motion of the \textit{JWST} reference ensemble, which affect our localisation and kinematic analysis of the target in the Milky Way. We therefore calibrated the absolute reference frame using both \textit{Gaia} stars and distant galaxies. The stellar component supplies the direct connection to the ICRS, whereas the galaxy component constrains the residual uncertainties with a larger-sample anchor.

We first searched {\it Gaia} DR3 \citep{GaiaCollaboration2016, GaiaCollaboration2023} for reference stars within the common F444W footprint. Six stars were found, of which one was rejected because its {\it JWST} centroid was compromised by the detector edge. The remaining five stars contributed 53 detections across the 50 F444W exposures. Their {\it Gaia} positions, proper motions and parallaxes, together with related uncertainties, were propagated from the {\it Gaia} reference epoch to the exact times of the {\it JWST} exposures using the parallax factors evaluated for the {\it JWST} observer. We measured their {\it JWST} positions with the hybrid PSF models (created as in Section~\ref{subsubsec:psf} but with a 4,500~K blackbody SED representative of a K-type star) and jointly fitted a common 2-D affine transformation.

The small number and \textit{Gaia}-motion uncertainties due to the faintness of the stars make a {\it Gaia}-only tie relatively noisy. We therefore supplemented the stellar tie with distant galaxies, for which the apparent motion should be zero. An extragalactic source was required to satisfy the previous reference-source standards and to appear in at least three epochs. It was also required to match either a reliable 3D-HST \citep{Brammer2012} source within $0.1\arcsec$, or a high-quality JADES \citep{CurtisLake2026} NIRSpec spectroscopic redshift within $0.2\arcsec$, and in either case with $z_{\rm spec}>0.1$. Candidates within $0.5\arcsec$ of a {\it Chandra} 7-Ms X-ray counterpart \citep{Luo2017} were removed to avoid variable Active Galactic Nuclei (AGN). The final sample contains 149 distant galaxies. Each galaxy was assigned its own unknown constant position, and the fit measured whether the ensemble retained a common apparent motion after relative alignment. Note that this is different from the previous relative calibration, because here we deliberately only focused on time-dependent motion to trace the residual motion from \textit{Gaia} reference stars, not the static imagery alignment. In this construction {\it Gaia} supplies the connection to the ICRS, whereas the galaxies constrain time-dependent terms (proper motion and parallax) in that connection and thereby reduce the uncertainty arising from the {\it Gaia} motion.

The \textit{Gaia} and extragalactic zero-point solutions are affected by the affine correction in the relative calibration, as it changes the relative position of reference sources between exposures and thus epochs. We subsequently incorporated this uncertainty in the bootstrapping approach detailed in Section~\ref{subsec:motion}. The derived median absolute calibration is $(+0.14\,\mathrm{mas~yr^{-1}},-0.21\,\mathrm{mas~yr^{-1}},+0.13\,\mathrm{mas})$ for the $\xi$- and $\eta$-direction proper motion, and parallax, respectively. Its marginal standard deviations are $(0.11,0.10,0.44)$ in the same units.

\subsection{Motion measurement}
\label{subsec:motion}

\begin{figure}
    \centering
    \includegraphics[width=\columnwidth]{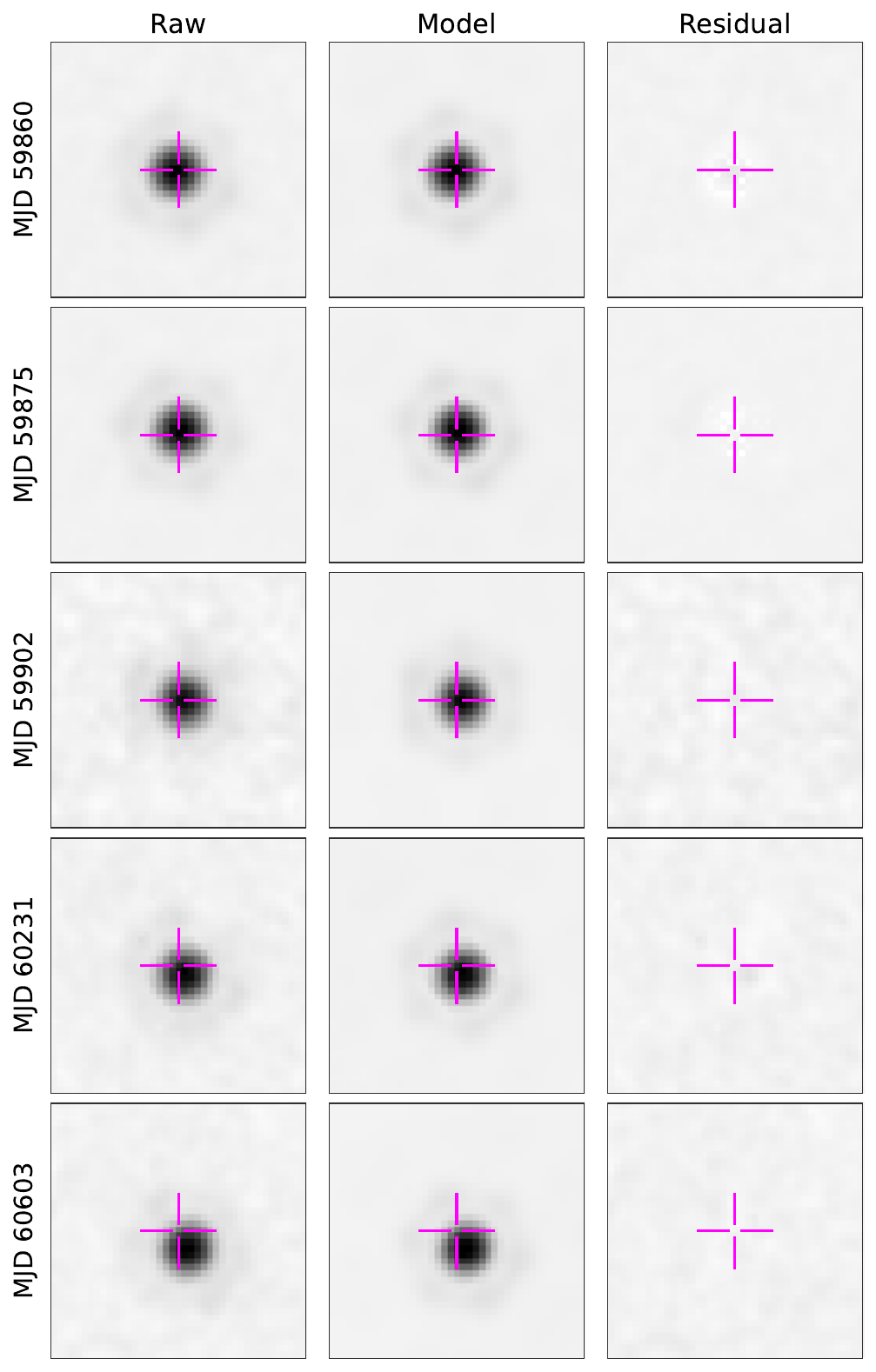}
    \caption{Raw \textit{JWST} images, hybrid PSF models, and the residual images for the F444W observations of GSBD5 in five epochs (rounded to the nearest day). Individual exposures are co-added to produce the $1.2\arcsec\times1.2\arcsec$ cutouts here. The magenta crosshairs mark the target position reported by \citet{hainline2026b} for GSBD5. All panels are plotted at the same colour scale. The hybrid PSF models provide excellent source extraction, and the centroids of GSBD5 display visible apparent motions.}
    \label{fig:stamps}
\end{figure}

\begin{figure*}
    \centering
    \includegraphics[width=0.415\linewidth]{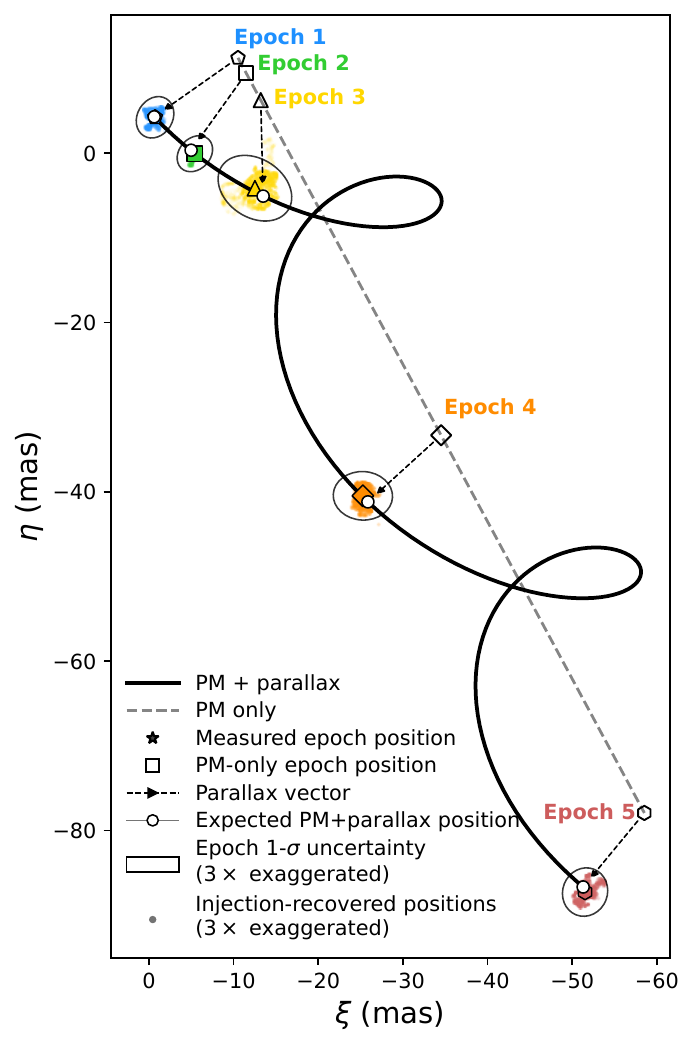}
    \includegraphics[width=0.575\linewidth]{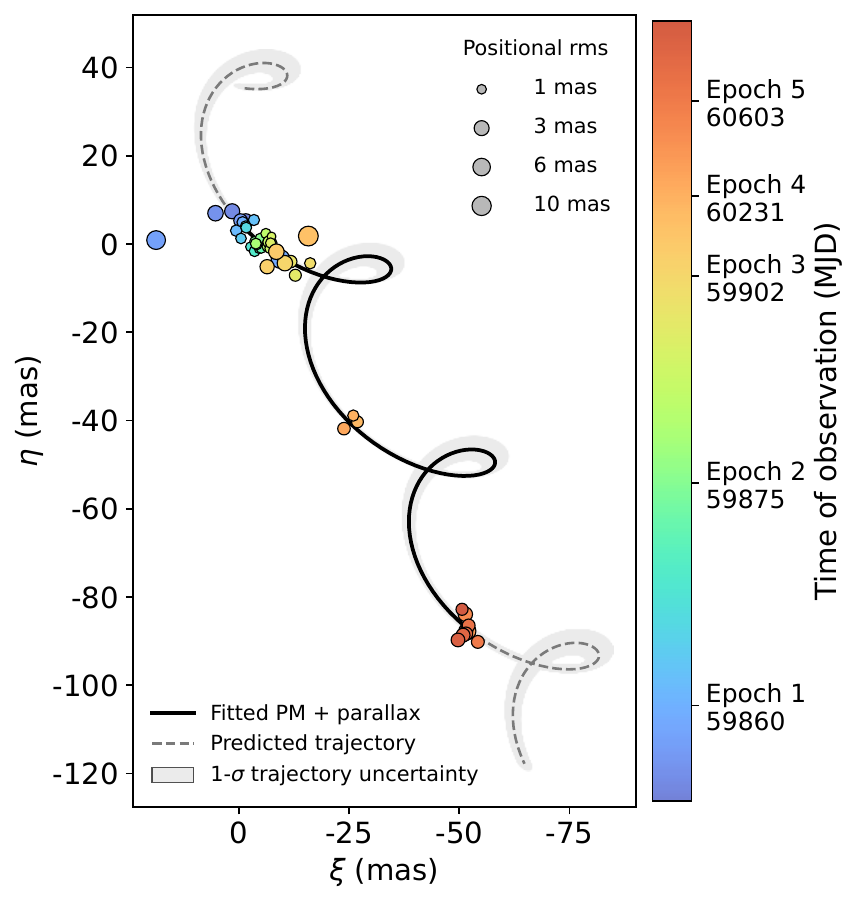}
    \caption{
    F444W positions and the fitted trajectory of JADES-GS-BD-5, after the relative and absolute reference-frame calibrations. \textit{Left:} Five-epoch astrometry, where exposures observed within a two-day gap are synthesized into one epoch. The solid black curve shows the best-fitting trajectory joining proper motion (PM) and parallax, while the dashed grey line shows the proper motion. The open symbols mark the PM-only positions and the colour-filled symbols mark the measured apparent centroid positions; matching marker shapes identify the same epoch. The dotted arrow shows the fitted parallax displacement, pointing to the trajectory-model positions (white-filled circles). The translucent coloured clouds show injection-recovery target positional offsets. For visibility, these clouds and the corresponding 1-$\sigma$ uncertainty envelopes (open black ellipses) are both exaggerated by a factor of 3 from the centroids. \textit{Right:} Per-exposure view of the same astrometric solution, showing all 50 calibrated centroids. Colours encode observation time and marker area increases with the radial rms, so higher-precision exposures are shown with smaller symbols. The trajectory remains highly consistent between different exposures, and reassuringly, the most extreme outliers are among the lowest-precision measurements. The grey shade shows the 1-$\sigma$ dispersion of the trajectory at each time point, and the dashed curves extend the trajectory outside the observed temporal baseline.
}
    \label{fig:trajectory}
\end{figure*}

\subsubsection{Hybrid PSF modelling}
\label{subsubsec:psf}

A key feature of GSBD5 compared to Capotauro \citep{Liu2026} is that it has significantly stronger observed S/Ns. Therefore, the PSF model used to extract its astrometry and photometry must match the level of significance, and thus requires a more careful treatment. Following the approach used by \citet{Liu2026} for Capotauro, which involved generating a simulated PSF model with \texttt{STPSF} v2.2.0 \citep{Perrin2012, Perrin2014}, we used the observed NIRSpec/PRISM spectrum of GSBD5 (see Section~\ref{sec:specphoto_analysis}) between 3.5 and 5.3~$\mu$m to weight a 41-wavelength polychromatic PSF and the wavefront-sensing optical-path-difference solution closest to the exposure date. Detector sampling and Fourier propagation were each oversampled by a factor of four, and the model was normalised to unit flux. We searched for GSBD5 around the position reported by \citet{hainline2026b}, allowing the centroid to move by up to five detector pixels ($\sim0.31''$) from its predicted position. All 50 exposures were measured independently; positions and covariances of images separated by less than two days were subsequently combined. 

The simulated PSF model may not perfectly describe the observed PSF. Indeed, we found negative holes in the residual image after PSF subtraction at $-8\%$ to $-2\%$ level compared to the PSF peak for F444W images in five epochs, implying the simulated PSF having a narrower core than realistic. This is a known tendency of simulated NIRCam PSFs being often slightly sharper \citep[e.g.,][]{Morishita2024, Weaver2024, Harvey2025}.
We therefore constructed a hybrid PSF (Figure~\ref{fig:stamps}), which combines both the simulation prediction and a Gaussian-convolution broadening kernel, best matched to the local measurement near GSBD5. We selected 14 point sources within $1'$ around GSBD5 by requiring a median $\mathrm{S/N}>20$ among all exposures, a minimum $\mathrm{S/N}>15$ in any single exposure and a point-source label by \textsc{sep}. We approximated their PSF weighting with a 4,500~K blackbody SED, though the change of SED models has been shown to have a minimal impact on the final PSF \citep{Liu2026}. We subsequently decided on a broadening convolution kernel by fitting the 14 point sources together with tilted local backgrounds and constructed the hybrid PSF for GSBD5, with the same kernel but the exact previous GSBD5 SED assumptions. The hybrid PSF reduced the $\chi_{\nu}^2$ of the GSBD5 fitting by $14\%$ on average, and 46 out of 50 frames were improved. The previous residual peaks were reduced by 50.0\% in absolute flux on average, and became $-3\%$ to $+3\%$ level across five epochs. These diagnostics motivated adoption of the hybrid model for all later analysis.

Furthermore, we tested a purely empirical F444W PSF which is a median stack of 25 bright stars in the \textit{JWST}/PRIMER survey over the Ultra Deep Survey (UDS) field. This PSF model has the advantage of $\sim2\times$ larger parent sample size and much brighter reference stars (and thus a more certain profile); however, collecting so many bright stars consequently requires a larger mosaic map, and thus naturally includes errors on local variation, drizzling and other uncalibrated offsets. Therefore, it is intuitively expected to be sub-optimal for our GSBD5 measurement. Indeed, this empirical PSF displayed virtually no improvement compared to the simulated PSF alone for both the residual and the fitted $\chi_{\nu}^2$. We therefore proceeded with the hybrid PSFs. The empirical PSF nevertheless provides a useful test of PSF sensitivity to our astrometric results, which we discuss in Section~\ref{subsubsec:binary_test}.

We also constructed hybrid PSFs for F356W, F410M, F560W and F770W, where residuals were visible with a simulated PSF. The hybrid PSFs were derived using point sources identified independently in each band, following the same criteria. The hybrid PSFs improved both the residuals and the target-fitting $\chi_{\nu}^2$. For the NIRCam LW~bands, the hybrid PSFs were later used for both astrometry and photometry, while F560W was used for photometry only. For other NIRCam and MIRI bands, the simulated PSF models agree well with the target for photometry, and we therefore retained the \texttt{STPSF} models.

\begin{figure*}
    \includegraphics[width=0.55\textwidth]{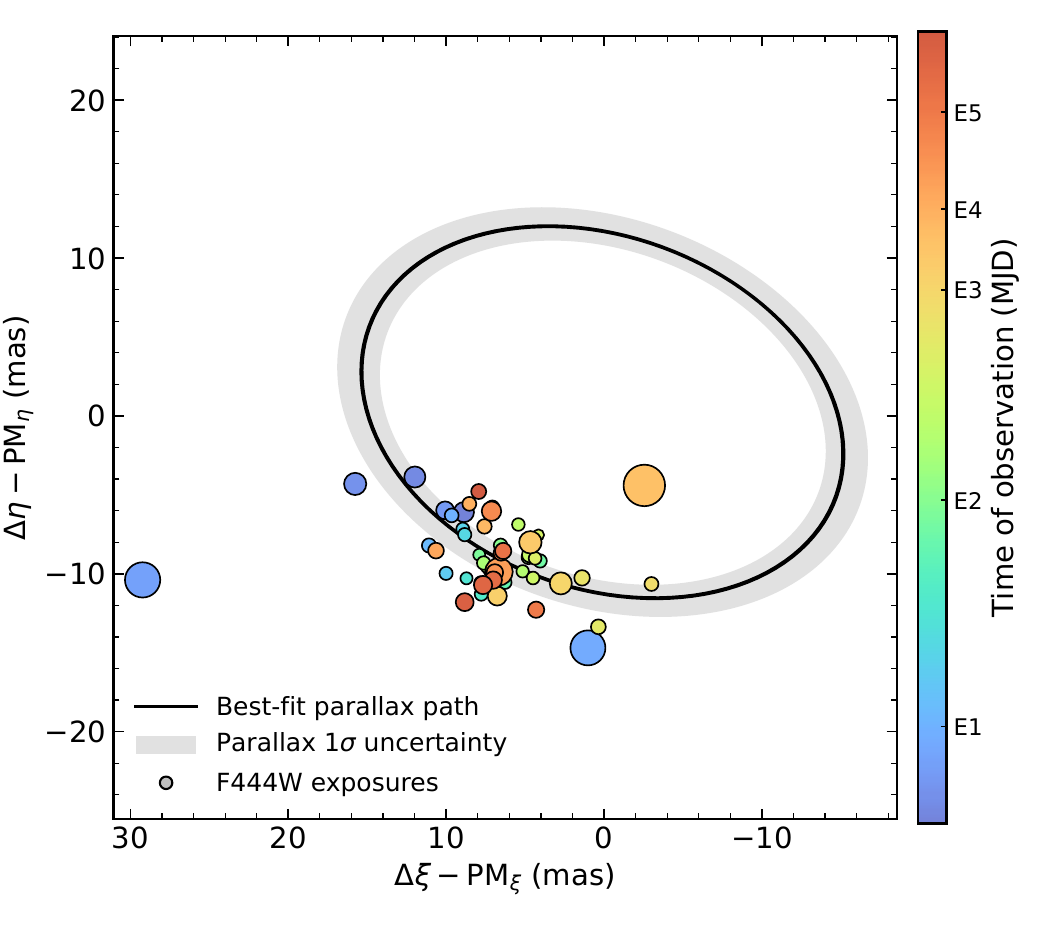}
    \includegraphics[width=0.44\textwidth]{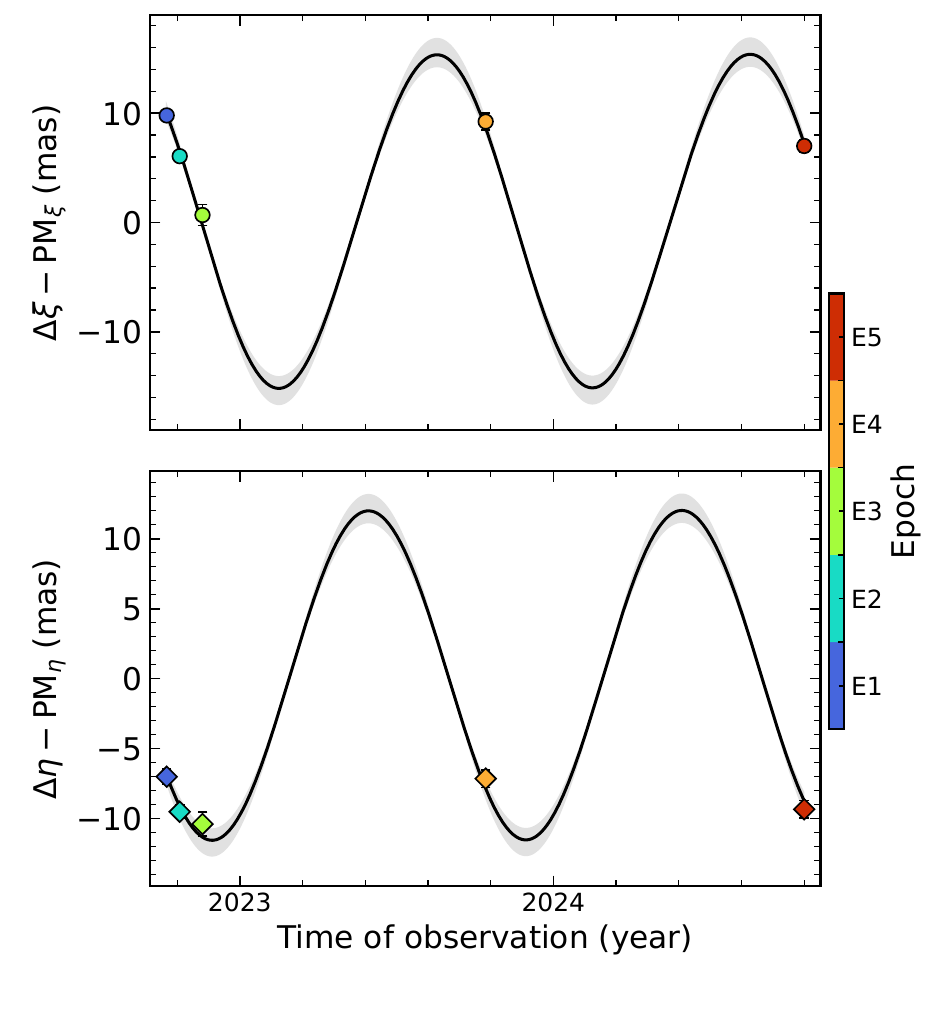}
    \caption{Parallax component of the GSBD5 astrometric motion.  \textit{Left:} Best-fit parallax solution shown in the east--north plane after subtracting the proper-motion component. Filled circles show the proper-motion-subtracted relative astrometry of GSBD5, colour-coded by observing epoch. Point sizes are proportional to the positional uncertainties, following the same scaling as in Figure~\ref{fig:trajectory}. The black curve shows the best-fit parallax path, while the grey curves show posterior samples of the parallax solution (panel \textit{d} of Figure~\ref{fig:histogram_residual}). The different exposures sample distinct phases of the projected annual parallax ellipse. \textit{Right:} The same proper-motion-subtracted parallax signal decomposed into eastward and northward components as a function of the time of observation. The residuals are shown in Figure~\ref{fig:histogram_residual}.}
    \label{fig:parallax}
\end{figure*}

\begin{figure*}
    \includegraphics[width=0.63\linewidth]{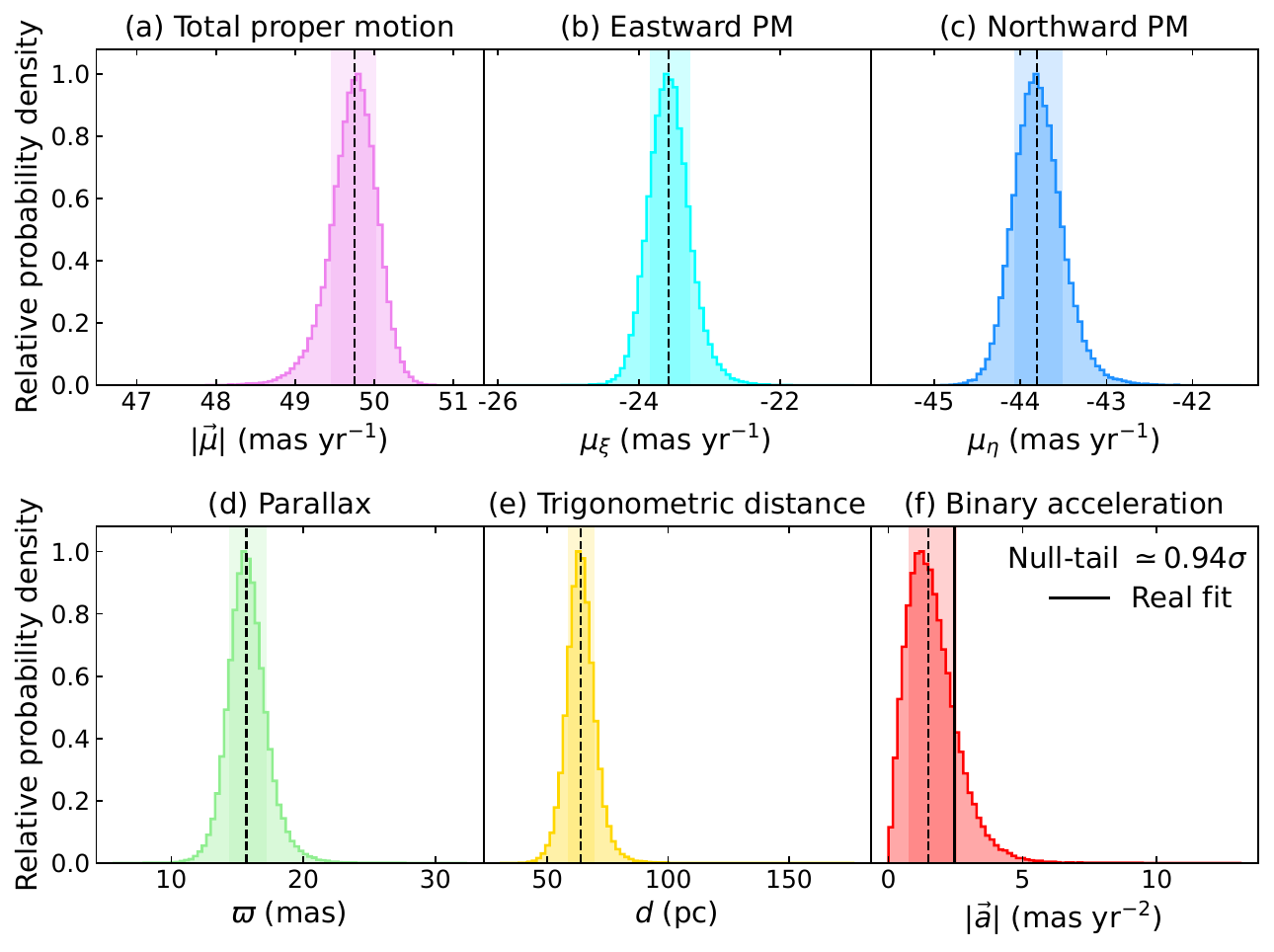}
    \includegraphics[width=0.35\linewidth]{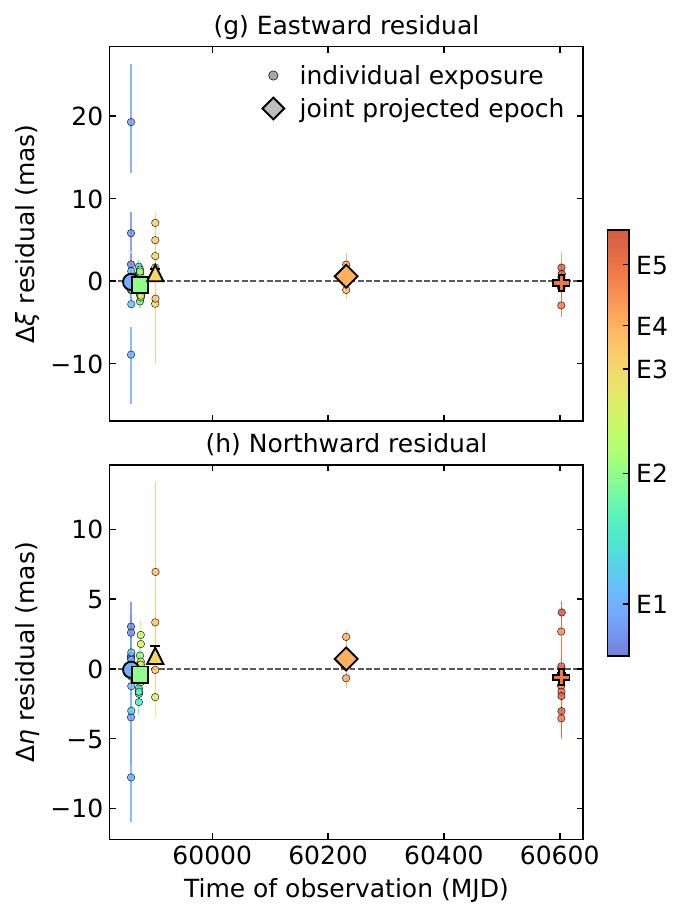}
    \caption{Five-parameter probability distributions and trajectory-fitting residuals. \textit{Left:} Probability distributions of all fitted parameters, with the black dashed line marking the median and the shaded regions indicating the $16\%-84\%$ ranges. Both proper motion and parallax are robustly detected, with their respective null values of zero lying well beyond the displayed ranges here. For binary acceleration, we show both the false-positive distribution (panel \textit{f}, red histograms, with the median marked by the dashed black line), obtained by forcing an acceleration fit to an injected trajectory with no acceleration, and the corresponding fit to the real data (solid black line). The real-data result lies within $\lesssim1\,\sigma$ of the false-positive median, indicating no significant evidence for binary acceleration. \textit{Right:} Residuals in both eastward and northward directions after subtracting the fitted trajectory from target positions at different times of observation. Larger markers, following the same style as in Figure~\ref{fig:trajectory}, represent the epoch-level residuals, while the residuals of individual exposures are shown in small circles; both are coloured by the time of observation. The positional residuals remain within $\lesssim1\,\sigma$ for all epochs and the vast majority of individual exposures. Epochs are coloured the same as in Figure~\ref{fig:trajectory}.}
    \label{fig:histogram_residual}
\end{figure*}

\begin{figure*}
    \includegraphics[width=\textwidth]{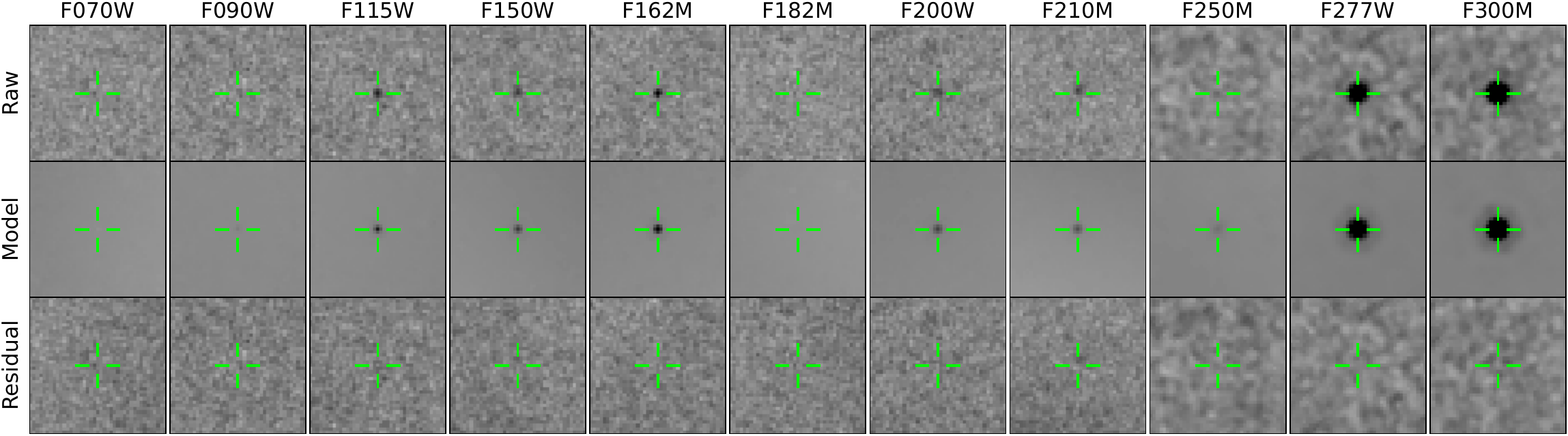}
    \vspace{0.03in}
    
    \includegraphics[width=\textwidth]{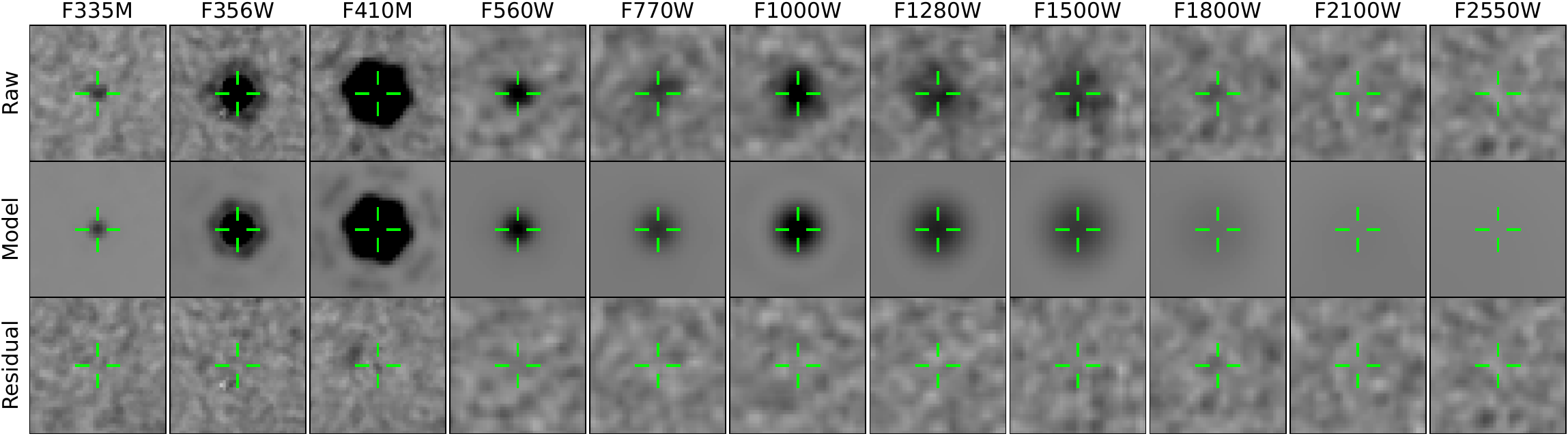}
    \caption{\textit{JWST} NIRCam and MIRI cutouts ($1.2\arcsec\times1.2\arcsec$) of JADES-GS-BD-5, stacked from individual exposures. In each band, the raw image is fitted simultaneously with a local PSF model and a tilted background. The F444W cutouts are presented separately in Figure~\ref{fig:stamps} and thus omitted here. As for F444W, locally calibrated hybrid PSFs are used for F410M, F560M and F770M, while simulated PSFs are used for all other NIRCam and MIRI bands. Crosshairs mark the fitted target centres after subtraction of the apparent motion.}
    \label{fig:cutouts}
\end{figure*}

\subsubsection{Proper motion and parallax}
\label{subsubsec:PM_parallax}

Using the hybrid PSFs, we measured GSBD5 independently in the 50 aligned F444W exposures. For each exposure, we fitted a $17\times17$ pixel cutout with a tilted background at the position reported by \citet{hainline2026b}. The same fit simultaneously derived GSBD5 photometry, as detailed in Section~\ref{sec:photometry}. The fitted detector centroid was transformed to the common $(\xi,\eta)$ tangent plane with the \textit{JWST}-pipeline WCS. The affine correction of the corresponding exposure was then evaluated and applied to its tangent-plane centroid.

The 50 exposures were grouped into five epochs where separation between exposures was shorter than two days. Within each epoch, the corrected exposure positions were combined, resulting in mean MJDs 59860, 59875, 59902, 60231, and 60603, as displayed in Figure~\ref{fig:trajectory}. For the astrometric fit, we first evaluated time and the \textit{JWST}-observer parallax factors at the exact MJD of every exposure and then passed this 50-exposure design through the same projector. Thus, the temporal and parallax information within a multi-day epoch was retained.
Specifically, we fitted the positions with a five-parameter model
\begin{align}\label{eq:model1}
\xi_i &= \xi_0+\mu_\xi(t_i-t_0)+\varpi P_{\xi,i},\\
\eta_i &= \eta_0+\mu_\eta(t_i-t_0)+\varpi P_{\eta,i},\label{eq:model2}
\end{align}
where $t_0=60094$ is the mean time measured in Julian years, $(\mu_\xi,\mu_\eta)$ are the eastward and northward proper-motion components, and a single signed parallax $\varpi$ is shared by the two coordinates. In Figure~\ref{fig:trajectory}, we decompose the proper motion and parallax components from the joint trajectory; the parallax-only measurements are displayed in Figure~\ref{fig:parallax}.

In addition to the pixel-fit covariance, the target position is also affected by local background structure, potential undetected faint neighbours, detector defects, and imperfect source subtraction. We therefore estimated these bias terms with 1,000 coherent injection-recovery tracks in the residual maps of the same 50 exposures. A preliminary five-parameter trajectory, as described by Equations~(\ref{eq:model1}) and (\ref{eq:model2}), was firstly fitted and used to subtract the source from the images; then we injected artificial sources coherently in the 50 residual images with the same trajectory model. Each track started from the first exposure in the temporal sequence, uniformly putting a random artificial source in area within $1\arcsec$ of the GSBD5 position. Injections in later exposures were defined by the trajectory at the exact exposure MJDs and the measured F444W flux of the corresponding image.
A track would be re-drawn if it put the artificial source onto an invalid pixel or a nearby bright source in one of the exposures. As a result, we obtained 1,000 successful tracks. 

Such injection recovery immediately revealed the different position-measurement uncertainties of individual exposures by comparing the injected positions with the recovered positions. In Figures~\ref{fig:trajectory} and \ref{fig:parallax}, we illustrate the positional scattering for each epoch, with exaggerated dispersions for better visibility. It can be seen that Epoch 3 has the largest uncertainty among the five epochs, which can be expected from its smaller number of exposures and their shallowness. In contrast, Epoch 2 provides much more certain positional constraint because of both a larger number of exposures and deeper integrations.

Critically, the injection recovery allowed us to combine uncertainties at different steps: the $20\arcsec$ reference-source blocks (Section~\ref{subsubsec:rela_calib}) propagated the relative-calibration uncertainty, and the background perturbation was measured from recovering the injected tracks. Specifically, the reference-source blocks were bootstrapped and every realisation was paired with a random choice of the injection-recovery track. The five parameters were then re-fitted to this combination, and a total of 100,000 bootstrap realisations formed the final five-parameter probability distributions.

As a result, we derived the relative trajectory parameters, the probability distributions of which are slightly non-Gaussian (Figure~\ref{fig:histogram_residual}):
\begin{align}
\mu_{\xi,{\rm rel}} &= -23.73^{+0.31}_{-0.28}~{\rm mas~yr^{-1}},\\
\mu_{\eta,{\rm rel}} &= -43.60^{+0.30}_{-0.26}~{\rm mas~yr^{-1}},\\
\varpi_{\rm rel} &= 15.6^{+1.4}_{-1.2}~{\rm mas},
\end{align}
with $\mu_{{\rm tot,rel}}=49.64^{+0.26}_{-0.32}$~mas~yr$^{-1}$. Its residual statistic is $\chi^2=6.4$ for five degrees of freedom, or $\chi^2_\nu=1.3$, suggesting an excellent agreement between the model and the data. Indeed, as displayed in Figure~\ref{fig:trajectory}, the temporal progression of the target position in the first three epochs exhibits very high consistency to the fitted trajectory. If fitting the trajectory with only proper motion, we would instead derive a significantly larger $\chi^2_{\nu}=18.6$, demonstrating that the trajectory cannot be approximated with a linear motion. Projected separately, the reference-frame term contributes $\sim70\%$ to the standard deviations of proper motion and parallax, whereas the target injection term contributes the remaining $\sim30\%$. This is expected, because for a bright source like GSBD5, the residual noise bias should be trivial.

Finally, we tied the relative result to the ICRS with the \textit{Gaia}-galaxy calibration. As described in Section~\ref{subsubsec:abso_calib}, the absolute calibration is also affected by relative calibration, and we thus paired the relative-calibration draw with the absolute-calibration solution. We derived the absolute trajectory:
\begin{align}
\mu_{\xi,{\rm abs}} &= -23.58^{+0.29}_{-0.27}~{\rm mas~yr^{-1}},\\
\mu_{\eta,{\rm abs}} &= -43.80^{+0.29}_{-0.26}~{\rm mas~yr^{-1}},\\
\varpi_{\rm abs} &= 15.7^{+1.4}_{-1.3}~{\rm mas},
\end{align}
with $\mu_{{\rm tot,abs}}=49.75^{+0.26}_{-0.30}$~mas~yr$^{-1}$. The total proper motion is consistent with the $50\pm20$~mas~yr$^{-1}$ value reported by \citet{Hainline2024a}, but has been now improved to $\mathrm{S/N\approx162}$. The $15.7^{+1.4}_{-1.3}~{\rm mas}$ parallax is detected at Gaussian-equivalent $\mathrm{S/N=11.5}$.
The slightly smaller uncertainty on the proper motion compared to the relatively calibrated result confirms that the \textit{Gaia}-galaxy anchor can improve how positional offset is explained by removing temporal motions in reference sources. Transforming the parallax draws separately, we obtained the parallax distance,
\begin{equation}
d=64^{+6}_{-5}~{\rm pc}.
\end{equation}
We stress that this distance is purely from trigonometric parallax and independent from physical property assumptions of GSBD5.

\subsubsection{Binary acceleration and other tests}
\label{subsubsec:binary_test}

To test whether GSBD5 is under significant perturbation of a (not yet known) companion object, we also fitted its trajectory with additional binary acceleration, i.e., the third term from Taylor expansion of a Kepler orbit:
\begin{align}\label{eq:acc1}
\xi_i &= \xi_0+\mu_\xi(t_i-t_0)+\varpi P_{\xi,i}+a_{\xi,i}(t_i-t_0)^2,\\
\eta_i &= \eta_0+\mu_\eta(t_i-t_0)+\varpi P_{\eta,i}+a_{\eta,i}(t_i-t_0)^2,\label{eq:acc2}
\end{align}
where $a_{\xi,i}$ and $a_{\eta,i}$ are the acceleration in two perpendicular directions. However, given our data only sparsely cover five epochs, these acceleration terms are expected to absorb the fitting residual and share degeneracy with the parallax. Indeed, while the previous fitting yielded a nominal total acceleration of 2.4~mas, the new fit has a reduced $\chi^2_{\nu}=0.06\ll1$, suggesting it is likely a false positive from over-fitting.

An injection-recovery test was then conducted to assess the false-positive rate. In this test, artificial F444W sources (with same fluxes as measured) were injected in the 50 residual images following a pre-known trajectory: the trajectory included only the pre-described fitted proper motion and parallax, with no acceleration term. We then repeated the full astrometric fitting procedure with the Equation--(\ref{eq:acc1}) and --(\ref{eq:acc2}) model on the injected images to determine whether a spurious acceleration could be recovered by chance. This test, whose corresponding distribution from 1,000 injections is presented in Figure~\ref{fig:histogram_residual}, showed that the previously fitted acceleration has only a Gaussian-equivalent significance of $\sim 0.94~\sigma$, and therefore we report a non-detection on binary acceleration. This is consistent with the absence of companions in our later search (Section~\ref{sec:discussion}).

We further performed a similar examination for parallax, i.e., forcing a parallax fit to an injected trajectory with only proper motion. No parallax at the same significance was found in 500,000 injections, corresponding to a parallax significance lower limit of at least $>4.7\,\sigma$, consistent with our measured parallax significance of $11.5\,\sigma$. While raising this floor to $\sim10\sigma$ is unpractical (which would require $\sim10^{29}$ injections), this test confirms the robustness of our detection of parallax for GSBD5. We also tested proper motion with static injections and reached the same significance floor. A larger test with more injections is unnecessary given the visibly secure detection. 

Another independent check verified the F444W-derived motion with other bands. We repeated the main measurement with both all 180 LW (i.e., F277W, F300M, F356W, F410M and F444W) exposures and the 130 non-F444W LW exposures, respectively. Each group used the same affine-network construction and the offsets were found generally larger between different bands during the relative calibration. The relative parallaxes are $13.9\pm1.7$ and $10.5\pm2.9$~mas, respectively. The corresponding proper motions are ($-24.0\pm0.4$, $-43.9\pm0.4$) and ($-27.1\pm2.9$, $-43.2\pm2.7$) mas~yr$^{-1}$. They are all largely consistent within uncertainties, though the non-F444W results were seriously worsened by the absence of final-epoch coverage and the additional uncertainty from wavelength-dependent morphology of reference sources. Therefore, we retain the F444W-only measurements as the final results.

\section{Photometric measurement}\label{sec:photometry}

\begin{table}
\centering
\caption{Photometry of GSBD5. Fluxes are measured from PSF fitting with injection-recovery uncertainties, stacked among all exposures. For non-detections ($\mathrm{S/N<2}$), we report both the 2-$\sigma$ upper limits and the raw measurements (in bracket). A 5\% flux uncertainty floor is applied during our analysis. The MIRI data are derived from concatenated mosaics and not used for motion measurement; therefore we do not specify their epochs here.}
\label{tab:photometry}
\setlength{\tabcolsep}{5pt}
\begin{tabularx}{\columnwidth}{@{}l l l l >{\raggedright\arraybackslash}X@{}}
\hline
Band & Flux & S/N & Epoch of \\
& (nJy) & & observation \\
\hline
F070W & $< 0.56~(-0.20 \pm 0.28)$ & 0.7 & E1 \\
F090W & $<0.38~(0.06 \pm 0.19)$ & 0.3 & E1, E2, E3 \\
F115W & $1.45 \pm 0.34$ & 4.3 & E1, E2, E3 \\
F150W & $1.39 \pm 0.21$ & 6.5 & E1, E2, E3, E4 \\
F162M & $3.77 \pm 0.30$ & 12.4 & E1, E2 \\
F182M & $<0.34~(0.06 \pm 0.17)$ & 0.4 & E1, E2, E3 \\
F200W & $1.60 \pm 0.37$ & 4.4 & E1, E2, E3, E4, E5 \\
F210M & $1.86 \pm 0.47$ & 3.9 & E1, E2, E3 \\
F250M & $<0.89~(0.77 \pm 0.45)$ & 1.7 & E1, E2 \\
F277W & $8.58 \pm 0.30$ & 29.0 & E1, E2, E3, E4 \\
F300M & $17.50 \pm 0.38$ & 46.1 & E1, E2 \\
F335M & $2.77 \pm 0.33$ & 8.5 & E1, E2, E3, E4 \\
F356W & $61.65 \pm 0.68$ & 90.8 & E1, E2, E3, E4 \\
F410M & $467.34 \pm 2.01$ & 232.3 & E1, E2 \\
F444W & $609.54 \pm 2.36$ & 258.2 & E1, E2, E3, E4, E5 \\
F560W & $360.67 \pm 49.98$ & 7.2 & -- \\
F770W & $393.54 \pm 42.04$ & 9.4 & -- \\
F1000W & $1422.17 \pm 80.91$ & 17.6 & -- \\
F1280W & $1675.82 \pm 152.18$ & 11.0 & -- \\
F1500W & $2037.08 \pm 182.45$ & 11.2 & -- \\
F1800W & $1780.58 \pm 480.79$ & 3.7 & -- \\
F2100W & $<1243.52~(1089.02 \pm 621.76)$ & 1.8 & -- \\
F2550W & $< 8651.33~(2422.85 \pm 4325.66)$ & 0.6 & -- \\
\hline
\end{tabularx}
\end{table}

We measured the photometry of GSBD5 by performing exposure-level PSF fitting, using the same PSF models and time-dependent trajectory framework as in Section~\ref{sec:astrometry}. For the five high-S/N detection bands, F277W, F300M, F356W, F410M, and F444W, we allowed the centroid to vary within a five-pixel radius. As described above, the resulting centroids were consistent with the trajectory derived from F444W. For the remaining 19 bands, we performed forced PSF photometry at the predicted position of GSBD5 in each exposure at its exact observation time. Photometric uncertainties were estimated from 1,000 injection-recovery trials on the residual maps.

After subtracting the best-fit PSF, we found no significant residuals in most bands (Figure~\ref{fig:cutouts}). The only exception is a faint source-like residual in the stack of all 30 F410M exposures, located \(0.25''\) east and \(0.09''\) north of GSBD5. A PSF fit to this residual gives \(F_\nu = 4.49 \pm 1.03~{\rm nJy}\), corresponding to \(\sim1\%\) of the GSBD5 flux in F410M, \(m_{\rm AB}=29.77\), and \(\mathrm{S/N}=4.34\). No comparable asymmetric residual was produced when the same hybrid PSF was applied to F410M \textit{Gaia} stars. Moreover, the signal is contributed nearly uniformly by 24 of the 30 F410M exposures, while the remaining six exposures are noise-dominated. These tests suggest that the residual is unlikely to be caused solely by PSF mismatch. Its non-detection in the other bands, especially F444W, argues against a normal Galactic stellar source. Possible interpretations are therefore either an image artefact, or a compact high-redshift emission-line galaxy.

The MIRI photometry is comparatively straightforward, because only the SMILES mosaic products are available and the exposure-level timing information is not retained. In addition, the broader MIRI PSFs make the measurements less sensitive to small positional uncertainties. For the NIRCam data, however, we measured photometry at three time resolutions: individual exposures, individual epochs constructed from exposure-level measurements, and the all-epoch combined photometry. Because brown dwarfs can be photometrically variable \citep[e.g.][]{Radigan2012, Apai2013, Metchev2015, Vos2020, Vos2022, Biller2018, Biller2024, Sutlieff2023, Sutlieff2024, Liu2024, McCarthy2024, McCarthy2025, Cushing2026}, combining measurements taken at different epochs may bias the final fluxes. We therefore tested for variability before adopting the stacked NIRCam photometry.

We searched for periodic photometric variability using a Lomb--Scargle \citep[LS,][]{Lomb1976, Scargle1982, VanderPlas2018} analysis of the per-exposure photometry within each epoch. For the high-S/N bands F277W, F300M, F356W, F410M, and F444W, we obtained 17 band--epoch combinations, with temporal baselines ranging from 0.3 to 83 hours and single-exposure times varying by up to an order of magnitude. Thirteen of these provide baselines longer than 2 hours and are therefore suitable for periodicity searches. We required the fitted semi-amplitude to exceed the uncertainty of the corresponding integrated photometry, and required the candidate period to be longer than the break-up limit for a \(20\,M_{\rm J}\) object, \(P>P_{\rm break}=0.66\)~h. Searching periods up to 20~h, we found no physically plausible solution with a significance of \(\geq3\sigma\). We further assessed false positives using a control sample of 25 bright, compact high-redshift (\(z>3\)) galaxies selected from the reference-source compilation. Owing to cosmological time dilation, non-blazar galaxies \citep[e.g., regular AGN/quasars or compact/bursty star-forming galaxies;][]{Winkler1992, MacLeod2016, Izotov2016, Ricci2023, Liu2024b, Liu2026a} at these redshifts are not expected to show intrinsic near- to mid-IR variability on the short timescales probed here, and therefore provide a useful empirical control sample for observational systematics. For all tentative LW-band variability signals with nominal significances of \(1.0\)--\(3.0\sigma\), the control galaxies produced equal or higher LS powers. We therefore attribute these tentative signals to observational systematics \citep[see e.g.][]{Alderson2023, Schlawin2023, Sutlieff2025, Zhou2026} and find no significant evidence for periodic photometric variability in GSBD5 within current sensitivity.

We also searched for long-term variability between epochs, and found no $>5\%$ semi-amplitude between epoch-integrated fluxes. The absence of significant variability justifies combining measurements from different exposures and epochs to obtain the final stacked fluxes reported in Table~\ref{tab:photometry}. The control-sample test, however, revealed artificial flux variations at the \(\sim1.5\%\)--\(3\%\) level, likely reflecting residual \textit{JWST} image-to-image zero-point differences in the data used here. We therefore imposed a \(5\%\) uncertainty floor on all \textit{JWST} fluxes in the subsequent analysis. This is consistent with the flux error floor used in \citet{hainline2026b}. Most variable brown dwarfs are observed to vary in the IR at a comparable amplitude \citep[0.2\% to 10\%, e.g.,][]{Metchev2015, Vos2022, Biller2024, Liu2024, Cushing2026}. Therefore, we cannot exclude GSBD5 having a weaker variability than that detectable with current flux uncertainties, although the data indeed rule out extreme variability amplitudes, which can sometimes reach as high as almost 40\% \citep[e.g.,][]{Radigan2012, Eriksson2019, B_Bowler2020, Zhou2022, Sutlieff2023, Sutlieff2024, Tan2025}.

We then used the NIRCam photometry to calibrate the PRISM spectrum. \textit{JWST}/NIRSpec PRISM spectra are known to suffer from wavelength-dependent baseline offsets and slit losses, which was particularly important for Capotauro \citep{Liu2026}. We therefore applied a second-order wavelength-dependent correction by matching synthetic PRISM photometry to the measured fluxes in the strong-detection ($\mathrm{S/N>5}$) bands F150W, F162M, F277W, F300M, F335M, F356W, F410M, and F444W using their filter profiles. The largest correction was required by F277W, at the \(\sim8\%\) level, whereas the remaining bands required corrections of only \(\lesssim3\%\). After this calibration, the PRISM spectrum is photometrically consistent with the NIRCam measurements. We note that the Galactic extinction is negligible for GSBD5. At its coordinates, we estimated maximum reddening ${E(B-V)=0.009}$\,mag \citep{Schlegel1998, Schlafly2011} with \texttt{IRSA} Galactic DUST Reddening and Extinction service\footnote{\href{https://irsa.ipac.caltech.edu/applications/DUST/}{https://irsa.ipac.caltech.edu/applications/DUST/}} for an extragalactic object, corresponding to about $1.4\%$ change for F070W flux and $0.1\%$ for F480M. Since GSBD5 is within the Milky Way, the extinction is even less.

\section{Spectrophotometric analysis}
\label{sec:specphoto_analysis}
\subsection{Spectral type classification}\label{subsec:spec_type}

\begin{figure}
    \centering
    \includegraphics[width=\linewidth]{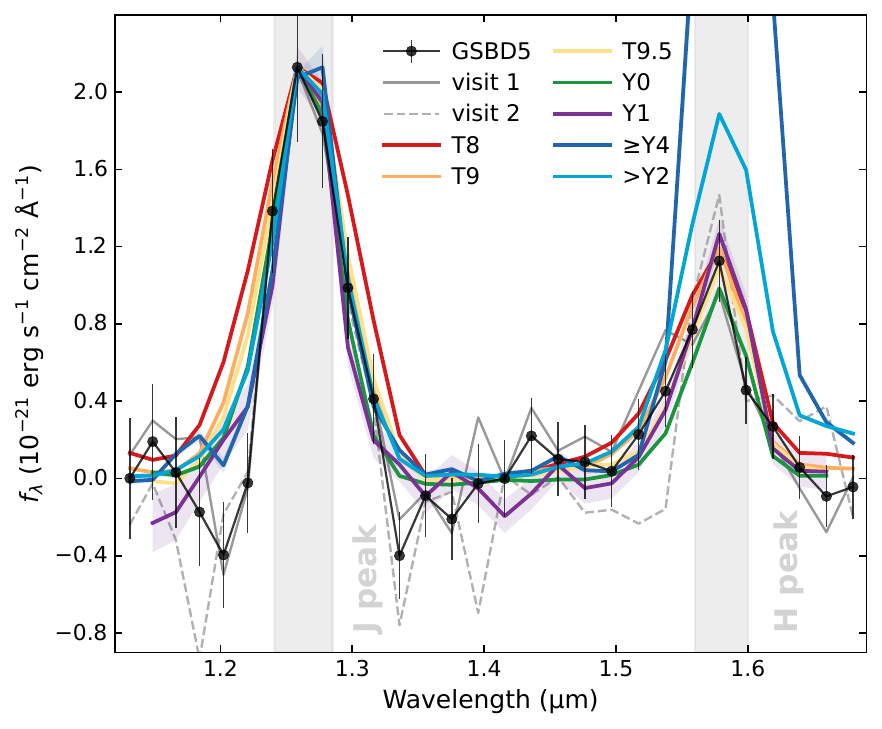}
    \vspace{-0.2in}
    \caption{Short-wavelength spectrum of GSBD5 in comparison with empirical average subtype templates. The integrated NIRSpec/PRISM spectrum is shown as black curve with scatters, with two separate visits plotted as solid and dashed grey curves, respectively. The subtype templates are plotted as coloured curves with shaded uncertainties, scaled to the integrated GSBD5 $J$-peak value. The $J$-peak profile broadly agrees with Y0--Y1, while the blue side clearly rejects T-type templates. We also show the $J$-peak scaled spectra of WISE~0855 ($\ge~$Y4) and WISE~1828 ($>~$Y2).}
    \label{fig:jhk_peaks}
\end{figure}

\begin{figure}
    \centering
    \includegraphics[width=\linewidth]{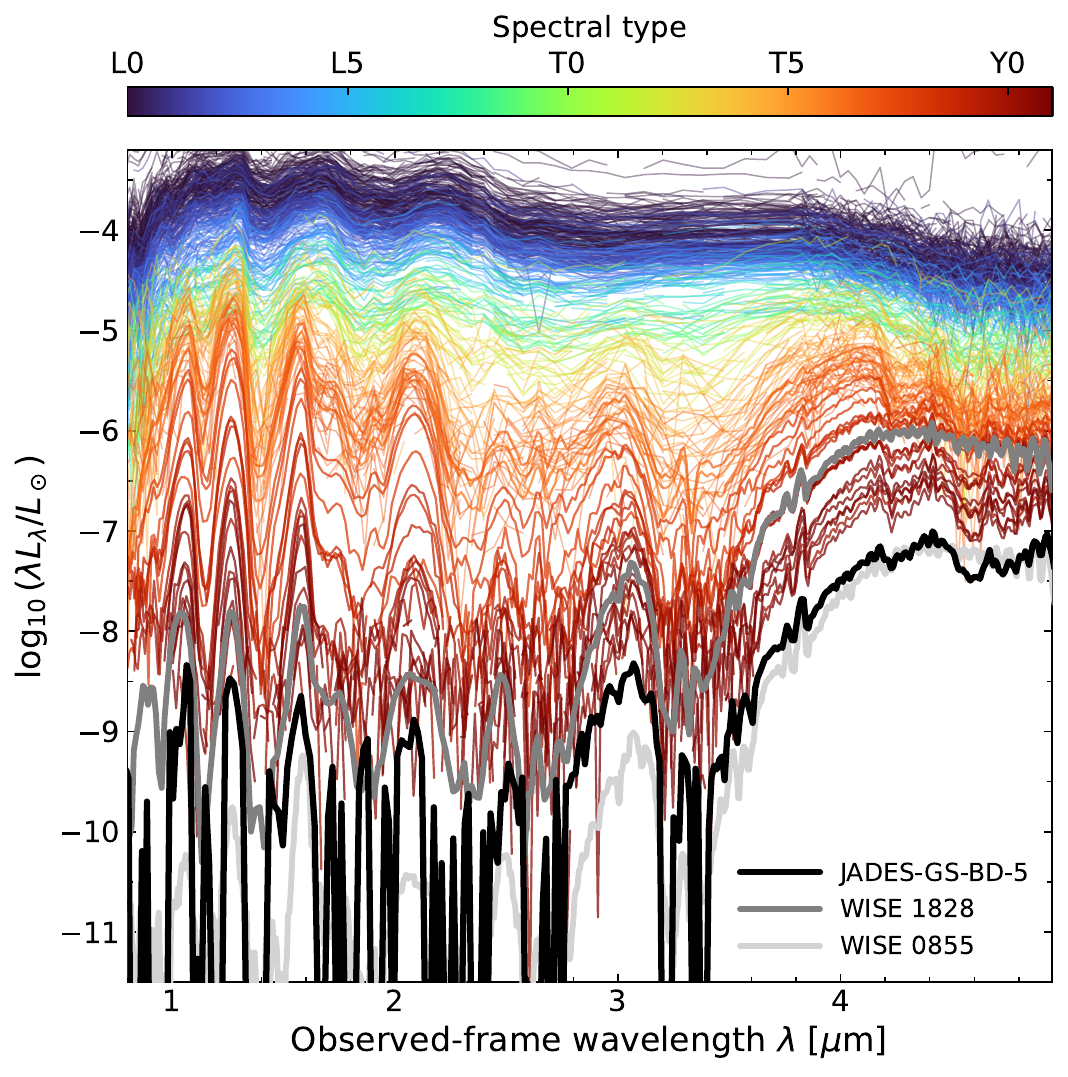}
    \vspace{-0.2in}
    \caption{Spectrum of GSBD5 compared with a sequence of L, T and Y dwarfs. All spectra are shown at their intrinsic luminosities, with spectral types indicated by the colour scale. GSBD5, shown in black, is among the least luminous brown dwarfs in the comparison sample of 512 L0--Y1 dwarfs displayed here, including the $\geq\mathrm{Y2}$ dwarf, WISE 1828. At SW, only the coldest dwarf WISE 0855 is fainter, while at LW GSBD5 has a luminosity comparable to that of WISE 0855.}
    \label{fig:spec_luminosity}
\end{figure}

\citet{Hainline2026c} classified GSBD5 as a Y0--Y1 dwarf from its PRISM spectrum, but noted a tension between the SW and LW spectral features. Using the PRISM spectrum recalibrated with our updated photometry, we revisit the spectral classification of GSBD5 in detail. 

We first assembled a set of reference spectra spanning late-T to early-Y dwarfs, including the T8 standard 2MASS J04151954$-$0935066 \citep[][data from \citealt{Burgasser2004}]{Burgasser2006}, the T9 standard UGPS J072227.51$-$054031.2 \citep[][data from the SpeX Prism Library; \citealt{Burgasser2014}]{Cushing2011}, the representative T9.5 dwarf WISEA J094306.00+360723.3 \citep[][data from the Cloud Atlas \textit{HST}/WFC3 library; \citealt{Schneider2015,Manjavacas2019}]{Schneider2015}, the Y0 standard WISEP J173835.52+273258.9 \citep[][data from the Cloud Atlas \textit{HST}/WFC3 library; \citealt{Schneider2015,Manjavacas2019}]{Cushing2011}, and the tentative Y1 standard WISE J035000.32$-$565830.2 \citep[][data from the Cloud Atlas \textit{HST}/WFC3 library; \citealt{Schneider2015,Manjavacas2019}]{Kirkpatrick2012}. To enable a direct comparison with the NIRSpec/PRISM observations, we convolved each reference spectrum to the corresponding instrumental resolution using the \texttt{msafit}\footnote{\href{https://github.com/annadeg/jwst-msafit}{https://github.com/annadeg/jwst-msafit}} forward model for a centred point source \citep{deGraaff2024}. The convolved spectra were then normalised to the $J$-band peak flux of GSBD5, as shown in Figure~\ref{fig:jhk_peaks}.

\begin{figure*}
    \centering
    \includegraphics[width=0.85\linewidth]{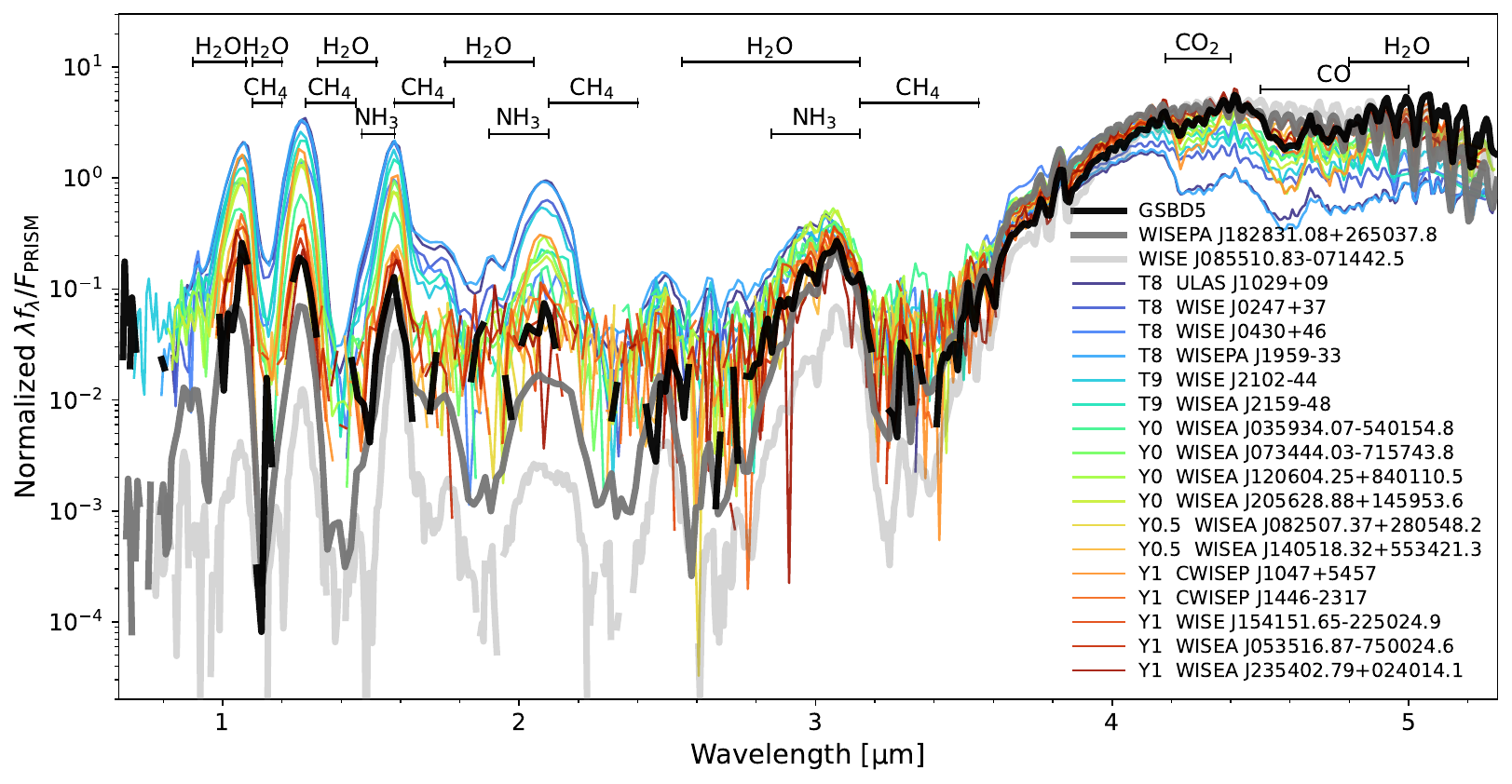}
    \vspace{-0.1in}
    \caption{NIRSpec/PRISM spectrum of GSBD5, shown in black, compared with a sequence of late-T and Y-dwarf benchmarks, shown by coloured curves. All spectra are normalized by their flux integrated over the shown wavelength range. GSBD5 lies near the faint lower envelope of the comparison sample over most of the near-IR range. It has an almost identical profile to the extremely cold WISE 1828, tentatively classified as $\geq\mathrm{Y2}$ and a binary candidate, and remains brighter only than the coldest benchmark WISE 0855, tentatively classified as Y4. The sequence shows a clear spectral gradient: the normalized flux decreases toward later spectral types at wavelengths shortward of $\sim4\,\mu$m, whereas beyond this wavelength the trend essentially reverses. Major molecular absorption ranges are marked. Negative values are left blank.}
    \label{fig:scaled_spec}
\end{figure*}

We first assessed the conventional approach of assigning spectral subtypes based on spectral morphology alone \citep[e.g.,][]{Cushing2011,Kirkpatrick2012}. Specifically, we fitted the $J$-band peak of GSBD5 using the convolved standard spectra. Among the standard templates, the Y1 standard provides the best fit, with a reduced chi-square of $\chi^2_{\nu}=1.2$. To assess the robustness of this result, we independently examined the $J$- and $H$-band peaks in the two NIRSpec visits (Figure~\ref{fig:jhk_peaks}), both of which show a consistent preference for the Y1 template. Nevertheless, the subtype assignment remains uncertain, as the Y1 template improves $\chi^2$ by only $\sim0.1$ relative to the next-best standard template, with five degrees of freedom (dofs). The $H$-band peak provides even weaker constraints: owing to its lower flux and larger relative uncertainties, the observed profile is consistent with all the reference subtypes considered here.

Notably, the blue wing of the GSBD5 $J$-band peak appears slightly narrower than that of the Y1 standard, a morphological characteristic generally associated with later spectral types. We therefore extended the comparison to the NIRSpec spectra of two of the coldest known brown dwarfs. One is WISE~J085510.83$-$071442.5, hereafter WISE~0855 (data from \citealt{Luhman2024}); this object is the coldest brown dwarf known to date ($T_{\rm eff}\sim290$~K) and has been tentatively, though somewhat arbitrarily, classified as $\mathrm{\geq Y4}$ \citep[e.g.,][]{Kirkpatrick2019}. Another is WISEPA~J182831.08$+$265037.8 \citep{Cushing2021}, hereafter WISE~1828. This object has an exceptionally red colour and once held the record for being the coldest brown dwarf \citep[$T_{\rm eff}\sim300$~K;][]{Kirkpatrick2012, Beichman2013}. Tentatively classified as $>\mathrm{Y2}$ \citep{Cushing2021}, it has received rich spectroscopic observations, including \textit{JWST} NIRSpec and MIRI \citep{DeFurio2023, Barrado2023}. However, interpretations of WISE 1828 must be treated with caution because it has been suspected to be an unresolved binary \citep{Leggett2025, Leggett2026}. Relative to the Y1 standard, WISE~1828 yields a further reduction in $\chi^2$ of $\sim0.9$ for five dofs, while WISE~0855 reduces $\chi^2$ by an additional $\sim0.5$. Thus, in terms of the minimum $\chi^2$, GSBD5 is most closely matched by WISE~0855, the coldest known brown dwarf. However, visual inspection of the $J$- and $H$-band profiles in Figure~\ref{fig:jhk_peaks} reveals that the Y0 and later-type spectra are largely indistinguishable within the observational uncertainties. Consequently, spectral morphology alone supports a Y-type classification for GSBD5, primarily based on the blue wing of the $J$-band peak, but does not allow a reliable determination of its subtype.

GSBD5 illustrates a fundamental limitation of spectral classification with \textit{JWST} NIRSpec/PRISM data. The instrument has its lowest resolving power at short wavelengths, reaching $R_{\rm min}\sim30$ near the $J$-band peak \citep{Jakobsen2022}. As the $J$-band peak becomes progressively fainter and narrower towards later spectral types because of the colder Planck function leaving little intrinsic emission in the Wien tail \citep{Schneider2016}, the combination of limited spectral resolution, increasing relative flux uncertainties, and fewer independent resolution elements makes it increasingly difficult to distinguish between subtypes. In the absence of established spectral standards later than Y1, the classification of objects with unusually narrow $J$-band peaks, such as GSBD5, is poorly defined. We therefore explore alternative empirical proxies for the spectral types of the coldest brown dwarfs, aiming to establish a more practical classification framework for \textit{JWST} observations and to identify a parameter space in which GSBD5 can be more distinctly characterised.

The absolute magnitude can provide an empirical tracer of spectral type \citep[e.g.,][]{Filippazzo2015}. We constructed a library of 539 dwarfs, including 25 M dwarfs from IRTF/SpeX spectral library \citep{Cushing2005, Rayner2009, Villaume2017}, 396 L dwarfs and 95 T dwarfs from the SUDA/SPHEREx QR2 \citep[10.26131/IRSA652;][]{Gagne2026}, and 11 T dwarfs and 12 Y dwarfs from \citet{Beiler2024}, excluding WISEA J220905.75+271143.6 because of its noisy SW spectrum. Each object is required to have an unambiguous standard spectral subtype, a measured parallax enabling a reliable flux--luminosity conversion, and a well-calibrated spectrum covering a similar range comparable to that of \textit{JWST} NIRSpec/PRISM. Additionally, we included the PRISM spectra of WISE~0855 and WISE~1828.

As illustrated in Figure~\ref{fig:spec_luminosity}, where only a representative subsample of L0--Y4 dwarfs is shown for clarity, the spectral luminosities follow a clear sequence of decreasing luminosities toward later types. Across the full PRISM wavelength range, GSBD5 is significantly fainter than all Y0--Y1 dwarfs in the comparison sample. At $\lesssim4\,\mu$m, it is only more luminous than WISE~0855, while at $\sim4-5\,\mu$m, the two objects have comparable luminosities. This clearly places GSBD5 at the boundary of, if not well beyond, the luminosity range current defined by Y0--Y1 standards. However, the comparison also reveals an important band dependence in luminosity-based typing. When evaluated at $\lesssim3.5\,\mu$m, GSBD5 lies between Y1 and the tentative Y4 classification of WISE 0855, but remains substantially more luminous than WISE~0855; when evaluated at $\gtrsim4.2\,\mu$m, its luminosity, or absolute magnitude, becomes comparable to that of WISE~0855. WISE~1828 also stands out from this sequence with its exceptionally high luminosity at $\gtrsim2.8\,\mu$m, which has led to the interpretation of an unresolved binary \citep{Leggett2025, Leggett2026}; conversely, the low luminosity of GSBD5 is in line with it being isolated (see Section~\ref{subsec:IPMO} for more discussion).

The overall spectral colour is also known to correlate with the spectral type \citep[e.g.,][]{Kirkpatrick2021, Kiman2026b}. To isolate spectral shape from luminosity, we normalised the 17 \textit{JWST} T8--Y1 dwarf spectra, as well as GSBD5, WISE~1828 and WISE~0855, by their total NIRSpec/PRISM-integrated flux. The resulting comparison is shown in Figure~\ref{fig:scaled_spec}. These cold-dwarf standards reveal a clear spectral sequence from late-T to Y types: the short-wavelength flux decreases systematically toward later spectral types, and this trend reverses at $\gtrsim4\,\mu$m, although with a more moderate amplitude. This evolutionary sequence is physically expected. As dwarfs cool toward later types, their effective temperatures decrease and the bulk of their emergent flux shifts to longer wavelengths. The increased opacity of H$_2$O, CH$_4$ and NH$_3$ further weakens the SW emission \citep{Faherty2026}, and the evolving $J$-peak profile is a by-product of this overall spectral evolution. Overall, GSBD5 lies near the coldest end of the T8--Y1 sequence of the comparison sample across most of the PRISM wavelength range. It shares similar spectral profiles to WISE~1828 at the $H$ peak and $\sim3\,\mu$m to $4.5\,\mu$m, strongly suggesting a subtype no earlier than Y1.

We also note that the spectral evolution becomes more gradual on the LW side toward the coldest objects, causing the later-Y spectra to converge more closely in this wavelength range. This partly reflects the reduced sensitivity of the normalised LW continuum to decreasing effective temperature, but it also indicates the growing importance of additional atmospheric processes at colder temperatures. Indeed, the diversity of spectral features in this region shows that, although effective temperature remains the dominant driver of the subtype sequence, the observed spectra are also shaped by CO$_2$ and CO absorption. The strengths of these features depend on metallicity, gravity, and the degree of vertical mixing in individual objects \citep{Beiler2024b}, producing diverse absorption troughs near $\sim4.3$, $4.6$, and $4.8\,\mu$m.

\begin{figure}
    \centering
    \includegraphics[width=\linewidth]{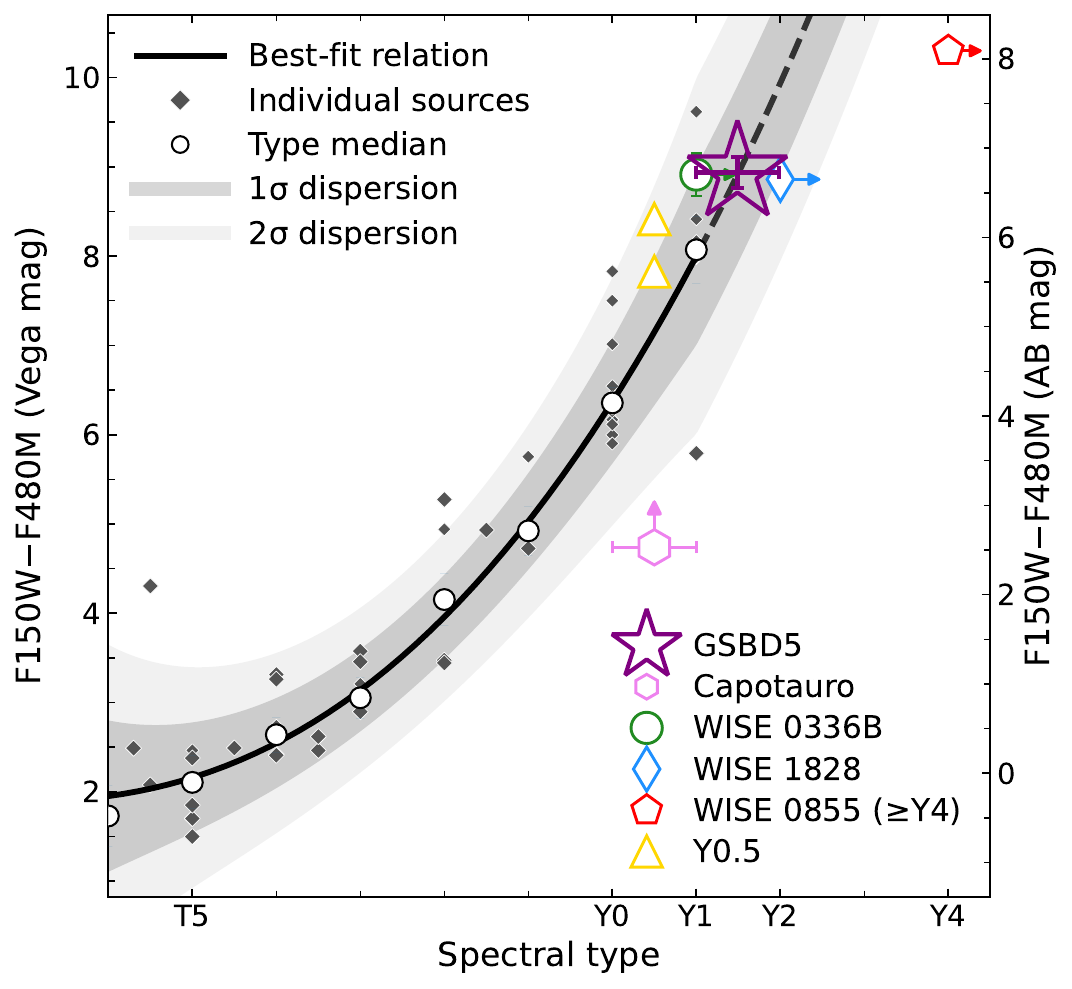}
    \vspace{-0.2in}
    \caption{F150W$-$F480M colour (AB magnitude) as a function of spectral type for late-T and Y dwarfs. The objects define a clear monotonic sequence, approximately described by a softened power law (black curve). GSBD5 and several sources previously suggested to be later than Y1 occupy the reddest, coldest end of the sequence.}
    \label{fig:y_tail_typing}
\end{figure}

Figure~\ref{fig:y_tail_typing} shows the F150W$-$F480M colour evolution with spectral type. A notable feature is that the late-T and Y dwarfs form a smooth, monotonic sequence, without abrupt breaks or reversals. Similar behaviour has been seen in pre-\textit{JWST} colour--spectral-type diagrams, e.g., \citet{Kirkpatrick2021}. We describe this late-T/Y tail phenomenologically with a softened power law,
\begin{flalign}
    & C_{\rm TY}(x) = A[S(x-k_1)]^p,\\
    & S(z) = w\ln\left(1+e^{z/w}\right), 
\end{flalign}
where C is the colour in AB magnitude, constant $w=0.65$ is the fixed smoothing width, $p$ is the power, $k_1$ is the horizontal offset, and $x=30+q$ for subtype Y$q$ and T$10+q$. To assess how the earlier-type sequence connects to this late-T/Y tail, we further modelled the joint M to T sequence, with a linear profile earlier than late L, and a softened turning curve between late L and early T,
\begin{flalign}
    & C_{\rm ML}(x) = a + b_1 x,\\
    & C_{\rm LT}(x) = (b_2-b_1)S(x-k_2),\\
    & C(x) = C_{\rm ML}(x)+C_{\rm LT}(x)+C_{\rm TY}(x),
\end{flalign}
where $x=0-9.5$ for M0--M9.5, $10-19$ for L0--L9, and $20-29$ for T0--T9. The main relation is fitted with all individual sources, with the 1- and 2-$\sigma$ shades marking the dispersion of their colour to the fitted relation for a given spectral type. Sources marked by coloured symbols are not involved in the fitting. We further explored colour combinations of different \textit{JWST} bands. The full set of colour--type fits, the used data, and the fitted parameters are introduced in Appendix~\ref{app:color_typing}.

This relation only reflects how the colour follows the existing subtype convention, instead of genuine physical correlations. Nevertheless, such a continuous parameterisation is useful because it provides a quantitative proxy for the spectral evolution underlying the classification sequence. \citet{Kirkpatrick2012} discussed which spectral quantity can best distil the progression of the $J$-peak sequence, but at the time the available data favoured classification through direct comparison of $J$-peak morphology. With \textit{JWST}, the F150W$-$F480M provides a complementary diagnostic: it traces the same broad late-T/Y spectral evolution as the $J$-peak sequence, while providing a simple and continuous quantity that can be applied uniformly to faint objects. It therefore represents a natural proxy to select late-type candidates before spectroscopy becomes available, with quantitative predictions into the colder regime beyond the current classification limit. Because of limited reference sources beyond Y1, we extrapolate the uncertainties with those at Y1 in Figure~\ref{fig:y_tail_typing}.

Applying the fitted relation to GSBD5 gives a colour-predicted spectral type of \textit{c}Y1.5$^{+0.5}_{-0.5}$ (\textit{c} for colour), assuming a 1-$\sigma$ equivalent spectral-type uncertainty. This is consistent with the conclusions from both its spectral luminosity and flux-normalised spectral profile. Interestingly, two sources classified beyond Y$1$, WISE~0336B and WISE~1828, have F150W$-$F480M colour nearly identical to that of GSBD5. Together, these three objects may indicate an emerging population beyond the Y1 standards. However, the colour--spectral-type relation remains poorly constrained at the Y0--Y1 end, with fewer than $\sim20$ reference objects, and its behaviour at later subtypes is currently unknown. As shown in Figure~\ref{fig:y_tail_typing}, the tentative Y4 object WISE~0855 does not follow a simple extrapolation of this sequence. Assigning precise spectral subtypes in this regime therefore remains premature. Nevertheless, the continuous colour relation provides an empirical framework for identifying and comparing objects beyond the current classification standards, which can be progressively refined as more ultracold dwarfs become available for spectroscopic characterisation.
We therefore adopt $>$Y1 as the formal spectral classification of GSBD5, a colour-based proxy of $c$Y1.5, and leave more robust spectral typing to future high-resolution spectroscopy.

\subsection{Bolometric luminosity}

\begin{figure}
    \centering
    \includegraphics[width=\linewidth]{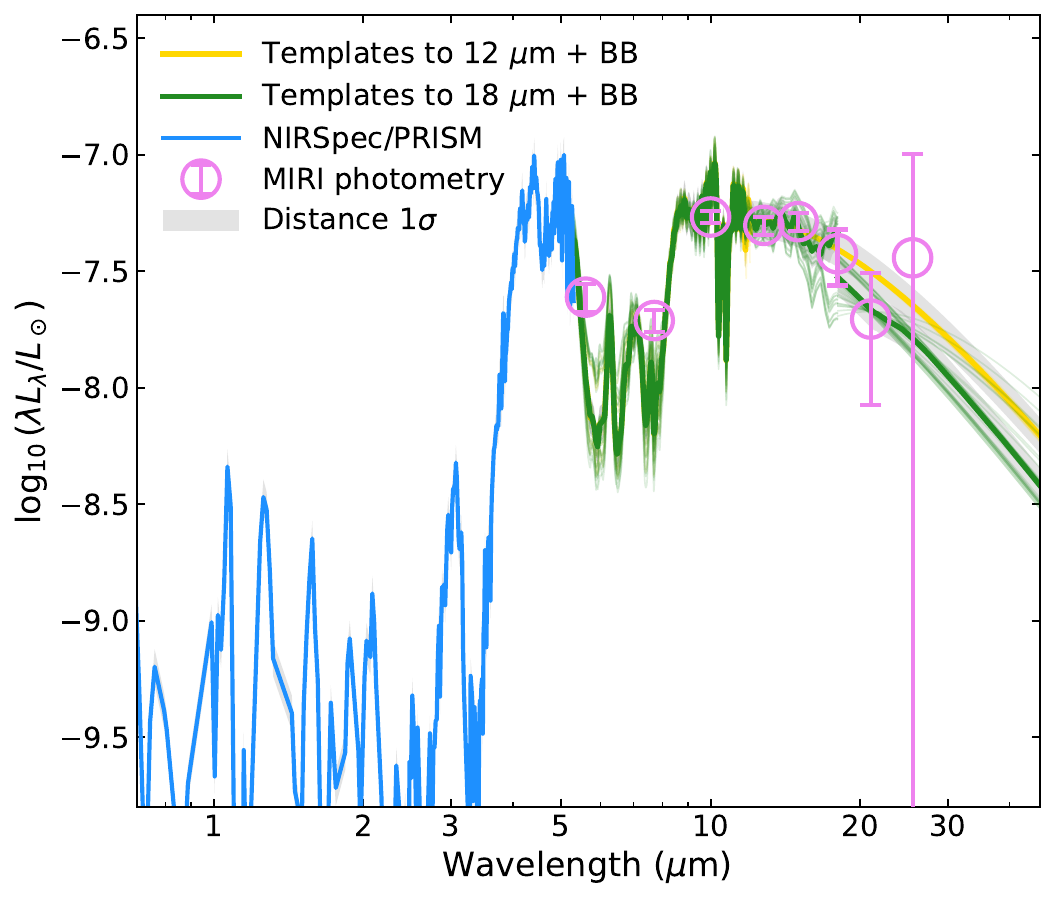}
    \vspace{-0.2in}
    \caption{SED of GSBD5 used to estimate its bolometric luminosity. The \textit{JWST} NIRSpec/PRISM spectrum (blue curve) is integrated directly, while the MIRI photometric data (pink circles) were fitted with Y-dwarf spectral templates and extended with a freely fitted blackbody tail. Fitted median SEDs are marked as thick curves, while individual templates are plotted as thin curves. We tested two extrapolations, using the templates out to $18\,\mu$m (green curve) or only to $12\,\mu$m (yellow curve) before appending the blackbody tail. The two estimates differ by only 5\%, indicating that the derived bolometric luminosity is robust against model assumptions in the long-wavelength extrapolation. The uncertainty from distance is marked by the grey shade.}
    \label{fig:sed_bolo}
\end{figure}

With the parallax measured and therefore distance inferred, the observed SED can be directly converted into an absolute luminosity scale, allowing the bolometric luminosity, $L_{\rm bol}$, to be estimated with essentially no dependence on physical-property assumptions. The flux shortward of 0.6$\,\mu$m is negligible (expected $<0.1\%$; \citealt{Beiler2023}). Over the NIRSpec/PRISM wavelength range, 0.6 to 5.3~$\mu$m, we directly integrated the observed spectrum. At longer wavelengths, from 5.3 to 12\,$\mu$m, we fitted the MIRI photometry using the 12 Y-dwarf templates from \citet{Beiler2024}, all of which have MIRI spectroscopic coverage. The measured MIRI fluxes, together with their uncertainties, were directly used in the fitting. We then adopted the median of the fitted templates as the fiducial SED. Beyond 12\,$\mu$m, only three templates remain available, i.e., the $>$Y1 candidates, WISE~1828 and WISE~0855, and the Y0 dwarf WISE~1738. Similarly, we fitted these templates to the MIRI photometry, and adopted their median. Beyond 18\,$\mu$m, we appended a blackbody tail by fitting it with the $12$--$18\,\mu$m median, allowing its temperature to vary between 175~K and 500~K. The SED profile is shown in Figure~\ref{fig:sed_bolo}. Uncertainties were drawn from Monte Carlo (MC) sampling of the extrapolated spectral templates, PRISM spectral uncertainties, a linear uniform prior of blackbody-tail temperature, and the probability distribution of the parallax.

This procedure gives a bolometric flux $\log_{10}\!\left( F_{\rm bol}/\mathrm{erg\,s^{-1}\,cm^{-2}} \right) = -15.18^{+0.03}_{-0.06} $, a bolometric luminosity $\mathrm{log}\left(L_{\rm bol}/L_{\odot}\right) = -7.08^{+0.09}_{-0.09}$ and a blackbody-tail temperature $T_{\rm BB}=371^{+91}_{-114}$~K. We note that $T_{\rm BB}$ describes only a component of GSBD5 and it is different from the global description, effective temperature $T_{\rm eff}$, although they are physically consistent, as in the models discussed below. The PRISM spectrum contributes $\sim22\%$ of the total luminosity, the MIRI-constrained region contributes $\sim58\%$, and the blackbody tail contributes the remaining $\sim20\%$. To assess the sensitivity of $L_{\rm bol}$ to the poorly constrained $12-18\,\mu$m SED shape, we repeated the calculation by omitting the three long-wavelength templates, and appended the blackbody tail directly after the \citet{Beiler2024} templates at 12$\,\mu$m. This approach yielded essentially the same $L_{\rm bol}$, only $\sim5\%$ higher than the fiducial value. We therefore conclude that the MIRI photometry provides sufficient leverage on the peak of the SED and thus our bolometric luminosity estimate.

\begin{figure*}
    \includegraphics[width=0.85\textwidth]{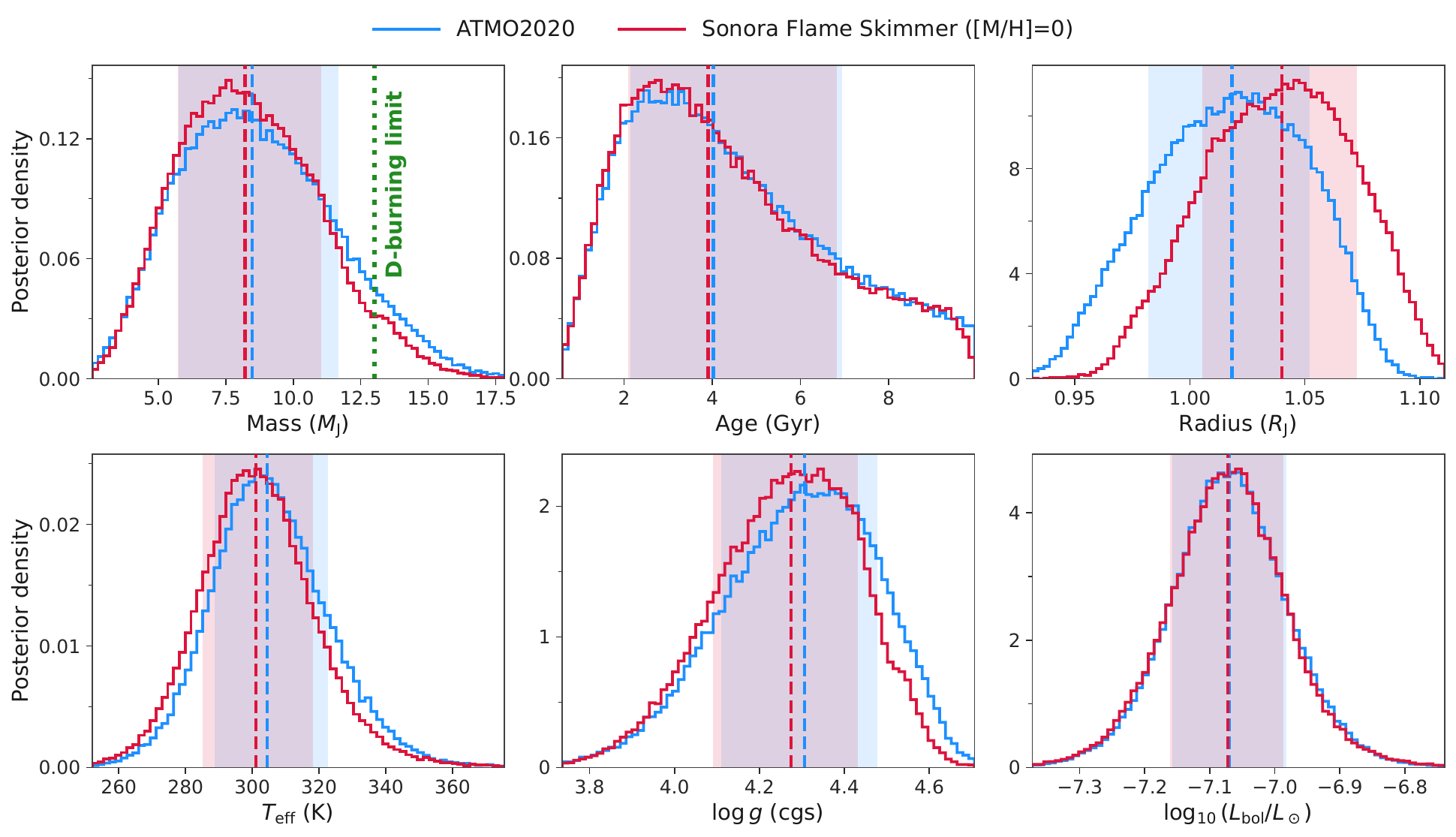}
    \vspace{-0.1in}
    \caption{Posterior distributions of the physical properties of GSBD5 inferred from the ATMO2020 and Sonora Flame Skimmer evolutionary models. Blue and red curves show the individual model results, the corresponding shaded regions indicate the central 68\% intervals, and dashed black lines mark the medians. The two model families give consistent results, with median values agreeing to within a few percent. $>90\%$ of the mass distribution from either model is below the deuterium-burning mass limit of 13 $M_{\rm J}$ \citep[dotted green line;][]{Spiegel2011}.}
    \label{fig:evolutionary_distribution}
\end{figure*}

\subsection{Evolutionary modelling}
\label{subsec:evolution_model}

We propagated the full $L_{\rm bol}$ distribution through two evolutionary models appropriate for cold dwarfs, ATMO2020 \citep{Phillips2020} and Sonora Flame Skimmer \citep{Mang2026}, to infer the physical properties of GSBD5. We adopted a uniform mass prior, $M\sim\mathrm{U}(0.1,30)\,M_{\rm J}$, and a log-normal age prior, $\ln(t/{\rm Gyr})\sim \mathcal{N}(\ln 3.5,[\ln 2]^2)$, truncated to $0<t<13.8$ Gyr. The age prior is motivated by the kinematically inferred ages of T dwarfs \citep{Hsu2021}, which provide the closest age-calibrated analogues to cold Y dwarfs. We verified that adopting a uniform age prior, $t\sim\mathrm{U}(0.1,10)$ Gyr, yields essentially unchanged results. Samples falling outside the valid parameter ranges of the evolutionary grids were discarded.
For ATMO2020, we evaluated the model luminosity along each $(M,t)$ evolutionary track against the input $L_{\rm bol}$ distribution by assigning a weight proportional to the luminosity likelihood, i.e., an importance weight. The Sonora Flame Skimmer models comprise 56 evolutionary grids, spanning equilibrium and disequilibrium chemistry, seven metallicities, and four C/O ratios. We only consider the solar-metallicity grids. We performed the same Bayesian analysis within each Sonora grid and assigned equal weight to all grids. The effective temperature $T_{\rm eff}$, surface gravity $\log g$, and radius $R$ were then obtained from the interpolated model properties. We report the results in Table~\ref{tab:properties}.

Our importance-weighting approach is mathematically equivalent to exact-likelihood rejection sampling \citep{Dupuy2023}: weighting prior draws by the luminosity likelihood and accepting them with probability proportional to that likelihood target the same posterior distribution. To test it, we conducted the traditional rejection sampling following the same steps as described in \citet{Miles2023}. The test yielded essentially the same posterior distributions.

The full posterior distributions of the inferred physical properties are shown in Figure~\ref{fig:evolutionary_distribution}. The ATMO2020 and Sonora Flame Skimmer models give broadly consistent results, with median values agreeing to within a few percent for all parameters. The inferred effective temperature is broadly consistent within $1\,\sigma$ with the atmospheric-model estimates of $T_{\rm eff}=322^{+5}_{-7}\,$K using Sonora Elf Owl and $330^{+7}_{-5}\,$K using ATMO2020 from \citet{hainline2026b}, although the latter used somewhat different photometry data. GSBD5 therefore remains the coldest brown dwarf identified in the \citet{Hainline2024a} JADES+CEERS sample. These results support the interpretation of GSBD5 as an ultracool $>$Y1-type brown dwarf. Its inferred properties are summarised in Table~\ref{tab:properties}, and we discuss its low mass and broader physical implications in Section~\ref{sec:discussion}.

\begin{table}
    \centering
    \caption{Physical properties of GSBD5 inferred using the Sonora Flame Skimmer and ATMO2020 evolutionary models.}
    \label{tab:properties}
    \renewcommand{\arraystretch}{1.25}
    \begin{tabular}{lccl}
    \hline
    Property & Sonora Flame Skimmer & ATMO2020 & Unit \\
    \hline
    $M$
        & $8.2^{+2.8}_{-2.5}$
        & $8.5^{+3.2}_{-2.7}$
        & $M_{\rm J}$ \\

    Age
        & $3.9^{+2.9}_{-1.8}$
        & $4.0^{+2.9}_{-1.9}$
        & Gyr \\

    $R$
        & $1.04^{+0.03}_{-0.04}$
        & $1.02^{+0.03}_{-0.04}$
        & $R_{\rm J}$ \\

    $T_{\rm eff}$
        & $301^{+17}_{-16}$
        & $305^{+18}_{-16}$
        & K \\

    $\log g$
        & $4.28^{+0.16}_{-0.18}$
        & $4.31^{+0.17}_{-0.20}$
        & cgs \\
    \hline
    \end{tabular}
\end{table}

\subsection{Atmospheric modelling}\label{subsec:free_grid}

We further explored atmospheric modelling. To explore possible physical scenarios for GSBD5, we collected seven major atmospheric forward model packages, including nine model families and spanning a range of physical treatments. They are:

\textit{i)} ATMO2020 \citep{Phillips2020}: a cloud-free grid including both chemical equilibrium (CEQ) and disequilibrium (DEQ) chemistry under different vertical mixing (VM) strengths. We adopted $T_{\rm eff}=200-600\,$K, $\log g=2.5-5.5$, and CEQ/DEQ+weak VM/DEQ+strong VM scenarios.

\textit{ii)} ATMO2020++ \citep{Leggett2021, Meisner2023}: an empirical extension of ATMO2020 to improve cold T/Y dwarf $4-15\,\mu$m SED fitting with DEQ chemistry and a non-adiabatic pressure-temperature relationship. We adopted the same $\log g$ and [M/H] range as for ATMO2020, and $T_{\rm eff}=200-1200\,$K.

\textit{iii)} Sonora Bobcat \citep{Marley2021}: a classic cloud-free, rainout-CEQ grid that provides a simple equilibrium baseline. We adopted $\log g=3.25-5.5$, $T_{\rm eff}=200-600\,$K, and [M/H]\,$=-0.5,\,0,\,0.5$.

\textit{iv)} Sonora Elf Owl \citep{Mukherjee2024}: a cloud-free DEQ model allowing VM, metallicity, and C/O. It was used as the most appropriate Sonora model in \citet{hainline2026b}. We adopted $\log g=3.0-5.5$, $T_{\rm eff}=275-600\,$K, and [M/H]\,$=-1$ to 1, $\mathrm{C/O}=0.5-2.5$, and $\log K_{zz}=2, 4, 7, 8, 9$.

\textit{v)} Sonora Flame Skimmer \citep{Mang2026}: the latest cloud-free Sonora model grid, extending to $T_{\rm eff}=50$ K and $\log g=2.0$, with CEQ/DEQ chemistry and a broad metallicity and C/O coverage. We adopted $\log g=2.0-5.5$, $T_{\rm eff}=200-600\,$K, and [M/H]\,$=-1$ to 2, $\mathrm{C/O}=0.5-2.5$, CEQ/DEQ, and $\log K_{zz}= 1-9$.

\textit{vi)} Water-cloud model from \citet{LacyBurrows2023}: this model specifically designed for Y dwarfs allows clear/thin/thick water clouds and EQ/DEQ with carbon and nitrogen molecules, although with a sparse grid. We adopted $\log g=3.5-5$, $T_{\rm eff}=200-450\,$K, and [M/H]\,$=-0.5,\,0,\,0.5$, and thin/thick clouds in combinations with CEQ/DEQ.

\textit{vii)} Exo-REM k26 \citep{Radcliffe2026}: this family includes three branches with different cloud treatments, i.e., with varying sedimentation efficiency factor $f_{\rm sed}$, simple microphysics, and cloud-free. The varying $f_{\rm sed}$ adjusts for condensate particle sedimentation efficiency, although the clouds here refer to silicate and iron clouds, usually at deeper layers of dwarfs and not necessarily dominate Y dwarfs. We adopted $\log g=3-5$, $T_{\rm eff}=200-500\,$K, $f_{\rm sed}=1-9$, $\mathrm{[M/H]=-0.5}$ to 2.0, $\mathrm{C/O}=0.1-0.8$ for the varying-$f_{\rm sed}$ grid, and remove unrelated parameters for the latter two.

Because the Exo-REM k26 family contains three separate branches, we essentially fitted nine model grids in total. 
All flux uncertainties include the 5\% error floor justified in Section~\ref{sec:photometry}. We used \texttt{species} \citep{Stolker2020} for grid interpolation and \texttt{PyMultiNest} \citep{Buchner2014, Buchner2016} for Bayesian sampling. First, we fitted to the NIRSpec/PRISM spectrum and the MIRI photometry, providing the highest spectral-resolution constraint available for this object, and the results are summarised in Appendix~\ref{app:forward_modelling_results} in Table~\ref{tab:gsbd5_freegrid_prism_miri}. Second, we used the full NIRCam+MIRI photometric SED, comprising 23 bands in total, with the results tabulated in Table~\ref{tab:gsbd5_freegrid_nircam_all_miri}. This separation allows us to distinguish failures in reproducing detailed spectral structure from failures in reproducing the broader SED shape, as discussed separately below.

\subsubsection{SED fitting with PRISM + MIRI}\label{subsubsec:prism_miri}

None of the joint NIRSpec/PRISM and MIRI spectrophotometric fits provides a statistically acceptable solution. Across all model families, the reduced chi-square values are large, with a median $\chi^2_{\nu}=23.2$, a mean $\chi^2_{\nu}=21.3$, a standard deviation of 4.4 and a minimum of 12.3. The poorest fits are obtained with Sonora Bobcat, Sonora Elf Owl, and cloudless Exo-REM, which yield the largest $\chi^2_{\nu}$ values and the lowest Bayesian evidence values, $\ln Z$ (see Table~\ref{tab:gsbd5_freegrid_prism_miri}). All three are cloud-free models.

The best fits are obtained with ATMO2020++ and Sonora Flame Skimmer. However, they imply distance-to-radius ratios, $d/R$, that correspond to equivalent distances of $d_{\rm 1R_J}\sim150$--$200$ pc for a typical Y-dwarf radius of $R=1\,R_{\rm J}$, in $>20\sigma$ tension with the parallax-derived value. We find a similar discrepancy compared to fits to the PRISM spectrum alone by \citet{Hainline2026c}, who obtained $\chi^2_{\nu}=19.63$ and $d=139$ pc with ATMO2020++, and $\chi^2_{\nu}=34.85$ and $d=188$ pc with Sonora Elf Owl. The recurrence of this discrepancy suggests that the problem is not specific to a single model family, but instead reflects a broader mismatch between current atmospheric models and the observed spectrum of GSBD5.

Another, perhaps more fundamental, feature is that the spectrophotometric fits systematically favour high effective temperatures (see a discussion of this feature in \citealt{Carter2023b}). The fitted temperatures have a mean of $T_{\rm eff}=434$ K and a median of 406 K, substantially higher than the evolutionary-model estimate of $\sim307$ K. A similar temperature bias has been reported for WISE~0855 \citep{Luhman2024}. Note that for a fixed observed bolometric flux and radius, the inferred distance scales as $d\propto T_{\rm eff}^2$; increasing $T_{\rm eff}$ from 300 K to 400 K alone therefore increases the inferred distance by a factor of $\sim1.8$. These cases indicate that forward modelling of high-quality spectra for such ultracool dwarfs can bias the inferred temperatures and consequently the distances to higher values, when a $\sim1\,R_{\rm J}$ radius is assumed. We discuss possible origins of this bias in Section~\ref{subsec:water_cloud}.

\begin{figure*}
    \centering
    \includegraphics[width=0.85\linewidth]{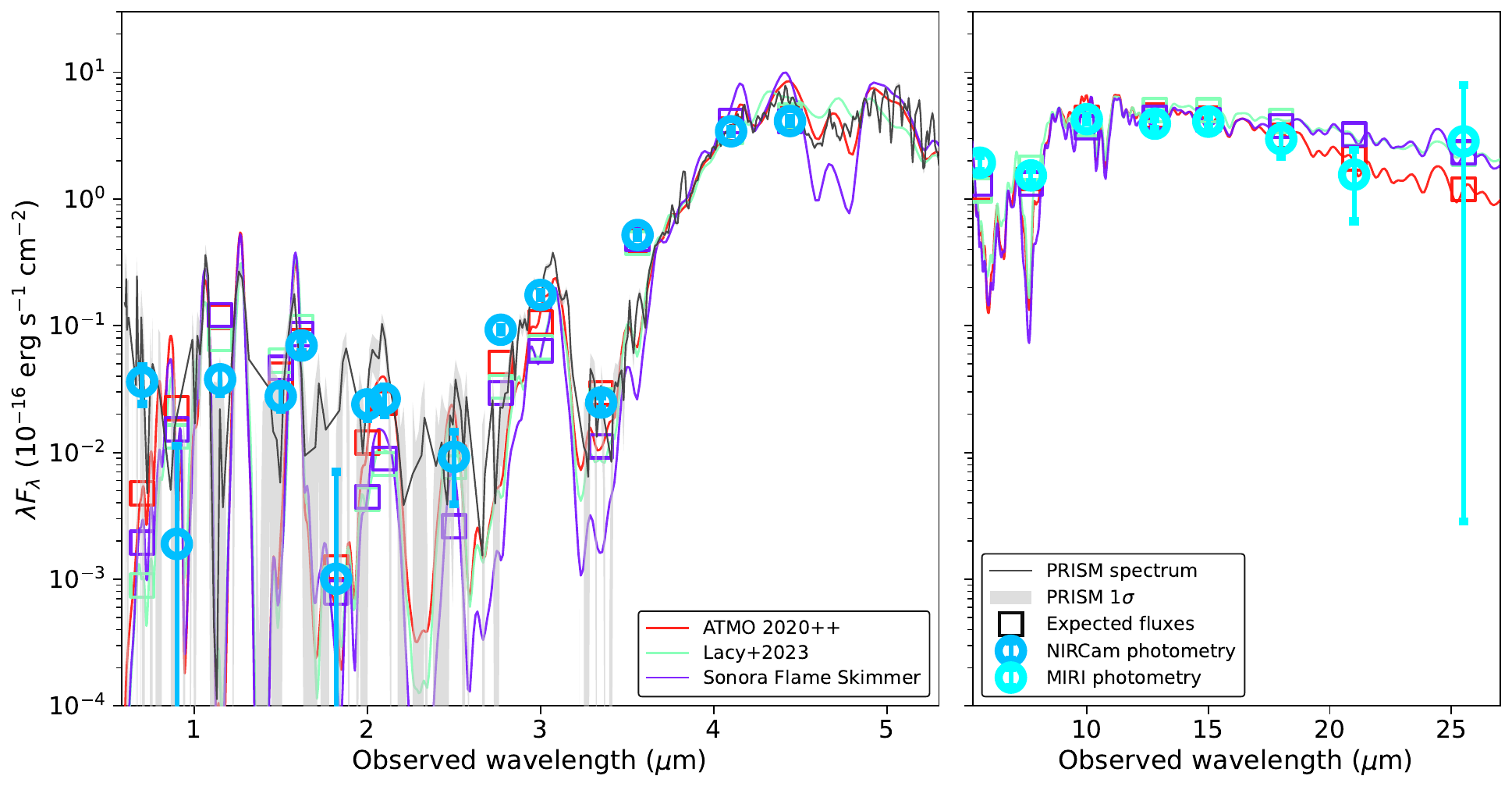}
    \includegraphics[width=0.95\linewidth]{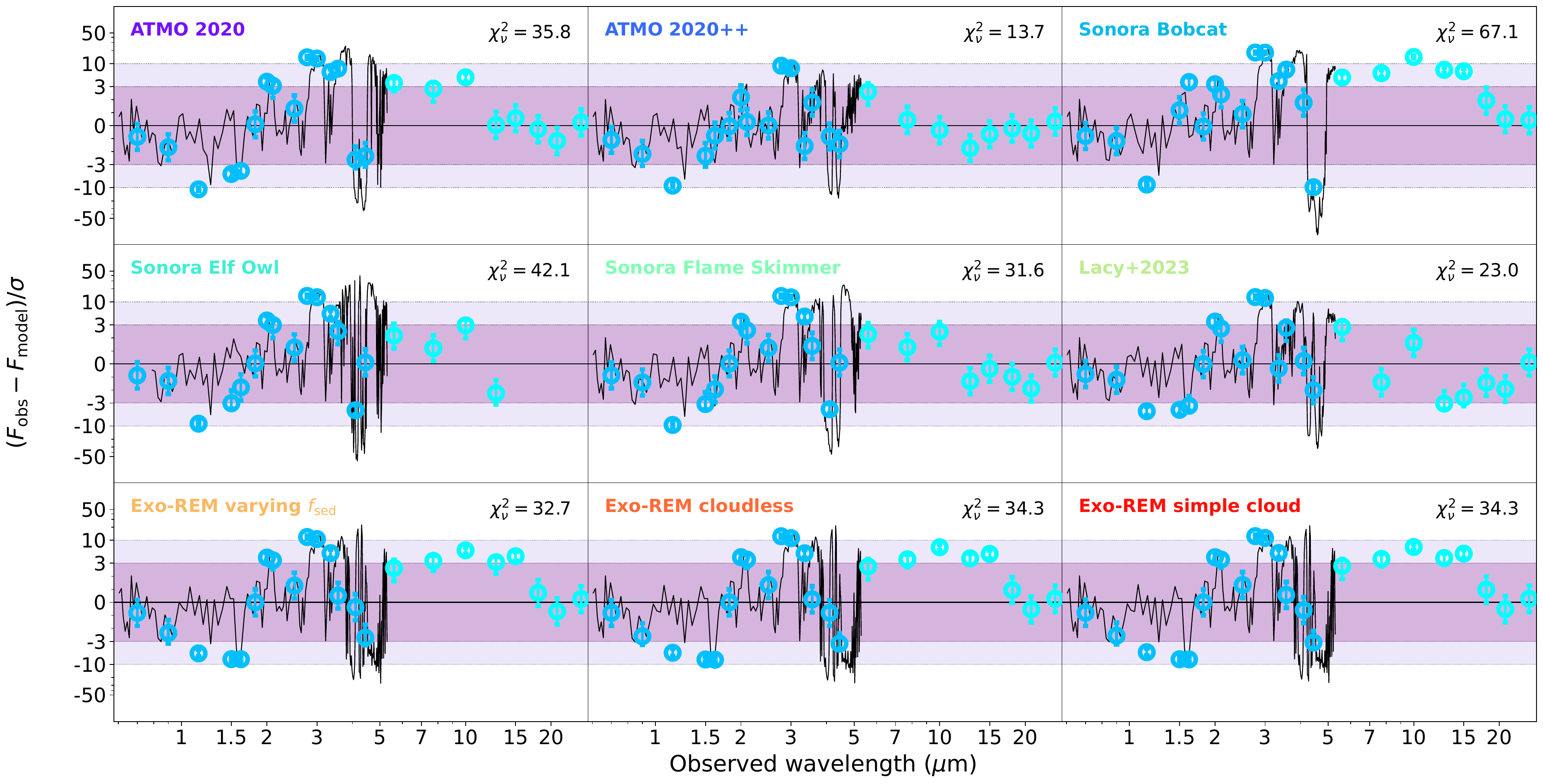}
    \caption{Best-fitting SED models for GSBD5 using evolutionary-model priors. \textit{Top:} NIRCam and MIRI photometry compared with the predicted photometric fluxes from the three best-fitting models across nine model families. The NIRSpec/PRISM spectrum and its uncertainty are overplotted for reference. The flux density is shown in units chosen to emphasise the short-wavelength spectral features, and the best-fitting model spectra are smoothed to five times the PRISM resolution for clarity. \textit{Bottom:} Residuals between the observed PRISM spectrum and the best-fitting model spectra for each model family. The shaded regions indicate residual amplitudes of $\leq3\sigma$ and $\leq10\sigma$. All fits are formally unacceptable, however the mismatch patterns are revealed by the residuals: significant discrepancies in F115W, F277W and F300M persist cross all fits, while model--data mismatch can be mitigated at various levels in other bands.}
    \label{fig:prior_sed}
\end{figure*}

\begin{figure}
    \centering
    \includegraphics[width=\linewidth]{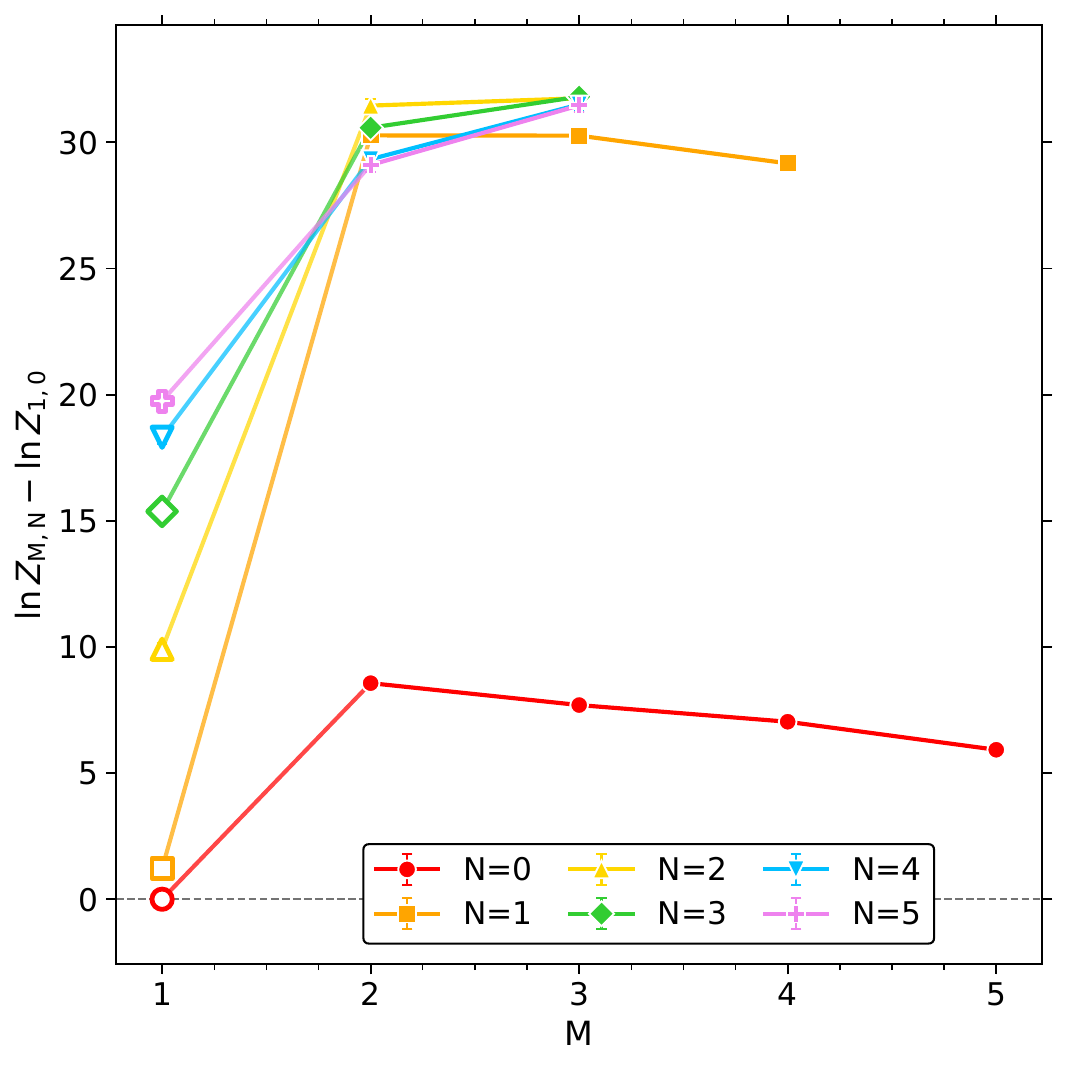}
    \vspace{-0.2in}
    \caption{Bayesian evidence in comparison with single-component no-correction forward modelling with NIRCam+MIRI photometric data and the Exo-REM varying-$f_{\rm sed}$ family. We show this difference as a function of component number, $M$, and polynomial order, $N$, for the wavelength-dependent weighting. The plateau occurs at $M=2$ and $N=1$, indicating that the GSBD5 photometric SED can be well reproduced by a two-component model with a linear wavelength-dependent redistribution before more complicated fine-tuning.}
    \label{fig:lnZ}
\end{figure}

\begin{figure*}
    \includegraphics[width=0.9\textwidth]{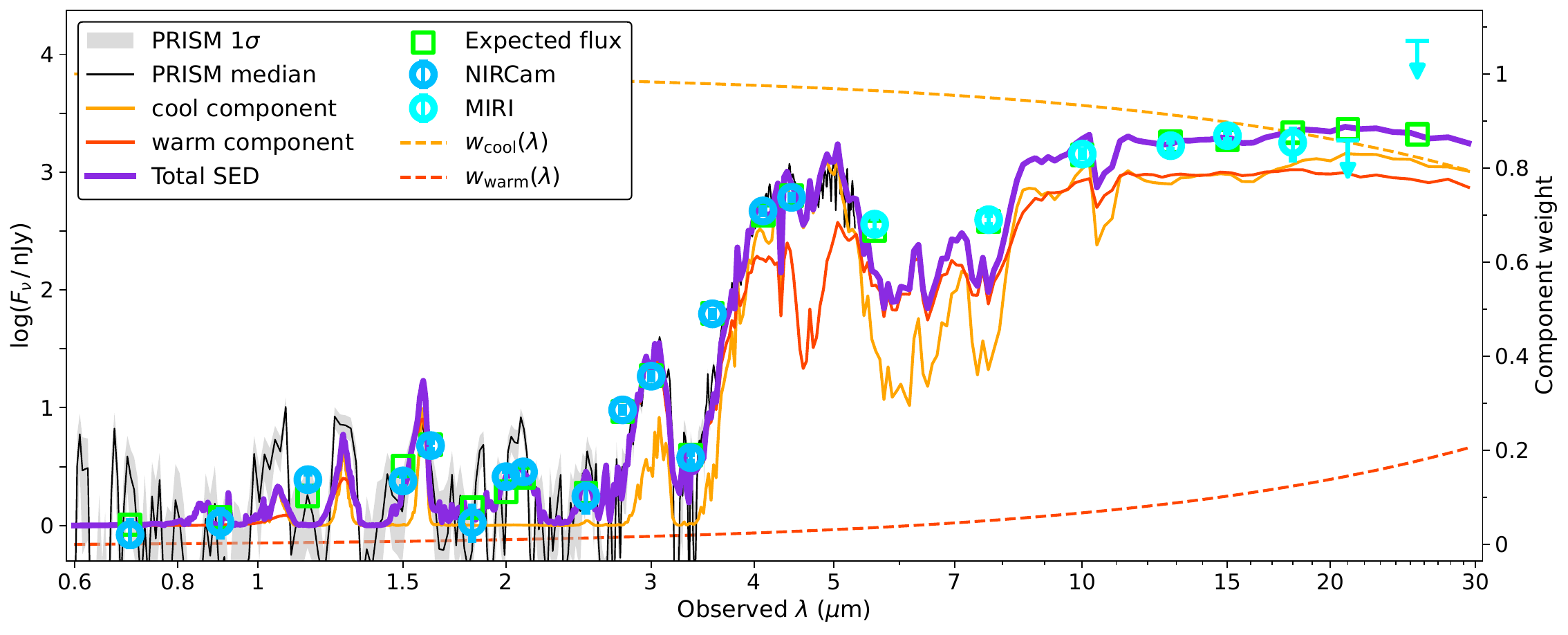}
    \includegraphics[width=0.9\textwidth]{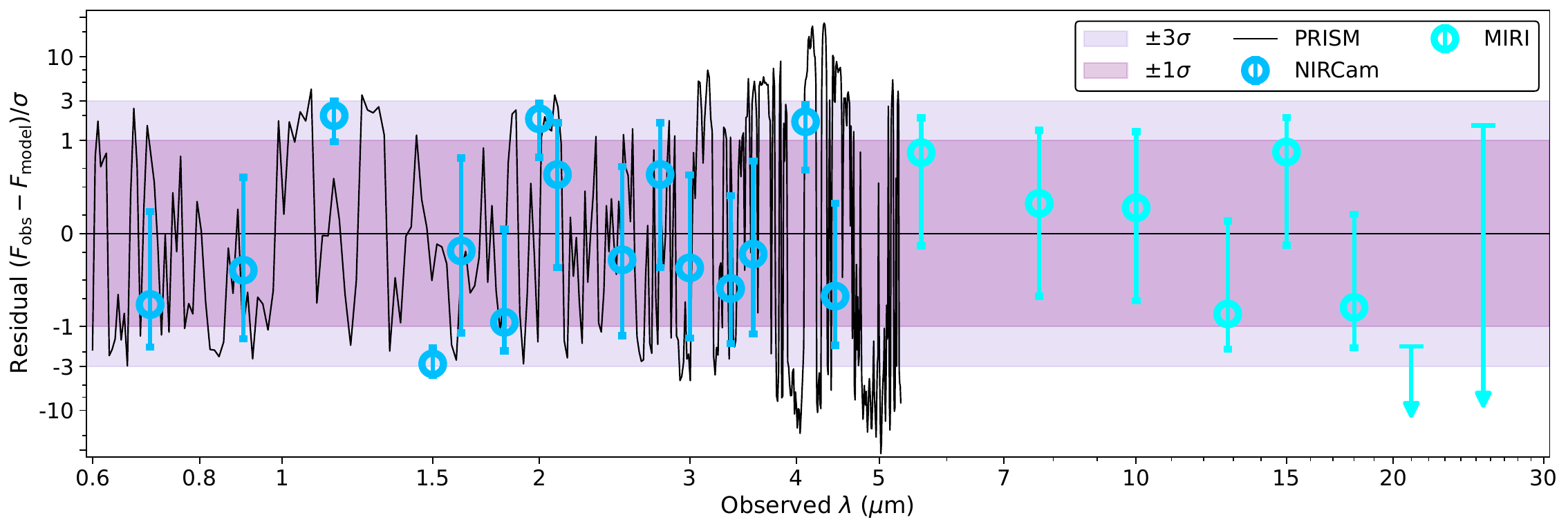}
    \caption{Best-fitting SED model for GSBD5. \textit{Top:} Two Exo-REM varying-$f_{\rm sed}$ components, with $T_{\rm eff}=250$\,K and $f_{\rm sed}=3$ (solid orange curve), and $T_{\rm eff}=450$\,K and $f_{\rm sed}=1$ (solid red curve), are combined to fit the NIRCam (blue circles) and MIRI (cyan circles) photometry. Each component is assigned a wavelength-dependent weight, shown by the dashed curve of the same colour. The NIRSpec/PRISM spectrum is overplotted for reference. \textit{Bottom:} Residuals of the photometric SED fit. Most photometric points are reproduced within the uncertainties, with the largest discrepancies in F115W and F410M at the $\sim2\,\sigma$ level. Compared with the best-fitting photometric model, the PRISM spectrum shows substantial residuals at long wavelengths, indicating a significant mismatch between the observed narrow spectral structure and the model prediction.}
    \label{fig:sed_fitting}
\end{figure*}

\subsubsection{SED fitting with NIRCam + MIRI}
\label{subsubsec:nircam_miri}

We fitted all 23 NIRCam+MIRI photometric bands listed in 
Table~\ref{tab:photometry} using the same nine model families with native flux measurements capped at the 5\% uncertainty floor. 
The resulting parameters are summarised in Table~\ref{tab:gsbd5_freegrid_nircam_all_miri}. 
Overall, the fits remain statistically poor: the reduced chi-square values have a median of $\chi^2_{\nu}=21.9$ and a mean of $\chi^2_{\nu}=21.3$. The large standard deviation of $\sigma_{\chi^2}=14.0$ is driven primarily by the Sonora Bobcat fit.

Nevertheless, several models reproduce the photometric SED substantially better than they reproduce the joint PRISM+MIRI spectrophotometry. The best solutions are obtained with two models, Exo-REM varying-$f_{\rm sed}$ and the Sonora Flame Skimmer DEQ, both with $T_{\rm eff}\sim300$ K and $d_{\rm 1R_J}\approx63$ pc. These inferred temperatures and flux scalings are broadly consistent with the parallax distance and with the effective temperature inferred from evolutionary models. These results are also similar to those reported in \citet{hainline2026b}. Fitting with selected NIRCam and MIRI bands, they estimated $d=70$ pc and $T_{\rm eff}=322$ K with Sonora Elf Owl, and $d=90$ pc and $T_{\rm eff}=330$ K with ATMO2020. Thus, while the models still fail to reproduce the full photometric SED at high precision, the photometric fits recover broadly reasonable global SED profile and flux scaling.

We then repeated the fitting after removing all NIRCam medium bands. The Exo-REM varying-$f_{\rm sed}$ and Sonora Flame Skimmer models again provide the best fits, with little change in fit quality. This indicates that the models perform similarly at the NIRCam medium+wide-band resolution, corresponding to an effective spectral resolution of $R\sim10$--15, and at the wide-band-only resolution of $R\sim4$. In other words, reducing the data to even broader photometric bands does not substantially alleviate the remaining mismatch.

The relative performance of different model families provides useful clues about the atmospheric physics favoured by GSBD5. Within the Exo-REM family, the varying-$f_{\rm sed}$ model fits much better than either the simple-cloud or cloud-free branches. The latter two also require less satisfactory flux scalings, with $d_{\rm 1R_J}\sim47$--48 pc. This suggests that adding clouds alone is not sufficient; rather, additional freedom in the cloud sedimentation efficiency, or a sedimentation-like modification of the vertical atmospheric structure, is necessary. The same preference for DEQ and/or vertical mixing is also confirmed by the significant improvement within the Sonora family from Bobcat to Elf Owl and finally Flame Skimmer.

\subsubsection{Prior-guided SED fitting}
\label{subsubsec:prior_sed_fitting}

The parallax measurement and evolutionary modelling provide independent constraints on the physical properties of GSBD5. A natural next step is therefore to use these constraints as priors in the SED fitting. However, since most atmospheric models only have very few (degenerate) parameters than those obtained through the evolutionary modelling, constraining all known properties via priors would compound parameter compensation in the fitting. We therefore only input the full Sonora Flame Skimmer $T_{\rm eff}$ and $\log g$ posteriors in Figure~\ref{fig:evolutionary_distribution} as priors, and allowed other properties to vary. Critically, the free-fitted flux scaling factor, $q=(R/d)^2$, can separate the spectral morphology from the overall flux scaling, allowing us to assess whether the fitted SED remains consistent with the parallax-based distance and expected radius.
We adopted the joint two-dimensional prior distribution obtained by equally combining the Sonora Flame Skimmer and ATMO2020 evolutionary-model posteriors (see Section~\ref{subsec:evolution_model} and Figure~\ref{fig:evolutionary_distribution}). We fitted all nine model branches to the 23-band NIRCam+MIRI photometry using the same fitting setup as above, except for the inclusion of the evolutionary-model priors. The resulting SEDs, together with the NIRCam and MIRI photometry and the PRISM spectrum for comparison, are shown in Figure~\ref{fig:prior_sed}. The fitted parameters are listed in Table~\ref{tab:gsbd5_evolution_guided_nircam_all_miri}.

The fitted global parameters in the prior-guided fits change in a physically meaningful way. Solar or subsolar metallicities stay largely unchanged, and the new flux scalings remain similar, corresponding to a median distance of $\sim61$ pc for $R=1\,R_{\rm J}$. The effective temperatures drop to $\sim281$ K (i.e., lower than the $\sim307$ K prior), and the gravity estimate is improved from the previous grid lower boundaries to a median of $\log g \sim 3.42$.

The large $\chi^2_{\nu}$ values are dominated by a small number of bands. As shown in Figure~\ref{fig:prior_sed}, F115W, F277W and F300M consistently show $\gtrsim10\,\sigma$ discrepancies across nearly all model branches, whereas the residuals in the other bands vary more strongly from model to model. For a typical number of degrees of freedom of $\sim20$, these three bands alone increase the reduced $\chi^2_{\nu}$ by about 15. Both F277W and F300M are assigned the 5\% flux-error floor because of their high native S/N. The large residuals therefore mainly reflect highly significant relative discrepancies with respect to the adopted uncertainties. This highlights the importance of high-S/N PSF photometry across a complete set of bands. With lower-S/N measurements, such as the $\sim3\times$ higher F277W noise adopted for GSBD5 by \citet{hainline2026b}, or with missing bands of any of the three, the statistical significance of this mismatch would be substantially reduced.

\subsubsection{Controlled SED-fitting comparison experiments}
\label{subsubsec:sed_experiment}

Because no single atmospheric model family provides a statistically acceptable fit to the SED of GSBD5, we performed a set of controlled $\chi^2$ comparison experiments to identify which additional model ingredients most effectively reduce the mismatch. These experiments are designed to compare the relative statistical performance of models that differ in specific assumptions and to identify missing ingredients in current models, a strategy that has been widely used in cold-dwarf studies when no fully adequate model is available \citep[e.g.,][]{Morley2014, Leggett2017, Miles2020, LacyBurrows2023, Beiler2024}.

The current best-performing free-grid model, Exo-REM with variable $f_{\rm sed}$, does not include water clouds. However, water clouds are expected to become important in atmospheres with $T_{\rm eff}\sim300$ K \citep[e.g.,][]{Morley2014, LacyBurrows2023}. We therefore tested the effect of water clouds using different branches of the \citet{LacyBurrows2023} model family. We fitted six combinations of clear, thin-cloud, and thick-cloud atmospheres, each with CEQ and DEQ chemistry, to the same NIRCam+MIRI photometry over the same parameter ranges. Within this model family, the best fit is obtained with the thick-cloud DEQ model, followed by the thin-cloud DEQ and thick-cloud CEQ models. All cloudy branches perform better than the corresponding clear branches, and the DEQ models generally outperform their equilibrium counterparts. The same qualitative preference is obtained when fitting the PRISM+MIRI data. Thus, within the \citet{LacyBurrows2023} framework, the SED of GSBD5 favours both water clouds and DEQ chemistry, plausibly associated with vertical mixing. However, the current SED fitting cannot uniquely identify if this preference comes from the the cloud composition or cloud structure or both.

We next tested whether horizontal inhomogeneity can improve the fit. Physically, this experiment represents an atmosphere composed of multiple columns with different local thermal or cloud properties \citep[e.g.,][]{X_Chen2025, OliverosGomez2026}. Using Exo-REM with variable $f_{\rm sed}$, we fitted the NIRCam+MIRI photometry with $M=1$--5 atmospheric components. The components were required to share the same $\log g$, [M/H], and C/O ratio, but were allowed to have different $T_{\rm eff}$ and $f_{\rm sed}$. Increasing the number of components produces only a modest improvement in Bayesian evidence, at the $\sim7\%$ level, with the best case occurring at $M=2$. Even this best multi-component fit still has $\chi^2_{\nu}>5$. We therefore find that horizontal patchiness alone cannot explain the SED mismatch.

We then explored a complementary possibility that the mismatch reflects an incomplete description of the vertical atmospheric structure. Different wavelengths probe different pressure levels \citep[e.g.][]{Buenzli2012, Biller2013, Biller2024, Ge2019, Vos2023, McCarthy2024, McCarthy2025, X_Chen2025, OliverosGomez2026}, so an imperfect pressure--temperature profile, opacity structure, or vertical mixing prescription could appear observationally as a wavelength-dependent redistribution of emergent flux. To test this phenomenologically, we introduced a smooth wavelength-dependent weighting function, $w(\lambda)$, that reshapes the relative SED brightness. We parameterised $w(\lambda)$ with polynomials of order $N=0$--5. The case $N=0$ corresponds to no wavelength-dependent correction beyond the usual flux scaling, while $N=1$ allows a linear tilt across wavelength. This experiment improves the Bayesian evidence by $\lesssim15\%$ across the tested polynomial orders (Figure~\ref{fig:lnZ}). Thus, a smooth wavelength-dependent correction moderately improves the fit, but higher-order corrections provide diminishing returns.

Finally, we combined horizontal multi-component structure with wavelength-dependent redistribution. We modelled the observed flux as
\begin{equation}
F_{\lambda}=q\sum^{M}_{m=1}w_m(\lambda)F_{\lambda,m},
\end{equation}
where $q=(R/d)^2$ is the overall flux-scaling factor, $w_m(\lambda)\geq0$, and $\sum^M_{m=1}w_m(\lambda)=1$ at each wavelength. Here, $F_{\lambda,m}$ represents the spectrum of the $m$-th atmospheric component, while $w_m(\lambda)$ describes its wavelength-dependent contribution to the disk-integrated emission. Combinations with excessively large $(M,N)$ were omitted when the number of free parameters exceeded the information content of the available photometric data. The resulting Bayesian evidences are shown in Figure~\ref{fig:lnZ}.

The comparison across $(M,N)$ combinations points to a relatively simple phenomenological solution. A large gain occurs at $M=2$ and $N=1$, corresponding to two atmospheric components whose relative contributions vary linearly with wavelength. This model gives $\chi^2=28.0$ for 13 degrees of freedom, or $\chi^2_{\nu}=2.2$, a substantial improvement over all single-component fits. Beyond this point, adding more components or higher-order wavelength dependence produces only marginal gains. For example, increasing the polynomial order changes $\ln Z$ by $<3\%$, while models with $M=4$ or 5 components have lower Bayesian evidence despite their additional flexibility. This plateau indicates that the improvement is not simply a numerical consequence of adding free parameters; rather, most of the recoverable mismatch is physically captured by a two-component model with a smooth wavelength-dependent redistribution. We show this preferred $M=2$, $N=1$ fit in Figure~\ref{fig:sed_fitting}, list the fitted parameters in Table~\ref{tab:two_components}, and discuss its physical implications in Section~\ref{subsec:water_cloud}.

\begin{table}
    \centering
    \caption{NIRCam and MIRI photometric SED Fitted results with two Exo-REM varying-$f_{\rm sed}$ components and wavelength-dependent weights. The total effective temperature is derived from the joint model-fitted SED.}
    \renewcommand{\arraystretch}{1.2}
    \begin{tabular}{lll}
    \hline
       Property  & Value & Unit \\
       \hline
        $T_{\rm cool}$ & $244^{+5}_{-8}$ & K\\
        $T_{\rm warm}$ & $425^{+14}_{-10}$ & K\\
        $T_{\rm total}$ & $284^{+10}_{-8}$  & K\\
        $f_{\rm sed,cool}$ & $5.0^{+2.7}_{-2.7}$ & --\\
        $f_{\rm sed,warm}$ & $1.03^{+0.04}_{-0.02}$  & -- \\
        $\log g$ & $3.21^{+0.17}_{-0.13}$  & cgs \\
        $[\mathrm{M/H}]$ & $-0.41^{+0.05}_{-0.05}$ & --\\
        ${\rm C/O}$ & $0.77^{+0.02}_{-0.03}$ & --\\
        $d_{\rm 1R_J}$ & $52.4^{+4.4}_{-3.6}$ & pc \\
        $\varpi_{\rm 1R_J}$ & $19.1^{+1.4}_{-1.5}$ & mas \\
        \hline
    \end{tabular}
    \label{tab:two_components}
\end{table}

\section{Discussion}\label{sec:discussion}

\subsection{An isolated planetary-mass object in the thin disk}\label{subsec:IPMO}

We searched for possible companions to GSBD5 over a wide range of projected separations. In the immediate vicinity of GSBD5, no plausible point-source companion is detected in the \textit{JWST} residual images. Given the analysed field of view, this rules out obvious companions over projected separations of $\lesssim7.5\times10^3$ AU, depending on direction (see Figure~\ref{fig:reference_frame}). Because the available imaging spans $0.7$--$25.5\,\mu$m and reaches very faint limits, a bound stellar companion that is undetected in all bands is strongly disfavoured. Indeed, we cannot rule out the possibility of a similarly cold, low-mass brown dwarf companion at a very small separation, given that the F444W PSF full-width-half-maximum (FWHM) corresponds to a projected distance of $\sim10~$AU, larger than a typical expected Y-dwarf binary separation of a few AU \citep[e.g.,][]{Fontanive2023}. However, the binary rate of such extremely cold, late-type dwarfs is statistically low \citep[e.g., $\lesssim10\%$;][]{Fontanive2018}.

We then searched for a possible wide stellar host. We queried the \textit{Gaia} DR3 catalogue \citep{GaiaCollaboration2016, GaiaCollaboration2023} within $1^\circ$ of GSBD5, corresponding to $2.3\times10^5$ AU at the distance of GSBD5. Among the 13 sources with parallaxes compatible with GSBD5, the closest proper-motion match differs by $\sim40.0\,{\rm mas\,yr^{-1}}$. Conversely, the closest overall proper-motion match has a parallax of $\varpi=5.137\pm0.013$ mas, in $8.2\sigma$ tension with GSBD5. We therefore find no evidence of a stellar host.

We further searched CatWISE2020 \citep{Marocco2021} and the All-Sky NOIRLab Source Catalogue DR2 \citep{Nidever2021} for fainter M/L/T/Y-dwarf companions. To reduce contamination from compact galaxies and other extragalactic sources, we cross-matched the candidates with the VIDEO DR5 \citep{Jarvis2013} and DES DR2 \citep{Abbott2021} catalogues. We required the proper motion to be consistent with that of GSBD5 within $3\sigma$ and restricted the search to a $5'$ radius, corresponding to $\sim2\times10^4$ AU at the distance of GSBD5. The only remaining candidate is CatWISE J033228.63$-$275029.5. However, its proper motion is constrained only at the $\sim1\sigma$ level, and its DES colours, compared with the colour templates of \citet{dalPonte2023}, are consistent with an M4 star. Assuming a conservative absolute magnitude of $M_J=9.5$ mag for an M4 dwarf, its photometry places it at $>300$ pc, ruling out physical association with GSBD5.
We therefore find no plausible dwarf companion to GSBD5. This is consistent with the non-detection of binary acceleration.

With the inferred mass of $\sim8\,M_{\rm J}$ (Section~\ref{subsec:evolution_model}), GSBD5 lies at the extreme low-mass end of the brown-dwarf population, far below the hydrogen-burning limit of $\sim75\,M_{\rm J}$ \citep[][]{Dieterich2014, Chabrier2023}. As a working threshold for the planetary-mass regime, we adopt $13\,M_{\rm J}$ \citep{Spiegel2011}, below which 95\% and 93\% of the GSBD5 mass posterior lies, from the evolutionary fits with Sonora Flame Skimmer and ATMO2020, respectively. Together with lack of known companions, this places GSBD5 in the regime of isolated planetary-mass objects, or sometimes termed as sub-brown dwarfs or free-floating planets \citep{MiretRoig2023}.

The coordinates place GSBD5 at Galactic coordinates of approximately \((l,b)=(223.62^{\circ},-54.47^{\circ})\). At the parallax distance of \(d\approx63.7\)~pc, this corresponds to a vertical offset of \(d\sin b\simeq -51.3\)~pc below the Galactic plane, well within the scale height of the Galactic thin disk. This spatial location is consistent with the current kinematic constraints: the measured proper motion implies a tangential velocity of \(v_{\rm tan}=15.0^{+1.4}_{-1.3}\,{\rm km\,s^{-1}}\), within, and toward the low-velocity end of, the thin-disk ultracool-dwarf tangential velocity distribution \citep[e.g.,][]{Faherty2009}. However, the current NIRSpec/PRISM spectrum does not provide a useful radial-velocity constraint, so the full three-dimensional space motion remains unknown. Future higher-resolution spectroscopy would be needed to constrain the full kinematics of GSBD5.

\begin{figure}
    \centering
    \includegraphics[width=\linewidth]{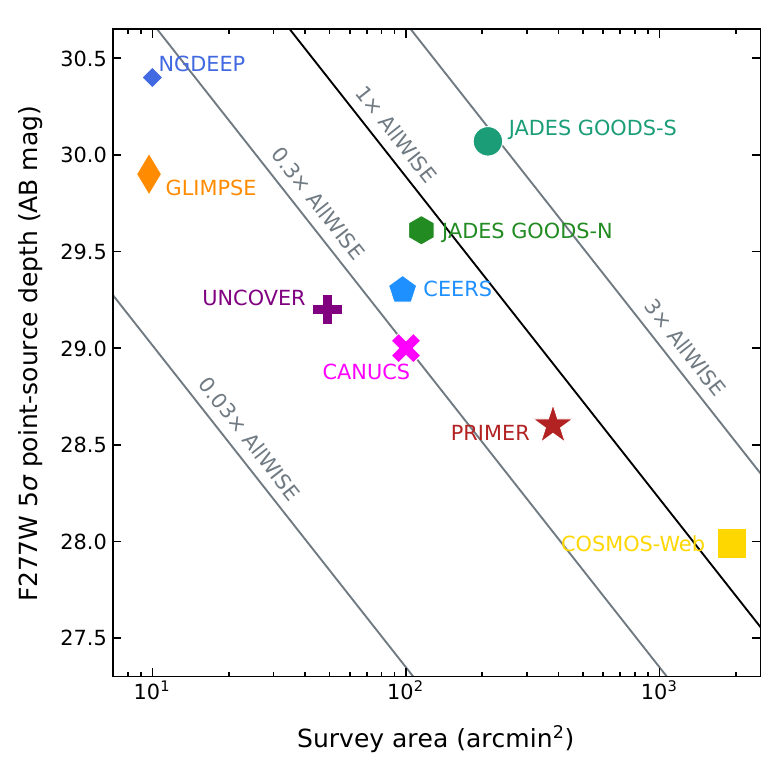}
    \vspace{-0.3in}
    \caption{Survey areas and depths of representative \textit{JWST} extragalactic surveys. Data of the two JADES fields and the COSMOS-Web field are described in the text. Depths of other surveys are taken from \citet{mcleod2026}, and their areas are obtained via \textit{JWST} MAST. The AllWISE-equivalent volumes are computed by comparing the F277W depth against the \textit{WISE} W1 depth (see text). Existing \textit{JWST} extragalactic fields provide large surveyed Galactic volumes, where extremely cold objects such as GSBD5 can be discovered.}
    \label{fig:surveys}
\end{figure}

\subsection{Expanding the astrometric frontier for the coldest brown dwarfs with \textit{JWST}}\label{subsec:frontier}

\begin{table*}
\centering
\caption{Astrometric and kinematic constraints of nine Y-dwarf candidates from a search in JADES and CEERS reported by \citet{Hainline2024a, hainline2026b}. PANO-GS-BD-18 has only received one epoch of observation and we cannot constrain its motion. The SED-based distances, $d_{\rm SED}$, are inferred from the median of the three models used by \citet{hainline2026b}, and the SED-based tangential velocities, $v_{\rm tan,\,SED}$, are derived from the SED-based distances and the newly obtained proper motions.}
\label{tab:astrometry_summary}
\setlength{\tabcolsep}{3pt}
\renewcommand{\arraystretch}{1.25}
\begin{tabular}{lcccccccccc}
\hline
ID &
$\mu_{\xi}$ &
$\mu_{\eta}$ &
$|\mu|$ &
S/N$_{|\mu|}$ &
$\varpi$ &
S/N$_{\rm\varpi}$ &
$d_{\varpi}$ &
$d_{\rm SED}$ &
$v_{\rm tan,\,\varpi}$ &
$v_{\rm tan,\,SED}$ \\
& (mas yr$^{-1}$) & (mas yr$^{-1}$) & (mas yr$^{-1}$) & & (mas) & & (pc) & (pc) & (km s$^{-1}$) & (km s$^{-1}$) \\
\hline
JADES-GN-BD-5 & $-21.1_{-1.2}^{+1.9}$ & $-5.2_{-2.3}^{+1.8}$ & $22.0_{-2.0}^{+0.7}$ & $20.2$ & $0.0_{-6.9}^{+4.9}$ & $0.0$ & $>203$ & $870_{-210}^{+210}$ & $>21.1_{-1.9}^{+0.7}$ & $88_{-21}^{+23}$ \\
JADES-GN-BD-8 & $-21.8_{-2.6}^{+2.6}$ & $7.4_{-3.7}^{+2.9}$ & $23.1_{-3.0}^{+2.9}$ & $7.7$ & $1.4_{-1.7}^{+1.9}$ & $0.8$ & $>309$ & $610_{-100}^{+80}$ & $>33.8_{-4.4}^{+4.3}$ & $66_{-13}^{+14}$ \\
JADES-GN-BD-9 & $-9.9_{-3.9}^{+1.9}$ & $1.9_{-4.4}^{+3.3}$ & $10.5_{-2.0}^{+5.2}$ & $3.4$ & $33_{-20}^{+11}$ & $2.0$ & $>23$ & $1190_{-370}^{+330}$ & $>1.1_{-0.2}^{+0.6}$ & $60_{-22}^{+31}$ \\
JADES-GN-BD-11 & $-19_{-12}^{+10}$ & $18_{-15}^{+11}$ & $28_{-8}^{+13}$ & $2.1$ & -- & -- & -- & $2090_{-1010}^{+950}$ & -- & $258_{-131}^{+203}$ \\
JADES-GN-BD-16 & $88_{-138}^{+125}$ & $213_{-95}^{+192}$ & $233_{-91}^{+192}$ & $1.7$ & -- & -- & -- & $960_{-310}^{+450}$ & -- & $1082_{-486}^{+982}$ \\
JADES-GS-BD-4 & $10.4_{-7.0}^{+4.8}$ & $-1.0_{-6.6}^{+5.1}$ & $11_{-6}^{+6}$ & $1.8$ & -- & -- & -- & $1380_{-420}^{+380}$ & -- & $66_{-36}^{+52}$ \\
JADES-GS-BD-8 & $12.8_{-2.4}^{+3.0}$ & $-11.3_{-1.6}^{+1.8}$ & $17.4_{-2.3}^{+2.2}$ & $8.5$ & $-0.9_{-2.0}^{+1.9}$ & $-0.4$ & $>963$ & $920_{-190}^{+150}$ & $>79_{-11}^{+10}$ & $74_{-16}^{+18}$ \\
JADES-GS-BD-21 & $9_{-13}^{+10}$ & $15_{-10}^{+9}$ & $23_{-14}^{+5}$ & $1.8$ & $-65_{-48}^{+62}$ & $-1.04$ & -- & $900_{-210}^{+310}$ & -- & $92_{-48}^{+48}$ \\
PANO-GS-BD-18 & -- & -- & -- & -- & -- & -- & -- & $440_{-140}^{+90}$ & -- & -- \\
\hline
\end{tabular}

\end{table*}

Until now, no brown dwarf in a \textit{JWST} extragalactic deep field has had a securely measured parallax, with previous attempts remaining unsuccessful \citep[e.g.,][]{Liu2026}. GSBD5 therefore provides the first such case, enabling a route to characterise distant brown dwarfs in extragalactic fields with parallaxes. In the pre-\textit{JWST} era, all-sky WISE searches provided a flux-limited Y-dwarf sample only out to $\sim10$ pc \citep{Kirkpatrick2014, Kirkpatrick2016, Wright2019, Kirkpatrick2021}. One of the most distant parallax-measured Y-dwarf candidates, the photometric Y0 candidate CWISEP J135937.65$-$435226.9, has $\varpi=41.1\pm4.3$ mas, corresponding to $d\sim24.3$ pc \citep{Marocco2021}. The previously most distant spectroscopically confirmed Y dwarf, WD 0806$-$661B, lies at a parallax-based distance of $\sim19.2$ pc, which was only discovered by targeted, deep \textit{Spitzer} observation \citep{Luhman2011}, and its parallax is measured from its brighter white-dwarf companion \citep{Subasavage2009, Luhman2012, Martin2018, Fontanive2026}. The parallax measurement of GSBD5 therefore extends geometric distance measurements of Y dwarfs by a factor of $\gtrsim3$.

\citet{Hainline2024a,hainline2026b} reported 10 Y-dwarf candidates from two of the deepest \textit{JWST} extragalactic fields (Fig.~\ref{fig:surveys}), the JADES GOODS-South and GOODS-North fields, including GSBD5. These candidates were selected through SED fitting with inferred $T_{\rm eff}<500$ K, but no astrometric motion was reported for any candidate other than GSBD5. Following the same approach used for GSBD5, we analysed the per-exposure photometry and astrometry of the remaining candidates using the reductions described in Section~\ref{sec:data}. Compared to \citet{hainline2026b}, we now have 10 new NIRCam bands (F070W, F162M, F182M, F187N, F210M, F250M, F430M, F460M, F480M) and slightly different flux values of existing bands, because of both a more complete collection of imaging data than JADES DR5 and the PSF photometry approach. The PSF-based photometric results are tabulated in Appendix~\ref{app:y_photo} and their astrometry is summarised in Table~\ref{tab:astrometry_summary}.

We robustly detect four additional proper motions at ${\rm S/N}>3$, for JADES-GN-BD-5, JADES-GN-BD-8, JADES-GN-BD-9, and JADES-GS-BD-8. These motions confirm that these sources are Galactic objects. We also obtain tentative proper-motion detections at $1.5<{\rm S/N}<3$ for JADES-GN-BD-11, JADES-GN-BD-16, JADES-GS-BD-21, and JADES-GS-BD-4. The only Y-dwarf candidate without a proper-motion constraint is PANO-GS-BD-18, which has only one available observing epoch.

No additional candidate shows a robust parallax detection with ${\rm S/N}>3$. The most suggestive case is JADES-GN-BD-9, with a tentative
${\rm S/N}\sim2$ parallax of $\varpi\sim33.3$ mas, corresponding to
$d\sim30$ pc. However, this parallax is driven almost entirely by the March 2023 epoch, which was observed in F410M rather than the F444W band used in the other epochs. Omitting one of the three exposures from this F410M epoch changes the inferred parallax by $\sim\pm1\sigma$. We therefore do not use this parallax as a direct distance constraint. Instead, using the 68\% intervals of all meaningful parallax constraints, we report trigonometric distance limits in Table~\ref{tab:astrometry_summary} for the four candidates whose parallaxes can be effectively constrained. These limits are broadly consistent with the SED-based distances inferred from the median of the three models used by \citet{hainline2026b}, although most of them remain relatively loose. The other five candidates either have too few temporal samplings to fit a proper motion + parallax model, or are too noisy for the parallax to be constrained.

Using the SED-fitted distances, the newly measured proper motions correspond to tangential velocities below $100~{\rm km~s^{-1}}$ for most sources, consistent with the tangential-velocity distribution of thin-disk cold dwarfs \citep[e.g.,][]{Faherty2009}. JADES-GN-BD-11 and JADES-GN-BD-16 have formally larger tangential velocities of $\sim260~{\rm km~s^{-1}}$ and $\sim1080~{\rm km~s^{-1}}$, respectively. However, these values have $\sim80\%-100\%$ uncertainties due to the uncertain SED-fitted distances. The current data therefore do not warrant a detailed kinematic interpretation of these two objects.

\begin{figure*}
    \includegraphics[width=0.48\textwidth]{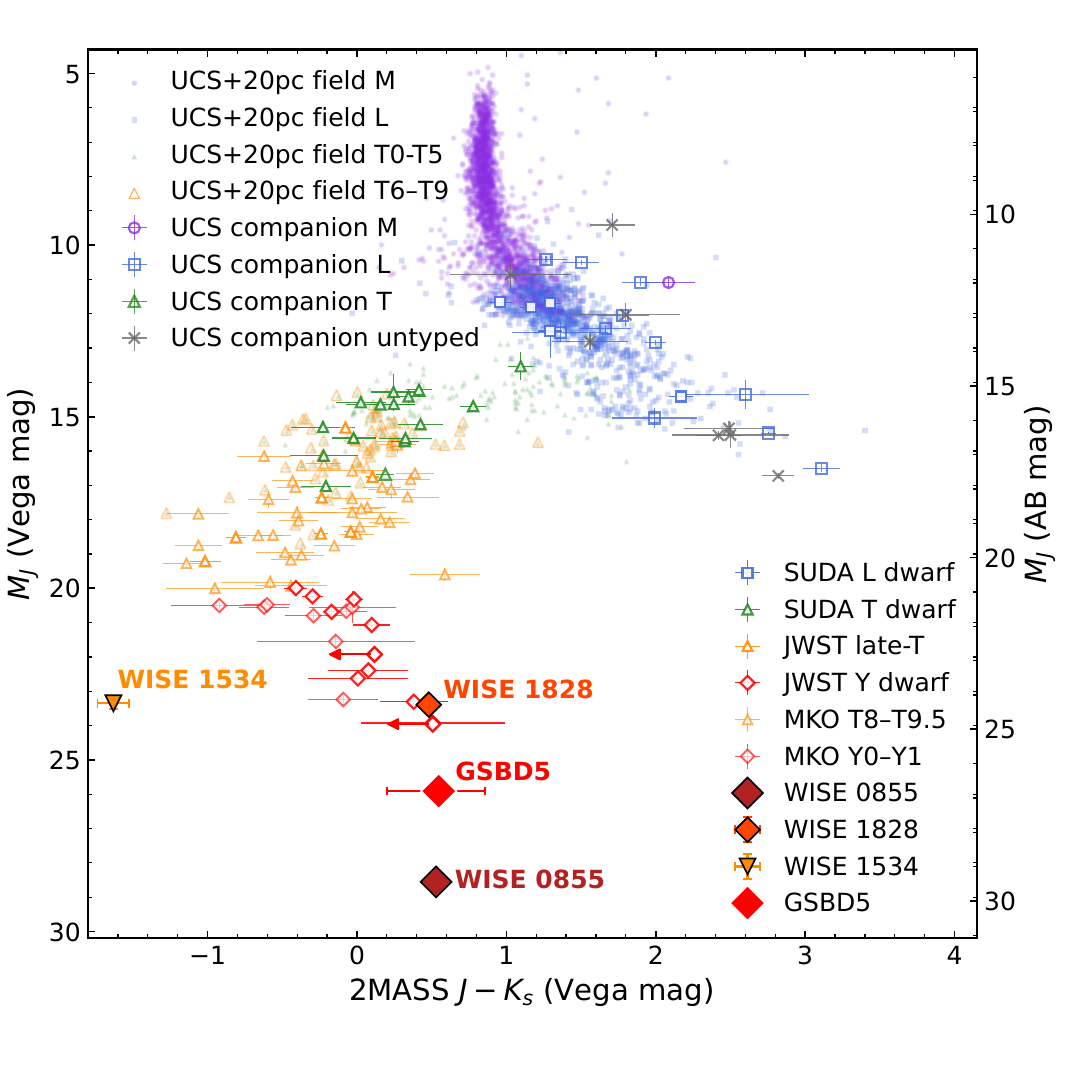}
    \includegraphics[width=0.48\textwidth]{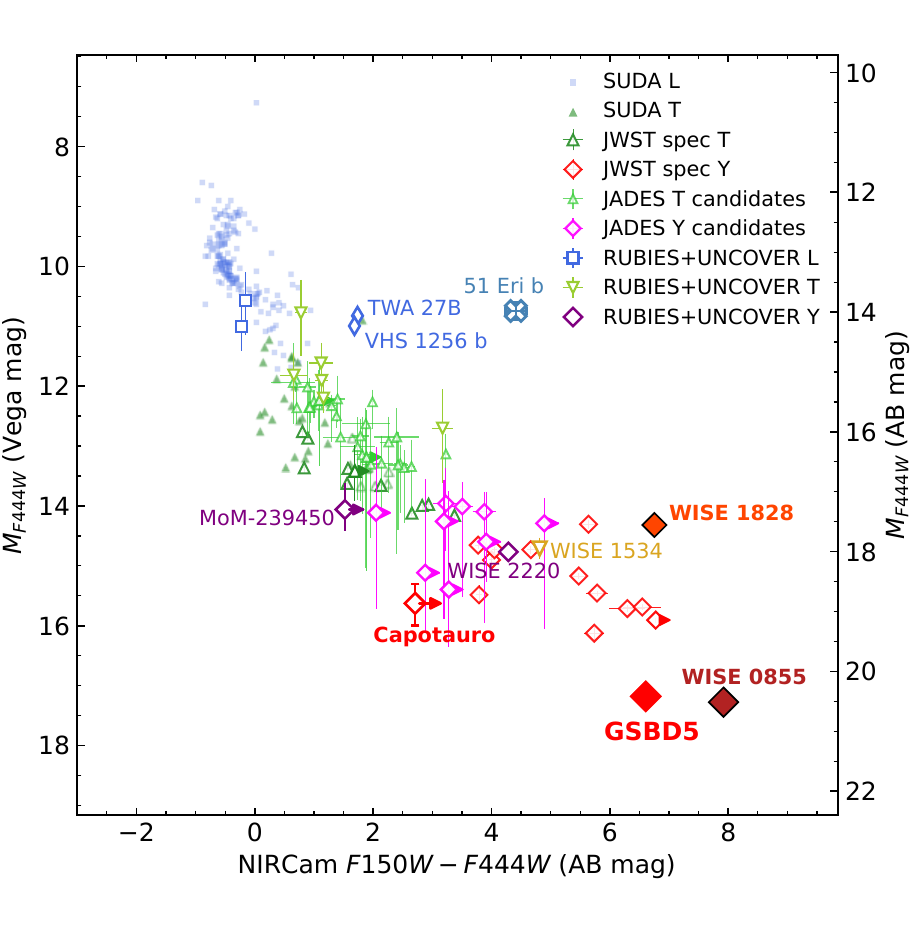}
    \includegraphics[width=0.46\textwidth]{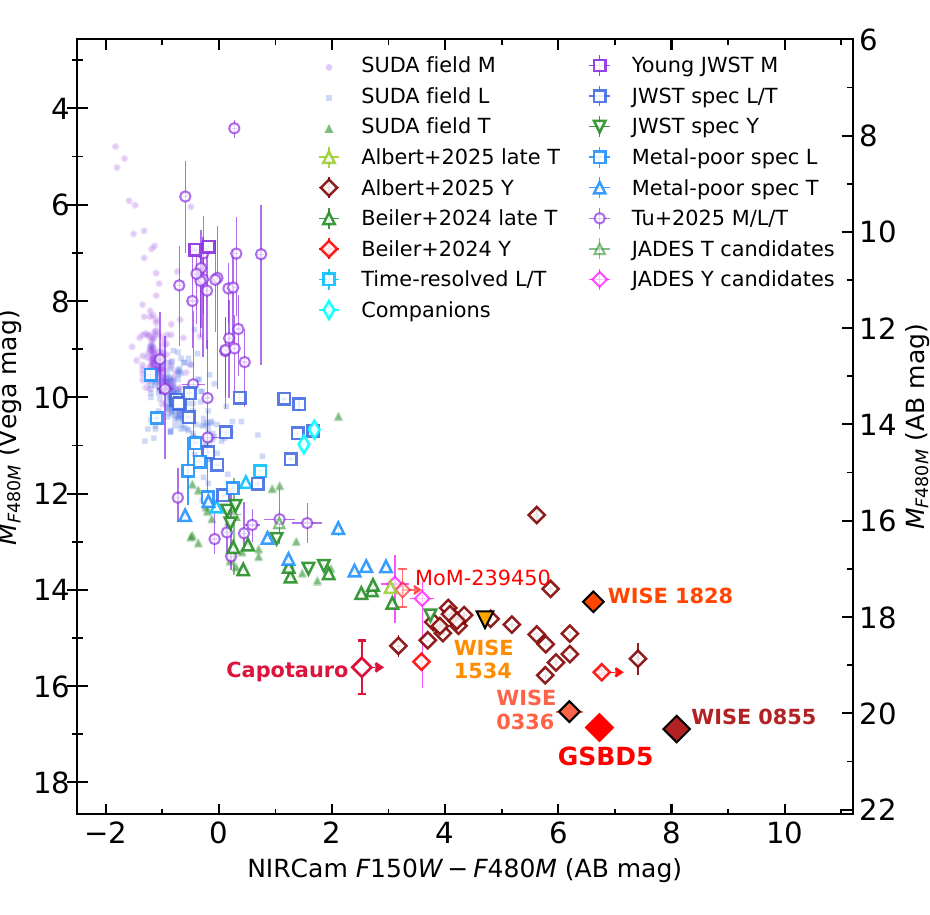}
    \includegraphics[width=0.49\textwidth]{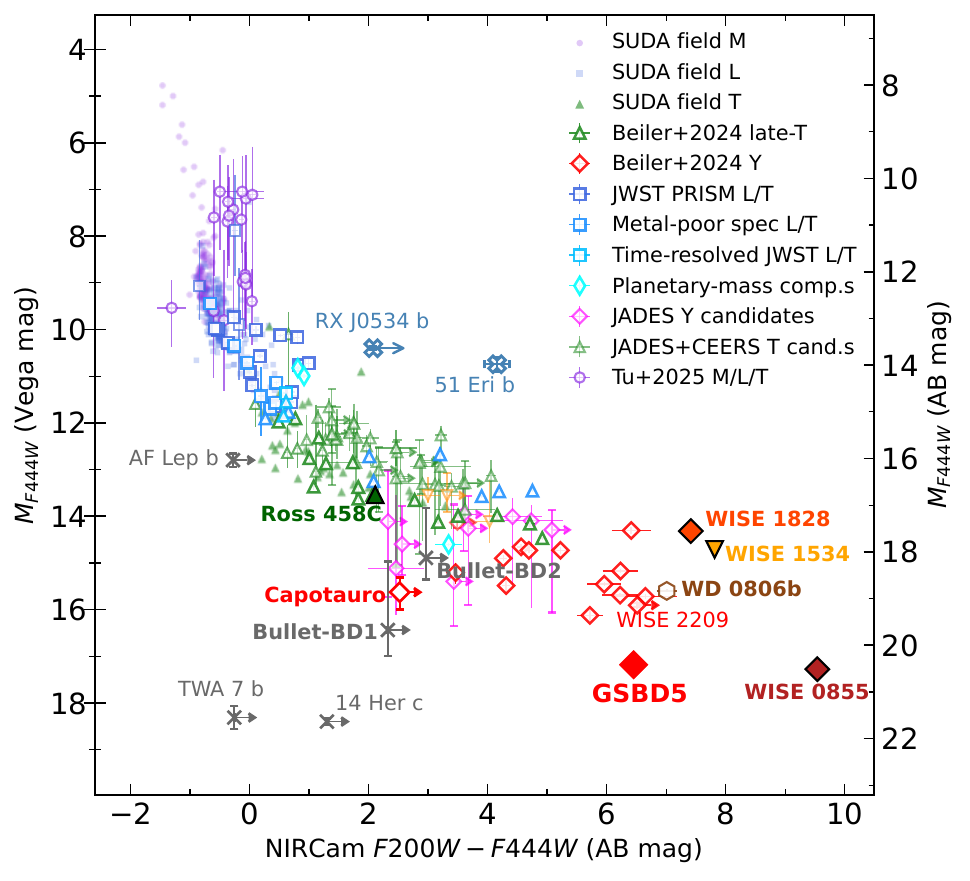}
    \caption{Colour-magnitude diagrams including many known brown dwarfs and exoplanets, with references detailed in Appendix~\ref{app:cmd_ref}. GSBD5 occupies the faintest, reddest corner of the parameter space, around the same location as other coldest known Y dwarfs. Other JADES-selected Y dwarf candidates are also among the late-T and Y-dwarf branch, however their general non-detections in the short-wavelength bands prevent further localisation. We highlight representative extremely cold brown dwarfs, cold exoplanets, and extragalactic-field brown dwarfs with special symbols and labels.}
    \label{fig:cmd}
\end{figure*}

We also fitted the PSF photometry of the nine candidates using late-T/Y \textit{JWST}/NIRSpec templates \citep{Beiler2024} over the entire NIRCam wavelength range. All candidates are formally best matched by Y-type templates (Table~\ref{tab:photometry_summary}). In the colour--magnitude diagrams shown in Figure~\ref{fig:cmd}, they lie on the branch occupied by late-T and early-Y dwarfs. However, most of these objects have only upper limits in F150W and F200W, affecting 6/9 and 7/9 sources, respectively. We therefore cannot locate their exact positions on the colour-magnitude diagrams or determine whether they are typical early-Y dwarfs or later-type objects similar to GSBD5. We further explored the colour proxy for spectral types using F356W and F444W, as shown in Appendix~\ref{fig:color_type}. These candidates span an approximate range from {\it c}T9.4 to {\it c}Y2, but with large uncertainties owing to the coarser spectral-type sensitivity of F356W-based colours.

More spectroscopy is needed to robustly classify this population. Two objects in this sample, JADES-GN-BD-5 and JADES-GN-BD-9, have been accepted for \textit{JWST}/NIRSpec observations in Cycle 5 GO~10264 (PI Finkelstein). Based on their F356W and F444W colours, they correspond to approximate continuous types of {\it c}T9.4 and {\it c}Y1.1, respectively. If these objects are late-T or early-Y dwarfs, their spectra, especially at shorter wavelengths, are unlikely to be informative as their NIRCam fluxes are $\sim20$ times fainter than those of GSBD5. We also expect them to be more massive, as their NIRCam fluxes indicate several times higher luminosity given the SED-fitted distances and the fact that the luminosity evolution of Y dwarfs depends more on mass than age \citep{Burrows2001}.

Regardless of whether these two dwarfs can be spectroscopically classified, \textit{JWST} extragalactic surveys are now revealing a growing population of ultracool brown dwarfs, e.g., Capotauro in the CEERS field \citep{gandolfi2026, Liu2026a} and two Y dwarfs recently discovered in the Bullet Cluster field \citep{bradac2026}. High Galactic-latitude extragalactic fields therefore provide a (perhaps unexpectedly) powerful discovery space for Galactic ultracool dwarfs: their exceptional depths, broad filter coverage, and frequent revisits together create favourable conditions for both selection and characterisation.

The main selection challenge in these fields is distinguishing ultracool dwarfs from compact high-redshift galaxies. Recently, cold dwarfs have been mostly confused with LRDs \citep[e.g.,][]{Tu2025a}. Indeed, GSBD5 itself satisfies all the colour-based LRD selection criteria in \citet{Kokorev2024}, \citet{Barro2024} and \citet{Labbe2025}. More robust selection techniques \citep[e.g.,][]{Kocevski2025} requires a photometric redshift check, which can only be obtained with at least two long-wavelength detections and informative non-detections at shorter wavelengths. In other words, an extragalactic survey must guarantee two detections of an object at $\gtrsim3~\mu$m to separate the coldest brown dwarfs effectively from galaxies and form a useful Galactic survey volume.

This criterion allows us to estimate the capability of existing \textit{JWST} surveys to select these objects. In the following estimate, we conservatively require a brown dwarf to be detected in F277W; for a typical Y dwarf in a \textit{JWST} deep extragalactic survey, this consequently suggests detections in F356W and F444W, i.e., three detections. NIRCam F277W is comparable to the blue side of the \textit{WISE} W1 band, and AllWISE, the major pre-\textit{JWST} survey yielding a large sample of brown dwarfs, reaches a point-source depth of $m_{\rm AB}=19.60$\,mag in $W1$ \citep{Wright2010, Cutri2013}. In JADES DR5 \citep{Johnson2026, Robertson2026}, the GOODS-South field covers $210~{\rm arcmin}^2$ to an aperture-corrected 5-$\sigma$ F277W point-source depth of 30.07 mag, while GOODS-North covers $115~{\rm arcmin}^2$ to a depth of 29.61 mag. Relative to the AllWISE W1 depth, these limits increase the distance reach by factors of $\sim100$--124. Thus, despite the small solid angle, the JADES search corresponds to a \textit{WISE}-equivalent survey volume of $\simeq1.43\times10^5~{\rm deg}^2$, or $\sim3.5$ times the AllWISE all-sky volume for comparable objects. Since \textit{WISE}-era all-sky searches identified only a couple of isolated planetary-mass brown dwarfs later than Y1, and a randomly placed source has a $\sim20\%$ probability of overlapping with a background galaxy in the JADES DR5 segmentation map, this simple volume-scaling argument predicts $\sim5.6\pm5.0$ such objects in the JADES volume, consistent with the discovery of GSBD5. Considering WISE~0855 is at $\sim2~$pc, such objects will have a characteristic distance of a few hundred pc in extragalactic fields.

Deep \textit{JWST} extragalactic surveys therefore provide an efficient route to expanding the rare sample of the coldest brown dwarfs, as shown in Figure~\ref{fig:surveys}. For example, PRIMER adds $\sim380~{\rm arcmin}^2$ across COSMOS and UDS with NIRCam, corresponding to roughly two thirds of the WISE all-sky volume at its F277W depth \citep{Liu2026a, mcleod2026}. COSMOS-Web \citep{Casey2023} and COSMOS-3D (GO~5893, PI Kakiichi) may contribute a further $\sim0.6$--$1.5$ WISE-equivalent volumes. Assuming a median F277W point-source limit of 28.2\,mag, the $0.63~{\rm deg}^2$ compilation of existing \textit{JWST} extragalactic fields from \citet{mcleod2026} corresponds to $\sim5$--10 AllWISE all-sky Y-dwarf volumes. These fields could plausibly yield $\sim10$--$20$ isolated extremely cold planetary-mass brown dwarfs. Such a sample would transform the currently sparse late-Y regime into a statistically useful population. This estimate is conservative, because survey volume scales steeply with limiting magnitude, $V\propto10^{0.6\Delta m}$, and the $\Delta m\sim3.5$ depth range among these surveys implies a $\gtrsim100$-fold variation in searchable volume per pointing.

The implication is not simply a larger search volume. Since \textit{WISE} Y dwarfs can be detected out to $\sim10$ pc, deep \textit{JWST} imaging can in principle detect comparable objects out to $\sim1,000$--2,000 pc. At high Galactic latitudes, this distance range corresponds to vertical heights beyond the thin-disk scale height. Although it suggests a reduced effective survey volume towards the Galactic outskirts, it allows a new opportunity to identify ultracool brown dwarfs in Galactic environments that were inaccessible in the \textit{WISE} era, including the transition between the thin and thick disks, and even the halo.

The main requirement for placing these distant objects on the same physical footing as nearby \textit{WISE} sources is robust astrometry, especially parallax. The \textit{JWST}/NIRCam F444W PSF \citep{Nardiello2022} is $\sim12$--14 times sharper than that of the \textit{Spitzer}/IRAC $4.5\,\mu$m channel \citep{Fazio2004}, partially compensating for the smaller parallax amplitudes at larger distances. In principle, with a follow-up strategy analogous to the \textit{Spitzer} parallax programs for \textit{WISE} discoveries, \textit{JWST} can extend parallax measurements to distances of order $\sim200-300$ pc for favourable targets. The six-month roll-angle flips commonly used in deep extragalactic fields naturally provide useful parallax baselines, and the serendipitous revisits of GSBD5 demonstrate the feasibility of this approach. In the most frequently revisited fields, parallaxes may already be measurable at larger distances than GSBD5, as will be explored in future work (Liu et al., in preparation). Future progress will therefore require three complementary efforts: larger \textit{JWST} NIRCam and MIRI deep-field imaging areas to build the sample, dedicated revisits of known candidates to constrain their parallaxes, and deep NIRSpec spectroscopic follow-up to reveal their physical conditions.

\subsection{The atmosphere of ultra-cool giant-planet analogues}\label{subsec:water_cloud}

\subsubsection{GSBD5 in the colour--magnitude diagram}

Figure~\ref{fig:cmd} places GSBD5 in colour--magnitude space together with a comprehensive compilation of M--Y dwarfs and directly imaged exoplanets. The first panel shows a traditional ground-based near-IR colour--magnitude diagram based on 2MASS $J$ and $K_s$ photometry, supplemented by synthetic photometry where spectra are available. In this diagram, late-T and Y dwarfs form a sharply turning sequence, with the Y dwarfs occupying a sparsely populated red/faint branch. GSBD5 lies around the faintest and reddest end of this branch, where data are very sparse.

The emergence of such ultracool planetary-mass objects is now becoming clearer in the \textit{JWST} era. This is illustrated by the three \textit{JWST} colour--magnitude diagrams shown in Figure~\ref{fig:cmd}: F150W$-$F444W and F200W$-$F444W, commonly used in extragalactic blind surveys \citep[and exoplanet imaging surveys, e.g.][]{Carter2023a, Carter2024, Biller2025, FerrerChavez2026}, and F150W$-$F480M, often used in \textit{JWST} ultracool brown-dwarf studies \citep[e.g.,][]{Calissendorff2023, Albert2025}. In these colour--magnitude spaces, cold substellar objects form clear, monotonic sequences. GSBD5 lies near the coolest end of the Y-dwarf population, more extreme than some other extremely cold brown dwarfs, such as WISE~1828, WISE~0336B, WD~0806B and WISE~1534, and close to WISE~0855. 
It is also similar to a few of the coldest directly imaged exoplanets or candidates: $\varepsilon$~Ind~A~b, with $T_{\rm eff}\simeq275\,{\rm K}$ and $M\simeq7.6\,M_{\rm J}$ \citep[][not illustrated here due to lack of data]{Matthews2024, Matthews2026}; 14~Her~c, with $T_{\rm eff}\simeq275$--$300\,{\rm K}$ and $M\simeq7.9\,M_{\rm J}$ \citep{BardalezGagliuffi2025}; and TWA~7~b, with $T_{\rm eff}\simeq320\,{\rm K}$ and $M\simeq0.3$--$0.4\,M_{\rm J}$ \citep{Lagrange2025, Crotts2025}. These exoplanets orbit bright K-type host stars at angular separations of only $\simeq4.1''$, $\simeq1.1''$, and $\simeq1.5''$, respectively; and the absence of a coronagraphic spectroscopic mode of \textit{JWST} means that light from the host star can strongly contaminate the planetary spectrum. Indeed, even for $\varepsilon$~Ind~A~b with the largest star--planet angular separation among the three, recent deep \textit{JWST} NIRSpec/IFU observations were not able to produce a separate spectrum of the planet \citep{Kravtchenko2026}. 

Isolated, self-luminous ultracool objects like GSBD5 avoid this contrast limitation and therefore provide a cleaner route to studying atmospheric conditions analogous to those of mature cold giant planets \citep[e.g.,][]{Dupuy2013, Beichman2014, Faherty2014, Skemer2016, Leggett2019}. Although GSBD5 need not have formed through a planet-like pathway, its physical properties place it close to the atmospheric parameter space of mature giant planets. With its high-quality NIRSpec spectrum and MIRI photometry, GSBD5 provides a precious isolated spectroscopic planetary-mass benchmark for studying $\sim300$~K hydrogen-rich super-Jovian atmospheres, where only no more than a few similar sources are known (e.g., WISE~0855; $T_{\rm eff}=285\,{\rm K}$, $M=6.5\pm3.5\,M_{\rm J}$; \citealt{Wilson2025}).

\subsubsection{Physical interpretations of GSBD5 atmosphere}

Previous studies of late-T and Y dwarfs have shown that atmospheres in this temperature regime differ substantially from those of warmer L and T dwarfs \citep[see the review by][]{Faherty2026}. As $T_{\rm eff}$ decreases, molecular opacity from H$_2$O, CH$_4$, and NH$_3$ becomes increasingly important, while the spectra remain highly sensitive to metallicity, gravity, vertical mixing, and cloud formation. DEQ chemistry can dredge CO and CO$_2$ from deeper, hotter layers into the observable atmosphere, strengthening their absorption features (see a discussion on the case of WISE~0855 in \citealt{Kuhnle2025}). At $T_{\rm eff}\lesssim300$ K, liquid water clouds, and potentially water-ice clouds, are also expected to emerge after silicate/iron clouds and subsequently salt/sulfide clouds have retreated to deeper atmospheric layers \citep[e.g.,][]{Morley2014, LacyBurrows2023}. WISE 0855 provides an important precedent in this regime, showing evidence for water depletion at mid-IR wavelengths, strong vertical mixing traced by CO \citep{Luhman2024}, and patchy water clouds \citep{Miles2026}.

Our fits provide several indications that vertical mixing is important in the atmosphere of GSBD5. Nearly all model comparisons prefer DEQ chemistry, whether the fits use PRISM+MIRI spectrophotometry, NIRCam+MIRI photometry, or evolutionary-model priors. The evidence for modified cloud sedimentation is more model-dependent, but within the Exo-REM family the fits favour variable-$f_{\rm sed}$ models over simpler cloud prescriptions. These trends suggest that both vertical chemical transport and cloud structure are important for reproducing the SED of GSBD5.

The spectral morphology supports this interpretation. In CEQ, carbon in the upper atmospheres of the coldest Y dwarfs should reside primarily in CH$_4$ \citep{Lodders2002, Zahnle2014}. Strong CO, and excess CO$_2$, absorption at observable pressures therefore requires DEQ chemistry, most naturally through vertical mixing \citep[e.g.][]{Visscher2011, Miles2020, Luhman2024, Beiler2024, Miles2026}. Consistent with this expectation, GSBD5 shows strong CO and CO$_2$ absorption features around $\sim4.2$--$5\,\mu$m (Figure~\ref{fig:scaled_spec}). At the same time, the main CH$_4$ absorption regions, near $\sim1.7\,\mu$m and $3.2$--$3.8\,\mu$m, appear relatively weakened.

However, it is unclear if H$_2$O, or indeed water clouds, plays a key role in the atmosphere of GSBD5. We see relatively stronger emission at 1.4, 1.85 and $2.5$--$3.0\,\mu$m in GSBD5 (Figure~\ref{fig:scaled_spec}), corresponding to water absorption peaks. These absorption troughs that are prominent in other cold Y dwarfs are partially filled in, and in some cases appear as local peaks, even compared to the coldest brown dwarf WISE~0855. However, the low resolution of the PRISM spectrum, as well as its high noise, makes it difficult to identify spectral features exclusively from water, as these bands are also primarily affected by CH$_4$ and NH$_3$. No absorption feature suggesting water ice clouds at $\sim2.8~\micron$ \citep{Morley2014} is identified in the PRISM spectrum. Therefore, we cannot conclude if there is reduced H$_2$O absorption. However, we note that water deficiency would help the SED-fitting of GSBD5, as it naturally produces the extra emission in F277W and F300M, mitigating the model--data discrepancies there found by our prior-guided fits (Section~\ref{subsubsec:prior_sed_fitting}).

In the meantime, the SED-fitting failures can also be mitigated by the two-component fitting approach, even without the involvement of water clouds (Section~\ref{subsubsec:sed_experiment}). In this scenario, the warmer component, with $T_{\rm eff}\sim425$ K, dominates the total flux at $\lesssim3.5\,\mu$m and helps reproduce the relatively weak H$_2$O and CH$_4$ absorption features. This component also favours very low $f_{\rm sed}$, i.e., inefficient sedimentation. However, a single warm, strongly mixed component would produce too much CO absorption at $\sim4$--$5\,\mu$m. The fit therefore shifts the long-wavelength contribution toward a cooler component with more moderate cloud sedimentation. Here, having two different-temperature atmospheric columns is not enough. The (M, N) test reveals the need to tweak the wavelength-dependent structure of these columns, compared to typical brown-dwarf DEQ models. 

Therefore, interestingly, here we witness a similar ambiguity previously seen in the interpretations of WISE~0855's atmosphere (see a summary in \citealt{Kuhnle2025}). On the one hand, water depletion, and even water ice clouds, were favoured by theoretical models for WISE~0855 \citep[e.g.,][]{Morley2014, LacyBurrows2023, Miles2026}. On the other hand, the same data could be similarly interpreted with cloudless models \citep[e.g.,][]{Luhman2024, Kothari2026}. Further investigating the two scenarios of GSBD5 requires bespoke modelling, which is beyond the scope of this paper, or indeed future deep spectroscopy \citep[e.g.,][]{Miles2026}.

\subsubsection{Causes of GSBD5 misinterpretations}

The features described in the section above provide a plausible explanation for the SED-fitting failures discussed in Section~\ref{subsubsec:prism_miri}. When the high-S/N PRISM spectrum dominates the likelihood, the atmospheric fits prefer hotter models, typically $T_{\rm eff}\sim350$--$400$ K, to align with the significant spectral features at NIRCam LW. Because the correspondingly boosted luminosity then needs to match the observed flux, these hotter solutions naturally lead to distances that are much larger than the parallax distance. By contrast, NIRCam photometric fitting averages over the detailed spectral structure and places more weight on the broad SED shape, leading to lower temperatures and more reasonable distance estimates. In the latter case, the remaining failures are more associated with limitations of the model grids, such as insufficient temperature coverage of the cold-Y regime \citep[e.g., the LOWZ models;][]{Meisner2021, hainline2026b} or the absence of DEQ chemistry to reproduce the carbon-bearing features at low temperature (e.g., Sonora Bobcat).

Several other fitted parameters can be understood as compensating for missing opacity. In the NIRCam+MIRI photometric fits, many models prefer the lowest surface gravities allowed by their grids, which reduces the photospheric pressure and weakens water and CH$_4$ absorption. They also prefer very high C/O ratios, which can reduce the available oxygen for H$_2$O formation; and high $\log K_{zz}$, which enhances vertical mixing, increases the abundance of CO and CO$_2$ and similarly reduces the available oxygen \citep{Visscher2011, Moses2013, Zahnle2014, Hobbs2019, Miles2020}. The extreme combinations in the fits are unlikely to represent the true physical state of GSBD5. For example, extremely low gravity is inconsistent with the evolutionary-model constraints, and the combination of very low metallicity and extremely high C/O is difficult to reconcile with a low-tangential-velocity, likely thin-disk object. A mildly subsolar metallicity may still be physical, as most models do not require the grid-edge solutions.

At present, however, we do not have a single self-consistent model grid that simultaneously includes DEQ chemistry, sufficiently low temperatures, a broad range of metallicities and C/O ratios, and multiple cloud species including water condensates. Therefore, interpreting physical properties of the coldest objects, especially those distant ones found in extragalactic fields without parallax measurements, needs particular caution. Nevertheless, our analysis provides a clear target for future model development: any successful model for GSBD5 must simultaneously reproduce its phenomenological features.

\section{Conclusions}
\label{section:conclusion}

We have presented a \textit{JWST} astrometric and spectrophotometric analysis of JADES-GS-BD-5, an extremely cold Y dwarf in the JADES/GOODS-South field. Our main conclusions are as follows:

\begin{enumerate}
    \item We obtained the first robust parallax measurement of a brown dwarf discovered in a \textit{JWST} extragalactic deep field. Using NIRCam images across five epochs spanning a 2.03-yr temporal baseline, we measure parallax $\varpi=15.7^{+1.4}_{-1.3}$~mas, detected at $\mathrm{S/N=11.5}$. This gives a model-independent trigonometric distance of $d=63^{+6}_{-5}$~pc. Its distance is roughly three times larger than that of the previously most distant Y dwarfs with reliable geometric distances. This demonstrates that \textit{JWST} deep-field imaging can provide highly precise astrometry for extremely faint Galactic substellar objects far beyond the local WISE-discovered sample. 

    \item JADES-GS-BD-5 is likely isolated and belongs to the Galactic thin disk. The absolute proper motion is measured with very high precision, $\mu_{\rm tot}=49.75^{+0.26}_{-0.30}$~mas~yr$^{-1}$, corresponding to a tangential velocity of $v_{\rm tan}=15.0^{+1.4}_{-1.3}$~km~s$^{-1}$. Its Galactic position and low tangential velocity are consistent with a thin-disk object. We find no evidence for a close companion in residual images, no significant binary acceleration, no convincing photometric variability and no plausible widely-separated stellar or substellar host. 
    
    \item JADES-GS-BD-5 is one of the coldest, faintest and latest-type Y dwarfs known to date. Although its short-wavelength \(J\)-peak morphology alone is broadly compatible with an early-Y classification, its absolute spectral luminosity, flux-normalized PRISM spectrum, and \textit{JWST} colour–type relation all place it beyond the current Y0–Y1 standards. We therefore adopt a formal classification of ${\rm >Y1}$, with a continuous colour-based proxy of {\it c}Y1.5.
    
    \item The parallax-calibrated luminosity places JADES-GS-BD-5 in the planetary mass regime. Integrating the NIRSpec/PRISM spectrum, MIRI-constrained mid-IR SED, and long-wavelength blackbody tail yields $\log(L_{\rm bol}/L_{\odot})=-7.08^{+0.09}_{-0.09}$. Evolutionary modelling subsequently gives $M\sim8~M_{\rm J}$, $T_{\rm eff} \sim 300$~K, $R\sim1.0~R_{\rm J}$ and $\log g\sim4.3$. More than 90\% of the mass posterior lies below 13 $M_{\rm J}$, placing the object at the extreme low-mass end of the brown-dwarf population. It is therefore a rare analogue of mature gas giant planets and one of the only isolated planetary-mass brown dwarf benchmarks to study their atmospheres.
    
    \item \textit{JWST} extragalactic surveys provide a new route to the coldest Galactic substellar population. Applying similar astrometric analyses to other nine JADES Y-dwarf candidates confirms eight additional proper motions, although no other robust parallax is yet detected. The combination of deep 1–5 \(\mu\)m imaging, repeated visits, and sharp astrometry makes \textit{JWST} deep fields uniquely powerful for discovering and characterising late-Y dwarfs and isolated planetary-mass objects beyond the local Solar neighbourhood. An ultracool dwarf sample several times larger than that in the \textit{WISE} era is expected in the existing \textit{JWST} extragalactic fields. Future dedicated observations will be crucial for turning these serendipitous discoveries into a statistically useful, parallax-calibrated sample.

    \item We investigated different physical scenarios to explain the challenge for current atmospheric models to reproduce its high-quality near-to-mid-IR SED. Free fits to the NIRSpec/PRISM spectrum and MIRI photometry produce formally unacceptable solutions and tend to overestimate the effective temperature and distance. Photometric fits with NIRCam and MIRI recover more plausible temperatures and flux scalings, but still show significant residuals. Prior-guided and variable-controlled fitting experiments, as well as spectral features, suggest that DEQ chemistry and vertical mixing are needed to explain the observed spectrum of this ultra-cool atmosphere. However, it is unclear if water deficiency happens in GSBD5, as similar SED behaviour can be reproduced by modifying non-water-cloud atmospheric models.
\end{enumerate}

\section*{acknowledgements}
DJM and JSD acknowledge the support of the Royal Society through the award of a Royal Society University Research Professorship to JSD. XC acknowledges funding from the China Scholarship Council (CSC) under grant CSC no. 202208170018. BJS and BAB acknowledge funding by the UK Science and Technology Facilities Council (STFC) grant nos. ST/V000594/1 and UKRI1196. HL and ACC acknowledge support from a UKRI Frontier Research Grantee Grant (PI Carnall; grant reference EP/Y037065/1).

This work made use of the University of Edinburgh's Institute for Astronomy research computing cluster, {\sl cuillin}, which is partially funded by the UK STFC. We would like to thank Eric Tittley for his work managing and maintaining {\sl cuillin}. This work is based in part on observations made with the NASA/ESA/CSA James Webb Space Telescope. The data were obtained from the Mikulski Archive for Space Telescopes at the Space Telescope Science Institute, which is operated by the Association of Universities for Research in Astronomy, Inc., under NASA contract NAS 5-03127 for JWST. These observations are associated with program nos. GTO~1180, GTO~1181, GTO~1189, GTO~1207, GTO~1210, GTO~1286, GTO~1275, GO~1895, GO~2198, GO~2514, GO~2674, GO~3215, GO~3486, GO~3577, GO~3670, GO~4539, GTO~4540, GO~4668, GO~5118, GO~5997, GO~6084, GO~6434, DD 6541, CAL/NIRCAM~6630, GO~7404 and CAL/NIRCAM~8882. The authors acknowledge the GO~1895, GO~2514, GO~3215, GO~3577, GO~6434, DD~6541, CAL/NIRCAM~6630, GO~7404, and CAL/NIRCAM~8882 teams for developing their observing programs with a zero-exclusive-access period. This work has made use of data from the European Space Agency (ESA) mission {\it Gaia} (\url{https://www.cosmos.esa.int/gaia}), processed by the {\it Gaia} Data Processing and Analysis Consortium (DPAC, \url{https://www.cosmos.esa.int/web/gaia/dpac/consortium}) \citep[][]{GaiaCollaboration2016, GaiaCollaboration2023}. Funding for the DPAC has been provided by national institutions, in particular the institutions participating in the {\it Gaia} Multilateral Agreement. This publication makes use of data products from the Spectro-Photometer for the History of the Universe, Epoch of Reionization and Ices Explorer (SPHEREx), which is a joint project of the Jet Propulsion Laboratory and the California Institute of Technology, and is funded by the National Aeronautics and Space Administration. This research has made use of the Spanish Virtual Observatory (SVO) Filter Profile Service ``Carlos Rodrigo" \url{https://svo2.cab.inta-csic.es/theory/fps/}, funded by MCIN/AEI/10.13039/501100011033/ through grant PID2023-146210NB-I00 \citep{Rodrigo2012, Rodrigo2020}. This research made use of SAOImageDS9, a tool for data visualization supported by the Chandra X-ray Science Center (CXC) and the High Energy Astrophysics Science Archive Center (HEASARC) with support from the JWST Mission office at the Space Telescope Science Institute for 3D visualization \citep{Joye2003}. Some of the data products presented herein were retrieved from the Dawn JWST Archive (DJA). DJA is an initiative of the Cosmic Dawn Center (DAWN), which is funded by the Danish National Research Foundation under grant DNRF140. This research has made use of the Astrophysics Data System, funded by NASA under Cooperative Agreement 80NSSC21M00561. This work makes use of the Python programming language\footnote{Python Software Foundation; \url{https://www.python.org/}}, in particular packages including \texttt{Astropy} \citep{astropy, Astropy2018, Astropy2022}, \texttt{Matplotlib} \citep{matplotlib}, \texttt{Numpy} \citep{numpy}, \texttt{Photutils} \citep{photutils}, \texttt{Scipy} \citep{scipy}, \texttt{species} \citep{Stolker2020}, \texttt{PyMultiNest} \citep{Buchner2014, Buchner2016}, \texttt{sep} \citep{Barbary2018}, and \texttt{STPSF} \citep{Perrin2012, Perrin2014}.

\bibliographystyle{mnras}
\bibliography{GSBD5_Liu_2026.bib}

@ARTICLE{Bertin1996,
       author = {{Bertin}, E. and {Arnouts}, S.},
        title = "{SExtractor: Software for source extraction.}",
      journal = {\aaps},
         year = 1996,
        month = jun,
       volume = {117},
        pages = {393-404},
          doi = {10.1051/aas:1996164},
       adsurl = {https://ui.adsabs.harvard.edu/abs/1996A&AS..117..393B}
}

@ARTICLE{bradac2026,
       author = {{Bradac}, Maru{\v{s}}a and {Willott}, Chris and {Asada}, Yoshihisa and {Albert}, Lo{\"\i}c and {Rihtar{\v{s}}i{\v{c}}}, Gregor and {Harshan}, Anishya and {Jude{\v{z}}}, Jon and {Martis}, Nicholas S. and {Ferrara}, Andrea and {Hainline}, Kevin and {Abdurro'uf} and {Allingham}, Joseph F.~V. and {Bromm}, Volker and {Chisholm}, John and {Coe}, Dan and {Desprez}, Guillaume and {Diego}, Jose M. and {Faisst}, Andreas L. and {Fujimoto}, Seiji and {Furtak}, Lukas J. and {Hsiao}, Tiger Yu-Yang and {Inayoshi}, Kohei and {Koekemoer}, Anton M. and {Kokorev}, Vasily and {Lemaux}, Brian C. and {Lopes}, Paulo A.~A. and {Lucas}, Ray A. and {Marchesini}, Danilo and {Markov}, Vladan and {Noirot}, Ga{\"e}l and {Pan}, Richard and {Randall}, Scott W. and {Richard}, Johan and {Robbins}, Luke and {Sarrouh}, Ghassan T.~E. and {Sawicki}, Marcin and {Schrabback}, Tim and {Tripodi}, Roberta and {Vanzella}, Eros and {Windhorst}, Rogier A.},
        title = "{Two Exciting High-redshift Galaxy Candidates Turn Out to Be Two Exciting Brown Dwarfs}",
      journal = {\apjl},
         year = 2026,
        month = oct,
       volume = {1010},
       number = {1},
          eid = {L13},
        pages = {L13},
          doi = {10.3847/2041-8213/ae9085},
       adsurl = {https://ui.adsabs.harvard.edu/abs/2026ApJ..1010L..13B}
}

@ARTICLE{eisenstein2023,
       author = {{Eisenstein}, Daniel J. and {Willott}, Chris and {Alberts}, Stacey and {Arribas}, Santiago and {Bonaventura}, Nina and {Bunker}, Andrew J. and {Cameron}, Alex J. and {Carniani}, Stefano and {Charlot}, Stephane and {Curtis-Lake}, Emma and {D'Eugenio}, Francesco and {Ferruit}, Pierre and {Giardino}, Giovanna and {Hainline}, Kevin and {Hausen}, Ryan and {Jakobsen}, Peter and {Johnson}, Benjamin D. and {Maiolino}, Roberto and {Rauscher}, Bernard J. and {Rieke}, Marcia and {Rieke}, George and {Rix}, Hans-Walter and {Robertson}, Brant and {Stark}, Daniel P. and {Tacchella}, Sandro and {Williams}, Christina C. and {Willmer}, Christopher N.~A. and {Baker}, William M. and {Baum}, Stefi and {Bhatawdekar}, Rachana and {Boyett}, Kristan and {Chen}, Zuyi and {Chevallard}, Jacopo and {Circosta}, Chiara and {Curti}, Mirko and {Danhaive}, A. Lola and {DeCoursey}, Christa and {Endsley}, Ryan and {de Graaff}, Anna and {Dressler}, Alan and {Egami}, Eiichi and {Helton}, Jakob M. and {Hviding}, Raphael E. and {Ji}, Zhiyuan and {Jones}, Gareth C. and {Kumari}, Nimisha and {L{\"u}tzgendorf}, Nora and {Laseter}, Isaac and {Looser}, Tobias J. and {Lyu}, Jianwei and {Maseda}, Michael V. and {Nelson}, Erica and {Parlanti}, Eleonora and {Perna}, Michele and {Pusk{\'a}s}, D{\'a}vid and {Rawle}, Tim and {Rodr{\'\i}guez Del Pino}, Bruno and {Rujopakarn}, Wiphu and {Sandles}, Lester and {Saxena}, Aayush and {Scholtz}, Jan and {Sharpe}, Katherine and {Shivaei}, Irene and {Silcock}, Maddie S. and {Simmonds}, Charlotte and {Skarbinski}, Maya and {Smit}, Renske and {Stone}, Meredith and {Suess}, Katherine A. and {Sun}, Fengwu and {Tang}, Mengtao and {Topping}, Michael W. and {{\"U}bler}, Hannah and {Villanueva}, Natalia C. and {Wallace}, Imaan E.~B. and {Whitler}, Lily and {Witstok}, Joris and {Woodrum}, Charity},
        title = "{Overview of the JWST Advanced Deep Extragalactic Survey (JADES)}",
      journal = {\apjs},
         year = 2026,
        month = mar,
       volume = {283},
       number = {1},
          eid = {6},
        pages = {6},
          doi = {10.3847/1538-4365/ae3163},
archivePrefix = {arXiv},
       eprint = {2306.02465},
 primaryClass = {astro-ph.GA},
       adsurl = {https://ui.adsabs.harvard.edu/abs/2026ApJS..283....6E}
}

@ARTICLE{eisenstein2025,
       author = {{Eisenstein}, Daniel J. and {Johnson}, Benjamin D. and {Robertson}, Brant and {Tacchella}, Sandro and {Hainline}, Kevin and {Jakobsen}, Peter and {Maiolino}, Roberto and {Bonaventura}, Nina and {Bunker}, Andrew J. and {Cameron}, Alex J. and {Cargile}, Phillip A. and {Curtis-Lake}, Emma and {Hausen}, Ryan and {Pusk{\'a}s}, D{\'a}vid and {Rieke}, Marcia and {Sun}, Fengwu and {Willmer}, Christopher N.~A. and {Willott}, Chris and {Alberts}, Stacey and {Arribas}, Santiago and {Baker}, William M. and {Baum}, Stefi and {Bhatawdekar}, Rachana and {Carniani}, Stefano and {Charlot}, Stephane and {Chen}, Zuyi and {Chevallard}, Jacopo and {Curti}, Mirko and {DeCoursey}, Christa and {D'Eugenio}, Francesco and {de Graaff}, Anna and {Egami}, Eiichi and {Helton}, Jakob M. and {Ji}, Zhiyuan and {Jones}, Gareth C. and {Kumari}, Nimisha and {L{\"u}tzgendorf}, Nora and {Laseter}, Isaac and {Looser}, Tobias J. and {Lyu}, Jianwei and {Maseda}, Michael V. and {Nelson}, Erica and {Parlanti}, Eleonora and {Rauscher}, Bernard J. and {Rawle}, Tim and {Rieke}, George and {Rix}, Hans-Walter and {Rujopakarn}, Wiphu and {Sandles}, Lester and {Saxena}, Aayush and {Scholtz}, Jan and {Sharpe}, Katherine and {Shivaei}, Irene and {Simmonds}, Charlotte and {Smit}, Renske and {Topping}, Michael W. and {{\"U}bler}, Hannah and {Venturi}, Giacomo and {Williams}, Christina C. and {Witstok}, Joris and {Woodrum}, Charity},
        title = "{The JADES Origins Field: A New JWST Deep Field in the JADES Second NIRCam Data Release}",
      journal = {\apjs},
         year = 2025,
        month = dec,
       volume = {281},
       number = {2},
          eid = {50},
        pages = {50},
          doi = {10.3847/1538-4365/ae1137},
archivePrefix = {arXiv},
       eprint = {2310.12340},
 primaryClass = {astro-ph.GA},
       adsurl = {https://ui.adsabs.harvard.edu/abs/2025ApJS..281...50E}
}

@ARTICLE{Finkelstein2023,
       author = {{Finkelstein}, Steven L. and {Bagley}, Micaela B. and {Ferguson}, Henry C. and {Wilkins}, Stephen M. and {Kartaltepe}, Jeyhan S. and {Papovich}, Casey and {Yung}, L.~Y. Aaron and {Arrabal Haro}, Pablo and {Behroozi}, Peter and {Dickinson}, Mark and {Kocevski}, Dale D. and {Koekemoer}, Anton M. and {Larson}, Rebecca L. and {Le Bail}, Aur{\'e}lien and {Morales}, Alexa M. and {P{\'e}rez-Gonz{\'a}lez}, Pablo G. and {Burgarella}, Denis and {Dav{\'e}}, Romeel and {Hirschmann}, Michaela and {Somerville}, Rachel S. and {Wuyts}, Stijn and {Bromm}, Volker and {Casey}, Caitlin M. and {Fontana}, Adriano and {Fujimoto}, Seiji and {Gardner}, Jonathan P. and {Giavalisco}, Mauro and {Grazian}, Andrea and {Grogin}, Norman A. and {Hathi}, Nimish P. and {Hutchison}, Taylor A. and {Jha}, Saurabh W. and {Jogee}, Shardha and {Kewley}, Lisa J. and {Kirkpatrick}, Allison and {Long}, Arianna S. and {Lotz}, Jennifer M. and {Pentericci}, Laura and {Pierel}, Justin D.~R. and {Pirzkal}, Nor and {Ravindranath}, Swara and {Ryan}, Russell E. and {Trump}, Jonathan R. and {Yang}, Guang and {Bhatawdekar}, Rachana and {Bisigello}, Laura and {Buat}, V{\'e}ronique and {Calabr{\`o}}, Antonello and {Castellano}, Marco and {Cleri}, Nikko J. and {Cooper}, M.~C. and {Croton}, Darren and {Daddi}, Emanuele and {Dekel}, Avishai and {Elbaz}, David and {Franco}, Maximilien and {Gawiser}, Eric and {Holwerda}, Benne W. and {Huertas-Company}, Marc and {Jaskot}, Anne E. and {Leung}, Gene C.~K. and {Lucas}, Ray A. and {Mobasher}, Bahram and {Pandya}, Viraj and {Tacchella}, Sandro and {Weiner}, Benjamin J. and {Zavala}, Jorge A.},
        title = "{CEERS Key Paper. I. An Early Look into the First 500 Myr of Galaxy Formation with JWST}",
      journal = {\apjl},
         year = 2023,
        month = mar,
       volume = {946},
       number = {1},
          eid = {L13},
        pages = {L13},
          doi = {10.3847/2041-8213/acade4},
archivePrefix = {arXiv},
       eprint = {2211.05792},
 primaryClass = {astro-ph.GA},
       adsurl = {https://ui.adsabs.harvard.edu/abs/2023ApJ...946L..13F}
}

@ARTICLE{gandolfi2026,
       author = {{Gandolfi}, G. and {Rodighiero}, G. and {Castellano}, M. and {Fontana}, A. and {Santini}, P. and {Dickinson}, M. and {Finkelstein}, S. and {Catone}, M. and {Calabr{\`o}}, A. and {Merlin}, E. and {Pentericci}, L. and {Bisigello}, L. and {Grazian}, A. and {Napolitano}, L. and {Vulcani}, B. and {Taylor}, A.~J. and {Arrabal Haro}, P. and {Kirkpatrick}, A. and {Backhaus}, B.~E. and {Holwerda}, B.~W. and {Giulietti}, M. and {Bianchetti}, A. and {Cassata}, P. and {Cleri}, N.~J. and {Daddi}, E. and {Ferguson}, H.~C. and {Girardi}, G. and {Hirschmann}, M. and {Koekemoer}, A.~M. and {Lapi}, A. and {Pacucci}, F. and {P{\'e}rez-Gonz{\'a}lez}, P.~G. and {de la Vega}, A. and {Vietri}, A. and {Wilkins}, S. and {Yung}, L.~Y.~A. and {Bagley}, M. and {Bhatawdekar}, R. and {Kartaltepe}, J. and {Papovich}, C. and {Pirzkal}, N.},
        title = "{Mysteries of Capotauro: Investigating the puzzling nature of an extreme F356W-dropout}",
      journal = {\aap},
         year = 2026,
        month = feb,
       volume = {706},
          eid = {A364},
        pages = {A364},
          doi = {10.1051/0004-6361/202557061},
archivePrefix = {arXiv},
       eprint = {2509.01664},
 primaryClass = {astro-ph.GA},
       adsurl = {https://ui.adsabs.harvard.edu/abs/2026A&A...706A.364G}
}

@ARTICLE{hainline2026b,
       author = {{Hainline}, Kevin N. and {Helton}, Jakob M. and {Miles}, Brittany E. and {Leisenring}, Jarron and {Marley}, Mark S. and {Mukherjee}, Sagnick and {Wogan}, Nicholas F. and {Bunker}, Andrew J. and {Johnson}, Benjamin D. and {Maiolino}, Roberto and {Rieke}, Marcia and {Rinaldi}, Pierluigi and {Robertson}, Brant and {Sun}, Fengwu and {Tacchella}, Sandro and {Williams}, Christina C. and {Willmer}, Christopher N.~A.},
        title = "{JADES: An Abundance of Ultradistant T and Y Dwarfs in Deep Extragalactic Data}",
      journal = {\apj},
         year = 2026,
        month = jun,
       volume = {1004},
       number = {2},
          eid = {223},
        pages = {223},
          doi = {10.3847/1538-4357/ae6ce0},
archivePrefix = {arXiv},
       eprint = {2510.00111},
 primaryClass = {astro-ph.SR},
       adsurl = {https://ui.adsabs.harvard.edu/abs/2026ApJ..1004..223H}
}

@ARTICLE{mcleod2026,
       author = {{McLeod}, D.~J. and {Dunlop}, J.~S. and {McLure}, R.~J. and {Donnan}, C.~T. and {Begley}, R. and {Antonogiannaki}, S. and {Magee}, D. and {Illingworth}, G.~D. and {Arrabal Haro}, P. and {Bondestam}, C. and {Carnall}, A.~C. and {Cullen}, F. and {Dickinson}, M. and {Ellis}, R.~S. and {Frye}, B.~L. and {Golawska}, H. and {Grogin}, N.~A. and {Holst}, I.~J.~B. and {Kamieneski}, P.~S. and {Leung}, H.-H. and {Liu}, F.-Y. and {Stanton}, T.~M. and {Tittley}, E.~R.},
        title = "{A search for the first galaxies across $>0.6$ deg$^2$ of JWST imaging: new evidence for a rapid decline in star-formation activity at $z>12$}",
      journal = {arXiv e-prints},
         year = 2026,
        month = apr,
          eid = {arXiv:2604.16666},
        pages = {arXiv:2604.16666},
          doi = {10.48550/arXiv.2604.16666},
archivePrefix = {arXiv},
       eprint = {2604.16666},
 primaryClass = {astro-ph.GA},
       adsurl = {https://ui.adsabs.harvard.edu/abs/2026arXiv260416666M}
}

@ARTICLE{Mukherjee2024,
       author = {{Mukherjee}, Sagnick and {Fortney}, Jonathan J. and {Morley}, Caroline V. and {Batalha}, Natasha E. and {Marley}, Mark S. and {Karalidi}, Theodora and {Visscher}, Channon and {Lupu}, Roxana and {Freedman}, Richard and {Gharib-Nezhad}, Ehsan},
        title = "{The Sonora Substellar Atmosphere Models. IV. Elf Owl: Atmospheric Mixing and Chemical Disequilibrium with Varying Metallicity and C/O Ratios}",
      journal = {\apj},
         year = 2024,
        month = mar,
       volume = {963},
       number = {1},
          eid = {73},
        pages = {73},
          doi = {10.3847/1538-4357/ad18c2},
archivePrefix = {arXiv},
       eprint = {2402.00756},
 primaryClass = {astro-ph.EP},
       adsurl = {https://ui.adsabs.harvard.edu/abs/2024ApJ...963...73M}
}

@ARTICLE{Muzzin2025,
       author = {{Muzzin}, Adam and {Suess}, Katherine A. and {Marchesini}, Danilo and {Robbins}, Luke and {Willott}, Chris J. and {Alberts}, Stacey and {Antwi-Danso}, Jacqueline and {Asada}, Yoshihisa and {Brammer}, Gabriel and {Cutler}, Sam E. and {Iyer}, Kartheik G. and {Labbe}, Ivo and {Martis}, Nicholas S. and {Miller}, Tim B. and {Mitsuhashi}, Ikki and {Pope}, Alexandra and {Sajina}, Anna and {Sarrouh}, Ghassan T.~E. and {Sharma}, Monu and {Stefanon}, Mauro and {Whitaker}, Katherine E. and {Abraham}, Roberto and {Atek}, Hakim and {Bradac}, Marusa and {Berek}, Samantha and {Bezanson}, Rachel and {Brown}, Westley and {Burgasser}, Adam J. and {Chicoine}, Nathalie and {Cloonan}, Aidan P. and {Cooper}, Olivia R. and {Dayal}, Pratika and {de Graaff}, Anna and {Desprez}, Guillaume and {Feldmann}, Robert and {Forrest}, Ben and {Franx}, Marijn and {Fudamoto}, Yoshinobu and {Fujimoto}, Seiji and {Furtak}, Lukas J. and {Glazebrook}, Karl and {Goovaerts}, Ilias and {Greene}, Jenny E. and {Jagga}, Naadiyah and {Jarvis}, William W.~H. and {Kriek}, Mariska and {Khullar}, Gourav and {La Torre}, Valentina and {Leja}, Joel and {Lin}, Jamie and {Lorenz}, Brian and {Lyon}, Daniel and {Markov}, Vladan and {Maseda}, Michael V. and {McConachie}, Ian and {Merchant}, Maya and {Merida}, Rosa M. and {Mowla}, Lamiya and {Myers}, Katherine and {Naidu}, Rohan P. and {Nanayakkara}, Themiya and {Nelson}, Erica J. and {Noirot}, Gael and {Oesch}, Pascal A. and {Omori}, Kiyoaki C. and {Pan}, Richard and {Porraz Barrera}, Natalia and {Price}, Sedona H. and {Ravindranath}, Swara and {Sawicki}, Marcin and {Setton}, David J. and {Smit}, Renske and {Sok}, Visal and {Speagle}, Joshua S. and {Taylor}, Edward N. and {Tan}, Vivian Yun Yan and {Tripodi}, Roberta and {van der Wel}, Arjen and {Perez Vidal}, Edgar and {Wang}, Bingjie and {Weaver}, John R. and {Williams}, Christina C. and {Withers}, Sunna and {Zaidi}, Kumail},
        title = "{MINERVA: A NIRCam Medium Band and MIRI Imaging Survey to Unlock the Hidden Gems of the Distant Universe}",
      journal = {arXiv e-prints},
         year = 2025,
        month = jul,
          eid = {arXiv:2507.19706},
        pages = {arXiv:2507.19706},
          doi = {10.48550/arXiv.2507.19706},
archivePrefix = {arXiv},
       eprint = {2507.19706},
 primaryClass = {astro-ph.GA},
       adsurl = {https://ui.adsabs.harvard.edu/abs/2025arXiv250719706M}
}

@ARTICLE{Oke1974,
       author = {{Oke}, J.~B.},
        title = "{Absolute Spectral Energy Distributions for White Dwarfs}",
      journal = {\apjs},
         year = 1974,
        month = feb,
       volume = {27},
        pages = {21},
          doi = {10.1086/190287},
       adsurl = {https://ui.adsabs.harvard.edu/abs/1974ApJS...27...21O}
}

@ARTICLE{Oke1983,
       author = {{Oke}, J.~B. and {Gunn}, J.~E.},
        title = "{Secondary standard stars for absolute spectrophotometry.}",
      journal = {\apj},
         year = 1983,
        month = mar,
       volume = {266},
        pages = {713-717},
          doi = {10.1086/160817},
       adsurl = {https://ui.adsabs.harvard.edu/abs/1983ApJ...266..713O}
}

@ARTICLE{rieke2023,
       author = {{Rieke}, Marcia J. and {Robertson}, Brant and {Tacchella}, Sandro and {Hainline}, Kevin and {Johnson}, Benjamin D. and {Hausen}, Ryan and {Ji}, Zhiyuan and {Willmer}, Christopher N.~A. and {Eisenstein}, Daniel J. and {Pusk{\'a}s}, D{\'a}vid and {Alberts}, Stacey and {Arribas}, Santiago and {Baker}, William M. and {Baum}, Stefi and {Bhatawdekar}, Rachana and {Bonaventura}, Nina and {Boyett}, Kristan and {Bunker}, Andrew J. and {Cameron}, Alex J. and {Carniani}, Stefano and {Charlot}, Stephane and {Chevallard}, Jacopo and {Chen}, Zuyi and {Curti}, Mirko and {Curtis-Lake}, Emma and {Danhaive}, A. Lola and {DeCoursey}, Christa and {Dressler}, Alan and {Egami}, Eiichi and {Endsley}, Ryan and {Helton}, Jakob M. and {Hviding}, Raphael E. and {Kumari}, Nimisha and {Looser}, Tobias J. and {Lyu}, Jianwei and {Maiolino}, Roberto and {Maseda}, Michael V. and {Nelson}, Erica J. and {Rieke}, George and {Rix}, Hans-Walter and {Sandles}, Lester and {Saxena}, Aayush and {Sharpe}, Katherine and {Shivaei}, Irene and {Skarbinski}, Maya and {Smit}, Renske and {Stark}, Daniel P. and {Stone}, Meredith and {Suess}, Katherine A. and {Sun}, Fengwu and {Topping}, Michael and {{\"U}bler}, Hannah and {Villanueva}, Natalia C. and {Wallace}, Imaan E.~B. and {Williams}, Christina C. and {Willott}, Chris and {Whitler}, Lily and {Witstok}, Joris and {Woodrum}, Charity},
        title = "{JADES Initial Data Release for the Hubble Ultra Deep Field: Revealing the Faint Infrared Sky with Deep JWST NIRCam Imaging}",
      journal = {\apjs},
         year = 2023,
        month = nov,
       volume = {269},
       number = {1},
          eid = {16},
        pages = {16},
          doi = {10.3847/1538-4365/acf44d},
archivePrefix = {arXiv},
       eprint = {2306.02466},
 primaryClass = {astro-ph.GA},
       adsurl = {https://ui.adsabs.harvard.edu/abs/2023ApJS..269...16R}
}

@MISC{Rodrigo2012,
       author = {{Rodrigo}, Carlos and {Solano}, Enrique and {Bayo}, Amelia},
        title = "{SVO Filter Profile Service Version 1.0}",
 howpublished = {IVOA Working Draft 15 October 2012},
         year = 2012,
        month = oct,
        pages = {1015},
          doi = {10.5479/ADS/bib/2012ivoa.rept.1015R},
       adsurl = {https://ui.adsabs.harvard.edu/abs/2012ivoa.rept.1015R}
}

@INPROCEEDINGS{Rodrigo2020,
       author = {{Rodrigo}, C. and {Solano}, E.},
        title = "{The SVO Filter Profile Service}",
    booktitle = {XIV.0 Scientific Meeting (virtual) of the Spanish Astronomical Society},
         year = 2020,
        month = jul,
          eid = {182},
        pages = {182},
       adsurl = {https://ui.adsabs.harvard.edu/abs/2020sea..confE.182R}
}

@ARTICLE{Faherty2009,
       author = {{Faherty}, Jacqueline K. and {Burgasser}, Adam J. and {Cruz}, Kelle L. and {Shara}, Michael M. and {Walter}, Frederick M. and {Gelino}, Christopher R.},
        title = "{The Brown Dwarf Kinematics Project I. Proper Motions and Tangential Velocities for a Large Sample of Late-Type M, L, and T Dwarfs}",
      journal = {\aj},
         year = 2009,
        month = jan,
       volume = {137},
       number = {1},
        pages = {1-18},
          doi = {10.1088/0004-6256/137/1/1},
archivePrefix = {arXiv},
       eprint = {0809.3008},
 primaryClass = {astro-ph},
       adsurl = {https://ui.adsabs.harvard.edu/abs/2009AJ....137....1F}
}

@INPROCEEDINGS{Joye2003,
       author = {{Joye}, W.~A. and {Mandel}, E.},
        title = "{New Features of SAOImage DS9}",
    booktitle = {Astronomical Data Analysis Software and Systems XII},
         year = 2003,
       editor = {{Payne}, H.~E. and {Jedrzejewski}, R.~I. and {Hook}, R.~N.},
       series = {Astronomical Society of the Pacific Conference Series},
       volume = {295},
        month = jan,
        pages = {489},
       adsurl = {https://ui.adsabs.harvard.edu/abs/2003ASPC..295..489J}
}

@ARTICLE{Hainline2024a,
       author = {{Hainline}, Kevin N. and {Helton}, Jakob M. and {Johnson}, Benjamin D. and {Sun}, Fengwu and {Topping}, Michael W. and {Leisenring}, Jarron M. and {Baker}, William M. and {Eisenstein}, Daniel J. and {Hausen}, Ryan and {Hviding}, Raphael E. and {Lyu}, Jianwei and {Robertson}, Brant and {Tacchella}, Sandro and {Williams}, Christina C. and {Willmer}, Christopher N.~A. and {Roellig}, Thomas L.},
        title = "{Brown Dwarf Candidates in the JADES and CEERS Extragalactic Surveys}",
      journal = {\apj},
         year = 2024,
        month = mar,
       volume = {964},
       number = {1},
          eid = {66},
        pages = {66},
          doi = {10.3847/1538-4357/ad20d1},
archivePrefix = {arXiv},
       eprint = {2309.03250},
 primaryClass = {astro-ph.SR},
       adsurl = {https://ui.adsabs.harvard.edu/abs/2024ApJ...964...66H}
}

@ARTICLE{Kirkpatrick2012,
       author = {{Kirkpatrick}, J. Davy and {Gelino}, Christopher R. and {Cushing}, Michael C. and {Mace}, Gregory N. and {Griffith}, Roger L. and {Skrutskie}, Michael F. and {Marsh}, Kenneth A. and {Wright}, Edward L. and {Eisenhardt}, Peter R. and {McLean}, Ian S. and {Mainzer}, Amanda K. and {Burgasser}, Adam J. and {Tinney}, C.~G. and {Parker}, Stephen and {Salter}, Graeme},
        title = "{Further Defining Spectral Type ``Y'' and Exploring the Low-mass End of the Field Brown Dwarf Mass Function}",
      journal = {\apj},
         year = 2012,
        month = jul,
       volume = {753},
       number = {2},
          eid = {156},
        pages = {156},
          doi = {10.1088/0004-637X/753/2/156},
archivePrefix = {arXiv},
       eprint = {1205.2122},
 primaryClass = {astro-ph.SR},
       adsurl = {https://ui.adsabs.harvard.edu/abs/2012ApJ...753..156K}
}

@ARTICLE{Cushing2011,
       author = {{Cushing}, Michael C. and {Kirkpatrick}, J. Davy and {Gelino}, Christopher R. and {Griffith}, Roger L. and {Skrutskie}, Michael F. and {Mainzer}, A. and {Marsh}, Kenneth A. and {Beichman}, Charles A. and {Burgasser}, Adam J. and {Prato}, Lisa A. and {Simcoe}, Robert A. and {Marley}, Mark S. and {Saumon}, D. and {Freedman}, Richard S. and {Eisenhardt}, Peter R. and {Wright}, Edward L.},
        title = "{The Discovery of Y Dwarfs using Data from the Wide-field Infrared Survey Explorer (WISE)}",
      journal = {\apj},
         year = 2011,
        month = dec,
       volume = {743},
       number = {1},
          eid = {50},
        pages = {50},
          doi = {10.1088/0004-637X/743/1/50},
archivePrefix = {arXiv},
       eprint = {1108.4678},
 primaryClass = {astro-ph.SR},
       adsurl = {https://ui.adsabs.harvard.edu/abs/2011ApJ...743...50C}
}

@ARTICLE{Leggett2021,
       author = {{Leggett}, S.~K. and {Tremblin}, Pascal and {Phillips}, Mark W. and {Dupuy}, Trent J. and {Marley}, Mark and {Morley}, Caroline and {Schneider}, Adam and {Caselden}, Dan and {Guillaume}, Colin and {Logsdon}, Sarah E.},
        title = "{Measuring and Replicating the 1-20 {\ensuremath{\mu}}m Energy Distributions of the Coldest Brown Dwarfs: Rotating, Turbulent, and Nonadiabatic Atmospheres}",
      journal = {\apj},
         year = 2021,
        month = sep,
       volume = {918},
       number = {1},
          eid = {11},
        pages = {11},
          doi = {10.3847/1538-4357/ac0cfe},
archivePrefix = {arXiv},
       eprint = {2107.00696},
 primaryClass = {astro-ph.SR},
       adsurl = {https://ui.adsabs.harvard.edu/abs/2021ApJ...918...11L}
}

@ARTICLE{Fontanive2026,
       author = {{Fontanive}, Cl{\'e}mence and {Bedin}, Luigi R. and {Phillips}, Mark W. and {Scalco}, Michele and {Albert}, Lo{\"\i}c and {Bardalez Gagliuffi}, Daniella C. and {Biller}, Beth},
        title = "{The Y-dwarf population with HST: unlocking the secrets of our coolest neighbours ─ III. Near-infrared photometry}",
      journal = {\mnras},
         year = 2026,
        month = mar,
       volume = {546},
       number = {4},
          eid = {stag284},
        pages = {stag284},
          doi = {10.1093/mnras/stag284},
archivePrefix = {arXiv},
       eprint = {2602.09160},
 primaryClass = {astro-ph.SR},
       adsurl = {https://ui.adsabs.harvard.edu/abs/2026MNRAS.546ag284F}
}

@ARTICLE{Kirkpatrick2021,
       author = {{Kirkpatrick}, J. Davy and {Gelino}, Christopher R. and {Faherty}, Jacqueline K. and {Meisner}, Aaron M. and {Caselden}, Dan and {Schneider}, Adam C. and {Marocco}, Federico and {Cayago}, Alfred J. and {Smart}, R.~L. and {Eisenhardt}, Peter R. and {Kuchner}, Marc J. and {Wright}, Edward L. and {Cushing}, Michael C. and {Allers}, Katelyn N. and {Bardalez Gagliuffi}, Daniella C. and {Burgasser}, Adam J. and {Gagn{\'e}}, Jonathan and {Logsdon}, Sarah E. and {Martin}, Emily C. and {Ingalls}, James G. and {Lowrance}, Patrick J. and {Abrahams}, Ellianna S. and {Aganze}, Christian and {Gerasimov}, Roman and {Gonzales}, Eileen C. and {Hsu}, Chih-Chun and {Kamraj}, Nikita and {Kiman}, Rocio and {Rees}, Jon and {Theissen}, Christopher and {Ammar}, Kareem and {Andersen}, Nikolaj Stevnbak and {Beaulieu}, Paul and {Colin}, Guillaume and {Elachi}, Charles A. and {Goodman}, Samuel J. and {Gramaize}, L{\'e}opold and {Hamlet}, Leslie K. and {Hong}, Justin and {Jonkeren}, Alexander and {Khalil}, Mohammed and {Martin}, David W. and {Pendrill}, William and {Pumphrey}, Benjamin and {Rothermich}, Austin and {Sainio}, Arttu and {Stenner}, Andres and {Tanner}, Christopher and {Th{\'e}venot}, Melina and {Voloshin}, Nikita V. and {Walla}, Jim and {W{\k{e}}dracki}, Zbigniew and {Backyard Worlds: Planet 9 Collaboration}},
        title = "{The Field Substellar Mass Function Based on the Full-sky 20 pc Census of 525 L, T, and Y Dwarfs}",
      journal = {\apjs},
         year = 2021,
        month = mar,
       volume = {253},
       number = {1},
          eid = {7},
        pages = {7},
          doi = {10.3847/1538-4365/abd107},
archivePrefix = {arXiv},
       eprint = {2011.11616},
 primaryClass = {astro-ph.SR},
       adsurl = {https://ui.adsabs.harvard.edu/abs/2021ApJS..253....7K}
}

@ARTICLE{Fontanive2025,
       author = {{Fontanive}, Cl{\'e}mence and {Bedin}, Luigi R. and {Albert}, Lo{\"\i}c and {Bardalez Gagliuffi}, Daniella C.},
        title = "{The Y dwarf population with HST: unlocking the secrets of our coolest neighbours - II. Parallaxes and proper motions}",
      journal = {\mnras},
         year = 2025,
        month = feb,
       volume = {537},
       number = {1},
        pages = {419-435},
          doi = {10.1093/mnras/staf032},
archivePrefix = {arXiv},
       eprint = {2412.16679},
 primaryClass = {astro-ph.SR},
       adsurl = {https://ui.adsabs.harvard.edu/abs/2025MNRAS.537..419F}
}

@ARTICLE{Matthews2024,
       author = {{Matthews}, E.~C. and {Carter}, A.~L. and {Pathak}, P. and {Morley}, C.~V. and {Phillips}, M.~W. and {P.~M.}, S. Krishanth and {Feng}, F. and {Bonse}, M.~J. and {Boogaard}, L.~A. and {Burt}, J.~A. and {Crossfield}, I.~J.~M. and {Douglas}, E.~S. and {Henning}, Th. and {Hom}, J. and {Ko}, C.-L. and {Kasper}, M. and {Lagrange}, A.-M. and {Petit dit de la Roche}, D. and {Philipot}, F.},
        title = "{A temperate super-Jupiter imaged with JWST in the mid-infrared}",
      journal = {\nat},
         year = 2024,
        month = sep,
       volume = {633},
       number = {8031},
        pages = {789-792},
          doi = {10.1038/s41586-024-07837-8},
archivePrefix = {arXiv},
       eprint = {2503.01599},
 primaryClass = {astro-ph.EP},
       adsurl = {https://ui.adsabs.harvard.edu/abs/2024Natur.633..789M}
}

@ARTICLE{Lagrange2025,
       author = {{Lagrange}, A.-M. and {Wilkinson}, C. and {M{\^a}lin}, M. and {Boccaletti}, A. and {Perrot}, C. and {Matr{\`a}}, L. and {Combes}, F. and {Beust}, H. and {Rouan}, D. and {Chomez}, A. and {Milli}, J. and {Charnay}, B. and {Mazevet}, S. and {Flasseur}, O. and {Olofsson}, J. and {Bayo}, A. and {Kral}, Q. and {Carter}, A. and {Crotts}, K.~A. and {Delorme}, P. and {Chauvin}, G. and {Thebault}, P. and {Rubini}, P. and {Kiefer}, F. and {Radcliffe}, A. and {Mazoyer}, J. and {Bodrito}, T. and {Stasevic}, S. and {Langlois}, M.},
        title = "{Evidence for a sub-Jovian planet in the young TWA 7 disk}",
      journal = {\nat},
         year = 2025,
        month = jun,
       volume = {642},
       number = {8069},
        pages = {905-908},
          doi = {10.1038/s41586-025-09150-4},
archivePrefix = {arXiv},
       eprint = {2502.15081},
 primaryClass = {astro-ph.EP},
       adsurl = {https://ui.adsabs.harvard.edu/abs/2025Natur.642..905L}
}

@ARTICLE{Crotts2025,
       author = {{Crotts}, Katie A. and {Carter}, Aarynn L. and {Lawson}, Kellen and {Mang}, James and {Biller}, Beth and {Booth}, Mark and {Ferrer-Chavez}, Rodrigo and {Girard}, Julien H. and {Lagrange}, Anne-Marie and {Liu}, Michael C. and {Marino}, Sebastian and {Millar-Blanchaer}, Maxwell A. and {Skemer}, Andy and {Strampelli}, Giovanni M. and {Wang}, Jason and {Absil}, Olivier and {Balmer}, William O. and {Bendahan-West}, Rapha{\"e}l and {Bogat}, Ellis and {Bowens-Rubin}, Rachel and {Chauvin}, Ga{\"e}l and {Fontanive}, Cl{\'e}mence and {Franson}, Kyle and {Kammerer}, Jens and {Leisenring}, Jarron and {Morley}, Caroline V. and {Rebollido}, Isabel and {Skaf}, Nour and {Sutlieff}, Ben J. and {Bruinsma}, Evelyn L. and {Hinkley}, Sasha and {Hoch}, Kielan and {James}, Andrew D. and {Kane}, Rohan and {Mawet}, Dimitri and {Meyer}, Michael R. and {Palatnick}, Skyler and {Perrin}, Marshall D. and {Ray}, Shrishmoy and {Rickman}, Emily and {Sanghi}, Aniket and {Stephenson}, Klaus Subbotina},
        title = "{Follow-up Exploration of the TWA 7 Planet─Disk System with JWST NIRCam}",
      journal = {\apjl},
         year = 2025,
        month = jul,
       volume = {987},
       number = {2},
          eid = {L41},
        pages = {L41},
          doi = {10.3847/2041-8213/ade798},
archivePrefix = {arXiv},
       eprint = {2506.19932},
 primaryClass = {astro-ph.EP},
       adsurl = {https://ui.adsabs.harvard.edu/abs/2025ApJ...987L..41C}
}

@ARTICLE{BardalezGagliuffi2025,
       author = {{Bardalez Gagliuffi}, Daniella C. and {Balmer}, William O. and {Pueyo}, Laurent and {Brandt}, Timothy D. and {Giovinazzi}, Mark R. and {Millholland}, Sarah and {Black}, Brennen and {Lu}, Tiger and {Rice}, Malena and {Mang}, James and {Morley}, Caroline and {Lacy}, Brianna and {Girard}, Julien H. and {Matthews}, Elisabeth C. and {Carter}, Aarynn L. and {Bowler}, Brendan P. and {Faherty}, Jacqueline K. and {Fontanive}, Clemence and {Rickman}, Emily},
        title = "{JWST Coronagraphic Images of 14 Her c: A Cold Giant Planet in a Dynamically Hot Multiplanet System}",
      journal = {\apjl},
         year = 2025,
        month = jul,
       volume = {988},
       number = {1},
          eid = {L18},
        pages = {L18},
          doi = {10.3847/2041-8213/ade30f},
archivePrefix = {arXiv},
       eprint = {2506.09201},
 primaryClass = {astro-ph.EP},
       adsurl = {https://ui.adsabs.harvard.edu/abs/2025ApJ...988L..18B}
}

@ARTICLE{Skemer2016,
       author = {{Skemer}, Andrew J. and {Morley}, Caroline V. and {Allers}, Katelyn N. and {Geballe}, Thomas R. and {Marley}, Mark S. and {Fortney}, Jonathan J. and {Faherty}, Jacqueline K. and {Bjoraker}, Gordon L. and {Lupu}, Roxana},
        title = "{The First Spectrum of the Coldest Brown Dwarf}",
      journal = {\apjl},
         year = 2016,
        month = aug,
       volume = {826},
       number = {2},
          eid = {L17},
        pages = {L17},
          doi = {10.3847/2041-8205/826/2/L17},
archivePrefix = {arXiv},
       eprint = {1605.04902},
 primaryClass = {astro-ph.EP},
       adsurl = {https://ui.adsabs.harvard.edu/abs/2016ApJ...826L..17S}
}

@ARTICLE{Leggett2019,
       author = {{Leggett}, Sandy and {Apai}, Daniel and {Burgasser}, Adam and {Cushing}, Michael and {Dupuy}, Trent and {Faherty}, Jackie and {Gizis}, John and {Kirkpatrick}, J. Davy and {Marley}, Mark and {Morley}, Caroline and {Schneider}, Adam and {Sousa-Silva}, Clara},
        title = "{Discovery of Cold Brown Dwarfs or Free-Floating Giant Planets Close to the Sun}",
      journal = {\baas},
         year = 2019,
        month = may,
       volume = {51},
       number = {3},
          eid = {95},
        pages = {95},
          doi = {10.48550/arXiv.1903.04686},
archivePrefix = {arXiv},
       eprint = {1903.04686},
 primaryClass = {astro-ph.SR},
       adsurl = {https://ui.adsabs.harvard.edu/abs/2019BAAS...51c..95L}
}

@ARTICLE{Faherty2014,
       author = {{Faherty}, Jacqueline K. and {Tinney}, C.~G. and {Skemer}, Andrew and {Monson}, Andrew J.},
        title = "{Indications of Water Clouds in the Coldest Known Brown Dwarf}",
      journal = {\apjl},
         year = 2014,
        month = sep,
       volume = {793},
       number = {1},
          eid = {L16},
        pages = {L16},
          doi = {10.1088/2041-8205/793/1/L16},
archivePrefix = {arXiv},
       eprint = {1408.4671},
 primaryClass = {astro-ph.SR},
       adsurl = {https://ui.adsabs.harvard.edu/abs/2014ApJ...793L..16F}
}

@ARTICLE{Dupuy2013,
       author = {{Dupuy}, Trent J. and {Kraus}, Adam L.},
        title = "{Distances, Luminosities, and Temperatures of the Coldest Known Substellar Objects}",
      journal = {Science},
         year = 2013,
        month = sep,
       volume = {341},
       number = {6153},
        pages = {1492-1495},
          doi = {10.1126/science.1241917},
archivePrefix = {arXiv},
       eprint = {1309.1422},
 primaryClass = {astro-ph.SR},
       adsurl = {https://ui.adsabs.harvard.edu/abs/2013Sci...341.1492D}
}

@ARTICLE{Beichman2014,
       author = {{Beichman}, C. and {Gelino}, Christopher R. and {Kirkpatrick}, J. Davy and {Cushing}, Michael C. and {Dodson-Robinson}, Sally and {Marley}, Mark S. and {Morley}, Caroline V. and {Wright}, E.~L.},
        title = "{WISE Y Dwarfs as Probes of the Brown Dwarf-Exoplanet Connection}",
      journal = {\apj},
         year = 2014,
        month = mar,
       volume = {783},
       number = {2},
          eid = {68},
        pages = {68},
          doi = {10.1088/0004-637X/783/2/68},
archivePrefix = {arXiv},
       eprint = {1401.1194},
 primaryClass = {astro-ph.SR},
       adsurl = {https://ui.adsabs.harvard.edu/abs/2014ApJ...783...68B}
}

@ARTICLE{LacyBurrows2023,
       author = {{Lacy}, Brianna and {Burrows}, Adam},
        title = "{Self-consistent Models of Y Dwarf Atmospheres with Water Clouds and Disequilibrium Chemistry}",
      journal = {\apj},
         year = 2023,
        month = jun,
       volume = {950},
       number = {1},
          eid = {8},
        pages = {8},
          doi = {10.3847/1538-4357/acc8cb},
archivePrefix = {arXiv},
       eprint = {2303.16295},
 primaryClass = {astro-ph.EP},
       adsurl = {https://ui.adsabs.harvard.edu/abs/2023ApJ...950....8L}
}

@ARTICLE{Mang2022,
       author = {{Mang}, James and {Gao}, Peter and {Hood}, Callie E. and {Fortney}, Jonathan J. and {Batalha}, Natasha and {Yu}, Xinting and {de Pater}, Imke},
        title = "{Microphysics of Water Clouds in the Atmospheres of Y Dwarfs and Temperate Giant Planets}",
      journal = {\apj},
         year = 2022,
        month = mar,
       volume = {927},
       number = {2},
          eid = {184},
        pages = {184},
          doi = {10.3847/1538-4357/ac51d3},
archivePrefix = {arXiv},
       eprint = {2202.01355},
 primaryClass = {astro-ph.EP},
       adsurl = {https://ui.adsabs.harvard.edu/abs/2022ApJ...927..184M}
}

@ARTICLE{Lee2025,
       author = {{Lee}, Elspeth K.~H. and {Ohno}, Kazumasa},
        title = "{Three-dimensional dynamical evolution of cloud particle microphysics in sub-stellar atmospheres: I. Description and exploring Y-dwarf atmospheric variability}",
      journal = {\aap},
         year = 2025,
        month = mar,
       volume = {695},
          eid = {A111},
        pages = {A111},
          doi = {10.1051/0004-6361/202452922},
archivePrefix = {arXiv},
       eprint = {2411.10305},
 primaryClass = {astro-ph.EP},
       adsurl = {https://ui.adsabs.harvard.edu/abs/2025A&A...695A.111L}
}

@ARTICLE{Morley2014,
       author = {{Morley}, Caroline V. and {Marley}, Mark S. and {Fortney}, Jonathan J. and {Lupu}, Roxana},
        title = "{Spectral Variability from the Patchy Atmospheres of T and Y Dwarfs}",
      journal = {\apjl},
         year = 2014,
        month = jul,
       volume = {789},
       number = {1},
          eid = {L14},
        pages = {L14},
          doi = {10.1088/2041-8205/789/1/L14},
archivePrefix = {arXiv},
       eprint = {1406.0863},
 primaryClass = {astro-ph.SR},
       adsurl = {https://ui.adsabs.harvard.edu/abs/2014ApJ...789L..14M}
}

@ARTICLE{Tan2025,
       author = {{Tan}, Xianyu and {Zhang}, Xi and {Marley}, Mark S. and {Zhou}, Yifan and {Lew}, Ben W.~P. and {Miles}, Brittany E. and {Batalha}, Natasha E. and {Biller}, Beth A. and {Chauvin}, Ga{\"e}l and {Hinkley}, Sasha and {Hoch}, Kielan K.~W. and {Manjavacas}, Elena and {Metchev}, Stanimir and {Petrus}, Simon and {Rickman}, Emily and {Skemer}, Andrew and {Su{\'a}rez}, Genaro and {Sutlieff}, Ben J. and {Vos}, Johanna M. and {Whiteford}, Niall},
        title = "{Large-amplitude variability driven by giant dust storms on a planetary-mass companion}",
      journal = {Science Advances},
         year = 2025,
        month = nov,
       volume = {11},
        pages = {22.3324},
          doi = {10.1126/sciadv.adv3324},
archivePrefix = {arXiv},
       eprint = {2511.23163},
 primaryClass = {astro-ph.EP},
       adsurl = {https://ui.adsabs.harvard.edu/abs/2025SciA...11v3324T}
}

@ARTICLE{Phillips2020,
       author = {{Phillips}, M.~W. and {Tremblin}, P. and {Baraffe}, I. and {Chabrier}, G. and {Allard}, N.~F. and {Spiegelman}, F. and {Goyal}, J.~M. and {Drummond}, B. and {H{\'e}brard}, E.},
        title = "{A new set of atmosphere and evolution models for cool T-Y brown dwarfs and giant exoplanets}",
      journal = {\aap},
         year = 2020,
        month = may,
       volume = {637},
          eid = {A38},
        pages = {A38},
          doi = {10.1051/0004-6361/201937381},
archivePrefix = {arXiv},
       eprint = {2003.13717},
 primaryClass = {astro-ph.SR},
       adsurl = {https://ui.adsabs.harvard.edu/abs/2020A&A...637A..38P}
}

@ARTICLE{Miles2020,
       author = {{Miles}, Brittany E. and {Skemer}, Andrew J.~I. and {Morley}, Caroline V. and {Marley}, Mark S. and {Fortney}, Jonathan J. and {Allers}, Katelyn N. and {Faherty}, Jacqueline K. and {Geballe}, Thomas R. and {Visscher}, Channon and {Schneider}, Adam C. and {Lupu}, Roxana and {Freedman}, Richard S. and {Bjoraker}, Gordon L.},
        title = "{Observations of Disequilibrium CO Chemistry in the Coldest Brown Dwarfs}",
      journal = {\aj},
         year = 2020,
        month = aug,
       volume = {160},
       number = {2},
          eid = {63},
        pages = {63},
          doi = {10.3847/1538-3881/ab9114},
archivePrefix = {arXiv},
       eprint = {2004.10770},
 primaryClass = {astro-ph.EP},
       adsurl = {https://ui.adsabs.harvard.edu/abs/2020AJ....160...63M}
}

@ARTICLE{Fontanive2021,
       author = {{Fontanive}, C. and {Bedin}, L.~R. and {Bardalez Gagliuffi}, D.~C.},
        title = "{The Y dwarf population with HST: unlocking the secrets of our coolest neighbours - I. Overview and first astrometric results}",
      journal = {\mnras},
         year = 2021,
        month = feb,
       volume = {501},
       number = {1},
        pages = {911-915},
          doi = {10.1093/mnras/staa3732},
archivePrefix = {arXiv},
       eprint = {2011.13873},
 primaryClass = {astro-ph.SR},
       adsurl = {https://ui.adsabs.harvard.edu/abs/2021MNRAS.501..911F}
}

@ARTICLE{Kirkpatrick2019,
       author = {{Kirkpatrick}, J. Davy and {Martin}, Emily C. and {Smart}, Richard L. and {Cayago}, Alfred J. and {Beichman}, Charles A. and {Marocco}, Federico and {Gelino}, Christopher R. and {Faherty}, Jacqueline K. and {Cushing}, Michael C. and {Schneider}, Adam C. and {Mace}, Gregory N. and {Tinney}, Christopher G. and {Wright}, Edward L. and {Lowrance}, Patrick J. and {Ingalls}, James G. and {Vrba}, Frederick J. and {Munn}, Jeffrey A. and {Dahm}, Scott E. and {McLean}, Ian S.},
        title = "{Preliminary Trigonometric Parallaxes of 184 Late-T and Y Dwarfs and an Analysis of the Field Substellar Mass Function into the {\textquotedblleft}Planetary{\textquotedblright} Mass Regime}",
      journal = {\apjs},
         year = 2019,
        month = feb,
       volume = {240},
       number = {2},
          eid = {19},
        pages = {19},
          doi = {10.3847/1538-4365/aaf6af},
archivePrefix = {arXiv},
       eprint = {1812.01208},
 primaryClass = {astro-ph.SR},
       adsurl = {https://ui.adsabs.harvard.edu/abs/2019ApJS..240...19K}
}

@ARTICLE{Morrissey2026,
       author = {{Morrissey}, Sara J. and {Burgasser}, Adam J. and {de Graaff}, Anna and {McConachie}, Ian and {Brammer}, Gabriel},
        title = "{Discovery of Seven Cold and Distant Brown Dwarfs with JWST RUBIES}",
      journal = {\aj},
         year = 2026,
        month = mar,
       volume = {171},
       number = {3},
          eid = {191},
        pages = {191},
          doi = {10.3847/1538-3881/ae40f1},
archivePrefix = {arXiv},
       eprint = {2511.01167},
 primaryClass = {astro-ph.SR},
       adsurl = {https://ui.adsabs.harvard.edu/abs/2026AJ....171..191M}
}

@ARTICLE{Li2026,
       author = {{Li}, D.~H. and {Zhang}, Z.~H. and {Peng}, H.~H. and {G{\'a}lvez-Ortiz}, M.~C. and {Zhou}, S.~Y. and {Jones}, H.~R.~A.},
        title = "{Two late-T dwarfs at kiloparsec distances revealed by JWST UNCOVER survey}",
      journal = {\mnras},
         year = 2026,
        month = mar,
       volume = {547},
       number = {1},
          eid = {stag227},
        pages = {stag227},
          doi = {10.1093/mnras/stag227},
archivePrefix = {arXiv},
       eprint = {2602.01894},
 primaryClass = {astro-ph.SR},
       adsurl = {https://ui.adsabs.harvard.edu/abs/2026MNRAS.547ag227L}
}

@ARTICLE{Burgasser2024,
       author = {{Burgasser}, Adam J. and {Bezanson}, Rachel and {Labbe}, Ivo and {Brammer}, Gabriel and {Cutler}, Sam E. and {Furtak}, Lukas J. and {Greene}, Jenny E. and {Gerasimov}, Roman and {Leja}, Joel and {Pan}, Richard and {Price}, Sedona H. and {Wang}, Bingjie and {Weaver}, John R. and {Whitaker}, Katherine E. and {Fujimoto}, Seiji and {Kokorev}, Vasily and {Dayal}, Pratika and {Nanayakkara}, Themiya and {Williams}, Christina C. and {Marchesini}, Danilo and {Zitrin}, Adi and {van Dokkum}, Pieter},
        title = "{UNCOVER: JWST Spectroscopy of Three Cold Brown Dwarfs at Kiloparsec-scale Distances}",
      journal = {\apj},
         year = 2024,
        month = feb,
       volume = {962},
       number = {2},
          eid = {177},
        pages = {177},
          doi = {10.3847/1538-4357/ad206f},
archivePrefix = {arXiv},
       eprint = {2308.12107},
 primaryClass = {astro-ph.SR},
       adsurl = {https://ui.adsabs.harvard.edu/abs/2024ApJ...962..177B}
}

@ARTICLE{Langeroodi2023,
       author = {{Langeroodi}, Danial and {Hjorth}, Jens},
        title = "{Little Red Dots or Brown Dwarfs? NIRSpec Discovery of Three Distant Brown Dwarfs Masquerading as NIRCam-selected Highly Reddened Active Galactic Nuclei}",
      journal = {\apjl},
         year = 2023,
        month = nov,
       volume = {957},
       number = {2},
          eid = {L27},
        pages = {L27},
          doi = {10.3847/2041-8213/acfeec},
archivePrefix = {arXiv},
       eprint = {2308.10900},
 primaryClass = {astro-ph.GA},
       adsurl = {https://ui.adsabs.harvard.edu/abs/2023ApJ...957L..27L}
}

@ARTICLE{Hainline2024b,
       author = {{Hainline}, Kevin N. and {D'Eugenio}, Francesco and {Sun}, Fengwu and {Helton}, Jakob M. and {Miles}, Brittany E. and {Marley}, Mark S. and {Lew}, Ben W.~P. and {Leisenring}, Jarron M. and {Bunker}, Andrew J. and {Cargile}, Phillip A. and {Carniani}, Stefano and {Eisenstein}, Daniel J. and {Juod{\v{z}}balis}, Ignas and {Johnson}, Benjamin D. and {Robertson}, Brant and {Tacchella}, Sandro and {Williams}, Christina C. and {Willmer}, Christopher N.~A.},
        title = "{JADES: Spectroscopic Confirmation and Proper Motion for a T-Dwarf at 2 kpc}",
      journal = {\apj},
         year = 2024,
        month = nov,
       volume = {975},
       number = {1},
          eid = {31},
        pages = {31},
          doi = {10.3847/1538-4357/ad76a7},
archivePrefix = {arXiv},
       eprint = {2407.08781},
 primaryClass = {astro-ph.GA},
       adsurl = {https://ui.adsabs.harvard.edu/abs/2024ApJ...975...31H}
}

@ARTICLE{Gardner2023,
       author = {{Gardner}, Jonathan P. and {Mather}, John C. and {Abbott}, Randy and {Abell}, James S. and {Abernathy}, Mark and {Abney}, Faith E. and {Abraham}, John G. and {Abraham}, Roberto and {Abul-Huda}, Yasin M. and {Acton}, Scott and {Adams}, Cynthia K. and {Adams}, Evan and {Adler}, David S. and {Adriaensen}, Maarten and {Aguilar}, Jonathan Albert and {Ahmed}, Mansoor and {Ahmed}, Nasif S. and {Ahmed}, Tanjira and {Albat}, R{\"u}deger and {Albert}, Lo{\"\i}c and {Alberts}, Stacey and {Aldridge}, David and {Allen}, Mary Marsha and {Allen}, Shaune S. and {Altenburg}, Martin and {Altunc}, Serhat and {Alvarez}, Jose Lorenzo and {{\'A}lvarez-M{\'a}rquez}, Javier and {Alves de Oliveira}, Catarina and {Ambrose}, Leslie L. and {Anandakrishnan}, Satya M. and {Andersen}, Gregory C. and {Anderson}, Harry James and {Anderson}, Jay and {Anderson}, Kristen and {Anderson}, Sara M. and {Aprea}, Julio and {Archer}, Benita J. and {Arenberg}, Jonathan W. and {Argyriou}, Ioannis and {Arribas}, Santiago and {Artigau}, {\'E}tienne and {Arvai}, Amanda Rose and {Atcheson}, Paul and {Atkinson}, Charles B. and {Averbukh}, Jesse and {Aymergen}, Cagatay and {Bacinski}, John J. and {Baggett}, Wayne E. and {Bagnasco}, Giorgio and {Baker}, Lynn L. and {Balzano}, Vicki Ann and {Banks}, Kimberly A. and {Baran}, David A. and {Barker}, Elizabeth A. and {Barrett}, Larry K. and {Barringer}, Bruce O. and {Barto}, Allison and {Bast}, William and {Baudoz}, Pierre and {Baum}, Stefi and {Beatty}, Thomas G. and {Beaulieu}, Mathilde and {Bechtold}, Kathryn and {Beck}, Tracy and {Beddard}, Megan M. and {Beichman}, Charles and {Bellagama}, Larry and {Bely}, Pierre and {Berger}, Timothy W. and {Bergeron}, Louis E. and {Bernier}, Antoine-Darveau and {Bertch}, Maria D. and {Beskow}, Charlotte and {Betz}, Laura E. and {Biagetti}, Carl P. and {Birkmann}, Stephan and {Bjorklund}, Kurt F. and {Blackwood}, James D. and {Blazek}, Ronald Paul and {Blossfeld}, Stephen and {Bluth}, Marcel and {Boccaletti}, Anthony and {Boegner}, Jr., Martin E. and {Bohlin}, Ralph C. and {Boia}, John Joseph and {B{\"o}ker}, Torsten and {Bonaventura}, N. and {Bond}, Nicholas A. and {Bosley}, Kari Ann and {Boucarut}, Rene A. and {Bouchet}, Patrice and {Bouwman}, Jeroen and {Bower}, Gary and {Bowers}, Ariel S. and {Bowers}, Charles W. and {Boyce}, Leslye A. and {Boyer}, Christine T. and {Boyer}, Martha L. and {Boyer}, Michael and {Boyer}, Robert and {Bradley}, Larry D. and {Brady}, Gregory R. and {Brandl}, Bernhard R. and {Brannen}, Judith L. and {Breda}, David and {Bremmer}, Harold G. and {Brennan}, David and {Bresnahan}, Pamela A. and {Bright}, Stacey N. and {Broiles}, Brian J. and {Bromenschenkel}, Asa and {Brooks}, Brian H. and {Brooks}, Keira J. and {Brown}, Bob and {Brown}, Bruce and {Brown}, Thomas M. and {Bruce}, Barry W. and {Bryson}, Jonathan G. and {Bujanda}, Edwin D. and {Bullock}, Blake M. and {Bunker}, A.~J. and {Bureo}, Rafael and {Burt}, Irving J. and {Bush}, James Aaron and {Bushouse}, Howard A. and {Bussman}, Marie C. and {Cabaud}, Olivier and {Cale}, Steven and {Calhoon}, Charles D. and {Calvani}, Humberto and {Canipe}, Alicia M. and {Caputo}, Francis M. and {Cara}, Mihai and {Carey}, Larkin and {Case}, Michael Eli and {Cesari}, Thaddeus and {Cetorelli}, Lee D. and {Chance}, Don R. and {Chandler}, Lynn and {Chaney}, Dave and {Chapman}, George N. and {Charlot}, S. and {Chayer}, Pierre and {Cheezum}, Jeffrey I. and {Chen}, Bin and {Chen}, Christine H. and {Cherinka}, Brian and {Chichester}, Sarah C. and {Chilton}, Zachary S. and {Chittiraibalan}, Dharini and {Clampin}, Mark and {Clark}, Charles R. and {Clark}, Kerry W. and {Clark}, Stephanie M. and {Claybrooks}, Edward E. and {Cleveland}, Keith A. and {Cohen}, Andrew L. and {Cohen}, Lester M. and {Col{\'o}n}, Knicole D. and {Coleman}, Benee L. and {Colina}, Luis and {Comber}, Brian J. and {Comeau}, Thomas M. and {Comer}, Thomas and {Conde Reis}, Alain and {Connolly}, Dennis C. and {Conroy}, Kyle E. and {Contos}, Adam R. and {Contreras}, James and {Cook}, Neil J. and {Cooper}, James L. and {Cooper}, Rachel Aviva and {Correia}, Michael F. and {Correnti}, Matteo and {Cossou}, Christophe and {Costanza}, Brian F. and {Coulais}, Alain and {Cox}, Colin R. and {Coyle}, Ray T. and {Cracraft}, Misty M. and {Crew}, Keith A. and {Curtis}, Gary J. and {Cusveller}, Bianca and {Da Costa Maciel}, Cleyciane and {Dailey}, Christopher T. and {Daugeron}, Fr{\'e}d{\'e}ric and {Davidson}, Greg S. and {Davies}, James E. and {Davis}, Katherine Anne and {Davis}, Michael S. and {Day}, Ratna and {de Chambure}, Daniel and {de Jong}, Pauline and {De Marchi}, Guido and {Dean}, Bruce H. and {Decker}, John E. and {Delisa}, Amy S. and {Dell}, Lawrence C. and {Dellagatta}, Gail},
        title = "{The James Webb Space Telescope Mission}",
      journal = {\pasp},
         year = 2023,
        month = jun,
       volume = {135},
       number = {1048},
          eid = {068001},
        pages = {068001},
          doi = {10.1088/1538-3873/acd1b5},
archivePrefix = {arXiv},
       eprint = {2304.04869},
 primaryClass = {astro-ph.IM},
       adsurl = {https://ui.adsabs.harvard.edu/abs/2023PASP..135f8001G}
}

@ARTICLE{GaiaCollaboration2016,
       author = {{Gaia Collaboration} and {Prusti}, T. and {de Bruijne}, J.~H.~J. and {Brown}, A.~G.~A. and {Vallenari}, A. and {Babusiaux}, C. and {Bailer-Jones}, C.~A.~L. and {Bastian}, U. and {Biermann}, M. and {Evans}, D.~W. and {Eyer}, L. and {Jansen}, F. and {Jordi}, C. and {Klioner}, S.~A. and {Lammers}, U. and {Lindegren}, L. and {Luri}, X. and {Mignard}, F. and {Milligan}, D.~J. and {Panem}, C. and {Poinsignon}, V. and {Pourbaix}, D. and {Randich}, S. and {Sarri}, G. and {Sartoretti}, P. and {Siddiqui}, H.~I. and {Soubiran}, C. and {Valette}, V. and {van Leeuwen}, F. and {Walton}, N.~A. and {Aerts}, C. and {Arenou}, F. and {Cropper}, M. and {Drimmel}, R. and {H{\o}g}, E. and {Katz}, D. and {Lattanzi}, M.~G. and {O'Mullane}, W. and {Grebel}, E.~K. and {Holland}, A.~D. and {Huc}, C. and {Passot}, X. and {Bramante}, L. and {Cacciari}, C. and {Casta{\~n}eda}, J. and {Chaoul}, L. and {Cheek}, N. and {De Angeli}, F. and {Fabricius}, C. and {Guerra}, R. and {Hern{\'a}ndez}, J. and {Jean-Antoine-Piccolo}, A. and {Masana}, E. and {Messineo}, R. and {Mowlavi}, N. and {Nienartowicz}, K. and {Ord{\'o}{\~n}ez-Blanco}, D. and {Panuzzo}, P. and {Portell}, J. and {Richards}, P.~J. and {Riello}, M. and {Seabroke}, G.~M. and {Tanga}, P. and {Th{\'e}venin}, F. and {Torra}, J. and {Els}, S.~G. and {Gracia-Abril}, G. and {Comoretto}, G. and {Garcia-Reinaldos}, M. and {Lock}, T. and {Mercier}, E. and {Altmann}, M. and {Andrae}, R. and {Astraatmadja}, T.~L. and {Bellas-Velidis}, I. and {Benson}, K. and {Berthier}, J. and {Blomme}, R. and {Busso}, G. and {Carry}, B. and {Cellino}, A. and {Clementini}, G. and {Cowell}, S. and {Creevey}, O. and {Cuypers}, J. and {Davidson}, M. and {De Ridder}, J. and {de Torres}, A. and {Delchambre}, L. and {Dell'Oro}, A. and {Ducourant}, C. and {Fr{\'e}mat}, Y. and {Garc{\'\i}a-Torres}, M. and {Gosset}, E. and {Halbwachs}, J.-L. and {Hambly}, N.~C. and {Harrison}, D.~L. and {Hauser}, M. and {Hestroffer}, D. and {Hodgkin}, S.~T. and {Huckle}, H.~E. and {Hutton}, A. and {Jasniewicz}, G. and {Jordan}, S. and {Kontizas}, M. and {Korn}, A.~J. and {Lanzafame}, A.~C. and {Manteiga}, M. and {Moitinho}, A. and {Muinonen}, K. and {Osinde}, J. and {Pancino}, E. and {Pauwels}, T. and {Petit}, J.-M. and {Recio-Blanco}, A. and {Robin}, A.~C. and {Sarro}, L.~M. and {Siopis}, C. and {Smith}, M. and {Smith}, K.~W. and {Sozzetti}, A. and {Thuillot}, W. and {van Reeven}, W. and {Viala}, Y. and {Abbas}, U. and {Abreu Aramburu}, A. and {Accart}, S. and {Aguado}, J.~J. and {Allan}, P.~M. and {Allasia}, W. and {Altavilla}, G. and {{\'A}lvarez}, M.~A. and {Alves}, J. and {Anderson}, R.~I. and {Andrei}, A.~H. and {Anglada Varela}, E. and {Antiche}, E. and {Antoja}, T. and {Ant{\'o}n}, S. and {Arcay}, B. and {Atzei}, A. and {Ayache}, L. and {Bach}, N. and {Baker}, S.~G. and {Balaguer-N{\'u}{\~n}ez}, L. and {Barache}, C. and {Barata}, C. and {Barbier}, A. and {Barblan}, F. and {Baroni}, M. and {Barrado y Navascu{\'e}s}, D. and {Barros}, M. and {Barstow}, M.~A. and {Becciani}, U. and {Bellazzini}, M. and {Bellei}, G. and {Bello Garc{\'\i}a}, A. and {Belokurov}, V. and {Bendjoya}, P. and {Berihuete}, A. and {Bianchi}, L. and {Bienaym{\'e}}, O. and {Billebaud}, F. and {Blagorodnova}, N. and {Blanco-Cuaresma}, S. and {Boch}, T. and {Bombrun}, A. and {Borrachero}, R. and {Bouquillon}, S. and {Bourda}, G. and {Bouy}, H. and {Bragaglia}, A. and {Breddels}, M.~A. and {Brouillet}, N. and {Br{\"u}semeister}, T. and {Bucciarelli}, B. and {Budnik}, F. and {Burgess}, P. and {Burgon}, R. and {Burlacu}, A. and {Busonero}, D. and {Buzzi}, R. and {Caffau}, E. and {Cambras}, J. and {Campbell}, H. and {Cancelliere}, R. and {Cantat-Gaudin}, T. and {Carlucci}, T. and {Carrasco}, J.~M. and {Castellani}, M. and {Charlot}, P. and {Charnas}, J. and {Charvet}, P. and {Chassat}, F. and {Chiavassa}, A. and {Clotet}, M. and {Cocozza}, G. and {Collins}, R.~S. and {Collins}, P. and {Costigan}, G.},
        title = "{The Gaia mission}",
      journal = {\aap},
         year = 2016,
        month = nov,
       volume = {595},
          eid = {A1},
        pages = {A1},
          doi = {10.1051/0004-6361/201629272},
archivePrefix = {arXiv},
       eprint = {1609.04153},
 primaryClass = {astro-ph.IM},
       adsurl = {https://ui.adsabs.harvard.edu/abs/2016A&A...595A...1G}
}

@ARTICLE{GaiaCollaboration2023,
       author = {{Gaia Collaboration} and {Vallenari}, A. and {Brown}, A.~G.~A. and {Prusti}, T. and {de Bruijne}, J.~H.~J. and {Arenou}, F. and {Babusiaux}, C. and {Biermann}, M. and {Creevey}, O.~L. and {Ducourant}, C. and {Evans}, D.~W. and {Eyer}, L. and {Guerra}, R. and {Hutton}, A. and {Jordi}, C. and {Klioner}, S.~A. and {Lammers}, U.~L. and {Lindegren}, L. and {Luri}, X. and {Mignard}, F. and {Panem}, C. and {Pourbaix}, D. and {Randich}, S. and {Sartoretti}, P. and {Soubiran}, C. and {Tanga}, P. and {Walton}, N.~A. and {Bailer-Jones}, C.~A.~L. and {Bastian}, U. and {Drimmel}, R. and {Jansen}, F. and {Katz}, D. and {Lattanzi}, M.~G. and {van Leeuwen}, F. and {Bakker}, J. and {Cacciari}, C. and {Casta{\~n}eda}, J. and {De Angeli}, F. and {Fabricius}, C. and {Fouesneau}, M. and {Fr{\'e}mat}, Y. and {Galluccio}, L. and {Guerrier}, A. and {Heiter}, U. and {Masana}, E. and {Messineo}, R. and {Mowlavi}, N. and {Nicolas}, C. and {Nienartowicz}, K. and {Pailler}, F. and {Panuzzo}, P. and {Riclet}, F. and {Roux}, W. and {Seabroke}, G.~M. and {Sordo}, R. and {Th{\'e}venin}, F. and {Gracia-Abril}, G. and {Portell}, J. and {Teyssier}, D. and {Altmann}, M. and {Andrae}, R. and {Audard}, M. and {Bellas-Velidis}, I. and {Benson}, K. and {Berthier}, J. and {Blomme}, R. and {Burgess}, P.~W. and {Busonero}, D. and {Busso}, G. and {C{\'a}novas}, H. and {Carry}, B. and {Cellino}, A. and {Cheek}, N. and {Clementini}, G. and {Damerdji}, Y. and {Davidson}, M. and {de Teodoro}, P. and {Nu{\~n}ez Campos}, M. and {Delchambre}, L. and {Dell'Oro}, A. and {Esquej}, P. and {Fern{\'a}ndez-Hern{\'a}ndez}, J. and {Fraile}, E. and {Garabato}, D. and {Garc{\'\i}a-Lario}, P. and {Gosset}, E. and {Haigron}, R. and {Halbwachs}, J.-L. and {Hambly}, N.~C. and {Harrison}, D.~L. and {Hern{\'a}ndez}, J. and {Hestroffer}, D. and {Hodgkin}, S.~T. and {Holl}, B. and {Jan{\ss}en}, K. and {Jevardat de Fombelle}, G. and {Jordan}, S. and {Krone-Martins}, A. and {Lanzafame}, A.~C. and {L{\"o}ffler}, W. and {Marchal}, O. and {Marrese}, P.~M. and {Moitinho}, A. and {Muinonen}, K. and {Osborne}, P. and {Pancino}, E. and {Pauwels}, T. and {Recio-Blanco}, A. and {Reyl{\'e}}, C. and {Riello}, M. and {Rimoldini}, L. and {Roegiers}, T. and {Rybizki}, J. and {Sarro}, L.~M. and {Siopis}, C. and {Smith}, M. and {Sozzetti}, A. and {Utrilla}, E. and {van Leeuwen}, M. and {Abbas}, U. and {{\'A}brah{\'a}m}, P. and {Abreu Aramburu}, A. and {Aerts}, C. and {Aguado}, J.~J. and {Ajaj}, M. and {Aldea-Montero}, F. and {Altavilla}, G. and {{\'A}lvarez}, M.~A. and {Alves}, J. and {Anders}, F. and {Anderson}, R.~I. and {Anglada Varela}, E. and {Antoja}, T. and {Baines}, D. and {Baker}, S.~G. and {Balaguer-N{\'u}{\~n}ez}, L. and {Balbinot}, E. and {Balog}, Z. and {Barache}, C. and {Barbato}, D. and {Barros}, M. and {Barstow}, M.~A. and {Bartolom{\'e}}, S. and {Bassilana}, J.-L. and {Bauchet}, N. and {Becciani}, U. and {Bellazzini}, M. and {Berihuete}, A. and {Bernet}, M. and {Bertone}, S. and {Bianchi}, L. and {Binnenfeld}, A. and {Blanco-Cuaresma}, S. and {Blazere}, A. and {Boch}, T. and {Bombrun}, A. and {Bossini}, D. and {Bouquillon}, S. and {Bragaglia}, A. and {Bramante}, L. and {Breedt}, E. and {Bressan}, A. and {Brouillet}, N. and {Brugaletta}, E. and {Bucciarelli}, B. and {Burlacu}, A. and {Butkevich}, A.~G. and {Buzzi}, R. and {Caffau}, E. and {Cancelliere}, R. and {Cantat-Gaudin}, T. and {Carballo}, R. and {Carlucci}, T. and {Carnerero}, M.~I. and {Carrasco}, J.~M. and {Casamiquela}, L. and {Castellani}, M. and {Castro-Ginard}, A. and {Chaoul}, L. and {Charlot}, P. and {Chemin}, L. and {Chiaramida}, V. and {Chiavassa}, A. and {Chornay}, N. and {Comoretto}, G. and {Contursi}, G. and {Cooper}, W.~J. and {Cornez}, T. and {Cowell}, S. and {Crifo}, F. and {Cropper}, M. and {Crosta}, M. and {Crowley}, C. and {Dafonte}, C. and {Dapergolas}, A. and {David}, M. and {David}, P. and {de Laverny}, P. and {De Luise}, F. and {De March}, R.},
        title = "{Gaia Data Release 3. Summary of the content and survey properties}",
      journal = {\aap},
         year = 2023,
        month = jun,
       volume = {674},
          eid = {A1},
        pages = {A1},
          doi = {10.1051/0004-6361/202243940},
archivePrefix = {arXiv},
       eprint = {2208.00211},
 primaryClass = {astro-ph.GA},
       adsurl = {https://ui.adsabs.harvard.edu/abs/2023A&A...674A...1G}
}

@ARTICLE{Metchev2015,
       author = {{Metchev}, Stanimir A. and {Heinze}, Aren and {Apai}, D{\'a}niel and {Flateau}, Davin and {Radigan}, Jacqueline and {Burgasser}, Adam and {Marley}, Mark S. and {Artigau}, {\'E}tienne and {Plavchan}, Peter and {Goldman}, Bertrand},
        title = "{Weather on Other Worlds. II. Survey Results: Spots are Ubiquitous on L and T Dwarfs}",
      journal = {\apj},
         year = 2015,
        month = feb,
       volume = {799},
       number = {2},
          eid = {154},
        pages = {154},
          doi = {10.1088/0004-637X/799/2/154},
archivePrefix = {arXiv},
       eprint = {1411.3051},
 primaryClass = {astro-ph.SR},
       adsurl = {https://ui.adsabs.harvard.edu/abs/2015ApJ...799..154M}
}

@ARTICLE{Vos2022,
       author = {{Vos}, Johanna M. and {Faherty}, Jacqueline K. and {Gagn{\'e}}, Jonathan and {Marley}, Mark and {Metchev}, Stanimir and {Gizis}, John and {Rice}, Emily L. and {Cruz}, Kelle},
        title = "{Let the Great World Spin: Revealing the Stormy, Turbulent Nature of Young Giant Exoplanet Analogs with the Spitzer Space Telescope}",
      journal = {\apj},
         year = 2022,
        month = jan,
       volume = {924},
       number = {2},
          eid = {68},
        pages = {68},
          doi = {10.3847/1538-4357/ac4502},
archivePrefix = {arXiv},
       eprint = {2201.04711},
 primaryClass = {astro-ph.EP},
       adsurl = {https://ui.adsabs.harvard.edu/abs/2022ApJ...924...68V}
}

@ARTICLE{Liu2024,
       author = {{Liu}, Pengyu and {Biller}, Beth A. and {Vos}, Johanna M. and {Whiteford}, Niall and {Zhang}, Zhoujian and {Liu}, Michael C. and {Fontanive}, Cl{\'e}mence and {Manjavacas}, Elena and {Henning}, Thomas and {Kenworthy}, Matthew A. and {Bonavita}, Mariangela and {Bonnefoy}, Micka{\"e}l and {Bubb}, Emma and {Petrus}, Simon and {Schlieder}, Joshua},
        title = "{A near-infrared variability survey of young planetary-mass objects}",
      journal = {\mnras},
         year = 2024,
        month = jan,
       volume = {527},
       number = {3},
        pages = {6624-6674},
          doi = {10.1093/mnras/stad3502},
archivePrefix = {arXiv},
       eprint = {2311.07218},
 primaryClass = {astro-ph.EP},
       adsurl = {https://ui.adsabs.harvard.edu/abs/2024MNRAS.527.6624L}
}

@ARTICLE{Cushing2026,
       author = {{Cushing}, Michael C. and {Trucks}, Jesica L. and {Hardegree-Ullman}, Kevin K. and {Burgasser}, Adam J. and {Carey}, Sean J. and {Fortney}, Jonathan J. and {Gelino}, Christopher R. and {Gizis}, John E. and {Kirkpatrick}, J. Davy and {Leggett}, Sandy and {Mace}, Gregory N. and {Marley}, Mark S. and {Morley}, Caroline V.},
        title = "{A Spitzer Space Telescope Exploration Science Program to Search for Y-dwarf Variability}",
      journal = {\aj},
         year = 2026,
        month = sep,
       volume = {172},
       number = {3},
          eid = {166},
        pages = {166},
          doi = {10.3847/1538-3881/ae8e7c},
archivePrefix = {arXiv},
       eprint = {2606.26411},
 primaryClass = {astro-ph.SR},
       adsurl = {https://ui.adsabs.harvard.edu/abs/2026AJ....172..166C}
}

@ARTICLE{Zhou2022,
       author = {{Zhou}, Yifan and {Bowler}, Brendan P. and {Apai}, D{\'a}niel and {Kataria}, Tiffany and {Morley}, Caroline V. and {Bryan}, Marta L. and {Skemer}, Andrew J. and {Benneke}, Bj{\"o}rn},
        title = "{Roaring Storms in the Planetary-mass Companion VHS 1256-1257 b: Hubble Space Telescope Multiepoch Monitoring Reveals Vigorous Evolution in an Ultracool Atmosphere}",
      journal = {\aj},
         year = 2022,
        month = dec,
       volume = {164},
       number = {6},
          eid = {239},
        pages = {239},
          doi = {10.3847/1538-3881/ac9905},
archivePrefix = {arXiv},
       eprint = {2210.02464},
 primaryClass = {astro-ph.EP},
       adsurl = {https://ui.adsabs.harvard.edu/abs/2022AJ....164..239Z}
}

@ARTICLE{Radigan2012,
       author = {{Radigan}, Jacqueline and {Jayawardhana}, Ray and {Lafreni{\`e}re}, David and {Artigau}, {\'E}tienne and {Marley}, Mark and {Saumon}, Didier},
        title = "{Large-amplitude Variations of an L/T Transition Brown Dwarf: Multi-wavelength Observations of Patchy, High-contrast Cloud Features}",
      journal = {\apj},
         year = 2012,
        month = may,
       volume = {750},
       number = {2},
          eid = {105},
        pages = {105},
          doi = {10.1088/0004-637X/750/2/105},
archivePrefix = {arXiv},
       eprint = {1201.3403},
 primaryClass = {astro-ph.SR},
       adsurl = {https://ui.adsabs.harvard.edu/abs/2012ApJ...750..105R}
}

@ARTICLE{B_Bowler2020,
       author = {{Bowler}, Brendan P. and {Zhou}, Yifan and {Morley}, Caroline V. and {Kataria}, Tiffany and {Bryan}, Marta L. and {Benneke}, Bj{\"o}rn and {Batygin}, Konstantin},
        title = "{Strong Near-infrared Spectral Variability of the Young Cloudy L Dwarf Companion VHS J1256-1257 b}",
      journal = {\apjl},
         year = 2020,
        month = apr,
       volume = {893},
       number = {2},
          eid = {L30},
        pages = {L30},
          doi = {10.3847/2041-8213/ab8197},
archivePrefix = {arXiv},
       eprint = {2004.05170},
 primaryClass = {astro-ph.EP},
       adsurl = {https://ui.adsabs.harvard.edu/abs/2020ApJ...893L..30B}
}

@ARTICLE{Eriksson2019,
       author = {{Eriksson}, Simon C. and {Janson}, Markus and {Calissendorff}, Per},
        title = "{Detection of new strongly variable brown dwarfs in the L/T transition}",
      journal = {\aap},
         year = 2019,
        month = sep,
       volume = {629},
          eid = {A145},
        pages = {A145},
          doi = {10.1051/0004-6361/201935671},
archivePrefix = {arXiv},
       eprint = {1910.02638},
 primaryClass = {astro-ph.EP},
       adsurl = {https://ui.adsabs.harvard.edu/abs/2019A&A...629A.145E}
}

@ARTICLE{Sutlieff2023,
       author = {{Sutlieff}, Ben J. and {Birkby}, Jayne L. and {Stone}, Jordan M. and {Doelman}, David S. and {Kenworthy}, Matthew A. and {Panwar}, Vatsal and {Bohn}, Alexander J. and {Ertel}, Steve and {Snik}, Frans and {Woodward}, Charles E. and {Skemer}, Andrew J. and {Leisenring}, Jarron M. and {Strassmeier}, Klaus G. and {Charbonneau}, David},
        title = "{Measuring the variability of directly imaged exoplanets using vector Apodizing Phase Plates combined with ground-based differential spectrophotometry}",
      journal = {\mnras},
         year = 2023,
        month = apr,
       volume = {520},
       number = {3},
        pages = {4235-4257},
          doi = {10.1093/mnras/stad249},
archivePrefix = {arXiv},
       eprint = {2301.08689},
 primaryClass = {astro-ph.EP},
       adsurl = {https://ui.adsabs.harvard.edu/abs/2023MNRAS.520.4235S}
}

@ARTICLE{Sutlieff2024,
       author = {{Sutlieff}, Ben J. and {Birkby}, Jayne L. and {Stone}, Jordan M. and {Derkink}, Annelotte and {Backs}, Frank and {Doelman}, David S. and {Kenworthy}, Matthew A. and {Bohn}, Alexander J. and {Ertel}, Steve and {Snik}, Frans and {Woodward}, Charles E. and {Ilyin}, Ilya and {Skemer}, Andrew J. and {Leisenring}, Jarron M. and {Strassmeier}, Klaus G. and {Wang}, Ji and {Charbonneau}, David and {Biller}, Beth A.},
        title = "{Exploring the directly imaged HD 1160 system through spectroscopic characterization and high-cadence variability monitoring}",
      journal = {\mnras},
         year = 2024,
        month = jun,
       volume = {531},
       number = {1},
        pages = {2168-2189},
          doi = {10.1093/mnras/stae1315},
archivePrefix = {arXiv},
       eprint = {2405.12271},
 primaryClass = {astro-ph.SR},
       adsurl = {https://ui.adsabs.harvard.edu/abs/2024MNRAS.531.2168S}
}

@ARTICLE{Biller2024,
       author = {{Biller}, Beth A. and {Vos}, Johanna M. and {Zhou}, Yifan and {McCarthy}, Allison M. and {Tan}, Xianyu and {Crossfield}, Ian J.~M. and {Whiteford}, Niall and {Suarez}, Genaro and {Faherty}, Jacqueline and {Manjavacas}, Elena and {Chen}, Xueqing and {Liu}, Pengyu and {Sutlieff}, Ben J. and {Limbach}, Mary Anne and {Molliere}, Paul and {Dupuy}, Trent J. and {Oliveros-Gomez}, Natalia and {Muirhead}, Philip S. and {Henning}, Thomas and {Mace}, Gregory and {Crouzet}, Nicolas and {Karalidi}, Theodora and {Morley}, Caroline V. and {Tremblin}, Pascal and {Kataria}, Tiffany},
        title = "{The JWST weather report from the nearest brown dwarfs I: multiperiod JWST NIRSpec + MIRI monitoring of the benchmark binary brown dwarf WISE 1049AB}",
      journal = {\mnras},
         year = 2024,
        month = aug,
       volume = {532},
       number = {2},
        pages = {2207-2233},
          doi = {10.1093/mnras/stae1602},
archivePrefix = {arXiv},
       eprint = {2407.09194},
 primaryClass = {astro-ph.SR},
       adsurl = {https://ui.adsabs.harvard.edu/abs/2024MNRAS.532.2207B}
}

@ARTICLE{Gardner2006,
       author = {{Gardner}, Jonathan P. and {Mather}, John C. and {Clampin}, Mark and {Doyon}, Rene and {Greenhouse}, Matthew A. and {Hammel}, Heidi B. and {Hutchings}, John B. and {Jakobsen}, Peter and {Lilly}, Simon J. and {Long}, Knox S. and {Lunine}, Jonathan I. and {McCaughrean}, Mark J. and {Mountain}, Matt and {Nella}, John and {Rieke}, George H. and {Rieke}, Marcia J. and {Rix}, Hans-Walter and {Smith}, Eric P. and {Sonneborn}, George and {Stiavelli}, Massimo and {Stockman}, H.~S. and {Windhorst}, Rogier A. and {Wright}, Gillian S.},
        title = "{The James Webb Space Telescope}",
      journal = {\ssr},
         year = 2006,
        month = apr,
       volume = {123},
       number = {4},
        pages = {485-606},
          doi = {10.1007/s11214-006-8315-7},
archivePrefix = {arXiv},
       eprint = {astro-ph/0606175},
 primaryClass = {astro-ph},
       adsurl = {https://ui.adsabs.harvard.edu/abs/2006SSRv..123..485G}
}

@INPROCEEDINGS{Rieke2003,
       author = {{Rieke}, Marcia J. and {Baum}, Stefi A. and {Beichman}, Charles A. and {Crampton}, David and {Doyon}, Rene and {Eisenstein}, Daniel and {Greene}, Thomas P. and {Hodapp}, Klaus-Werner and {Horner}, Scott D. and {Johnstone}, Doug and {Lesyna}, Lawrence and {Lilly}, Simon and {Meyer}, Michael and {Martin}, Peter and {McCarthy}, Jr., Donald W. and {Rieke}, George H. and {Roellig}, Thomas L. and {Stauffer}, John and {Trauger}, John T. and {Young}, Erick T.},
        title = "{NGST NIRCam Scientific Program and Design Concept}",
    booktitle = {IR Space Telescopes and Instruments},
         year = 2003,
       editor = {{Mather}, John C.},
       series = {Society of Photo-Optical Instrumentation Engineers (SPIE) Conference Series},
       volume = {4850},
        month = mar,
        pages = {478-485},
          doi = {10.1117/12.489103},
       adsurl = {https://ui.adsabs.harvard.edu/abs/2003SPIE.4850..478R}
}

@ARTICLE{Rieke2023a,
       author = {{Rieke}, Marcia J. and {Kelly}, Douglas M. and {Misselt}, Karl and {Stansberry}, John and {Boyer}, Martha and {Beatty}, Thomas and {Egami}, Eiichi and {Florian}, Michael and {Greene}, Thomas P. and {Hainline}, Kevin and {Leisenring}, Jarron and {Roellig}, Thomas and {Schlawin}, Everett and {Sun}, Fengwu and {Tinnin}, Lee and {Williams}, Christina C. and {Willmer}, Christopher N.~A. and {Wilson}, Debra and {Clark}, Charles R. and {Rohrbach}, Scott and {Brooks}, Brian and {Canipe}, Alicia and {Correnti}, Matteo and {DiFelice}, Audrey and {Gennaro}, Mario and {Girard}, Julien H. and {Hartig}, George and {Hilbert}, Bryan and {Koekemoer}, Anton M. and {Nikolov}, Nikolay K. and {Pirzkal}, Norbert and {Rest}, Armin and {Robberto}, Massimo and {Sunnquist}, Ben and {Telfer}, Randal and {Wu}, Chi Rai and {Ferry}, Malcolm and {Lewis}, Dan and {Baum}, Stefi and {Beichman}, Charles and {Doyon}, Ren{\'e} and {Dressler}, Alan and {Eisenstein}, Daniel J. and {Ferrarese}, Laura and {Hodapp}, Klaus and {Horner}, Scott and {Jaffe}, Daniel T. and {Johnstone}, Doug and {Krist}, John and {Martin}, Peter and {McCarthy}, Donald W. and {Meyer}, Michael and {Rieke}, George H. and {Trauger}, John and {Young}, Erick T.},
        title = "{Performance of NIRCam on JWST in Flight}",
      journal = {\pasp},
         year = 2023,
        month = feb,
       volume = {135},
       number = {1044},
          eid = {028001},
        pages = {028001},
          doi = {10.1088/1538-3873/acac53},
archivePrefix = {arXiv},
       eprint = {2212.12069},
 primaryClass = {astro-ph.IM},
       adsurl = {https://ui.adsabs.harvard.edu/abs/2023PASP..135b8001R}
}

@ARTICLE{Jakobsen2022,
       author = {{Jakobsen}, P. and {Ferruit}, P. and {Alves de Oliveira}, C. and {Arribas}, S. and {Bagnasco}, G. and {Barho}, R. and {Beck}, T.~L. and {Birkmann}, S. and {B{\"o}ker}, T. and {Bunker}, A.~J. and {Charlot}, S. and {de Jong}, P. and {de Marchi}, G. and {Ehrenwinkler}, R. and {Falcolini}, M. and {Fels}, R. and {Franx}, M. and {Franz}, D. and {Funke}, M. and {Giardino}, G. and {Gnata}, X. and {Holota}, W. and {Honnen}, K. and {Jensen}, P.~L. and {Jentsch}, M. and {Johnson}, T. and {Jollet}, D. and {Karl}, H. and {Kling}, G. and {K{\"o}hler}, J. and {Kolm}, M.-G. and {Kumari}, N. and {Lander}, M.~E. and {Lemke}, R. and {L{\'o}pez-Caniego}, M. and {L{\"u}tzgendorf}, N. and {Maiolino}, R. and {Manjavacas}, E. and {Marston}, A. and {Maschmann}, M. and {Maurer}, R. and {Messerschmidt}, B. and {Moseley}, S.~H. and {Mosner}, P. and {Mott}, D.~B. and {Muzerolle}, J. and {Pirzkal}, N. and {Pittet}, J.-F. and {Plitzke}, A. and {Posselt}, W. and {Rapp}, B. and {Rauscher}, B.~J. and {Rawle}, T. and {Rix}, H.-W. and {R{\"o}del}, A. and {Rumler}, P. and {Sabbi}, E. and {Salvignol}, J.-C. and {Schmid}, T. and {Sirianni}, M. and {Smith}, C. and {Strada}, P. and {te Plate}, M. and {Valenti}, J. and {Wettemann}, T. and {Wiehe}, T. and {Wiesmayer}, M. and {Willott}, C.~J. and {Wright}, R. and {Zeidler}, P. and {Zincke}, C.},
        title = "{The Near-Infrared Spectrograph (NIRSpec) on the James Webb Space Telescope. I. Overview of the instrument and its capabilities}",
      journal = {\aap},
         year = 2022,
        month = may,
       volume = {661},
          eid = {A80},
        pages = {A80},
          doi = {10.1051/0004-6361/202142663},
archivePrefix = {arXiv},
       eprint = {2202.03305},
 primaryClass = {astro-ph.IM},
       adsurl = {https://ui.adsabs.harvard.edu/abs/2022A&A...661A..80J}
}

@ARTICLE{Boker2023,
       author = {{B{\"o}ker}, T. and {Beck}, T.~L. and {Birkmann}, S.~M. and {Giardino}, G. and {Keyes}, C. and {Kumari}, N. and {Muzerolle}, J. and {Rawle}, T. and {Zeidler}, P. and {Abul-Huda}, Y. and {Alves de Oliveira}, C. and {Arribas}, S. and {Bechtold}, K. and {Bhatawdekar}, R. and {Bonaventura}, N. and {Bunker}, A.~J. and {Cameron}, A.~J. and {Carniani}, S. and {Charlot}, S. and {Curti}, M. and {Espinoza}, N. and {Ferruit}, P. and {Franx}, M. and {Jakobsen}, P. and {Karakla}, D. and {L{\'o}pez-Caniego}, M. and {L{\"u}tzgendorf}, N. and {Maiolino}, R. and {Manjavacas}, E. and {Marston}, A.~P. and {Moseley}, S.~H. and {Ogle}, P. and {Perna}, M. and {Pe{\~n}a-Guerrero}, M. and {Pirzkal}, N. and {Plesha}, R. and {Proffitt}, C.~R. and {Rauscher}, B.~J. and {Rix}, H.-W. and {Rodr{\'\i}guez del Pino}, B. and {Rustamkulov}, Z. and {Sabbi}, E. and {Sing}, D.~K. and {Sirianni}, M. and {te Plate}, M. and {{\'U}beda}, L. and {Wahlgren}, G.~M. and {Wislowski}, E. and {Wu}, R. and {Willott}, Chris J.},
        title = "{In-orbit Performance of the Near-infrared Spectrograph NIRSpec on the James Webb Space Telescope}",
      journal = {\pasp},
         year = 2023,
        month = mar,
       volume = {135},
       number = {1045},
          eid = {038001},
        pages = {038001},
          doi = {10.1088/1538-3873/acb846},
archivePrefix = {arXiv},
       eprint = {2301.13766},
 primaryClass = {astro-ph.IM},
       adsurl = {https://ui.adsabs.harvard.edu/abs/2023PASP..135c8001B}
}

@INPROCEEDINGS{Perrin2012,
       author = {{Perrin}, Marshall D. and {Soummer}, R{\'e}mi and {Elliott}, Erin M. and {Lallo}, Matthew D. and {Sivaramakrishnan}, Anand},
        title = "{Simulating point spread functions for the James Webb Space Telescope with WebbPSF}",
    booktitle = {Space Telescopes and Instrumentation 2012: Optical, Infrared, and Millimeter Wave},
         year = 2012,
       editor = {{Clampin}, Mark C. and {Fazio}, Giovanni G. and {MacEwen}, Howard A. and {Oschmann}, Jr., Jacobus M.},
       series = {Society of Photo-Optical Instrumentation Engineers (SPIE) Conference Series},
       volume = {8442},
        month = sep,
          eid = {84423D},
        pages = {84423D},
          doi = {10.1117/12.925230},
       adsurl = {https://ui.adsabs.harvard.edu/abs/2012SPIE.8442E..3DP}
}

@INPROCEEDINGS{Perrin2014,
       author = {{Perrin}, Marshall D. and {Sivaramakrishnan}, Anand and {Lajoie}, Charles-Philippe and {Elliott}, Erin and {Pueyo}, Laurent and {Ravindranath}, Swara and {Albert}, Lo{\"\i}c.},
        title = "{Updated point spread function simulations for JWST with WebbPSF}",
    booktitle = {Space Telescopes and Instrumentation 2014: Optical, Infrared, and Millimeter Wave},
         year = 2014,
       editor = {{Oschmann}, Jr., Jacobus M. and {Clampin}, Mark and {Fazio}, Giovanni G. and {MacEwen}, Howard A.},
       series = {Society of Photo-Optical Instrumentation Engineers (SPIE) Conference Series},
       volume = {9143},
        month = aug,
          eid = {91433X},
        pages = {91433X},
          doi = {10.1117/12.2056689},
       adsurl = {https://ui.adsabs.harvard.edu/abs/2014SPIE.9143E..3XP}
}

@ARTICLE{Gandolfi2026b,
       author = {{Gandolfi}, G. and {Rodighiero}, G. and {Bisigello}, L. and {Grazian}, A. and {Finkelstein}, S.~L. and {Dickinson}, M. and {Castellano}, M. and {Merlin}, E. and {Calabr{\`o}}, A. and {Papovich}, C. and {Bianchetti}, A. and {Ba{\~n}ados}, E. and {Benotto}, P. and {Catone}, M. and {Buitrago}, F. and {Daddi}, E. and {Girardi}, G. and {Giulietti}, M. and {Hirschmann}, M. and {Holwerda}, B.~W. and {Arrabal Haro}, P. and {Lapi}, A. and {Lucas}, R.~A. and {Lyu}, Y. and {Massardi}, M. and {Pacucci}, F. and {P{\'e}rez-Gonz{\'a}lez}, P.~G. and {Ronconi}, T. and {Tarrasse}, M. and {Wilkins}, S. and {Vulcani}, B. and {Yung}, L.~Y.~A. and {Zavala}, J.~A. and {Backhaus}, B. and {Bagley}, M. and {Buat}, V. and {Burgarella}, D. and {Kartaltepe}, J. and {Khusanova}, Y. and {Kirkpatrick}, A. and {Kocevski}, D. and {Koekemoer}, A.~M. and {Lambrides}, E. and {Pirzkal}, N. and {Yang}, G.},
        title = "{Ultrahigh-redshift or closer-by, dust-obscured galaxies?: Deciphering the nature of faint, previously missed F200W dropouts in CEERS}",
      journal = {\aap},
         year = 2026,
        month = apr,
       volume = {708},
          eid = {A195},
        pages = {A195},
          doi = {10.1051/0004-6361/202554009},
archivePrefix = {arXiv},
       eprint = {2502.02637},
 primaryClass = {astro-ph.GA},
       adsurl = {https://ui.adsabs.harvard.edu/abs/2026A&A...708A.195G}
}

@ARTICLE{Burningham2013,
       author = {{Burningham}, Ben and {Cardoso}, C.~V. and {Smith}, L. and {Leggett}, S.~K. and {Smart}, R.~L. and {Mann}, A.~W. and {Dhital}, S. and {Lucas}, P.~W. and {Tinney}, C.~G. and {Pinfield}, D.~J. and {Zhang}, Z. and {Morley}, C. and {Saumon}, D. and {Aller}, K. and {Littlefair}, S.~P. and {Homeier}, D. and {Lodieu}, N. and {Deacon}, N. and {Marley}, M.~S. and {van Spaandonk}, L. and {Baker}, D. and {Allard}, F. and {Andrei}, A.~H. and {Canty}, J. and {Clarke}, J. and {Day-Jones}, A.~C. and {Dupuy}, T. and {Fortney}, J.~J. and {Gomes}, J. and {Ishii}, M. and {Jones}, H.~R.~A. and {Liu}, M. and {Magazz{\'u}}, A. and {Marocco}, F. and {Murray}, D.~N. and {Rojas-Ayala}, B. and {Tamura}, M.},
        title = "{76 T dwarfs from the UKIDSS LAS: benchmarks, kinematics and an updated space density}",
      journal = {\mnras},
         year = 2013,
        month = jul,
       volume = {433},
       number = {1},
        pages = {457-497},
          doi = {10.1093/mnras/stt740},
archivePrefix = {arXiv},
       eprint = {1304.7246},
 primaryClass = {astro-ph.SR},
       adsurl = {https://ui.adsabs.harvard.edu/abs/2013MNRAS.433..457B}
}

@ARTICLE{Luhman2011,
       author = {{Luhman}, K.~L. and {Burgasser}, A.~J. and {Bochanski}, J.~J.},
        title = "{Discovery of a Candidate for the Coolest Known Brown Dwarf}",
      journal = {\apjl},
         year = 2011,
        month = mar,
       volume = {730},
       number = {1},
          eid = {L9},
        pages = {L9},
          doi = {10.1088/2041-8205/730/1/L9},
archivePrefix = {arXiv},
       eprint = {1102.5411},
 primaryClass = {astro-ph.GA},
       adsurl = {https://ui.adsabs.harvard.edu/abs/2011ApJ...730L...9L}
}

@ARTICLE{Beiler2024,
       author = {{Beiler}, Samuel A. and {Cushing}, Michael C. and {Kirkpatrick}, J. Davy and {Schneider}, Adam C. and {Mukherjee}, Sagnick and {Marley}, Mark S. and {Marocco}, Federico and {Smart}, Richard L.},
        title = "{Precise Bolometric Luminosities and Effective Temperatures of 23 Late-T and Y Dwarfs Obtained with JWST}",
      journal = {\apj},
         year = 2024,
        month = oct,
       volume = {973},
       number = {2},
          eid = {107},
        pages = {107},
          doi = {10.3847/1538-4357/ad6301},
archivePrefix = {arXiv},
       eprint = {2407.08518},
 primaryClass = {astro-ph.SR},
       adsurl = {https://ui.adsabs.harvard.edu/abs/2024ApJ...973..107B}
}

@ARTICLE{Beiler2024b,
       author = {{Beiler}, Samuel A. and {Mukherjee}, Sagnick and {Cushing}, Michael C. and {Kirkpatrick}, J. Davy and {Schneider}, Adam C. and {Kothari}, Harshil and {Marley}, Mark S. and {Visscher}, Channon},
        title = "{A Tale of Two Molecules: The Underprediction of CO$_{2}$ and Overprediction of PH$_{3}$ in Late T and Y Dwarf Atmospheric Models}",
      journal = {\apj},
         year = 2024,
        month = sep,
       volume = {973},
       number = {1},
          eid = {60},
        pages = {60},
          doi = {10.3847/1538-4357/ad6759},
archivePrefix = {arXiv},
       eprint = {2407.15950},
 primaryClass = {astro-ph.EP},
       adsurl = {https://ui.adsabs.harvard.edu/abs/2024ApJ...973...60B}
}

@ARTICLE{Albert2025,
       author = {{Albert}, Lo{\"\i}c and {Leggett}, Sandy K. and {Calissendorff}, Per and {Vandal}, Thomas and {Kirkpatrick}, J. Davy and {Bardalez Gagliuffi}, Daniella C. and {De Furio}, Matthew and {Meyer}, Michael and {Beichman}, Charles A. and {Burgasser}, Adam J. and {Cushing}, Michael C. and {Faherty}, Jacqueline Kelly and {Fontanive}, Cl{\'e}mence and {Gelino}, Christopher R. and {Gizis}, John E. and {Greenbaum}, Alexandra Z. and {Martinache}, Frantz and {N'Diaye}, Mamadou and {Pope}, Benjamin J.~S. and {Roellig}, Thomas L. and {Sahlmann}, Johannes and {Sivaramakrishnan}, Anand and {Ygouf}, Marie},
        title = "{JWST 1.5 {\ensuremath{\mu}}m and 4.8 {\ensuremath{\mu}}m Photometry of Y Dwarfs}",
      journal = {\aj},
         year = 2025,
        month = mar,
       volume = {169},
       number = {3},
          eid = {163},
        pages = {163},
          doi = {10.3847/1538-3881/adadf9},
archivePrefix = {arXiv},
       eprint = {2501.14100},
 primaryClass = {astro-ph.SR},
       adsurl = {https://ui.adsabs.harvard.edu/abs/2025AJ....169..163A}
}

@ARTICLE{Dupuy2012,
       author = {{Dupuy}, Trent J. and {Liu}, Michael C.},
        title = "{The Hawaii Infrared Parallax Program. I. Ultracool Binaries and the L/T Transition}",
      journal = {\apjs},
         year = 2012,
        month = aug,
       volume = {201},
       number = {2},
          eid = {19},
        pages = {19},
          doi = {10.1088/0067-0049/201/2/19},
archivePrefix = {arXiv},
       eprint = {1201.2465},
 primaryClass = {astro-ph.SR},
       adsurl = {https://ui.adsabs.harvard.edu/abs/2012ApJS..201...19D}
}

@ARTICLE{Glasse2015,
       author = {{Glasse}, Alistair and {Rieke}, G.~H. and {Bauwens}, E. and {Garc{\'\i}a-Mar{\'\i}n}, Macarena and {Ressler}, M.~E. and {Rost}, Steffen and {Tikkanen}, T.~V. and {Vandenbussche}, B. and {Wright}, G.~S.},
        title = "{The Mid-Infrared Instrument for the James Webb Space Telescope, IX: Predicted Sensitivity}",
      journal = {\pasp},
         year = 2015,
        month = jul,
       volume = {127},
       number = {953},
        pages = {686},
          doi = {10.1086/682259},
archivePrefix = {arXiv},
       eprint = {1508.02427},
 primaryClass = {astro-ph.IM},
       adsurl = {https://ui.adsabs.harvard.edu/abs/2015PASP..127..686G}
}

@ARTICLE{Wright2023,
       author = {{Wright}, Gillian S. and {Rieke}, George H. and {Glasse}, Alistair and {Ressler}, Michael and {Garc{\'\i}a Mar{\'\i}n}, Macarena and {Aguilar}, Jonathan and {Alberts}, Stacey and {{\'A}lvarez-M{\'a}rquez}, Javier and {Argyriou}, Ioannis and {Banks}, Kimberly and {Baudoz}, Pierre and {Boccaletti}, Anthony and {Bouchet}, Patrice and {Bouwman}, Jeroen and {Brandl}, Bernard R. and {Breda}, David and {Bright}, Stacey and {Cale}, Steven and {Colina}, Luis and {Cossou}, Christophe and {Coulais}, Alain and {Cracraft}, Misty and {De Meester}, Wim and {Dicken}, Daniel and {Engesser}, Michael and {Etxaluze}, Mireya and {Fox}, Ori D. and {Friedman}, Scott and {Fu}, Henry and {Gasman}, Danny and {G{\'a}sp{\'a}r}, Andr{\'a}s and {Gastaud}, Ren{\'e} and {Geers}, Vincent and {Glauser}, Adrian Michael and {Gordon}, Karl D. and {Greene}, Thomas and {Greve}, Thomas R. and {Grundy}, Timothy and {G{\"u}del}, Manuel and {Guillard}, Pierre and {Haderlein}, Peter and {Hashimoto}, Ryan and {Henning}, Thomas and {Hines}, Dean and {Holler}, Bryan and {Detre}, {\"O}rs Hunor and {Jahromi}, Amir and {James}, Bryan and {Jones}, Olivia C. and {Justtanont}, Kay and {Kavanagh}, Patrick and {Kendrew}, Sarah and {Klaassen}, Pamela and {Krause}, Oliver and {Labiano}, Alvaro and {Lagage}, Pierre-Olivier and {Lambros}, Scott and {Larson}, Kirsten and {Law}, David and {Lee}, David and {Libralato}, Mattia and {Lorenzo Alverez}, Jose and {Meixner}, Margaret and {Morrison}, Jane and {Mueller}, Migo and {Murray}, Katherine and {Mycroft}, Matthew and {Myers}, Richard and {Nayak}, Omnarayani and {Naylor}, Bret and {Nickson}, Bryony and {Noriega-Crespo}, Alberto and {{\"O}stlin}, G{\"o}ran and {O'Sullivan}, Brian and {Ottens}, Richard and {Patapis}, Polychronis and {Penanen}, Konstantin and {Pietraszkiewicz}, Martin and {Ray}, Tom and {Regan}, Michael and {Roteliuk}, Anthony and {Royer}, Pierre and {Samara-Ratna}, Piyal and {Samuelson}, Bridget and {Sargent}, Beth A. and {Scheithauer}, Silvia and {Schneider}, Analyn and {Schreiber}, J{\"u}rgen and {Shaughnessy}, Bryan and {Sheehan}, Evan and {Shivaei}, Irene and {Sloan}, G.~C. and {Tamas}, Laszlo and {Teague}, Kelly and {Temim}, Tea and {Tikkanen}, Tuomo and {Tustain}, Samuel and {van Dishoeck}, Ewine F. and {Vandenbussche}, Bart and {Weilert}, Mark and {Whitehouse}, Paul and {Wolff}, Schuyler},
        title = "{The Mid-infrared Instrument for JWST and Its In-flight Performance}",
      journal = {\pasp},
         year = 2023,
        month = apr,
       volume = {135},
       number = {1046},
          eid = {048003},
        pages = {048003},
          doi = {10.1088/1538-3873/acbe66},
       adsurl = {https://ui.adsabs.harvard.edu/abs/2023PASP..135d8003W}
}

@software{Bushouse2025,
  author       = {Bushouse, Howard and
                  Eisenhamer, Jonathan and
                  Dencheva, Nadia and
                  Davies, James and
                  Greenfield, Perry and
                  Morrison, Jane and
                  Hodge, Phil and
                  Simon, Bernie and
                  Grumm, David and
                  Droettboom, Michael and
                  Slavich, Edward and
                  Sosey, Megan and
                  Pauly, Tyler and
                  Miller, Todd and
                  Jedrzejewski, Robert and
                  Hack, Warren and
                  Davis, David and
                  Crawford, Steven and
                  Law, David and
                  Gordon, Karl and
                  Regan, Michael and
                  Cara, Mihai and
                  MacDonald, Ken and
                  Bradley, Larry and
                  Shanahan, Clare and
                  Jamieson, William and
                  Teodoro, Mairan and
                  Williams, Thomas and
                  Pena-Guerrero, Maria and
                  Graham, Brett and
                  Molter, Edward and
                  Brandt, Timothy and
                  Hayes, Christian and
                  Cooper, Rachel and
                  Clarke, Melanie and
                  Filippazzo, Joseph},
  title        = {JWST Calibration Pipeline},
  month        = nov,
  year         = 2025,
  publisher    = {Zenodo},
  version      = {1.20.2},
  doi          = {10.5281/zenodo.17515973},
  url          = {https://doi.org/10.5281/zenodo.17515973},
  swhid        = {swh:1:dir:fc8e0b17375bd6292e6f2fe6b758a3e44b81aa01
                   ;origin=https://doi.org/10.5281/zenodo.6984365;vis
                   it=swh:1:snp:ee0c72d562544a3226903308be9ab3969e139
                   858;anchor=swh:1:rel:4164a7a6a89b54a00ff5ed5b74c67
                   5e41f26b3fc;path=spacetelescope-jwst-ee52a96
                  },
}

@ARTICLE{Leggett2025,
       author = {{Leggett}, S.~K. and {Phillips}, Mark W. and {Tremblin}, Pascal},
        title = "{Exploration of the Near-infrared Colors of Cold Y Dwarfs from the Ground and Space}",
      journal = {\apj},
         year = 2025,
        month = oct,
       volume = {991},
       number = {2},
          eid = {193},
        pages = {193},
          doi = {10.3847/1538-4357/adf21f},
archivePrefix = {arXiv},
       eprint = {2509.05514},
 primaryClass = {astro-ph.SR},
       adsurl = {https://ui.adsabs.harvard.edu/abs/2025ApJ...991..193L}
}

@ARTICLE{Leggett2026,
       author = {{Leggett}, S.~K.},
        title = "{The Coldest Known Y Dwarfs: Estimates of Their Effective Temperatures}",
      journal = {\apj},
         year = 2026,
        month = may,
       volume = {1002},
       number = {1},
          eid = {113},
        pages = {113},
          doi = {10.3847/1538-4357/ae5814},
archivePrefix = {arXiv},
       eprint = {2603.24740},
 primaryClass = {astro-ph.SR},
       adsurl = {https://ui.adsabs.harvard.edu/abs/2026ApJ..1002..113L}
}

@ARTICLE{astropy,
       author = {{Astropy Collaboration} and {Robitaille}, Thomas P. and {Tollerud}, Erik J. and {Greenfield}, Perry and {Droettboom}, Michael and {Bray}, Erik and {Aldcroft}, Tom and {Davis}, Matt and {Ginsburg}, Adam and {Price-Whelan}, Adrian M. and {Kerzendorf}, Wolfgang E. and {Conley}, Alexander and {Crighton}, Neil and {Barbary}, Kyle and {Muna}, Demitri and {Ferguson}, Henry and {Grollier}, Fr{\'e}d{\'e}ric and {Parikh}, Madhura M. and {Nair}, Prasanth H. and {Unther}, Hans M. and {Deil}, Christoph and {Woillez}, Julien and {Conseil}, Simon and {Kramer}, Roban and {Turner}, James E.~H. and {Singer}, Leo and {Fox}, Ryan and {Weaver}, Benjamin A. and {Zabalza}, Victor and {Edwards}, Zachary I. and {Azalee Bostroem}, K. and {Burke}, D.~J. and {Casey}, Andrew R. and {Crawford}, Steven M. and {Dencheva}, Nadia and {Ely}, Justin and {Jenness}, Tim and {Labrie}, Kathleen and {Lim}, Pey Lian and {Pierfederici}, Francesco and {Pontzen}, Andrew and {Ptak}, Andy and {Refsdal}, Brian and {Servillat}, Mathieu and {Streicher}, Ole},
        title = "{Astropy: A community Python package for astronomy}",
      journal = {\aap},
         year = 2013,
        month = oct,
       volume = {558},
          eid = {A33},
        pages = {A33},
          doi = {10.1051/0004-6361/201322068},
archivePrefix = {arXiv},
       eprint = {1307.6212},
 primaryClass = {astro-ph.IM},
       adsurl = {https://ui.adsabs.harvard.edu/abs/2013A&A...558A..33A}
}

@ARTICLE{matplotlib,
       author = {{Hunter}, John D.},
        title = "{Matplotlib: A 2D Graphics Environment}",
      journal = {Computing in Science and Engineering},
         year = 2007,
        month = may,
       volume = {9},
       number = {3},
        pages = {90-95},
          doi = {10.1109/MCSE.2007.55},
       adsurl = {https://ui.adsabs.harvard.edu/abs/2007CSE.....9...90H}
}

@ARTICLE{numpy,
       author = {{Harris}, Charles R. and {Millman}, K. Jarrod and {van der Walt}, St{\'e}fan J. and {Gommers}, Ralf and {Virtanen}, Pauli and {Cournapeau}, David and {Wieser}, Eric and {Taylor}, Julian and {Berg}, Sebastian and {Smith}, Nathaniel J. and {Kern}, Robert and {Picus}, Matti and {Hoyer}, Stephan and {van Kerkwijk}, Marten H. and {Brett}, Matthew and {Haldane}, Allan and {del R{\'\i}o}, Jaime Fern{\'a}ndez and {Wiebe}, Mark and {Peterson}, Pearu and {G{\'e}rard-Marchant}, Pierre and {Sheppard}, Kevin and {Reddy}, Tyler and {Weckesser}, Warren and {Abbasi}, Hameer and {Gohlke}, Christoph and {Oliphant}, Travis E.},
        title = "{Array programming with NumPy}",
      journal = {\nat},
         year = 2020,
        month = sep,
       volume = {585},
       number = {7825},
        pages = {357-362},
          doi = {10.1038/s41586-020-2649-2},
archivePrefix = {arXiv},
       eprint = {2006.10256},
 primaryClass = {cs.MS},
       adsurl = {https://ui.adsabs.harvard.edu/abs/2020Natur.585..357H}
}

@ARTICLE{scipy,
       author = {{Virtanen}, Pauli and {Gommers}, Ralf and {Oliphant}, Travis E. and {Haberland}, Matt and {Reddy}, Tyler and {Cournapeau}, David and {Burovski}, Evgeni and {Peterson}, Pearu and {Weckesser}, Warren and {Bright}, Jonathan and {van der Walt}, St{\'e}fan J. and {Brett}, Matthew and {Wilson}, Joshua and {Millman}, K. Jarrod and {Mayorov}, Nikolay and {Nelson}, Andrew R.~J. and {Jones}, Eric and {Kern}, Robert and {Larson}, Eric and {Carey}, C.~J. and {Polat}, {\.I}lhan and {Feng}, Yu and {Moore}, Eric W. and {VanderPlas}, Jake and {Laxalde}, Denis and {Perktold}, Josef and {Cimrman}, Robert and {Henriksen}, Ian and {Quintero}, E.~A. and {Harris}, Charles R. and {Archibald}, Anne M. and {Ribeiro}, Ant{\^o}nio H. and {Pedregosa}, Fabian and {van Mulbregt}, Paul and {SciPy 1. 0 Contributors}},
        title = "{SciPy 1.0: fundamental algorithms for scientific computing in Python}",
      journal = {Nature Methods},
         year = 2020,
        month = feb,
       volume = {17},
        pages = {261-272},
          doi = {10.1038/s41592-019-0686-2},
archivePrefix = {arXiv},
       eprint = {1907.10121},
 primaryClass = {cs.MS},
       adsurl = {https://ui.adsabs.harvard.edu/abs/2020NatMe..17..261V}
}

@ARTICLE{Astropy2018,
       author = {{Astropy Collaboration} and {Price-Whelan}, A.~M. and {Sip{\H{o}}cz}, B.~M. and {G{\"u}nther}, H.~M. and {Lim}, P.~L. and {Crawford}, S.~M. and {Conseil}, S. and {Shupe}, D.~L. and {Craig}, M.~W. and {Dencheva}, N. and {Ginsburg}, A. and {VanderPlas}, J.~T. and {Bradley}, L.~D. and {P{\'e}rez-Su{\'a}rez}, D. and {de Val-Borro}, M. and {Aldcroft}, T.~L. and {Cruz}, K.~L. and {Robitaille}, T.~P. and {Tollerud}, E.~J. and {Ardelean}, C. and {Babej}, T. and {Bach}, Y.~P. and {Bachetti}, M. and {Bakanov}, A.~V. and {Bamford}, S.~P. and {Barentsen}, G. and {Barmby}, P. and {Baumbach}, A. and {Berry}, K.~L. and {Biscani}, F. and {Boquien}, M. and {Bostroem}, K.~A. and {Bouma}, L.~G. and {Brammer}, G.~B. and {Bray}, E.~M. and {Breytenbach}, H. and {Buddelmeijer}, H. and {Burke}, D.~J. and {Calderone}, G. and {Cano Rodr{\'\i}guez}, J.~L. and {Cara}, M. and {Cardoso}, J.~V.~M. and {Cheedella}, S. and {Copin}, Y. and {Corrales}, L. and {Crichton}, D. and {D'Avella}, D. and {Deil}, C. and {Depagne}, {\'E}. and {Dietrich}, J.~P. and {Donath}, A. and {Droettboom}, M. and {Earl}, N. and {Erben}, T. and {Fabbro}, S. and {Ferreira}, L.~A. and {Finethy}, T. and {Fox}, R.~T. and {Garrison}, L.~H. and {Gibbons}, S.~L.~J. and {Goldstein}, D.~A. and {Gommers}, R. and {Greco}, J.~P. and {Greenfield}, P. and {Groener}, A.~M. and {Grollier}, F. and {Hagen}, A. and {Hirst}, P. and {Homeier}, D. and {Horton}, A.~J. and {Hosseinzadeh}, G. and {Hu}, L. and {Hunkeler}, J.~S. and {Ivezi{\'c}}, {\v{Z}}. and {Jain}, A. and {Jenness}, T. and {Kanarek}, G. and {Kendrew}, S. and {Kern}, N.~S. and {Kerzendorf}, W.~E. and {Khvalko}, A. and {King}, J. and {Kirkby}, D. and {Kulkarni}, A.~M. and {Kumar}, A. and {Lee}, A. and {Lenz}, D. and {Littlefair}, S.~P. and {Ma}, Z. and {Macleod}, D.~M. and {Mastropietro}, M. and {McCully}, C. and {Montagnac}, S. and {Morris}, B.~M. and {Mueller}, M. and {Mumford}, S.~J. and {Muna}, D. and {Murphy}, N.~A. and {Nelson}, S. and {Nguyen}, G.~H. and {Ninan}, J.~P. and {N{\"o}the}, M. and {Ogaz}, S. and {Oh}, S. and {Parejko}, J.~K. and {Parley}, N. and {Pascual}, S. and {Patil}, R. and {Patil}, A.~A. and {Plunkett}, A.~L. and {Prochaska}, J.~X. and {Rastogi}, T. and {Reddy Janga}, V. and {Sabater}, J. and {Sakurikar}, P. and {Seifert}, M. and {Sherbert}, L.~E. and {Sherwood-Taylor}, H. and {Shih}, A.~Y. and {Sick}, J. and {Silbiger}, M.~T. and {Singanamalla}, S. and {Singer}, L.~P. and {Sladen}, P.~H. and {Sooley}, K.~A. and {Sornarajah}, S. and {Streicher}, O. and {Teuben}, P. and {Thomas}, S.~W. and {Tremblay}, G.~R. and {Turner}, J.~E.~H. and {Terr{\'o}n}, V. and {van Kerkwijk}, M.~H. and {de la Vega}, A. and {Watkins}, L.~L. and {Weaver}, B.~A. and {Whitmore}, J.~B. and {Woillez}, J. and {Zabalza}, V. and {Astropy Contributors}},
        title = "{The Astropy Project: Building an Open-science Project and Status of the v2.0 Core Package}",
      journal = {\aj},
         year = 2018,
        month = sep,
       volume = {156},
       number = {3},
          eid = {123},
        pages = {123},
          doi = {10.3847/1538-3881/aabc4f},
archivePrefix = {arXiv},
       eprint = {1801.02634},
 primaryClass = {astro-ph.IM},
       adsurl = {https://ui.adsabs.harvard.edu/abs/2018AJ....156..123A}
}

@ARTICLE{Astropy2022,
       author = {{Astropy Collaboration} and {Price-Whelan}, Adrian M. and {Lim}, Pey Lian and {Earl}, Nicholas and {Starkman}, Nathaniel and {Bradley}, Larry and {Shupe}, David L. and {Patil}, Aarya A. and {Corrales}, Lia and {Brasseur}, C.~E. and {N{\"o}the}, Maximilian and {Donath}, Axel and {Tollerud}, Erik and {Morris}, Brett M. and {Ginsburg}, Adam and {Vaher}, Eero and {Weaver}, Benjamin A. and {Tocknell}, James and {Jamieson}, William and {van Kerkwijk}, Marten H. and {Robitaille}, Thomas P. and {Merry}, Bruce and {Bachetti}, Matteo and {G{\"u}nther}, H. Moritz and {Aldcroft}, Thomas L. and {Alvarado-Montes}, Jaime A. and {Archibald}, Anne M. and {B{\'o}di}, Attila and {Bapat}, Shreyas and {Barentsen}, Geert and {Baz{\'a}n}, Juanjo and {Biswas}, Manish and {Boquien}, M{\'e}d{\'e}ric and {Burke}, D.~J. and {Cara}, Daria and {Cara}, Mihai and {Conroy}, Kyle E. and {Conseil}, Simon and {Craig}, Matthew W. and {Cross}, Robert M. and {Cruz}, Kelle L. and {D'Eugenio}, Francesco and {Dencheva}, Nadia and {Devillepoix}, Hadrien A.~R. and {Dietrich}, J{\"o}rg P. and {Eigenbrot}, Arthur Davis and {Erben}, Thomas and {Ferreira}, Leonardo and {Foreman-Mackey}, Daniel and {Fox}, Ryan and {Freij}, Nabil and {Garg}, Suyog and {Geda}, Robel and {Glattly}, Lauren and {Gondhalekar}, Yash and {Gordon}, Karl D. and {Grant}, David and {Greenfield}, Perry and {Groener}, Austen M. and {Guest}, Steve and {Gurovich}, Sebastian and {Handberg}, Rasmus and {Hart}, Akeem and {Hatfield-Dodds}, Zac and {Homeier}, Derek and {Hosseinzadeh}, Griffin and {Jenness}, Tim and {Jones}, Craig K. and {Joseph}, Prajwel and {Kalmbach}, J. Bryce and {Karamehmetoglu}, Emir and {Ka{\l}uszy{\'n}ski}, Miko{\l}aj and {Kelley}, Michael S.~P. and {Kern}, Nicholas and {Kerzendorf}, Wolfgang E. and {Koch}, Eric W. and {Kulumani}, Shankar and {Lee}, Antony and {Ly}, Chun and {Ma}, Zhiyuan and {MacBride}, Conor and {Maljaars}, Jakob M. and {Muna}, Demitri and {Murphy}, N.~A. and {Norman}, Henrik and {O'Steen}, Richard and {Oman}, Kyle A. and {Pacifici}, Camilla and {Pascual}, Sergio and {Pascual-Granado}, J. and {Patil}, Rohit R. and {Perren}, Gabriel I. and {Pickering}, Timothy E. and {Rastogi}, Tanuj and {Roulston}, Benjamin R. and {Ryan}, Daniel F. and {Rykoff}, Eli S. and {Sabater}, Jose and {Sakurikar}, Parikshit and {Salgado}, Jes{\'u}s and {Sanghi}, Aniket and {Saunders}, Nicholas and {Savchenko}, Volodymyr and {Schwardt}, Ludwig and {Seifert-Eckert}, Michael and {Shih}, Albert Y. and {Jain}, Anany Shrey and {Shukla}, Gyanendra and {Sick}, Jonathan and {Simpson}, Chris and {Singanamalla}, Sudheesh and {Singer}, Leo P. and {Singhal}, Jaladh and {Sinha}, Manodeep and {Sip{\H{o}}cz}, Brigitta M. and {Spitler}, Lee R. and {Stansby}, David and {Streicher}, Ole and {{\v{S}}umak}, Jani and {Swinbank}, John D. and {Taranu}, Dan S. and {Tewary}, Nikita and {Tremblay}, Grant R. and {de Val-Borro}, Miguel and {Van Kooten}, Samuel J. and {Vasovi{\'c}}, Zlatan and {Verma}, Shresth and {de Miranda Cardoso}, Jos{\'e} Vin{\'\i}cius and {Williams}, Peter K.~G. and {Wilson}, Tom J. and {Winkel}, Benjamin and {Wood-Vasey}, W.~M. and {Xue}, Rui and {Yoachim}, Peter and {Zhang}, Chen and {Zonca}, Andrea and {Astropy Project Contributors}},
        title = "{The Astropy Project: Sustaining and Growing a Community-oriented Open-source Project and the Latest Major Release (v5.0) of the Core Package}",
      journal = {\apj},
         year = 2022,
        month = aug,
       volume = {935},
       number = {2},
          eid = {167},
        pages = {167},
          doi = {10.3847/1538-4357/ac7c74},
archivePrefix = {arXiv},
       eprint = {2206.14220},
 primaryClass = {astro-ph.IM},
       adsurl = {https://ui.adsabs.harvard.edu/abs/2022ApJ...935..167A}
}

@software{photutils,
  author       = {Bradley, Larry and
                  Sipőcz, Brigitta M. and
                  Robitaille, T. P. and
                  Tollerud, E. J. and
                  Vinícius, Zé and
                  Deil, Christoph and
                  Barbary, Kyle and
                  Wilson, Tom J. and
                  Busko, Ivo and
                  Donath, Axel and
                  Günther, Hans Moritz and
                  Cara, Mihai and
                  Lim, P. L. and
                  Meßlinger, Sebastian and
                  Conseil, Simon and
                  Droettboom, Michael and
                  Bostroem, K. Azalee and
                  Bray, E. M. and
                  Bratholm, Lars Andersen and
                  Burnett, Zach and
                  Jamieson, William and
                  Ginsburg, Adam and
                  Taranu, Dan and
                  Barentsen, Geert and
                  Craig, Matthew W. and
                  Morris, Brett M. and
                  Perrin, Marshall and
                  Rathi, Shivangee},
  title        = {Photutils},
  month        = apr,
  year         = 2026,
  publisher    = {Zenodo},
  version      = {3.0.0},
  doi          = {10.5281/zenodo.19636730},
  url          = {https://doi.org/10.5281/zenodo.19636730},
  swhid        = {swh:1:dir:5ea90432cd987336a26af201c1b50343b12e2daa
                   ;origin=https://doi.org/10.5281/zenodo.596036;visi
                   t=swh:1:snp:25dd84d95353ec3813baa8785c248017da7192
                   33;anchor=swh:1:rel:253cbde27398f9aea9d7368f386661
                   8c00fe16f9;path=astropy-photutils-3322558
                  },
}

@ARTICLE{Luhman2014,
       author = {{Luhman}, K.~L.},
        title = "{Discovery of a \raisebox{-0.5ex}\textasciitilde250 K Brown Dwarf at 2 pc from the Sun}",
      journal = {\apjl},
         year = 2014,
        month = may,
       volume = {786},
       number = {2},
          eid = {L18},
        pages = {L18},
          doi = {10.1088/2041-8205/786/2/L18},
archivePrefix = {arXiv},
       eprint = {1404.6501},
 primaryClass = {astro-ph.GA},
       adsurl = {https://ui.adsabs.harvard.edu/abs/2014ApJ...786L..18L}
}

@ARTICLE{Luhman2024,
       author = {{Luhman}, K.~L. and {Tremblin}, P. and {Alves de Oliveira}, C. and {Birkmann}, S.~M. and {Baraffe}, I. and {Chabrier}, G. and {Manjavacas}, E. and {Parker}, R.~J. and {Valenti}, J.},
        title = "{JWST/NIRSpec Observations of the Coldest Known Brown Dwarf}",
      journal = {\aj},
         year = 2024,
        month = jan,
       volume = {167},
       number = {1},
          eid = {5},
        pages = {5},
          doi = {10.3847/1538-3881/ad0b72},
archivePrefix = {arXiv},
       eprint = {2311.17316},
 primaryClass = {astro-ph.SR},
       adsurl = {https://ui.adsabs.harvard.edu/abs/2024AJ....167....5L}
}

@INPROCEEDINGS{Dunlop2013,
       author = {{Dunlop}, James S.},
        title = "{Observing the First Galaxies}",
    booktitle = {The First Galaxies},
         year = 2013,
       editor = {{Wiklind}, Tommy and {Mobasher}, Bahram and {Bromm}, Volker},
       series = {Astrophysics and Space Science Library},
       volume = {396},
        month = jan,
        pages = {223},
          doi = {10.1007/978-3-642-32362-1_5},
archivePrefix = {arXiv},
       eprint = {1205.1543},
 primaryClass = {astro-ph.CO},
       adsurl = {https://ui.adsabs.harvard.edu/abs/2013ASSL..396..223D}
}

@ARTICLE{Burrows2001,
       author = {{Burrows}, Adam and {Hubbard}, W.~B. and {Lunine}, J.~I. and {Liebert}, James},
        title = "{The theory of brown dwarfs and extrasolar giant planets}",
      journal = {Reviews of Modern Physics},
         year = 2001,
        month = jul,
       volume = {73},
       number = {3},
        pages = {719-765},
          doi = {10.1103/RevModPhys.73.719},
archivePrefix = {arXiv},
       eprint = {astro-ph/0103383},
 primaryClass = {astro-ph},
       adsurl = {https://ui.adsabs.harvard.edu/abs/2001RvMP...73..719B}
}

@ARTICLE{Chabrier2023,
       author = {{Chabrier}, Gilles and {Baraffe}, Isabelle and {Phillips}, Mark and {Debras}, Florian},
        title = "{Impact of a new H/He equation of state on the evolution of massive brown dwarfs. New determination of the hydrogen burning limit}",
      journal = {\aap},
         year = 2023,
        month = mar,
       volume = {671},
          eid = {A119},
        pages = {A119},
          doi = {10.1051/0004-6361/202243832},
archivePrefix = {arXiv},
       eprint = {2212.07153},
 primaryClass = {astro-ph.SR},
       adsurl = {https://ui.adsabs.harvard.edu/abs/2023A&A...671A.119C}
}

@ARTICLE{Best2021,
       author = {{Best}, William M.~J. and {Liu}, Michael C. and {Magnier}, Eugene A. and {Dupuy}, Trent J.},
        title = "{A Volume-limited Sample of Ultracool Dwarfs. I. Construction, Space Density, and a Gap in the L/T Transition}",
      journal = {\aj},
         year = 2021,
        month = jan,
       volume = {161},
       number = {1},
          eid = {42},
        pages = {42},
          doi = {10.3847/1538-3881/abc893},
archivePrefix = {arXiv},
       eprint = {2010.15853},
 primaryClass = {astro-ph.SR},
       adsurl = {https://ui.adsabs.harvard.edu/abs/2021AJ....161...42B}
}

@ARTICLE{Kirkpatrick2024,
       author = {{Kirkpatrick}, J. Davy and {Marocco}, Federico and {Gelino}, Christopher R. and {Raghu}, Yadukrishna and {Faherty}, Jacqueline K. and {Bardalez Gagliuffi}, Daniella C. and {Schurr}, Steven D. and {Apps}, Kevin and {Schneider}, Adam C. and {Meisner}, Aaron M. and {Kuchner}, Marc J. and {Caselden}, Dan and {Smart}, R.~L. and {Casewell}, S.~L. and {Raddi}, Roberto and {Kesseli}, Aurora and {Stevnbak Andersen}, Nikolaj and {Antonini}, Edoardo and {Beaulieu}, Paul and {Bickle}, Thomas P. and {Bilsing}, Martin and {Chieng}, Raymond and {Colin}, Guillaume and {Deen}, Sam and {Dereveanco}, Alexandru and {Doll}, Katharina and {Durantini Luca}, Hugo A. and {Frazer}, Anya and {Gantier}, Jean Marc and {Gramaize}, L{\'e}opold and {Grant}, Kristin and {Hamlet}, Leslie K. and {Higashimura}, Hiro and {Hyogo}, Michiharu and {Ja{\l}owiczor}, Peter A. and {Jonkeren}, Alexander and {Kabatnik}, Martin and {Kiwy}, Frank and {Martin}, David W. and {Michaels}, Marianne N. and {Pendrill}, William and {Pessanha Machado}, Celso and {Pumphrey}, Benjamin and {Rothermich}, Austin and {Russwurm}, Rebekah and {Sainio}, Arttu and {Sanchez}, John and {Sapelkin-Tambling}, Fyodor Theo and {Sch{\"u}mann}, J{\"o}rg and {Selg-Mann}, Karl and {Singh}, Harshdeep and {Stenner}, Andres and {Sun}, Guoyou and {Tanner}, Christopher and {Th{\'e}venot}, Melina and {Ventura}, Maurizio and {Voloshin}, Nikita V. and {Walla}, Jim and {W{\k{e}}dracki}, Zbigniew and {Adorno}, Jose I. and {Aganze}, Christian and {Allers}, Katelyn N. and {Brooks}, Hunter and {Burgasser}, Adam J. and {Calamari}, Emily and {Connor}, Thomas and {Costa}, Edgardo and {Eisenhardt}, Peter R. and {Gagn{\'e}}, Jonathan and {Gerasimov}, Roman and {Gonzales}, Eileen C. and {Hsu}, Chih-Chun and {Kiman}, Rocio and {Li}, Guodong and {Low}, Ryan and {Mamajek}, Eric and {Pantoja}, Blake M. and {Popinchalk}, Mark and {Rees}, Jon M. and {Stern}, Daniel and {Su{\'a}rez}, Genaro and {Theissen}, Christopher and {Tsai}, Chao-Wei and {Vos}, Johanna M. and {Zurek}, David and {The Backyard Worlds: Planet 9 Collaboration}},
        title = "{The Initial Mass Function Based on the Full-sky 20 pc Census of {\ensuremath{\sim}}3600 Stars and Brown Dwarfs}",
      journal = {\apjs},
         year = 2024,
        month = apr,
       volume = {271},
       number = {2},
          eid = {55},
        pages = {55},
          doi = {10.3847/1538-4365/ad24e2},
archivePrefix = {arXiv},
       eprint = {2312.03639},
 primaryClass = {astro-ph.SR},
       adsurl = {https://ui.adsabs.harvard.edu/abs/2024ApJS..271...55K}
}

@ARTICLE{Marley2021,
       author = {{Marley}, Mark S. and {Saumon}, Didier and {Visscher}, Channon and {Lupu}, Roxana and {Freedman}, Richard and {Morley}, Caroline and {Fortney}, Jonathan J. and {Seay}, Christopher and {Smith}, Adam J.~R.~W. and {Teal}, D.~J. and {Wang}, Ruoyan},
        title = "{The Sonora Brown Dwarf Atmosphere and Evolution Models. I. Model Description and Application to Cloudless Atmospheres in Rainout Chemical Equilibrium}",
      journal = {\apj},
         year = 2021,
        month = oct,
       volume = {920},
       number = {2},
          eid = {85},
        pages = {85},
          doi = {10.3847/1538-4357/ac141d},
archivePrefix = {arXiv},
       eprint = {2107.07434},
 primaryClass = {astro-ph.SR},
       adsurl = {https://ui.adsabs.harvard.edu/abs/2021ApJ...920...85M}
}

@ARTICLE{Marsh2013,
       author = {{Marsh}, Kenneth A. and {Wright}, Edward L. and {Kirkpatrick}, J. Davy and {Gelino}, Christopher R. and {Cushing}, Michael C. and {Griffith}, Roger L. and {Skrutskie}, Michael F. and {Eisenhardt}, Peter R.},
        title = "{Parallaxes and Proper Motions of Ultracool Brown Dwarfs of Spectral Types Y and Late T}",
      journal = {\apj},
         year = 2013,
        month = jan,
       volume = {762},
       number = {2},
          eid = {119},
        pages = {119},
          doi = {10.1088/0004-637X/762/2/119},
archivePrefix = {arXiv},
       eprint = {1211.6977},
 primaryClass = {astro-ph.SR},
       adsurl = {https://ui.adsabs.harvard.edu/abs/2013ApJ...762..119M}
}

@ARTICLE{Martin2018,
       author = {{Martin}, Emily C. and {Kirkpatrick}, J. Davy and {Beichman}, Charles A. and {Smart}, Richard L. and {Faherty}, Jacqueline K. and {Gelino}, Christopher R. and {Cushing}, Michael C. and {Schneider}, Adam C. and {Wright}, Edward L. and {Lowrance}, Patrick and {Ingalls}, James and {Tinney}, C.~G. and {McLean}, Ian S. and {Logsdon}, Sarah E. and {Lebreton}, J{\'e}r{\'e}my},
        title = "{Y Dwarf Trigonometric Parallaxes from the Spitzer Space Telescope}",
      journal = {\apj},
         year = 2018,
        month = nov,
       volume = {867},
       number = {2},
          eid = {109},
        pages = {109},
          doi = {10.3847/1538-4357/aae1af},
archivePrefix = {arXiv},
       eprint = {1809.06479},
 primaryClass = {astro-ph.SR},
       adsurl = {https://ui.adsabs.harvard.edu/abs/2018ApJ...867..109M}
}

@ARTICLE{Casey2023,
       author = {{Casey}, Caitlin M. and {Kartaltepe}, Jeyhan S. and {Drakos}, Nicole E. and {Franco}, Maximilien and {Harish}, Santosh and {Paquereau}, Louise and {Ilbert}, Olivier and {Rose}, Caitlin and {Cox}, Isabella G. and {Nightingale}, James W. and {Robertson}, Brant E. and {Silverman}, John D. and {Koekemoer}, Anton M. and {Massey}, Richard and {McCracken}, Henry Joy and {Rhodes}, Jason and {Akins}, Hollis B. and {Allen}, Natalie and {Amvrosiadis}, Aristeidis and {Arango-Toro}, Rafael C. and {Bagley}, Micaela B. and {Bongiorno}, Angela and {Capak}, Peter L. and {Champagne}, Jaclyn B. and {Chartab}, Nima and {Ch{\'a}vez Ortiz}, {\'O}scar A. and {Chworowsky}, Katherine and {Cooke}, Kevin C. and {Cooper}, Olivia R. and {Darvish}, Behnam and {Ding}, Xuheng and {Faisst}, Andreas L. and {Finkelstein}, Steven L. and {Fujimoto}, Seiji and {Gentile}, Fabrizio and {Gillman}, Steven and {Gould}, Katriona M.~L. and {Gozaliasl}, Ghassem and {Hayward}, Christopher C. and {He}, Qiuhan and {Hemmati}, Shoubaneh and {Hirschmann}, Michaela and {Jahnke}, Knud and {Jin}, Shuowen and {Khostovan}, Ali Ahmad and {Kokorev}, Vasily and {Lambrides}, Erini and {Laigle}, Clotilde and {Larson}, Rebecca L. and {Leung}, Gene C.~K. and {Liu}, Daizhong and {Liaudat}, Tobias and {Long}, Arianna S. and {Magdis}, Georgios and {Mahler}, Guillaume and {Mainieri}, Vincenzo and {Manning}, Sinclaire M. and {Maraston}, Claudia and {Martin}, Crystal L. and {McCleary}, Jacqueline E. and {McKinney}, Jed and {McPartland}, Conor J.~R. and {Mobasher}, Bahram and {Pattnaik}, Rohan and {Renzini}, Alvio and {Rich}, R. Michael and {Sanders}, David B. and {Sattari}, Zahra and {Scognamiglio}, Diana and {Scoville}, Nick and {Sheth}, Kartik and {Shuntov}, Marko and {Sparre}, Martin and {Suzuki}, Tomoko L. and {Talia}, Margherita and {Toft}, Sune and {Trakhtenbrot}, Benny and {Urry}, C. Megan and {Valentino}, Francesco and {Vanderhoof}, Brittany N. and {Vardoulaki}, Eleni and {Weaver}, John R. and {Whitaker}, Katherine E. and {Wilkins}, Stephen M. and {Yang}, Lilan and {Zavala}, Jorge A.},
        title = "{COSMOS-Web: An Overview of the JWST Cosmic Origins Survey}",
      journal = {\apj},
         year = 2023,
        month = sep,
       volume = {954},
       number = {1},
          eid = {31},
        pages = {31},
          doi = {10.3847/1538-4357/acc2bc},
archivePrefix = {arXiv},
       eprint = {2211.07865},
 primaryClass = {astro-ph.GA},
       adsurl = {https://ui.adsabs.harvard.edu/abs/2023ApJ...954...31C}
}

@ARTICLE{Bezanson2024,
       author = {{Bezanson}, Rachel and {Labbe}, Ivo and {Whitaker}, Katherine E. and {Leja}, Joel and {Price}, Sedona H. and {Franx}, Marijn and {Brammer}, Gabriel and {Marchesini}, Danilo and {Zitrin}, Adi and {Wang}, Bingjie and {Weaver}, John R. and {Furtak}, Lukas J. and {Atek}, Hakim and {Coe}, Dan and {Cutler}, Sam E. and {Dayal}, Pratika and {van Dokkum}, Pieter and {Feldmann}, Robert and {F{\"o}rster Schreiber}, Natascha M. and {Fujimoto}, Seiji and {Geha}, Marla and {Glazebrook}, Karl and {de Graaff}, Anna and {Greene}, Jenny E. and {Juneau}, St{\'e}phanie and {Kassin}, Susan and {Kriek}, Mariska and {Khullar}, Gourav and {Maseda}, Michael and {Mowla}, Lamiya A. and {Muzzin}, Adam and {Nanayakkara}, Themiya and {Nelson}, Erica J. and {Oesch}, Pascal A. and {Pacifici}, Camilla and {Pan}, Richard and {Papovich}, Casey and {Setton}, David J. and {Shapley}, Alice E. and {Smit}, Renske and {Stefanon}, Mauro and {Taylor}, Edward N. and {Williams}, Christina C.},
        title = "{The JWST UNCOVER Treasury Survey: Ultradeep NIRSpec and NIRCam Observations before the Epoch of Reionization}",
      journal = {\apj},
         year = 2024,
        month = oct,
       volume = {974},
       number = {1},
          eid = {92},
        pages = {92},
          doi = {10.3847/1538-4357/ad66cf},
archivePrefix = {arXiv},
       eprint = {2212.04026},
 primaryClass = {astro-ph.GA},
       adsurl = {https://ui.adsabs.harvard.edu/abs/2024ApJ...974...92B}
}

@ARTICLE{Nonino2023,
       author = {{Nonino}, Mario and {Glazebrook}, Karl and {Burgasser}, Adam J. and {Polenta}, Gianluca and {Morishita}, Takahiro and {Lepinzan}, Marius and {Castellano}, Marco and {Fontana}, Adriano and {Merlin}, Emiliano and {Bonchi}, Andrea and {Paris}, Diego and {Treu}, Tommaso and {Vulcani}, Benedetta and {Wang}, Xin and {Santini}, Paola and {Vanzella}, Eros and {Nanayakkara}, Themiya and {Mercurio}, Amata and {Rosati}, Piero and {Grillo}, Claudio and {Bradac}, Marusa},
        title = "{Early Results from GLASS-JWST. XIII. A Faint, Distant, and Cold Brown Dwarf}",
      journal = {\apjl},
         year = 2023,
        month = jan,
       volume = {942},
       number = {2},
          eid = {L29},
        pages = {L29},
          doi = {10.3847/2041-8213/ac8e5f},
archivePrefix = {arXiv},
       eprint = {2207.14802},
 primaryClass = {astro-ph.SR},
       adsurl = {https://ui.adsabs.harvard.edu/abs/2023ApJ...942L..29N}
}

@ARTICLE{Wang2023,
       author = {{Wang}, Po-Ya and {Goto}, Tomotsugu and {Ho}, Simon C.-C. and {Lin}, Yu-Wei and {Wu}, Cossas K.-W. and {Ling}, Chih-Teng and {Hashimoto}, Tetsuya and {Kim}, Seong Jin and {Hsiao}, Tiger Y.-Y.},
        title = "{A T-dwarf candidate from JWST early release NIRCam data}",
      journal = {\mnras},
         year = 2023,
        month = aug,
       volume = {523},
       number = {3},
        pages = {4534-4538},
          doi = {10.1093/mnras/stad1679},
archivePrefix = {arXiv},
       eprint = {2306.02663},
 primaryClass = {astro-ph.SR},
       adsurl = {https://ui.adsabs.harvard.edu/abs/2023MNRAS.523.4534W}
}

@ARTICLE{A_Chen2025,
       author = {{Chen}, Amos Y.~A. and {Goto}, Tomotsugu and {Wu}, Cossas K.-W. and {Ling}, Chih-Teng and {Kim}, Seong Jin and {Ho}, Simon C.-C. and {Kilerci}, Ece and {Uno}, Yuri and {Phan}, Terry Long and {Lin}, Yu-Wei and {Yang}, Tsung-Ching and {Hashimoto}, Tetsuya},
        title = "{Brown dwarf number density in the JWST COSMOS-Web field}",
      journal = {\pasa},
         year = 2025,
        month = jan,
       volume = {42},
          eid = {e042},
        pages = {e042},
          doi = {10.1017/pasa.2025.25},
archivePrefix = {arXiv},
       eprint = {2503.18510},
 primaryClass = {astro-ph.SR},
       adsurl = {https://ui.adsabs.harvard.edu/abs/2025PASA...42...42C}
}

@ARTICLE{Tu2025a,
       author = {{Tu}, Zhijun and {Wang}, Shu and {Chen}, Xiaodian and {Liu}, Jifeng},
        title = "{Three Brown Dwarfs Masquerading as High-redshift Galaxies in JWST Observations}",
      journal = {\apj},
         year = 2025,
        month = feb,
       volume = {980},
       number = {2},
          eid = {230},
        pages = {230},
          doi = {10.3847/1538-4357/adaf9f},
archivePrefix = {arXiv},
       eprint = {2501.16648},
 primaryClass = {astro-ph.SR},
       adsurl = {https://ui.adsabs.harvard.edu/abs/2025ApJ...980..230T}
}

@ARTICLE{Tu2025b,
       author = {{Tu}, Zhijun and {Wang}, Shu and {Chen}, Xiaodian and {Liu}, Jifeng},
        title = "{A Large Sample of JWST/NIRSpec Brown Dwarfs: New Distant Discoveries}",
      journal = {\apjs},
         year = 2025,
        month = dec,
       volume = {281},
       number = {2},
          eid = {49},
        pages = {49},
          doi = {10.3847/1538-4365/ae112e},
archivePrefix = {arXiv},
       eprint = {2510.02026},
 primaryClass = {astro-ph.SR},
       adsurl = {https://ui.adsabs.harvard.edu/abs/2025ApJS..281...49T}
}

@ARTICLE{Pirzkal2005,
       author = {{Pirzkal}, N. and {Sahu}, K.~C. and {Burgasser}, A. and {Moustakas}, L.~A. and {Xu}, C. and {Malhotra}, S. and {Rhoads}, J.~E. and {Koekemoer}, A.~M. and {Nelan}, E.~P. and {Windhorst}, R.~A. and {Panagia}, N. and {Gronwall}, C. and {Pasquali}, A. and {Walsh}, J.~R.},
        title = "{Stars in the Hubble Ultra Deep Field}",
      journal = {\apj},
         year = 2005,
        month = mar,
       volume = {622},
       number = {1},
        pages = {319-332},
          doi = {10.1086/427896},
archivePrefix = {arXiv},
       eprint = {astro-ph/0412097},
 primaryClass = {astro-ph},
       adsurl = {https://ui.adsabs.harvard.edu/abs/2005ApJ...622..319P}
}

@ARTICLE{Aganze2022,
       author = {{Aganze}, Christian and {Burgasser}, Adam J. and {Malkan}, Mathew and {Theissen}, Christopher A. and {Tejada Arevalo}, Roberto A. and {Hsu}, Chih-Chun and {Bardalez Gagliuffi}, Daniella C. and {Ryan}, Russell E. and {Holwerda}, Benne},
        title = "{Beyond the Local Volume. I. Surface Densities of Ultracool Dwarfs in Deep HST/WFC3 Parallel Fields}",
      journal = {\apj},
         year = 2022,
        month = jan,
       volume = {924},
       number = {2},
          eid = {114},
        pages = {114},
          doi = {10.3847/1538-4357/ac35ea},
archivePrefix = {arXiv},
       eprint = {2110.07672},
 primaryClass = {astro-ph.SR},
       adsurl = {https://ui.adsabs.harvard.edu/abs/2022ApJ...924..114A}
}

@ARTICLE{Liu2026,
       author = {{Liu}, F.-Y.~F. and {McLeod}, D.~J. and {Sutlieff}, B.~J. and {Dupuy}, T.~J. and {Dunlop}, J.~S. and {Chen}, X. and {McLure}, R.~J. and {Leung}, H.-H. and {Antonogiannaki}, S. and {Bondestam}, C. and {Laplace}, E. and {Begley}, R. and {Carnall}, A.~C. and {Cullen}, F. and {Donnan}, C.~T. and {Ellis}, R.~S.},
        title = "{A clear detection of proper motion confirms that the claimed z ≃ 32 galaxy candidate, ``Capotauro'', is a Y-type brown dwarf}",
      journal = {\mnras},
         year = 2026,
        month = oct,
          doi = {10.1093/mnras/stag2055},
       adsurl = {https://ui.adsabs.harvard.edu/abs/2026MNRAS.tmp.1893L}
}

@ARTICLE{Leung2026a,
       author = {{Leung}, Ho-Hin and {Carnall}, Adam C. and {Taylor}, Elizabeth and {Stevenson}, Struan D. and {Beverage}, Aliza G. and {Cullen}, Fergus and {Dunlop}, James S. and {McLeod}, Derek J. and {McLure}, Ross J. and {Begley}, Ryan and {Almaini}, Omar and {Antonogiannaki}, Stella and {Arellano-C{\'o}rdova}, Karla Z. and {Barrufet}, Laia and {Bondestam}, Cecilia and {Donnan}, Callum T. and {Holst}, Isaac J.~B. and {Liu}, Feng-Yuan F. and {Rowlands}, Kate and {Sanders}, Ryan L. and {Scholte}, Dirk and {Skarbinski}, Maya and {Stanton}, Thomas M. and {Wild}, Vivienne},
        title = "{The JWST EXCELS survey: the ages and abundances of 3 < z < 5 massive quiescent galaxies show that downsizing was already in place by z ≃ 4}",
      journal = {\mnras},
         year = 2026,
        month = jun,
       volume = {549},
       number = {1},
          eid = {stag827},
        pages = {stag827},
          doi = {10.1093/mnras/stag827},
archivePrefix = {arXiv},
       eprint = {2602.05934},
 primaryClass = {astro-ph.GA},
       adsurl = {https://ui.adsabs.harvard.edu/abs/2026MNRAS.549ag827L}
}

@ARTICLE{Horne1986,
       author = {{Horne}, K.},
        title = "{An optimal extraction algorithm for CCD spectroscopy.}",
      journal = {\pasp},
         year = 1986,
        month = jun,
       volume = {98},
        pages = {609-617},
          doi = {10.1086/131801},
       adsurl = {https://ui.adsabs.harvard.edu/abs/1986PASP...98..609H}
}

@ARTICLE{Barrufet2025,
       author = {{Barrufet}, L. and {Oesch}, P.~A. and {Marques-Chaves}, R. and {Arellano-Cordova}, K. and {Baggen}, J.~F.~W. and {Carnall}, A.~C. and {Cullen}, F. and {Dunlop}, J.~S. and {Gottumukkala}, R. and {Fudamoto}, Y. and {Illingworth}, G.~D. and {Magee}, D. and {McLure}, R.~J. and {McLeod}, D.~J. and {Micha{\l}owski}, M.~J. and {Stefanon}, M. and {van Dokkum}, P.~G. and {Weibel}, A.},
        title = "{Quiescent or dusty? Unveiling the nature of extremely red galaxies at z > 3}",
      journal = {\mnras},
         year = 2025,
        month = mar,
       volume = {537},
       number = {4},
        pages = {3453-3469},
          doi = {10.1093/mnras/staf013},
archivePrefix = {arXiv},
       eprint = {2404.08052},
 primaryClass = {astro-ph.GA},
       adsurl = {https://ui.adsabs.harvard.edu/abs/2025MNRAS.537.3453B}
}

@ARTICLE{oesch2023,
       author = {{Oesch}, P.~A. and {Brammer}, G. and {Naidu}, R.~P. and {Bouwens}, R.~J. and {Chisholm}, J. and {Illingworth}, G.~D. and {Matthee}, J. and {Nelson}, E. and {Qin}, Y. and {Reddy}, N. and {Shapley}, A. and {Shivaei}, I. and {van Dokkum}, P. and {Weibel}, A. and {Whitaker}, K. and {Wuyts}, S. and {Covelo-Paz}, A. and {Endsley}, R. and {Fudamoto}, Y. and {Giovinazzo}, E. and {Herard-Demanche}, T. and {Kerutt}, J. and {Kramarenko}, I. and {Labbe}, I. and {Leonova}, E. and {Lin}, J. and {Magee}, D. and {Marchesini}, D. and {Maseda}, M. and {Mason}, C. and {Matharu}, J. and {Meyer}, R.~A. and {Neufeld}, C. and {Prieto Lyon}, G. and {Schaerer}, D. and {Sharma}, R. and {Shuntov}, M. and {Smit}, R. and {Stefanon}, M. and {Wyithe}, J.~S.~B. and {Xiao}, M.},
        title = "{The JWST FRESCO survey: legacy NIRCam/grism spectroscopy and imaging in the two GOODS fields}",
      journal = {\mnras},
         year = 2023,
        month = oct,
       volume = {525},
       number = {2},
        pages = {2864-2874},
          doi = {10.1093/mnras/stad2411},
archivePrefix = {arXiv},
       eprint = {2304.02026},
 primaryClass = {astro-ph.GA},
       adsurl = {https://ui.adsabs.harvard.edu/abs/2023MNRAS.525.2864O}
}

@ARTICLE{williams2025,
       author = {{Williams}, Christina C. and {Oesch}, Pascal A. and {Weibel}, Andrea and {Brammer}, Gabriel and {Cloonan}, Aidan P. and {Whitaker}, Katherine E. and {Barrufet}, Laia and {Bezanson}, Rachel and {Bowler}, Rebecca A.~A. and {Dayal}, Pratika and {Franx}, Marijn and {Greene}, Jenny E. and {Hutter}, Anne and {Ji}, Zhiyuan and {Labb{\'e}}, Ivo and {Manning}, Sinclaire M. and {Maseda}, Michael V. and {Xiao}, Mengyuan},
        title = "{The PANORAMIC Survey: Pure Parallel Wide Area Legacy Imaging with JWST/NIRCam}",
      journal = {\apj},
         year = 2025,
        month = feb,
       volume = {979},
       number = {2},
          eid = {140},
        pages = {140},
          doi = {10.3847/1538-4357/ad97bc},
archivePrefix = {arXiv},
       eprint = {2410.01875},
 primaryClass = {astro-ph.GA},
       adsurl = {https://ui.adsabs.harvard.edu/abs/2025ApJ...979..140W}
}

@ARTICLE{Mang2026,
       author = {{Mang}, James and {Chachan}, Yayaati and {Morley}, Caroline V. and {Batalha}, Natasha E. and {Wogan}, Nicholas F. and {Mukherjee}, Sagnick and {Fortney}, Jonathan J. and {Marley}, Mark S. and {Visscher}, Channon and {Gharib-Nezhad}, Ehsan},
        title = "{The Sonora Substellar Atmosphere Models. VII. Flame Skimmer: Cloud-free Atmospheric and Evolutionary Models for the Coldest Substellar Objects}",
      journal = {\apj},
         year = 2026,
        month = sep,
       volume = {1008},
       number = {2},
          eid = {254},
        pages = {254},
          doi = {10.3847/1538-4357/ae9609},
archivePrefix = {arXiv},
       eprint = {2608.06454},
 primaryClass = {astro-ph.EP},
       adsurl = {https://ui.adsabs.harvard.edu/abs/2026ApJ..1008..254M}
}

@ARTICLE{Bowler2016,
       author = {{Bowler}, Brendan P.},
        title = "{Imaging Extrasolar Giant Planets}",
      journal = {\pasp},
         year = 2016,
        month = oct,
       volume = {128},
       number = {968},
        pages = {102001},
          doi = {10.1088/1538-3873/128/968/102001},
archivePrefix = {arXiv},
       eprint = {1605.02731},
 primaryClass = {astro-ph.EP},
       adsurl = {https://ui.adsabs.harvard.edu/abs/2016PASP..128j2001B}
}

@ARTICLE{Dupuy2017,
       author = {{Dupuy}, Trent J. and {Liu}, Michael C.},
        title = "{Individual Dynamical Masses of Ultracool Dwarfs}",
      journal = {\apjs},
         year = 2017,
        month = aug,
       volume = {231},
       number = {2},
          eid = {15},
        pages = {15},
          doi = {10.3847/1538-4365/aa5e4c},
archivePrefix = {arXiv},
       eprint = {1703.05775},
 primaryClass = {astro-ph.SR},
       adsurl = {https://ui.adsabs.harvard.edu/abs/2017ApJS..231...15D}
}

@ARTICLE{Fontanive2023,
       author = {{Fontanive}, Cl{\'e}mence and {Bedin}, Luigi R. and {De Furio}, Matthew and {Biller}, Beth and {Anderson}, Jay and {Bonavita}, Mariangela and {Allers}, Katelyn and {Pantoja}, Blake},
        title = "{An HST survey of 33 T8 to Y1 brown dwarfs: NIR photometry and multiplicity of the coldest isolated objects}",
      journal = {\mnras},
         year = 2023,
        month = dec,
       volume = {526},
       number = {2},
        pages = {1783-1798},
          doi = {10.1093/mnras/stad2870},
archivePrefix = {arXiv},
       eprint = {2309.09923},
 primaryClass = {astro-ph.SR},
       adsurl = {https://ui.adsabs.harvard.edu/abs/2023MNRAS.526.1783F}
}

@ARTICLE{Faherty2012,
       author = {{Faherty}, Jacqueline K. and {Burgasser}, Adam J. and {Walter}, Frederick M. and {Van der Bliek}, Nicole and {Shara}, Michael M. and {Cruz}, Kelle L. and {West}, Andrew A. and {Vrba}, Frederick J. and {Anglada-Escud{\'e}}, Guillem},
        title = "{The Brown Dwarf Kinematics Project (BDKP). III. Parallaxes for 70 Ultracool Dwarfs}",
      journal = {\apj},
         year = 2012,
        month = jun,
       volume = {752},
       number = {1},
          eid = {56},
        pages = {56},
          doi = {10.1088/0004-637X/752/1/56},
archivePrefix = {arXiv},
       eprint = {1203.5543},
 primaryClass = {astro-ph.SR},
       adsurl = {https://ui.adsabs.harvard.edu/abs/2012ApJ...752...56F}
}

@ARTICLE{Radcliffe2026,
       author = {{Radcliffe}, Alice and {Charnay}, Benjamin and {Lagrange}, Anne-Marie and {Kiefer}, Flavien and {B{\'e}zard}, Bruno and {Petrus}, Simon and {Palma-Bifani}, Paulina and {Ravet}, Matthieu and {Leconte}, J{\'e}r{\'e}my and {Marleau}, Gabriel-Dominique},
        title = "{Next-generation Exo-REM atmospheric models Application to VHS 1256 b to emulate patchy clouds}",
      journal = {\aap},
         year = 2026,
        month = jul,
       volume = {711},
          eid = {A257},
        pages = {A257},
          doi = {10.1051/0004-6361/202659006},
archivePrefix = {arXiv},
       eprint = {2605.29070},
 primaryClass = {astro-ph.EP},
       adsurl = {https://ui.adsabs.harvard.edu/abs/2026A&A...711A.257R}
}

@ARTICLE{Hainline2026c,
       author = {{Hainline}, Kevin N. and {Beiler}, Samuel A. and {Burgasser}, Adam J. and {Chen}, Evan S. and {Helton}, Jakob M. and {Leisenring}, Jarron and {Miles}, Brittany E. and {Marley}, Mark S. and {Mukherjee}, Sagnick and {Baum}, Stefi and {Bunker}, Andrew J. and {Carniani}, Stefano and {D'Eugenio}, Francesco and {Egami}, Eiichi and {Looser}, Tobias J. and {Rinaldi}, Pierluigi and {Willmer}, Christopher N.~A.},
        title = "{JWST/NIRSpec Spectra for Three Ultracool Brown Dwarfs Detected in Extragalactic Surveys: Further Evidence for Phosphine Absorption}",
      journal = {arXiv e-prints},
         year = 2026,
        month = aug,
          eid = {arXiv:2608.14786},
        pages = {arXiv:2608.14786},
archivePrefix = {arXiv},
       eprint = {2608.14786},
 primaryClass = {astro-ph.SR},
       adsurl = {https://ui.adsabs.harvard.edu/abs/2026arXiv260814786H}
}

@ARTICLE{Arias1995,
       author = {{Arias}, E.~F. and {Charlot}, P. and {Feissel}, M. and {Lestrade}, J.-F.},
        title = "{The extragalactic reference system of the International Earth Rotation Service, ICRS.}",
      journal = {\aap},
         year = 1995,
        month = nov,
       volume = {303},
        pages = {604-608},
       adsurl = {https://ui.adsabs.harvard.edu/abs/1995A&A...303..604A}
}

@software{Barbary2018,
       author = {{Barbary}, Kyle},
        title = "{SEP: Source Extraction and Photometry}",
 howpublished = {Astrophysics Source Code Library, record ascl:1811.004},
         year = 2018,
        month = nov,
          eid = {ascl:1811.004},
archivePrefix = {ascl},
       eprint = {1811.004},
       adsurl = {https://ui.adsabs.harvard.edu/abs/2018ascl.soft11004B}
}

@ARTICLE{CurtisLake2026,
       author = {{Curtis-Lake}, Emma and {Cameron}, Alex J. and {Bunker}, Andrew J. and {Scholtz}, Jan and {Carniani}, Stefano and {Parlanti}, Eleonora and {D'Eugenio}, Francesco and {Jakobsen}, Peter and {Willmer}, Christopher N.~A. and {Arribas}, Santiago and {Baker}, William M. and {Charlot}, St{\'e}phane and {Chevallard}, Jacopo and {Circosta}, Chiara and {Curti}, Mirko and {Duan}, Qiao and {Eisenstein}, Daniel J. and {Hainline}, Kevin and {Ji}, Zhiyuan and {Johnson}, Benjamin D. and {Jones}, Gareth C. and {Maiolino}, Roberto and {Maseda}, Michael V. and {Perna}, Michele and {P{\'e}rez-Gonz{\'a}lez}, Pablo G. and {Rawle}, Tim and {Rieke}, Marcia and {Rinaldi}, Pierluigi and {Robertson}, Brant and {Rodr{\'\i}guez Del Pino}, Bruno and {Saxena}, Aayush and {Shivaei}, Irene and {Smit}, Renske and {Tacchella}, Sandro and {{\"U}bler}, Hannah and {Venturi}, Giacomo and {Williams}, Christina C. and {Willott}, Chris},
        title = "{JADES data release 4 ─ Paper I. Sample selection, observing strategy and redshifts of the complete spectroscopic sample}",
      journal = {\mnras},
         year = 2026,
        month = jul,
       volume = {549},
       number = {4},
          eid = {stag836},
        pages = {stag836},
          doi = {10.1093/mnras/stag836},
archivePrefix = {arXiv},
       eprint = {2510.01033},
 primaryClass = {astro-ph.GA},
       adsurl = {https://ui.adsabs.harvard.edu/abs/2026MNRAS.549ag836C}
}

@ARTICLE{Brammer2012,
       author = {{Brammer}, Gabriel B. and {van Dokkum}, Pieter G. and {Franx}, Marijn and {Fumagalli}, Mattia and {Patel}, Shannon and {Rix}, Hans-Walter and {Skelton}, Rosalind E. and {Kriek}, Mariska and {Nelson}, Erica and {Schmidt}, Kasper B. and {Bezanson}, Rachel and {da Cunha}, Elisabete and {Erb}, Dawn K. and {Fan}, Xiaohui and {F{\"o}rster Schreiber}, Natascha and {Illingworth}, Garth D. and {Labb{\'e}}, Ivo and {Leja}, Joel and {Lundgren}, Britt and {Magee}, Dan and {Marchesini}, Danilo and {McCarthy}, Patrick and {Momcheva}, Ivelina and {Muzzin}, Adam and {Quadri}, Ryan and {Steidel}, Charles C. and {Tal}, Tomer and {Wake}, David and {Whitaker}, Katherine E. and {Williams}, Anna},
        title = "{3D-HST: A Wide-field Grism Spectroscopic Survey with the Hubble Space Telescope}",
      journal = {\apjs},
         year = 2012,
        month = jun,
       volume = {200},
       number = {2},
          eid = {13},
        pages = {13},
          doi = {10.1088/0067-0049/200/2/13},
archivePrefix = {arXiv},
       eprint = {1204.2829},
 primaryClass = {astro-ph.CO},
       adsurl = {https://ui.adsabs.harvard.edu/abs/2012ApJS..200...13B}
}

@ARTICLE{Luo2017,
       author = {{Luo}, B. and {Brandt}, W.~N. and {Xue}, Y.~Q. and {Lehmer}, B. and {Alexander}, D.~M. and {Bauer}, F.~E. and {Vito}, F. and {Yang}, G. and {Basu-Zych}, A.~R. and {Comastri}, A. and {Gilli}, R. and {Gu}, Q.-S. and {Hornschemeier}, A.~E. and {Koekemoer}, A. and {Liu}, T. and {Mainieri}, V. and {Paolillo}, M. and {Ranalli}, P. and {Rosati}, P. and {Schneider}, D.~P. and {Shemmer}, O. and {Smail}, I. and {Sun}, M. and {Tozzi}, P. and {Vignali}, C. and {Wang}, J.-X.},
        title = "{The Chandra Deep Field-South Survey: 7 Ms Source Catalogs}",
      journal = {\apjs},
         year = 2017,
        month = jan,
       volume = {228},
       number = {1},
          eid = {2},
        pages = {2},
          doi = {10.3847/1538-4365/228/1/2},
archivePrefix = {arXiv},
       eprint = {1611.03501},
 primaryClass = {astro-ph.GA},
       adsurl = {https://ui.adsabs.harvard.edu/abs/2017ApJS..228....2L}
}

@ARTICLE{Morishita2024,
       author = {{Morishita}, Takahiro and {Stiavelli}, Massimo and {Chary}, Ranga-Ram and {Trenti}, Michele and {Bergamini}, Pietro and {Chiaberge}, Marco and {Leethochawalit}, Nicha and {Roberts-Borsani}, Guido and {Shen}, Xuejian and {Treu}, Tommaso},
        title = "{Enhanced Subkiloparsec-scale Star Formation: Results from a JWST Size Analysis of 341 Galaxies at 5 < z < 14}",
      journal = {\apj},
         year = 2024,
        month = mar,
       volume = {963},
       number = {1},
          eid = {9},
        pages = {9},
          doi = {10.3847/1538-4357/ad1404},
archivePrefix = {arXiv},
       eprint = {2308.05018},
 primaryClass = {astro-ph.GA},
       adsurl = {https://ui.adsabs.harvard.edu/abs/2024ApJ...963....9M}
}

@ARTICLE{Weaver2024,
       author = {{Weaver}, John R. and {Cutler}, Sam E. and {Pan}, Richard and {Whitaker}, Katherine E. and {Labb{\'e}}, Ivo and {Price}, Sedona H. and {Bezanson}, Rachel and {Brammer}, Gabriel and {Marchesini}, Danilo and {Leja}, Joel and {Wang}, Bingjie and {Furtak}, Lukas J. and {Zitrin}, Adi and {Atek}, Hakim and {Chemerynska}, Iryna and {Coe}, Dan and {Dayal}, Pratika and {van Dokkum}, Pieter and {Feldmann}, Robert and {F{\"o}rster Schreiber}, Natascha M. and {Franx}, Marijn and {Fujimoto}, Seiji and {Fudamoto}, Yoshinobu and {Glazebrook}, Karl and {de Graaff}, Anna and {Greene}, Jenny E. and {Juneau}, St{\'e}phanie and {Kassin}, Susan and {Kriek}, Mariska and {Khullar}, Gourav and {Maseda}, Michael V. and {Mowla}, Lamiya A. and {Muzzin}, Adam and {Nanayakkara}, Themiya and {Nelson}, Erica J. and {Oesch}, Pascal A. and {Pacifici}, Camilla and {Papovich}, Casey and {Setton}, David J. and {Shapley}, Alice E. and {Shipley}, Heath V. and {Smit}, Renske and {Stefanon}, Mauro and {Taylor}, Edward N. and {Weibel}, Andrea and {Williams}, Christina C.},
        title = "{The UNCOVER Survey: A First-look HST + JWST Catalog of 60,000 Galaxies near A2744 and beyond}",
      journal = {\apjs},
         year = 2024,
        month = jan,
       volume = {270},
       number = {1},
          eid = {7},
        pages = {7},
          doi = {10.3847/1538-4365/ad07e0},
archivePrefix = {arXiv},
       eprint = {2301.02671},
 primaryClass = {astro-ph.GA},
       adsurl = {https://ui.adsabs.harvard.edu/abs/2024ApJS..270....7W}
}

@ARTICLE{Harvey2025,
       author = {{Harvey}, Thomas and {Conselice}, Christopher J. and {Adams}, Nathan J. and {Austin}, Duncan and {Li}, Qiong and {Rusakov}, Vadim and {Westcott}, Lewi and {Goolsby}, Caio M. and {Lovell}, Christopher C. and {Cochrane}, Rachel K. and {Vijayan}, Aswin P. and {Trussler}, James},
        title = "{Behind the spotlight: a systematic assessment of outshining using NIRCam medium bands in the JADES Origins Field}",
      journal = {\mnras},
         year = 2025,
        month = oct,
       volume = {542},
       number = {4},
        pages = {2998-3027},
          doi = {10.1093/mnras/staf1396},
archivePrefix = {arXiv},
       eprint = {2504.05244},
 primaryClass = {astro-ph.GA},
       adsurl = {https://ui.adsabs.harvard.edu/abs/2025MNRAS.542.2998H}
}

@ARTICLE{VanderPlas2018,
       author = {{VanderPlas}, Jacob T.},
        title = "{Understanding the Lomb-Scargle Periodogram}",
      journal = {\apjs},
         year = 2018,
        month = may,
       volume = {236},
       number = {1},
          eid = {16},
        pages = {16},
          doi = {10.3847/1538-4365/aab766},
archivePrefix = {arXiv},
       eprint = {1703.09824},
 primaryClass = {astro-ph.IM},
       adsurl = {https://ui.adsabs.harvard.edu/abs/2018ApJS..236...16V}
}

@ARTICLE{Manjavacas2019,
       author = {{Manjavacas}, Elena and {Apai}, D{\'a}niel and {Zhou}, Yifan and {Lew}, Ben W.~P. and {Schneider}, Glenn and {Metchev}, Stan and {Miles-P{\'a}ez}, Paulo A. and {Radigan}, Jacqueline and {Marley}, Mark S. and {Cowan}, Nicolas and {Karalidi}, Theodora and {Burgasser}, Adam J. and {Bedin}, Luigi R. and {Lowrance}, Patrick J. and {Kauffmann}, Parker},
        title = "{Cloud Atlas: Hubble Space Telescope Near-infrared Spectral Library of Brown Dwarfs, Planetary-mass Companions, and Hot Jupiters}",
      journal = {\aj},
         year = 2019,
        month = mar,
       volume = {157},
       number = {3},
          eid = {101},
        pages = {101},
          doi = {10.3847/1538-3881/aaf88f},
archivePrefix = {arXiv},
       eprint = {1812.03963},
 primaryClass = {astro-ph.SR},
       adsurl = {https://ui.adsabs.harvard.edu/abs/2019AJ....157..101M}
}

@ARTICLE{Cushing2005,
       author = {{Cushing}, Michael C. and {Rayner}, John T. and {Vacca}, William D.},
        title = "{An Infrared Spectroscopic Sequence of M, L, and T Dwarfs}",
      journal = {\apj},
         year = 2005,
        month = apr,
       volume = {623},
       number = {2},
        pages = {1115-1140},
          doi = {10.1086/428040},
archivePrefix = {arXiv},
       eprint = {astro-ph/0412313},
 primaryClass = {astro-ph},
       adsurl = {https://ui.adsabs.harvard.edu/abs/2005ApJ...623.1115C}
}

@ARTICLE{Rayner2009,
       author = {{Rayner}, John T. and {Cushing}, Michael C. and {Vacca}, William D.},
        title = "{The Infrared Telescope Facility (IRTF) Spectral Library: Cool Stars}",
      journal = {\apjs},
         year = 2009,
        month = dec,
       volume = {185},
       number = {2},
        pages = {289-432},
          doi = {10.1088/0067-0049/185/2/289},
archivePrefix = {arXiv},
       eprint = {0909.0818},
 primaryClass = {astro-ph.SR},
       adsurl = {https://ui.adsabs.harvard.edu/abs/2009ApJS..185..289R}
}

@ARTICLE{Villaume2017,
       author = {{Villaume}, Alexa and {Conroy}, Charlie and {Johnson}, Benjamin and {Rayner}, John and {Mann}, Andrew W. and {van Dokkum}, Pieter},
        title = "{The Extended IRTF Spectral Library: Expanded Coverage in Metallicity, Temperature, and Surface Gravity}",
      journal = {\apjs},
         year = 2017,
        month = jun,
       volume = {230},
       number = {2},
          eid = {23},
        pages = {23},
          doi = {10.3847/1538-4365/aa72ed},
archivePrefix = {arXiv},
       eprint = {1705.08906},
 primaryClass = {astro-ph.SR},
       adsurl = {https://ui.adsabs.harvard.edu/abs/2017ApJS..230...23V}
}

@ARTICLE{Gagne2026,
       author = {{Gagn{\'e}}, Jonathan and {Faherty}, Jacqueline K. and {Ruiz Diaz}, Azul and {Coulombe}, Louis-Philippe and {Bickle}, Thomas P. and {Schneider}, Adam C. and {Kirkpatrick}, J. Davy and {Kuchner}, Marc J. and {Meisner}, Aaron M. and {Caselden}, Dan and {Burgasser}, Adam J. and {Casewell}, Sarah and {Honaker}, Easton J. and {Kiwy}, Frank and {Marocco}, Federico and {Bardalez Gagliuffi}, Daniella C. and {Stevnbak Andersen}, Nikolaj and {Ruiz Arroyo}, Lizzeth and {Baller}, Bruce and {Beaulieu}, Paul and {Bell}, John and {Bilsing}, Martin and {Bohling}, Troy K. and {Colin}, Guillaume and {Colombo}, Giovanni and {Deen}, Sam and {Dereveanco}, Alexandru and {Dixon}, Kevin and {Durantini Luca}, Hugo A. and {Flores}, Deiby and {Franck}, Christoph and {Fulvi}, Christopher and {Gallmann}, Michael and {Gantier}, Jean Marc and {Glebov}, Konstantin and {Gramaize}, L{\'e}opold and {Hamlet}, Leslie K. and {Hinckley}, Ken and {Jablonski}, Kevin and {Ja{\l}owiczor}, Peter A. and {Kabatnik}, Martin and {Kasprowitz}, Peter and {Ly}, K. and {Martin}, David W. and {Marzak}, Naoufel and {McColgan}, Alexander and {McEwan}, Neil J. and {Michaels}, Marianne N. and {Pendrill}, William and {Perlin}, St{\'e}phane and {Pumphrey}, Benjamin and {Rabe}, James and {Raway}, Henry and {Robledo}, Walter Ruben and {Roser}, David and {Roy}, Animesh and {Sainio}, Arttu and {Schindler}, Vincent and {Schonau}, Manfred and {Sch{\"u}mann}, J{\"o}rg and {Selg-Mann}, Karl and {Serio}, Andrea and {Smith}, Patrick and {Stenner}, Andres and {Tanner}, Christopher and {Th{\'e}venot}, Melina and {Thakur}, Vinod and {Torres Guerrero}, Mayahuel and {Ventura}, Maurizio and {Voloshin}, Nikita V. and {Walla}, Jim and {W{\c{e}}dracki}, Zbigniew and {Weyandt}, Bailey and {Wilhite}, Breck and {Zitouni}, Spartacus},
        title = "{A SPHEREx Pipeline and Spectral Library for Ultracool Dwarfs}",
      journal = {\apj},
         year = 2026,
        month = sep,
       volume = {1009},
       number = {1},
          eid = {15},
        pages = {15},
          doi = {10.3847/1538-4357/ae8aed},
       adsurl = {https://ui.adsabs.harvard.edu/abs/2026ApJ..1009...15G}
}

@ARTICLE{Pritchard2026,
       author = {{Pritchard}, Evan and {Loritsch}, Marylin and {Haynes}, Julia and {Morrissey}, Sara and {Softich}, Emma and {Burgasser}, Adam J.},
        title = "{JWST/NIRSpec Spectral Standards for M, L, and T Dwarfs and Subdwarfs}",
      journal = {Research Notes of the American Astronomical Society},
         year = 2026,
        month = may,
       volume = {10},
       number = {5},
          eid = {144},
        pages = {144},
          doi = {10.3847/2515-5172/ae736f},
archivePrefix = {arXiv},
       eprint = {2606.23928},
 primaryClass = {astro-ph.SR},
       adsurl = {https://ui.adsabs.harvard.edu/abs/2026RNAAS..10..144P}
}

@ARTICLE{Rieke2024,
       author = {{Rieke}, G.~H. and {Alberts}, Stacey and {Shivaei}, Irene and {Lyu}, Jianwei and {Willmer}, Christopher N.~A. and {P{\'e}rez-Gonz{\'a}lez}, Pablo and {Williams}, Christina C.},
        title = "{SMILES: A Prototype JWST Multiband Mid-infrared Survey}",
      journal = {\apj},
         year = 2024,
        month = nov,
       volume = {975},
       number = {1},
          eid = {83},
        pages = {83},
          doi = {10.3847/1538-4357/ad6cd2},
archivePrefix = {arXiv},
       eprint = {2406.03518},
 primaryClass = {astro-ph.GA},
       adsurl = {https://ui.adsabs.harvard.edu/abs/2024ApJ...975...83R}
}

@ARTICLE{Alberts2024,
       author = {{Alberts}, Stacey and {Lyu}, Jianwei and {Shivaei}, Irene and {Rieke}, George H. and {P{\'e}rez-Gonz{\'a}lez}, Pablo G. and {Bonaventura}, Nina and {Zhu}, Yongda and {Helton}, Jakob M. and {Ji}, Zhiyuan and {Morrison}, Jane and {Robertson}, Brant E. and {Stone}, Meredith A. and {Sun}, Yang and {Williams}, Christina C. and {Willmer}, Christopher N.~A.},
        title = "{SMILES Initial Data Release: Unveiling the Obscured Universe with MIRI Multiband Imaging}",
      journal = {\apj},
         year = 2024,
        month = dec,
       volume = {976},
       number = {2},
          eid = {224},
        pages = {224},
          doi = {10.3847/1538-4357/ad7396},
archivePrefix = {arXiv},
       eprint = {2405.15972},
 primaryClass = {astro-ph.GA},
       adsurl = {https://ui.adsabs.harvard.edu/abs/2024ApJ...976..224A}
}

@ARTICLE{Ferruit2022,
       author = {{Ferruit}, P. and {Jakobsen}, P. and {Giardino}, G. and {Rawle}, T. and {Alves de Oliveira}, C. and {Arribas}, S. and {Beck}, T.~L. and {Birkmann}, S. and {B{\"o}ker}, T. and {Bunker}, A.~J. and {Charlot}, S. and {de Marchi}, G. and {Franx}, M. and {Henry}, A. and {Karakla}, D. and {Kassin}, S.~A. and {Kumari}, N. and {L{\'o}pez-Caniego}, M. and {L{\"u}tzgendorf}, N. and {Maiolino}, R. and {Manjavacas}, E. and {Marston}, A. and {Moseley}, S.~H. and {Muzerolle}, J. and {Pirzkal}, N. and {Rauscher}, B. and {Rix}, H.-W. and {Sabbi}, E. and {Sirianni}, M. and {te Plate}, M. and {Valenti}, J. and {Willott}, C.~J. and {Zeidler}, P.},
        title = "{The Near-Infrared Spectrograph (NIRSpec) on the James Webb Space Telescope. II. Multi-object spectroscopy (MOS)}",
      journal = {\aap},
         year = 2022,
        month = may,
       volume = {661},
          eid = {A81},
        pages = {A81},
          doi = {10.1051/0004-6361/202142673},
archivePrefix = {arXiv},
       eprint = {2202.03306},
 primaryClass = {astro-ph.IM},
       adsurl = {https://ui.adsabs.harvard.edu/abs/2022A&A...661A..81F}
}

@ARTICLE{Beichman2013,
       author = {{Beichman}, C. and {Gelino}, Christopher R. and {Kirkpatrick}, J. Davy and {Barman}, Travis S. and {Marsh}, Kenneth A. and {Cushing}, Michael C. and {Wright}, E.~L.},
        title = "{The Coldest Brown Dwarf (or Free-floating Planet)?: The Y Dwarf WISE 1828+2650}",
      journal = {\apj},
         year = 2013,
        month = feb,
       volume = {764},
       number = {1},
          eid = {101},
        pages = {101},
          doi = {10.1088/0004-637X/764/1/101},
archivePrefix = {arXiv},
       eprint = {1301.1669},
 primaryClass = {astro-ph.SR},
       adsurl = {https://ui.adsabs.harvard.edu/abs/2013ApJ...764..101B}
}

@ARTICLE{DeFurio2023,
       author = {{De Furio}, Matthew and {Lew}, Ben and {Beichman}, Charles and {Roellig}, Thomas and {Bryden}, Geoffrey and {Ciardi}, David and {Meyer}, Michael and {Rieke}, Marcia and {Greenbaum}, Alexandra and {Leisenring}, Jarron and {Llop-Sayson}, Jorge and {Ygouf}, Marie and {Albert}, Loic and {Boyer}, Martha and {Eisenstein}, Daniel and {Hodapp}, Klaus and {Horner}, Scott and {Johnstone}, Doug and {Kelly}, Doug and {Misselt}, Karl and {Rieke}, George and {Stansberry}, John and {Young}, Erick},
        title = "{JWST Observations of the Enigmatic Y-Dwarf WISE 1828+2650. I. Limits to a Binary Companion}",
      journal = {\apj},
         year = 2023,
        month = may,
       volume = {948},
       number = {2},
          eid = {92},
        pages = {92},
          doi = {10.3847/1538-4357/acbf1e},
archivePrefix = {arXiv},
       eprint = {2302.12723},
 primaryClass = {astro-ph.SR},
       adsurl = {https://ui.adsabs.harvard.edu/abs/2023ApJ...948...92D}
}

@ARTICLE{Barrado2023,
       author = {{Barrado}, David and {Molli{\`e}re}, Paul and {Patapis}, Polychronis and {Min}, Michiel and {Tremblin}, Pascal and {Ardevol Martinez}, Francisco and {Whiteford}, Niall and {Vasist}, Malavika and {Argyriou}, Ioannis and {Samland}, Matthias and {Lagage}, Pierre-Olivier and {Decin}, Leen and {Waters}, Rens and {Henning}, Thomas and {Morales-Calder{\'o}n}, Mar{\'\i}a and {Guedel}, Manuel and {Vandenbussche}, Bart and {Absil}, Olivier and {Baudoz}, Pierre and {Boccaletti}, Anthony and {Bouwman}, Jeroen and {Cossou}, Christophe and {Coulais}, Alain and {Crouzet}, Nicolas and {Gastaud}, Ren{\'e} and {Glasse}, Alistair and {Glauser}, Adrian M. and {Kamp}, Inga and {Kendrew}, Sarah and {Krause}, Oliver and {Lahuis}, Fred and {Mueller}, Michael and {Olofsson}, G{\"o}ran and {Pye}, John and {Rouan}, Daniel and {Royer}, Pierre and {Scheithauer}, Silvia and {Waldmann}, Ingo and {Colina}, Luis and {van Dishoeck}, Ewine F. and {Ray}, Tom and {{\"O}stlin}, G{\"o}ran and {Wright}, Gillian},
        title = "{$^{15}$NH$_{3}$ in the atmosphere of a cool brown dwarf}",
      journal = {\nat},
         year = 2023,
        month = dec,
       volume = {624},
       number = {7991},
        pages = {263-266},
          doi = {10.1038/s41586-023-06813-y},
archivePrefix = {arXiv},
       eprint = {2311.08054},
 primaryClass = {astro-ph.EP},
       adsurl = {https://ui.adsabs.harvard.edu/abs/2023Natur.624..263B}
}

@ARTICLE{Nardiello2022,
       author = {{Nardiello}, D. and {Bedin}, L.~R. and {Burgasser}, A. and {Salaris}, M. and {Cassisi}, S. and {Griggio}, M. and {Scalco}, M.},
        title = "{Photometry and astrometry with JWST - I. NIRCam point spread functions and the first JWST colour-magnitude diagrams of a globular cluster}",
      journal = {\mnras},
         year = 2022,
        month = nov,
       volume = {517},
       number = {1},
        pages = {484-497},
          doi = {10.1093/mnras/stac2659},
archivePrefix = {arXiv},
       eprint = {2209.06547},
 primaryClass = {astro-ph.SR},
       adsurl = {https://ui.adsabs.harvard.edu/abs/2022MNRAS.517..484N}
}

@ARTICLE{Fazio2004,
       author = {{Fazio}, G.~G. and {Hora}, J.~L. and {Allen}, L.~E. and {Ashby}, M.~L.~N. and {Barmby}, P. and {Deutsch}, L.~K. and {Huang}, J.-S. and {Kleiner}, S. and {Marengo}, M. and {Megeath}, S.~T. and {Melnick}, G.~J. and {Pahre}, M.~A. and {Patten}, B.~M. and {Polizotti}, J. and {Smith}, H.~A. and {Taylor}, R.~S. and {Wang}, Z. and {Willner}, S.~P. and {Hoffmann}, W.~F. and {Pipher}, J.~L. and {Forrest}, W.~J. and {McMurty}, C.~W. and {McCreight}, C.~R. and {McKelvey}, M.~E. and {McMurray}, R.~E. and {Koch}, D.~G. and {Moseley}, S.~H. and {Arendt}, R.~G. and {Mentzell}, J.~E. and {Marx}, C.~T. and {Losch}, P. and {Mayman}, P. and {Eichhorn}, W. and {Krebs}, D. and {Jhabvala}, M. and {Gezari}, D.~Y. and {Fixsen}, D.~J. and {Flores}, J. and {Shakoorzadeh}, K. and {Jungo}, R. and {Hakun}, C. and {Workman}, L. and {Karpati}, G. and {Kichak}, R. and {Whitley}, R. and {Mann}, S. and {Tollestrup}, E.~V. and {Eisenhardt}, P. and {Stern}, D. and {Gorjian}, V. and {Bhattacharya}, B. and {Carey}, S. and {Nelson}, B.~O. and {Glaccum}, W.~J. and {Lacy}, M. and {Lowrance}, P.~J. and {Laine}, S. and {Reach}, W.~T. and {Stauffer}, J.~A. and {Surace}, J.~A. and {Wilson}, G. and {Wright}, E.~L. and {Hoffman}, A. and {Domingo}, G. and {Cohen}, M.},
        title = "{The Infrared Array Camera (IRAC) for the Spitzer Space Telescope}",
      journal = {\apjs},
         year = 2004,
        month = sep,
       volume = {154},
       number = {1},
        pages = {10-17},
          doi = {10.1086/422843},
archivePrefix = {arXiv},
       eprint = {astro-ph/0405616},
 primaryClass = {astro-ph},
       adsurl = {https://ui.adsabs.harvard.edu/abs/2004ApJS..154...10F}
}

@ARTICLE{Liu2026a,
       author = {{Liu}, Feng-Yuan and {Dunlop}, James S. and {McLure}, Ross J. and {McLeod}, Derek J. and {Barrufet}, Laia and {Carnall}, Adam C. and {Begley}, Ryan and {P{\'e}rez-Gonz{\'a}lez}, Pablo G. and {Donnan}, Callum T. and {Ellis}, Richard S. and {Grogin}, Norman A. and {Magee}, Dan and {Illingworth}, Garth D. and {Cullen}, Fergus and {Stevenson}, Struan D. and {Koekemoer}, Anton M. and {Fontana}, Adriano and {Bowler}, Rebecca A.~A.},
        title = "{JWST PRIMER: a deep JWST study of all ALMA-detected galaxies in PRIMER COSMOS ─ dust-obscured star formation history back to z ≃ 7}",
      journal = {\mnras},
         year = 2026,
        month = jan,
       volume = {545},
       number = {2},
          eid = {staf1961},
        pages = {staf1961},
          doi = {10.1093/mnras/staf1961},
archivePrefix = {arXiv},
       eprint = {2503.07774},
 primaryClass = {astro-ph.GA},
       adsurl = {https://ui.adsabs.harvard.edu/abs/2026MNRAS.545f1961L}
}

@ARTICLE{Wright2010,
       author = {{Wright}, Edward L. and {Eisenhardt}, Peter R.~M. and {Mainzer}, Amy K. and {Ressler}, Michael E. and {Cutri}, Roc M. and {Jarrett}, Thomas and {Kirkpatrick}, J. Davy and {Padgett}, Deborah and {McMillan}, Robert S. and {Skrutskie}, Michael and {Stanford}, S.~A. and {Cohen}, Martin and {Walker}, Russell G. and {Mather}, John C. and {Leisawitz}, David and {Gautier}, III, Thomas N. and {McLean}, Ian and {Benford}, Dominic and {Lonsdale}, Carol J. and {Blain}, Andrew and {Mendez}, Bryan and {Irace}, William R. and {Duval}, Valerie and {Liu}, Fengchuan and {Royer}, Don and {Heinrichsen}, Ingolf and {Howard}, Joan and {Shannon}, Mark and {Kendall}, Martha and {Walsh}, Amy L. and {Larsen}, Mark and {Cardon}, Joel G. and {Schick}, Scott and {Schwalm}, Mark and {Abid}, Mohamed and {Fabinsky}, Beth and {Naes}, Larry and {Tsai}, Chao-Wei},
        title = "{The Wide-field Infrared Survey Explorer (WISE): Mission Description and Initial On-orbit Performance}",
      journal = {\aj},
         year = 2010,
        month = dec,
       volume = {140},
       number = {6},
        pages = {1868-1881},
          doi = {10.1088/0004-6256/140/6/1868},
archivePrefix = {arXiv},
       eprint = {1008.0031},
 primaryClass = {astro-ph.IM},
       adsurl = {https://ui.adsabs.harvard.edu/abs/2010AJ....140.1868W}
}

@MISC{Cutri2013,
       author = {{Cutri}, R.~M. and {Wright}, E.~L. and {Conrow}, T. and {Fowler}, J.~W. and {Eisenhardt}, P.~R.~M. and {Grillmair}, C. and {Kirkpatrick}, J.~D. and {Masci}, F. and {McCallon}, H.~L. and {Wheelock}, S.~L. and {Fajardo-Acosta}, S. and {Yan}, L. and {Benford}, D. and {Harbut}, M. and {Jarrett}, T. and {Lake}, S. and {Leisawitz}, D. and {Ressler}, M.~E. and {Stanford}, S.~A. and {Tsai}, C.~W. and {Liu}, F. and {Helou}, G. and {Mainzer}, A. and {Gettings}, D. and {Gonzalez}, A. and {Hoffman}, D. and {Marsh}, K.~A. and {Padgett}, D. and {Skrutskie}, M.~F. and {Beck}, R.~P. and {Papin}, M. and {Wittman}, M.},
        title = "{Explanatory Supplement to the AllWISE Data Release Products}",
 howpublished = {Explanatory Supplement to the AllWISE Data Release Products, by R. M. Cutri et al.},
         year = 2013,
        month = nov,
        pages = {1},
       adsurl = {https://ui.adsabs.harvard.edu/abs/2013wise.rept....1C}
}

@ARTICLE{Johnson2026,
       author = {{Johnson}, Benjamin D. and {Robertson}, Brant E. and {Eisenstein}, Daniel J. and {Tacchella}, Sandro and {Pusk{\'a}s}, D{\'a}vid and {Duan}, Qiao and {Wu}, Zihao and {Hainline}, Kevin and {Rieke}, Marcia and {Willott}, Chris and {Willmer}, Christopher N.~A. and {Trussler}, James A.~A. and {Alberts}, Stacey and {Arribas}, Santiago and {Baker}, William M. and {Bunker}, Andrew J. and {Cameron}, Alex J. and {Carniani}, Stefano and {Carreira}, Courtney and {Cargile}, Phillip A. and {Curtis-Lake}, Emma and {Egami}, Eiichi and {Hausen}, Ryan and {Helton}, Jakob M. and {Ji}, Zhiyuan and {Maiolino}, Roberto and {P{\'e}rez-Gonz{\'a}lez}, Pablo G. and {Rinaldi}, Pierluigi and {Sun}, Fengwu and {Sun}, Yang and {Villanueva}, Natalia C. and {Williams}, Christina C. and {Zhu}, Yongda},
        title = "{JWST Advanced Deep Extragalactic Survey (JADES) Data Release 5: NIRCam Imaging in GOODS-S and GOODS-N}",
      journal = {arXiv e-prints},
         year = 2026,
        month = jan,
          eid = {arXiv:2601.15954},
        pages = {arXiv:2601.15954},
          doi = {10.48550/arXiv.2601.15954},
archivePrefix = {arXiv},
       eprint = {2601.15954},
 primaryClass = {astro-ph.GA},
       adsurl = {https://ui.adsabs.harvard.edu/abs/2026arXiv260115954J}
}

@ARTICLE{Hsu2021,
       author = {{Hsu}, Chih-Chun and {Burgasser}, Adam J. and {Theissen}, Christopher A. and {Gelino}, Christopher R. and {Birky}, Jessica L. and {Diamant}, Sharon J.~M. and {Bardalez Gagliuffi}, Daniella C. and {Aganze}, Christian and {Blake}, Cullen H. and {Faherty}, Jacqueline K.},
        title = "{The Brown Dwarf Kinematics Project (BDKP). V. Radial and Rotational Velocities of T Dwarfs from Keck/NIRSPEC High-resolution Spectroscopy}",
      journal = {\apjs},
         year = 2021,
        month = dec,
       volume = {257},
       number = {2},
          eid = {45},
        pages = {45},
          doi = {10.3847/1538-4365/ac1c7d},
archivePrefix = {arXiv},
       eprint = {2107.01222},
 primaryClass = {astro-ph.SR},
       adsurl = {https://ui.adsabs.harvard.edu/abs/2021ApJS..257...45H}
}

@ARTICLE{Spiegel2011,
       author = {{Spiegel}, David S. and {Burrows}, Adam and {Milsom}, John A.},
        title = "{The Deuterium-burning Mass Limit for Brown Dwarfs and Giant Planets}",
      journal = {\apj},
         year = 2011,
        month = jan,
       volume = {727},
       number = {1},
          eid = {57},
        pages = {57},
          doi = {10.1088/0004-637X/727/1/57},
archivePrefix = {arXiv},
       eprint = {1008.5150},
 primaryClass = {astro-ph.EP},
       adsurl = {https://ui.adsabs.harvard.edu/abs/2011ApJ...727...57S}
}

@ARTICLE{Meisner2023,
       author = {{Meisner}, Aaron M. and {Leggett}, S.~K. and {Logsdon}, Sarah E. and {Schneider}, Adam C. and {Tremblin}, Pascal and {Phillips}, Mark},
        title = "{Exploring the Extremes: Characterizing a New Population of Old and Cold Brown Dwarfs}",
      journal = {\aj},
         year = 2023,
        month = aug,
       volume = {166},
       number = {2},
          eid = {57},
        pages = {57},
          doi = {10.3847/1538-3881/acdb68},
archivePrefix = {arXiv},
       eprint = {2301.09817},
 primaryClass = {astro-ph.SR},
       adsurl = {https://ui.adsabs.harvard.edu/abs/2023AJ....166...57M}
}

@ARTICLE{Marocco2021,
       author = {{Marocco}, Federico and {Eisenhardt}, Peter R.~M. and {Fowler}, John W. and {Kirkpatrick}, J. Davy and {Meisner}, Aaron M. and {Schlafly}, Edward F. and {Stanford}, S.~A. and {Garcia}, Nelson and {Caselden}, Dan and {Cushing}, Michael C. and {Cutri}, Roc M. and {Faherty}, Jacqueline K. and {Gelino}, Christopher R. and {Gonzalez}, Anthony H. and {Jarrett}, Thomas H. and {Koontz}, Renata and {Mainzer}, Amanda and {Marchese}, Elijah J. and {Mobasher}, Bahram and {Schlegel}, David J. and {Stern}, Daniel and {Teplitz}, Harry I. and {Wright}, Edward L.},
        title = "{The CatWISE2020 Catalog}",
      journal = {\apjs},
         year = 2021,
        month = mar,
       volume = {253},
       number = {1},
          eid = {8},
        pages = {8},
          doi = {10.3847/1538-4365/abd805},
archivePrefix = {arXiv},
       eprint = {2012.13084},
 primaryClass = {astro-ph.IM},
       adsurl = {https://ui.adsabs.harvard.edu/abs/2021ApJS..253....8M}
}

@ARTICLE{Jarvis2013,
       author = {{Jarvis}, Matt J. and {Bonfield}, D.~G. and {Bruce}, V.~A. and {Geach}, J.~E. and {McAlpine}, K. and {McLure}, R.~J. and {Gonz{\'a}lez-Solares}, E. and {Irwin}, M. and {Lewis}, J. and {Yoldas}, A. Kupcu and {Andreon}, S. and {Cross}, N.~J.~G. and {Emerson}, J.~P. and {Dalton}, G. and {Dunlop}, J.~S. and {Hodgkin}, S.~T. and {Le}, F{\`e}vre O. and {Karouzos}, M. and {Meisenheimer}, K. and {Oliver}, S. and {Rawlings}, S. and {Simpson}, C. and {Smail}, I. and {Smith}, D.~J.~B. and {Sullivan}, M. and {Sutherland}, W. and {White}, S.~V. and {Zwart}, J.~T.~L.},
        title = "{The VISTA Deep Extragalactic Observations (VIDEO) survey}",
      journal = {\mnras},
         year = 2013,
        month = jan,
       volume = {428},
       number = {2},
        pages = {1281-1295},
          doi = {10.1093/mnras/sts118},
archivePrefix = {arXiv},
       eprint = {1206.4263},
 primaryClass = {astro-ph.CO},
       adsurl = {https://ui.adsabs.harvard.edu/abs/2013MNRAS.428.1281J}
}

@ARTICLE{Nidever2021,
       author = {{Nidever}, David L. and {Dey}, Arjun and {Fasbender}, Katie and {Juneau}, St{\'e}phanie and {Meisner}, Aaron M. and {Wishart}, Joseph and {Scott}, Adam and {Matt}, Kyle and {Nikutta}, Robert and {Pucha}, Ragadeepika},
        title = "{Second Data Release of the All-sky NOIRLab Source Catalog}",
      journal = {\aj},
         year = 2021,
        month = apr,
       volume = {161},
       number = {4},
          eid = {192},
        pages = {192},
          doi = {10.3847/1538-3881/abd6e1},
archivePrefix = {arXiv},
       eprint = {2011.08868},
 primaryClass = {astro-ph.GA},
       adsurl = {https://ui.adsabs.harvard.edu/abs/2021AJ....161..192N}
}

@ARTICLE{Abbott2021,
       author = {{Abbott}, T.~M.~C. and {Adam{\'o}w}, M. and {Aguena}, M. and {Allam}, S. and {Amon}, A. and {Annis}, J. and {Avila}, S. and {Bacon}, D. and {Banerji}, M. and {Bechtol}, K. and {Becker}, M.~R. and {Bernstein}, G.~M. and {Bertin}, E. and {Bhargava}, S. and {Bridle}, S.~L. and {Brooks}, D. and {Burke}, D.~L. and {Carnero Rosell}, A. and {Carrasco Kind}, M. and {Carretero}, J. and {Castander}, F.~J. and {Cawthon}, R. and {Chang}, C. and {Choi}, A. and {Conselice}, C. and {Costanzi}, M. and {Crocce}, M. and {da Costa}, L.~N. and {Davis}, T.~M. and {De Vicente}, J. and {DeRose}, J. and {Desai}, S. and {Diehl}, H.~T. and {Dietrich}, J.~P. and {Drlica-Wagner}, A. and {Eckert}, K. and {Elvin-Poole}, J. and {Everett}, S. and {Evrard}, A.~E. and {Ferrero}, I. and {Fert{\'e}}, A. and {Flaugher}, B. and {Fosalba}, P. and {Friedel}, D. and {Frieman}, J. and {Garc{\'\i}a-Bellido}, J. and {Gaztanaga}, E. and {Gelman}, L. and {Gerdes}, D.~W. and {Giannantonio}, T. and {Gill}, M.~S.~S. and {Gruen}, D. and {Gruendl}, R.~A. and {Gschwend}, J. and {Gutierrez}, G. and {Hartley}, W.~G. and {Hinton}, S.~R. and {Hollowood}, D.~L. and {Honscheid}, K. and {Huterer}, D. and {James}, D.~J. and {Jeltema}, T. and {Johnson}, M.~D. and {Kent}, S. and {Kron}, R. and {Kuehn}, K. and {Kuropatkin}, N. and {Lahav}, O. and {Li}, T.~S. and {Lidman}, C. and {Lin}, H. and {MacCrann}, N. and {Maia}, M.~A.~G. and {Manning}, T.~A. and {Maloney}, J.~D. and {March}, M. and {Marshall}, J.~L. and {Martini}, P. and {Melchior}, P. and {Menanteau}, F. and {Miquel}, R. and {Morgan}, R. and {Myles}, J. and {Neilsen}, E. and {Ogando}, R.~L.~C. and {Palmese}, A. and {Paz-Chinch{\'o}n}, F. and {Petravick}, D. and {Pieres}, A. and {Plazas}, A.~A. and {Pond}, C. and {Rodriguez-Monroy}, M. and {Romer}, A.~K. and {Roodman}, A. and {Rykoff}, E.~S. and {Sako}, M. and {Sanchez}, E. and {Santiago}, B. and {Scarpine}, V. and {Serrano}, S. and {Sevilla-Noarbe}, I. and {Smith}, J. Allyn and {Smith}, M. and {Soares-Santos}, M. and {Suchyta}, E. and {Swanson}, M.~E.~C. and {Tarle}, G. and {Thomas}, D. and {To}, C. and {Tremblay}, P.~E. and {Troxel}, M.~A. and {Tucker}, D.~L. and {Turner}, D.~J. and {Varga}, T.~N. and {Walker}, A.~R. and {Wechsler}, R.~H. and {Weller}, J. and {Wester}, W. and {Wilkinson}, R.~D. and {Yanny}, B. and {Zhang}, Y. and {Nikutta}, R. and {Fitzpatrick}, M. and {Jacques}, A. and {Scott}, A. and {Olsen}, K. and {Huang}, L. and {Herrera}, D. and {Juneau}, S. and {Nidever}, D. and {Weaver}, B.~A. and {Adean}, C. and {Correia}, V. and {de Freitas}, M. and {Freitas}, F.~N. and {Singulani}, C. and {Vila-Verde}, G. and {Linea Science Server}},
        title = "{The Dark Energy Survey Data Release 2}",
      journal = {\apjs},
         year = 2021,
        month = aug,
       volume = {255},
       number = {2},
          eid = {20},
        pages = {20},
          doi = {10.3847/1538-4365/ac00b3},
archivePrefix = {arXiv},
       eprint = {2101.05765},
 primaryClass = {astro-ph.IM},
       adsurl = {https://ui.adsabs.harvard.edu/abs/2021ApJS..255...20A}
}

@ARTICLE{dalPonte2023,
       author = {{dal Ponte}, M. and {Santiago}, B. and {Carnero Rosell}, A. and {De Paris}, L. and {Pace}, A.~B. and {Bechtol}, K. and {Abbott}, T.~M.~C. and {Aguena}, M. and {Allam}, S. and {Alves}, O. and {Bacon}, D. and {Bertin}, E. and {Bocquet}, S. and {Brooks}, D. and {Burke}, D.~L. and {Carrasco Kind}, M. and {Carretero}, J. and {Conselice}, C. and {Costanzi}, M. and {Desai}, S. and {De Vicente}, J. and {Doel}, P. and {Everett}, S. and {Ferrero}, I. and {Flaugher}, B. and {Frieman}, J. and {Garc{\'\i}a-Bellido}, J. and {Gerdes}, D.~W. and {Gruendl}, R.~A. and {Gruen}, D. and {Gutierrez}, G. and {Hinton}, S.~R. and {Hollowood}, D.~L. and {James}, D.~J. and {Kuehn}, K. and {Kuropatkin}, N. and {Marshall}, J.~L. and {Mena-Fern{\'a}ndez}, J. and {Menanteau}, F. and {Miquel}, R. and {Ogando}, R.~L.~C. and {Palmese}, A. and {Paz-Chinch{\'o}n}, F. and {Pereira}, M.~E.~S. and {Plazas Malag{\'o}n}, A.~A. and {Pieres}, A. and {Raveri}, M. and {Rodriguez-Monroy}, M. and {Sanchez}, E. and {Scarpine}, V. and {Schubnell}, M. and {Sevilla-Noarbe}, I. and {Smith}, M. and {Soares-Santos}, M. and {Suchyta}, E. and {Swanson}, M.~E.~C. and {Tarle}, G. and {Thomas}, D. and {To}, C. and {Weaverdyck}, N. and {DES Collaboration}},
        title = "{Ultracool dwarfs candidates based on 6 yr of the Dark Energy Survey data}",
      journal = {\mnras},
         year = 2023,
        month = jun,
       volume = {522},
       number = {2},
        pages = {1951-1967},
          doi = {10.1093/mnras/stad955},
archivePrefix = {arXiv},
       eprint = {2303.15156},
 primaryClass = {astro-ph.SR},
       adsurl = {https://ui.adsabs.harvard.edu/abs/2023MNRAS.522.1951D}
}

@ARTICLE{MiretRoig2023,
       author = {{Miret-Roig}, N{\'u}ria},
        title = "{The origin of free-floating planets}",
      journal = {\apss},
         year = 2023,
        month = mar,
       volume = {368},
       number = {3},
          eid = {17},
        pages = {17},
          doi = {10.1007/s10509-023-04175-5},
archivePrefix = {arXiv},
       eprint = {2303.05522},
 primaryClass = {astro-ph.EP},
       adsurl = {https://ui.adsabs.harvard.edu/abs/2023Ap&SS.368...17M}
}

@ARTICLE{Lomb1976,
       author = {{Lomb}, N.~R.},
        title = "{Least-Squares Frequency Analysis of Unequally Spaced Data}",
      journal = {\apss},
         year = 1976,
        month = feb,
       volume = {39},
       number = {2},
        pages = {447-462},
          doi = {10.1007/BF00648343},
       adsurl = {https://ui.adsabs.harvard.edu/abs/1976Ap&SS..39..447L}
}

@ARTICLE{Scargle1982,
       author = {{Scargle}, J.~D.},
        title = "{Studies in astronomical time series analysis. II. Statistical aspects of spectral analysis of unevenly spaced data.}",
      journal = {\apj},
         year = 1982,
        month = dec,
       volume = {263},
        pages = {835-853},
          doi = {10.1086/160554},
       adsurl = {https://ui.adsabs.harvard.edu/abs/1982ApJ...263..835S}
}

@ARTICLE{Dupuy2023,
       author = {{Dupuy}, Trent J. and {Liu}, Michael C. and {Evans}, Elise L. and {Best}, William M.~J. and {Pearce}, Logan A. and {Sanghi}, Aniket and {Phillips}, Mark W. and {Bardalez Gagliuffi}, Daniella C.},
        title = "{On the masses, age, and architecture of the VHS J1256-1257AB b system}",
      journal = {\mnras},
         year = 2023,
        month = feb,
       volume = {519},
       number = {2},
        pages = {1688-1694},
          doi = {10.1093/mnras/stac3557},
archivePrefix = {arXiv},
       eprint = {2208.08448},
 primaryClass = {astro-ph.EP},
       adsurl = {https://ui.adsabs.harvard.edu/abs/2023MNRAS.519.1688D}
}

@ARTICLE{Miles2023,
       author = {{Miles}, Brittany E. and {Biller}, Beth A. and {Patapis}, Polychronis and {Worthen}, Kadin and {Rickman}, Emily and {Hoch}, Kielan K.~W. and {Skemer}, Andrew and {Perrin}, Marshall D. and {Whiteford}, Niall and {Chen}, Christine H. and {Sargent}, B. and {Mukherjee}, Sagnick and {Morley}, Caroline V. and {Moran}, Sarah E. and {Bonnefoy}, Mickael and {Petrus}, Simon and {Carter}, Aarynn L. and {Choquet}, Elodie and {Hinkley}, Sasha and {Ward-Duong}, Kimberly and {Leisenring}, Jarron M. and {Millar-Blanchaer}, Maxwell A. and {Pueyo}, Laurent and {Ray}, Shrishmoy and {Sallum}, Steph and {Stapelfeldt}, Karl R. and {Stone}, Jordan M. and {Wang}, Jason J. and {Absil}, Olivier and {Balmer}, William O. and {Boccaletti}, Anthony and {Bonavita}, Mariangela and {Booth}, Mark and {Bowler}, Brendan P. and {Chauvin}, Gael and {Christiaens}, Valentin and {Currie}, Thayne and {Danielski}, Camilla and {Fortney}, Jonathan J. and {Girard}, Julien H. and {Grady}, Carol A. and {Greenbaum}, Alexandra Z. and {Henning}, Thomas and {Hines}, Dean C. and {Janson}, Markus and {Kalas}, Paul and {Kammerer}, Jens and {Kennedy}, Grant M. and {Kenworthy}, Matthew A. and {Kervella}, Pierre and {Lagage}, Pierre-Olivier and {Lew}, Ben W.~P. and {Liu}, Michael C. and {Macintosh}, Bruce and {Marino}, Sebastian and {Marley}, Mark S. and {Marois}, Christian and {Matthews}, Elisabeth C. and {Matthews}, Brenda C. and {Mawet}, Dimitri and {McElwain}, Michael W. and {Metchev}, Stanimir and {Meyer}, Michael R. and {Molliere}, Paul and {Pantin}, Eric and {Quirrenbach}, Andreas and {Rebollido}, Isabel and {Ren}, Bin B. and {Schneider}, Glenn and {Vasist}, Malavika and {Wyatt}, Mark C. and {Zhou}, Yifan and {Briesemeister}, Zackery W. and {Bryan}, Marta L. and {Calissendorff}, Per and {Cantalloube}, Faustine and {Cugno}, Gabriele and {De Furio}, Matthew and {Dupuy}, Trent J. and {Factor}, Samuel M. and {Faherty}, Jacqueline K. and {Fitzgerald}, Michael P. and {Franson}, Kyle and {Gonzales}, Eileen C. and {Hood}, Callie E. and {Howe}, Alex R. and {Kraus}, Adam L. and {Kuzuhara}, Masayuki and {Lagrange}, Anne-Marie and {Lawson}, Kellen and {Lazzoni}, Cecilia and {Liu}, Pengyu and {Llop-Sayson}, Jorge and {Lloyd}, James P. and {Martinez}, Raquel A. and {Mazoyer}, Johan and {Quanz}, Sascha P. and {Redai}, Jea Adams and {Samland}, Matthias and {Schlieder}, Joshua E. and {Tamura}, Motohide and {Tan}, Xianyu and {Uyama}, Taichi and {Vigan}, Arthur and {Vos}, Johanna M. and {Wagner}, Kevin and {Wolff}, Schuyler G. and {Ygouf}, Marie and {Zhang}, Xi and {Zhang}, Keming and {Zhang}, Zhoujian},
        title = "{The JWST Early-release Science Program for Direct Observations of Exoplanetary Systems II: A 1 to 20 {\ensuremath{\mu}}m Spectrum of the Planetary-mass Companion VHS 1256-1257 b}",
      journal = {\apjl},
         year = 2023,
        month = mar,
       volume = {946},
       number = {1},
          eid = {L6},
        pages = {L6},
          doi = {10.3847/2041-8213/acb04a},
archivePrefix = {arXiv},
       eprint = {2209.00620},
 primaryClass = {astro-ph.EP},
       adsurl = {https://ui.adsabs.harvard.edu/abs/2023ApJ...946L...6M}
}

@ARTICLE{Liu2024b,
       author = {{Liu}, Feng-Yuan and {Dai}, Y. Sophia and {Omont}, Alain and {Liu}, Daizhong and {Cox}, Pierre and {Neri}, Roberto and {Krips}, Melanie and {Yang}, Chentao and {Wu}, Xue-Bing and {Huang}, Jia-Sheng},
        title = "{Dust and Cold Gas Properties of Starburst HyLIRG Quasars at z {\ensuremath{\sim}} 2.5}",
      journal = {\apj},
         year = 2024,
        month = apr,
       volume = {964},
       number = {2},
          eid = {136},
        pages = {136},
          doi = {10.3847/1538-4357/ad24fe},
archivePrefix = {arXiv},
       eprint = {2402.03909},
 primaryClass = {astro-ph.GA},
       adsurl = {https://ui.adsabs.harvard.edu/abs/2024ApJ...964..136L}
}

@ARTICLE{MacLeod2016,
       author = {{MacLeod}, Chelsea L. and {Ross}, Nicholas P. and {Lawrence}, Andy and {Goad}, Mike and {Horne}, Keith and {Burgett}, William and {Chambers}, Ken C. and {Flewelling}, Heather and {Hodapp}, Klaus and {Kaiser}, Nick and {Magnier}, Eugene and {Wainscoat}, Richard and {Waters}, Christopher},
        title = "{A systematic search for changing-look quasars in SDSS}",
      journal = {\mnras},
         year = 2016,
        month = mar,
       volume = {457},
       number = {1},
        pages = {389-404},
          doi = {10.1093/mnras/stv2997},
archivePrefix = {arXiv},
       eprint = {1509.08393},
 primaryClass = {astro-ph.GA},
       adsurl = {https://ui.adsabs.harvard.edu/abs/2016MNRAS.457..389M}
}

@ARTICLE{Izotov2016,
       author = {{Izotov}, Y.~I. and {Guseva}, N.~G. and {Fricke}, K.~J. and {Henkel}, C.},
        title = "{The bursting nature of star formation in compact star-forming galaxies from the Sloan Digital Sky Survey}",
      journal = {\mnras},
         year = 2016,
        month = nov,
       volume = {462},
       number = {4},
        pages = {4427-4434},
          doi = {10.1093/mnras/stw1973},
archivePrefix = {arXiv},
       eprint = {1608.01523},
 primaryClass = {astro-ph.GA},
       adsurl = {https://ui.adsabs.harvard.edu/abs/2016MNRAS.462.4427I}
}

@ARTICLE{Ricci2023,
       author = {{Ricci}, Claudio and {Trakhtenbrot}, Benny},
        title = "{Changing-look active galactic nuclei}",
      journal = {Nature Astronomy},
         year = 2023,
        month = nov,
       volume = {7},
        pages = {1282-1294},
          doi = {10.1038/s41550-023-02108-4},
archivePrefix = {arXiv},
       eprint = {2211.05132},
 primaryClass = {astro-ph.GA},
       adsurl = {https://ui.adsabs.harvard.edu/abs/2023NatAs...7.1282R}
}

@ARTICLE{Winkler1992,
       author = {{Winkler}, H. and {Glass}, I.~S. and {van Wyk}, F. and {Marang}, F. and {Jones}, J.~H.~S. and {Buckley}, D.~A.~H. and {Sekiguchi}, K.},
        title = "{Variability studies of seyfert galaxies - I. Broad-band optical photometry .}",
      journal = {\mnras},
         year = 1992,
        month = aug,
       volume = {257},
        pages = {659-676},
          doi = {10.1093/mnras/257.4.659},
       adsurl = {https://ui.adsabs.harvard.edu/abs/1992MNRAS.257..659W}
}

@ARTICLE{Kuhnle2025,
       author = {{K{\"u}hnle}, H. and {Patapis}, P. and {Molli{\`e}re}, P. and {Tremblin}, P. and {Matthews}, E. and {Glauser}, A.~M. and {Whiteford}, N. and {Vasist}, M. and {Absil}, O. and {Barrado}, D. and {Min}, M. and {Lagage}, P.-O. and {Waters}, L.~B.~F.~M. and {Guedel}, M. and {Henning}, Th. and {Vandenbussche}, B. and {Baudoz}, P. and {Decin}, L. and {Pye}, J.~P. and {Royer}, P. and {van Dishoeck}, E.~F. and {{\"O}stlin}, G. and {Ray}, T.~P. and {Wright}, G.},
        title = "{Water depletion and $^{15}$NH$_{3}$ in the atmosphere of the coldest brown dwarf observed with JWST/MIRI}",
      journal = {\aap},
         year = 2025,
        month = mar,
       volume = {695},
          eid = {A224},
        pages = {A224},
          doi = {10.1051/0004-6361/202452547},
archivePrefix = {arXiv},
       eprint = {2410.10933},
 primaryClass = {astro-ph.EP},
       adsurl = {https://ui.adsabs.harvard.edu/abs/2025A&A...695A.224K}
}

@ARTICLE{Wilson2025,
       author = {{Wilson}, Mikayla J. and {Limbach}, Mary Anne and {Skemer}, Andrew J. and {Vos}, Johanna M. and {Miles}, Brittany E. and {Rowland}, Melanie J. and {Vanderburg}, Andrew and {Schneider}, Adam C. and {Morley}, Caroline and {Kotten}, Brooke and {Householder}, Andrew and {Lupu}, Roxana and {Mang}, James and {Freedman}, Richard},
        title = "{A Deep Search for Exomoons around WISE 0855 with JWST}",
      journal = {\aj},
         year = 2025,
        month = dec,
       volume = {170},
       number = {6},
          eid = {357},
        pages = {357},
          doi = {10.3847/1538-3881/ae17ca},
archivePrefix = {arXiv},
       eprint = {2510.24575},
 primaryClass = {astro-ph.EP},
       adsurl = {https://ui.adsabs.harvard.edu/abs/2025AJ....170..357W}
}

@ARTICLE{Stolker2020,
       author = {{Stolker}, T. and {Quanz}, S.~P. and {Todorov}, K.~O. and {K{\"u}hn}, J. and {Molli{\`e}re}, P. and {Meyer}, M.~R. and {Currie}, T. and {Daemgen}, S. and {Lavie}, B.},
        title = "{MIRACLES: atmospheric characterization of directly imaged planets and substellar companions at 4-5 {\ensuremath{\mu}}m. I. Photometric analysis of {\ensuremath{\beta}} Pic b, HIP 65426 b, PZ Tel B, and HD 206893 B}",
      journal = {\aap},
         year = 2020,
        month = mar,
       volume = {635},
          eid = {A182},
        pages = {A182},
          doi = {10.1051/0004-6361/201937159},
archivePrefix = {arXiv},
       eprint = {1912.13316},
 primaryClass = {astro-ph.EP},
       adsurl = {https://ui.adsabs.harvard.edu/abs/2020A&A...635A.182S}
}

@ARTICLE{Buchner2014,
       author = {{Buchner}, J. and {Georgakakis}, A. and {Nandra}, K. and {Hsu}, L. and {Rangel}, C. and {Brightman}, M. and {Merloni}, A. and {Salvato}, M. and {Donley}, J. and {Kocevski}, D.},
        title = "{X-ray spectral modelling of the AGN obscuring region in the CDFS: Bayesian model selection and catalogue}",
      journal = {\aap},
         year = 2014,
        month = apr,
       volume = {564},
          eid = {A125},
        pages = {A125},
          doi = {10.1051/0004-6361/201322971},
archivePrefix = {arXiv},
       eprint = {1402.0004},
 primaryClass = {astro-ph.HE},
       adsurl = {https://ui.adsabs.harvard.edu/abs/2014A&A...564A.125B}
}

@ARTICLE{Meynardie2025,
       author = {{Meynardie}, William W. and {Meyer}, Michael R. and {MacDonald}, Ryan J. and {Calissendorff}, Per and {Mullens}, Elijah and {Zarazua}, Gabriel Munoz and {Roy}, Anuranj and {Ganta}, Hansica and {Gonzales}, Eileen C. and {Adams}, Arthur and {Lewis}, Nikole and {Hong}, Yucian and {Lunine}, Jonathan},
        title = "{Ross 458 C: Gas Giant or Brown Dwarf?}",
      journal = {\apj},
         year = 2025,
        month = dec,
       volume = {994},
       number = {2},
          eid = {237},
        pages = {237},
          doi = {10.3847/1538-4357/ae0ad0},
archivePrefix = {arXiv},
       eprint = {2509.22803},
 primaryClass = {astro-ph.EP},
       adsurl = {https://ui.adsabs.harvard.edu/abs/2025ApJ...994..237M}
}

@ARTICLE{Schlegel1998,
       author = {{Schlegel}, David J. and {Finkbeiner}, Douglas P. and {Davis}, Marc},
        title = "{Maps of Dust Infrared Emission for Use in Estimation of Reddening and Cosmic Microwave Background Radiation Foregrounds}",
      journal = {\apj},
         year = 1998,
        month = jun,
       volume = {500},
       number = {2},
        pages = {525-553},
          doi = {10.1086/305772},
archivePrefix = {arXiv},
       eprint = {astro-ph/9710327},
 primaryClass = {astro-ph},
       adsurl = {https://ui.adsabs.harvard.edu/abs/1998ApJ...500..525S}
}

@ARTICLE{Schlafly2011,
       author = {{Schlafly}, Edward F. and {Finkbeiner}, Douglas P.},
        title = "{Measuring Reddening with Sloan Digital Sky Survey Stellar Spectra and Recalibrating SFD}",
      journal = {\apj},
         year = 2011,
        month = aug,
       volume = {737},
       number = {2},
          eid = {103},
        pages = {103},
          doi = {10.1088/0004-637X/737/2/103},
archivePrefix = {arXiv},
       eprint = {1012.4804},
 primaryClass = {astro-ph.GA},
       adsurl = {https://ui.adsabs.harvard.edu/abs/2011ApJ...737..103S}
}

@ARTICLE{Tu2026,
       author = {{Tu}, Zhijun and {Wang}, Shu and {Huang}, Haomiao and {Chen}, Xiaodian and {Liu}, Jifeng},
        title = "{SPHEREx Ultracool Dwarf Spectrophotometric Atlas (SUDA): Atmospheric and Fundamental Parameters of Ultracool Dwarfs}",
      journal = {\apj},
         year = 2026,
        month = sep,
       volume = {1008},
       number = {2},
          eid = {250},
        pages = {250},
          doi = {10.3847/1538-4357/ae9082},
archivePrefix = {arXiv},
       eprint = {2604.26771},
 primaryClass = {astro-ph.SR},
       adsurl = {https://ui.adsabs.harvard.edu/abs/2026ApJ..1008..250T}
}

@ARTICLE{Stephens2004,
       author = {{Stephens}, D.~C. and {Leggett}, S.~K.},
        title = "{JHK Magnitudes for L and T Dwarfs and Infrared Photometric Systems}",
      journal = {\pasp},
         year = 2004,
        month = jan,
       volume = {116},
       number = {815},
        pages = {9-21},
          doi = {10.1086/381135},
archivePrefix = {arXiv},
       eprint = {astro-ph/0311229},
 primaryClass = {astro-ph},
       adsurl = {https://ui.adsabs.harvard.edu/abs/2004PASP..116....9S}
}

@ARTICLE{Faherty2025,
       author = {{Faherty}, Jacqueline K. and {Meisner}, Aaron M. and {Burningham}, Ben and {Visscher}, Channon and {Line}, Michael and {Su{\'a}rez}, Genaro and {Gagn{\'e}}, Jonathan and {Alejandro Merchan}, Sherelyn and {Rothermich}, Austin James and {Burgasser}, Adam J. and {Schneider}, Adam C. and {Caselden}, Dan and {Kirkpatrick}, J. Davy and {Kuchner}, Marc Jason and {Bardalez Gagliuffi}, Daniella Carolina and {Eisenhardt}, Peter and {Gelino}, Christopher R. and {Gonzales}, Eileen C. and {Marocco}, Federico and {Leggett}, Sandy and {Lodieu}, Nicolas and {Casewell}, Sarah L. and {Tremblin}, Pascal and {Cushing}, Michael and {Zapatero Osorio}, Maria Rosa and {B{\'e}jar}, V{\'\i}ctor J.~S. and {Gauza}, Bartosz and {Wright}, Edward and {Phillips}, Mark W. and {Zhang}, Jun-Yan and {Martin}, Eduardo L.},
        title = "{Silicate precursor silane detected in cold low-metallicity brown dwarf}",
      journal = {\nat},
         year = 2025,
        month = sep,
       volume = {645},
       number = {8079},
        pages = {62-66},
          doi = {10.1038/s41586-025-09369-1},
archivePrefix = {arXiv},
       eprint = {2509.18635},
 primaryClass = {astro-ph.EP},
       adsurl = {https://ui.adsabs.harvard.edu/abs/2025Natur.645...62F}
}

@ARTICLE{Luhman2023,
       author = {{Luhman}, K.~L. and {Tremblin}, P. and {Birkmann}, S.~M. and {Manjavacas}, E. and {Valenti}, J. and {Alves de Oliveira}, C. and {Beck}, T.~L. and {Giardino}, G. and {L{\"u}tzgendorf}, N. and {Rauscher}, B.~J. and {Sirianni}, M.},
        title = "{JWST/NIRSpec Observations of the Planetary Mass Companion TWA 27B}",
      journal = {\apjl},
         year = 2023,
        month = jun,
       volume = {949},
       number = {2},
          eid = {L36},
        pages = {L36},
          doi = {10.3847/2041-8213/acd635},
archivePrefix = {arXiv},
       eprint = {2305.18603},
 primaryClass = {astro-ph.EP},
       adsurl = {https://ui.adsabs.harvard.edu/abs/2023ApJ...949L..36L}
}

@ARTICLE{X_Chen2025,
       author = {{Chen}, Xueqing and {Biller}, Beth A. and {Tan}, Xianyu and {Vos}, Johanna M. and {Zhou}, Yifan and {Su{\'a}rez}, Genaro and {McCarthy}, Allison M. and {Morley}, Caroline V. and {Whiteford}, Niall and {Dupuy}, Trent J. and {Faherty}, Jacqueline and {Sutlieff}, Ben J. and {Oliveros-Gomez}, Natalia and {Manjavacas}, Elena and {Limbach}, Mary Anne and {Lee}, Elspeth K.~H. and {Karalidi}, Theodora and {Crossfield}, Ian J.~M. and {Liu}, Pengyu and {Molliere}, Paul and {Muirhead}, Philip S. and {Henning}, Thomas and {Mace}, Gregory and {Crouzet}, Nicolas and {Kataria}, Tiffany},
        title = "{The JWST weather report from the nearest brown dwarfs II: consistent variability mechanisms over 7 months revealed by 1─14 {\ensuremath{\mu}}m NIRSpec + MIRI monitoring of WISE 1049AB}",
      journal = {\mnras},
         year = 2025,
        month = jun,
       volume = {539},
       number = {4},
        pages = {3758-3777},
          doi = {10.1093/mnras/staf737},
archivePrefix = {arXiv},
       eprint = {2505.00794},
 primaryClass = {astro-ph.SR},
       adsurl = {https://ui.adsabs.harvard.edu/abs/2025MNRAS.539.3758C}
}

@ARTICLE{OliverosGomez2026,
       author = {{Oliveros-Gomez}, Natalia and {Manjavacas}, Elena and {Karalidi}, Theodora and {Phillippe}, Myrla and {Campos Estrada}, Beatriz and {Biller}, Beth and {Vos}, Johanna M. and {Faherty}, Jacqueline and {Chen}, Xueqing and {Dupuy}, Trent J. and {Henning}, Thomas and {McCarthy}, Allison M. and {Muirhead}, Philip S. and {Lee}, Elspeth K.~H. and {Tremblin}, Pascal and {Ramirez}, Jasmine and {Suarez}, Genaro and {Sutlieff}, Ben J. and {Tan}, Xianyu and {Crouzet}, Nicolas},
        title = "{The JWST Weather Report from the Nearest Brown Dwarfs. III. Heterogeneous Clouds and Thermochemical Instabilities as Possible Drivers of WISE 1049AB's Spectroscopic Variability}",
      journal = {\apj},
         year = 2026,
        month = feb,
       volume = {997},
       number = {2},
          eid = {136},
        pages = {136},
          doi = {10.3847/1538-4357/ae21d9},
archivePrefix = {arXiv},
       eprint = {2511.15667},
 primaryClass = {astro-ph.SR},
       adsurl = {https://ui.adsabs.harvard.edu/abs/2026ApJ...997..136O}
}

@ARTICLE{Nasedkin2025,
       author = {{Nasedkin}, E. and {Schrader}, M. and {Vos}, J.~M. and {Biller}, B. and {Burningham}, B. and {Cowan}, N.~B. and {Faherty}, J.~K. and {Gonzales}, E. and {Lam}, M.~B. and {McCarthy}, A.~M. and {Muirhead}, P.~S. and {O'Toole}, C. and {Plummer}, M.~K. and {Su{\'a}rez}, G. and {Tan}, X. and {Visscher}, C. and {Whiteford}, N. and {Zhou}, Y.},
        title = "{The JWST weather report: Retrieving temperature variations, auroral heating, and static cloud coverage on SIMP-0136}",
      journal = {\aap},
         year = 2025,
        month = oct,
       volume = {702},
          eid = {A1},
        pages = {A1},
          doi = {10.1051/0004-6361/202555370},
archivePrefix = {arXiv},
       eprint = {2507.07772},
 primaryClass = {astro-ph.EP},
       adsurl = {https://ui.adsabs.harvard.edu/abs/2025A&A...702A...1N}
}

@ARTICLE{Manjavacas2024,
       author = {{Manjavacas}, Elena and {Tremblin}, Pascal and {Birkmann}, Stephan and {Valenti}, Jeff and {Alves de Oliveira}, Catarina and {Beck}, Tracy L. and {Giardino}, G. and {L{\"u}tzgendorf}, N. and {Rauscher}, B.~J. and {Sirianni}, M.},
        title = "{Medium-resolution 0.97─5.3 {\ensuremath{\mu}}m Spectra of Very Young Benchmark Brown Dwarfs with NIRSpec on Board the James Webb Space Telescope}",
      journal = {\aj},
         year = 2024,
        month = apr,
       volume = {167},
       number = {4},
          eid = {168},
        pages = {168},
          doi = {10.3847/1538-3881/ad2938},
archivePrefix = {arXiv},
       eprint = {2402.04230},
 primaryClass = {astro-ph.SR},
       adsurl = {https://ui.adsabs.harvard.edu/abs/2024AJ....167..168M}
}

@ARTICLE{Lew2026,
       author = {{Lew}, Ben W.~P. and {Roellig}, Thomas and {Batalha}, Natasha E. and {Wogan}, Nicholas F. and {Greene}, Thomas and {Marley}, Mark S. and {Fortney}, Jonathan J. and {Leisenring}, Jarron and {Johnstone}, Doug and {De Furio}, Matthew and {Hodapp}, Klaus and {Beichman}, Charles and {Rieke}, Marcia},
        title = "{JWST Spectral Retrieval of Cold Directly Imaged Planet WD 0806 b and the First Measurement of Altitude-dependent K$_{zz}$ in Exoplanet Atmospheres}",
      journal = {\aj},
         year = 2026,
        month = apr,
       volume = {171},
       number = {4},
          eid = {227},
        pages = {227},
          doi = {10.3847/1538-3881/ae4747},
archivePrefix = {arXiv},
       eprint = {2602.08200},
 primaryClass = {astro-ph.EP},
       adsurl = {https://ui.adsabs.harvard.edu/abs/2026AJ....171..227L}
}

@ARTICLE{Kiman2026a,
       author = {{Kiman}, Rocio and {Beichman}, Charles A. and {Ruiz Diaz}, Azul and {Faherty}, Jacqueline K. and {Lacy}, Brianna and {Su{\'a}rez}, Genaro and {Marocco}, Federico and {Kirkpatrick}, J. Davy and {Gagn{\'e}}, Jonathan and {Copeland}, Jessica and {Burningham}, Ben and {Whiteford}, Niall and {Rowland}, Melanie J. and {Bardalez Gagliuffi}, Daniella C. and {Vos}, Johanna M. and {Schneider}, Adam C. and {Gonzales}, Eileen C. and {Alejandro Merchan}, Sherelyn and {Rothermich}, Austin and {Smart}, Richard and {Costa}, Edgardo and {Mendez}, Rene A.},
        title = "{The Diversity of Cold Worlds: Age and Characterization of the Exoplanet COCONUTS-2 b}",
      journal = {\aj},
         year = 2026,
        month = feb,
       volume = {171},
       number = {2},
          eid = {60},
        pages = {60},
          doi = {10.3847/1538-3881/ae230f},
archivePrefix = {arXiv},
       eprint = {2511.20923},
 primaryClass = {astro-ph.EP},
       adsurl = {https://ui.adsabs.harvard.edu/abs/2026AJ....171...60K}
}

@ARTICLE{Faherty2026,
       author = {{Faherty}, Jacqueline K. and {Vos}, Johanna M.},
        title = "{Observed Characteristics of Brown Dwarf Atmospheres}",
      journal = {\araa},
         year = 2026,
        month = aug,
       volume = {64},
       number = {1},
        pages = {409-447},
          doi = {10.1146/annurev-astro-052722-103322},
       adsurl = {https://ui.adsabs.harvard.edu/abs/2026ARA&A..64..409F}
}

\section*{Data Availability}
The data underlying this article will be shared on reasonable request to the corresponding author.

\appendix
\section{Colour-driven spectral typing}
\label{app:color_typing}

We constructed the empirical colour--spectral-type relations from an enlarged parent library of 899 spectroscopically classified objects. The parent library comprises 25 M0--M9.5 spectra from the IRTF/SpeX Spectral Library \citep{Rayner2009}; 845 L0--T7 SPHEREx Quick Release 2 spectrophotometric spectra from the SPHEREx Ultracool Dwarf spectrophotometric Atlas (SUDA; \citealt{Tu2026}); 23 T6--Y1 JWST spectra from \citet{Beiler2024}; three T5--T8 JWST/NIRSpec spectra from the RUBIES brown-dwarf sample \citep{Morrissey2026}; two independently classified T5 and T8 JWST/NIRSpec standards from \citet{Pritchard2026}, for which we adopted the original DAWN v4.4 reductions; and direct F150W and F480M photometry of 20 Y dwarfs and the T9 standard UGPS~J072227.51$-$054031.2 from \citet{Albert2025} (NIRCam photometry is used in priority to the synthesized photometry).

Synthetic AB photometry was calculated by integrating each spectrum through the JWST/NIRCam system-throughput curves obtained from the SVO Filter Profile Service \citep{Rodrigo2012, Rodrigo2024}. The sample was selected independently for each colour: an object was retained only when it had positive measured fluxes, raw S/N$>2$ in both constituent bands, and sufficient spectral coverage in both filters. Consequently, the fitted sample varies between panels, from 216 to 425 objects, while the union of all objects entering at least one relation contains 720 unique sources.

\begin{figure*}
    \centering
    \includegraphics[width=\linewidth]{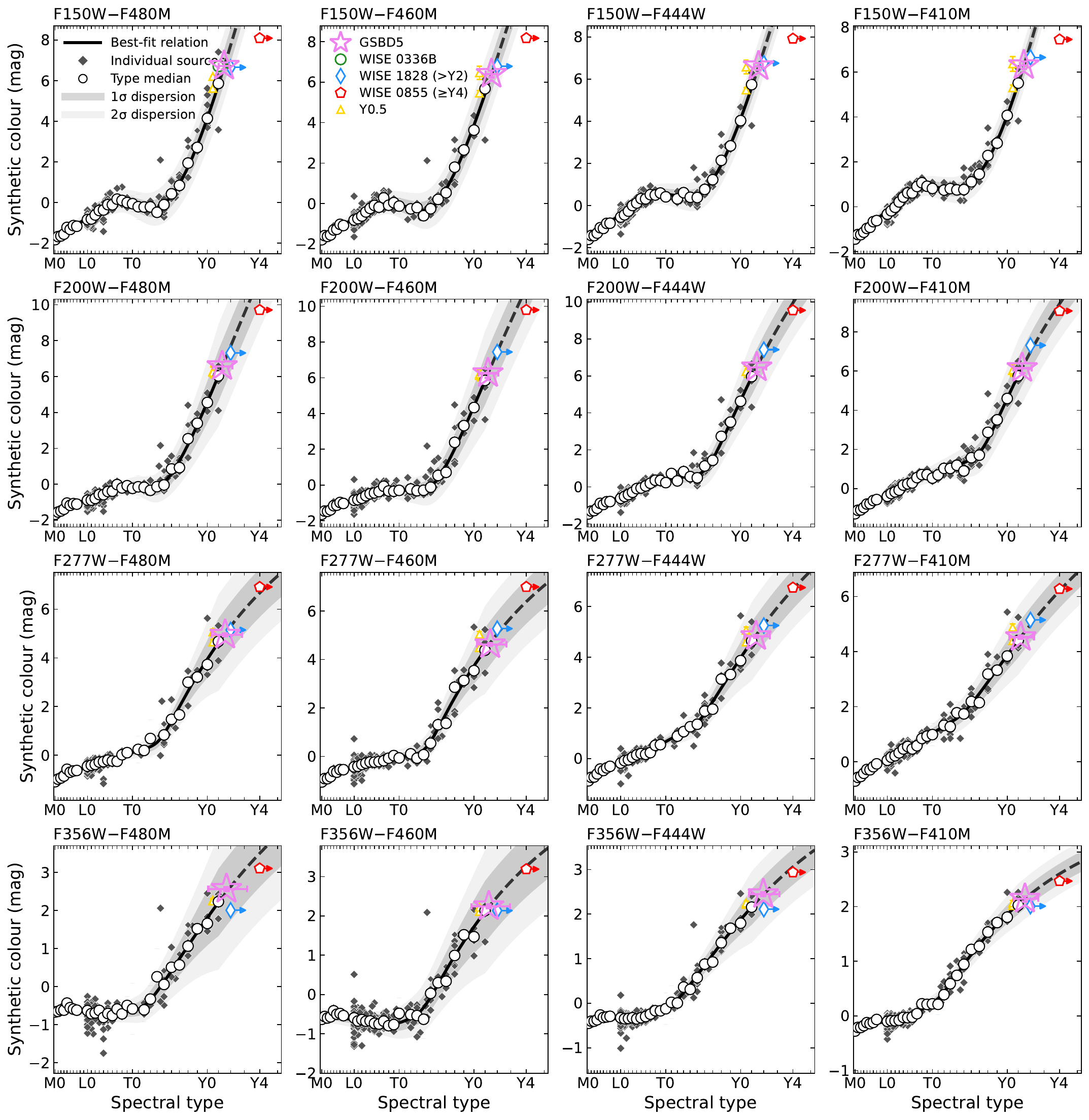}
    \vspace{-0.1in}
    \caption{Synthetic or observed colours, in AB magnitudes, as a function of spectral type. The two Y0.5 sources and the $\geq$Y2 sources are excluded from the fits. All SW--LW colour combinations show consistent trends that can be described by a power law in the late-T to Y regime. The grey bands indicate the 1$\sigma$ and 2$\sigma$ dispersions about the corresponding relation. Among all colours considered, F150W--F480M is the most sensitive subtype diagnostic owing to its steep colour--type gradient. Shaded regions mark 1-$\sigma$ and 2-$\sigma$ equivalent intervals of the fitted relations.}
    \label{fig:color_type}
\end{figure*}

The two objects historically assigned the intermediate Y0.5 subtype in the \citet{Beiler2024} sample were excluded from all fits and shown only as external diagnostics. WISE~J033605.05$-$014350.4B \citep{Albert2025}, WISEPA~J182831.08+265037.8 \citep{Cushing2021, DeFurio2023}, WISE~J085510.83$-$071442.5 \citep{Luhman2024}, and GSBD5 \citep{hainline2026b} were likewise displayed only for comparison and did not constrain the fitted relations. The WISE~1828+2650 colours were synthesized from its public JWST/NIRSpec PRISM spectrum obtained under program GTO~1189 (PI: T. Roellig), whereas the GSBD5 colours were derived from the recalibrated PRISM spectrum presented in this work. We illustrate the relations in Figure~\ref{fig:color_type}, tabulate the fitted parameters in Table~\ref{tab:colour_type_parameters}, and summarise the criteria for different spectral types in Table~\ref{tab:T6_Y4_diagnostic_colours_mnras}.

\begin{table*}
\centering
\caption{Parameters of the colour--spectral-type relations. The model is $C(x)=a+s_1x+(s_2-s_1)S_{0.65}(x-k_1)+A[S_{0.65}(x-k_2)]^p$, where $S_w(z)=w\ln(1+\exp[z/w])$ and the numerical type coordinate is M0 $=0$, L0 $=10$, T0 $=20$, and Y0 $=30$. Each entry gives the fit to the observed subtype medians followed by the $[16,84]$ percentile bootstrap interval.}
\label{tab:colour_type_parameters}

\setlength{\tabcolsep}{4pt}
\renewcommand{\arraystretch}{1.0}

\begin{tabular*}{0.8\textwidth}{@{\extracolsep{\fill}}lrrrr@{}}
\toprule
Colour & $N$ & $a$ & $s_1$ & $s_2$ \\
\midrule
$\mathrm{F150W}-\mathrm{F480M}$ & 245 & $-1.839\,[-1.871,-1.811]$ & $0.108\,[0.104,0.111]$ & $-0.129\,[-0.160,-0.057]$ \\
$\mathrm{F150W}-\mathrm{F460M}$ & 313 & $-1.790\,[-1.812,-1.749]$ & $0.105\,[0.103,0.109]$ & $-0.097\,[-0.148,-0.068]$ \\
$\mathrm{F150W}-\mathrm{F444W}$ & 309 & $-1.647\,[-1.672,-1.614]$ & $0.123\,[0.119,0.126]$ & $-0.043\,[-0.100,-0.025]$ \\
$\mathrm{F150W}-\mathrm{F410M}$ & 372 & $-1.512\,[-1.535,-1.478]$ & $0.135\,[0.130,0.137]$ & $-0.056\,[-0.094,-0.031]$ \\
$\mathrm{F200W}-\mathrm{F480M}$ & 291 & $-1.629\,[-1.648,-1.593]$ & $0.079\,[0.074,0.080]$ & $-0.136\,[-0.092,0.036]$ \\
$\mathrm{F200W}-\mathrm{F460M}$ & 388 & $-1.572\,[-1.584,-1.524]$ & $0.075\,[0.070,0.077]$ & $-0.005\,[-0.033,0.026]$ \\
$\mathrm{F200W}-\mathrm{F444W}$ & 358 & $-1.434\,[-1.451,-1.402]$ & $0.093\,[0.089,0.094]$ & $0.039\,[-0.001,0.075]$ \\
$\mathrm{F200W}-\mathrm{F410M}$ & 425 & $-1.234\,[-1.289,-1.235]$ & $0.094\,[0.093,0.103]$ & $0.111\,[0.042,0.114]$ \\
$\mathrm{F277W}-\mathrm{F480M}$ & 216 & $-1.005\,[-1.034,-0.973]$ & $0.053\,[0.051,0.057]$ & $0.069\,[0.020,0.080]$ \\
$\mathrm{F277W}-\mathrm{F460M}$ & 289 & $-0.973\,[-0.990,-0.931]$ & $0.054\,[0.049,0.056]$ & $0.035\,[-0.084,0.033]$ \\
$\mathrm{F277W}-\mathrm{F444W}$ & 277 & $-0.796\,[-0.826,-0.781]$ & $0.064\,[0.064,0.069]$ & $0.122\,[0.092,0.131]$ \\
$\mathrm{F277W}-\mathrm{F410M}$ & 336 & $-0.653\,[-0.676,-0.634]$ & $0.075\,[0.073,0.077]$ & $0.127\,[0.102,0.142]$ \\
$\mathrm{F356W}-\mathrm{F480M}$ & 221 & $-0.580\,[-0.601,-0.548]$ & $-0.008\,[-0.011,-0.006]$ & $0.023\,[0.009,0.175]$ \\
$\mathrm{F356W}-\mathrm{F460M}$ & 294 & $-0.516\,[-0.550,-0.500]$ & $-0.012\,[-0.013,-0.008]$ & $0.070\,[-0.023,0.107]$ \\
$\mathrm{F356W}-\mathrm{F444W}$ & 295 & $-0.375\,[-0.396,-0.350]$ & $0.004\,[0.001,0.006]$ & $0.050\,[0.046,0.130]$ \\
$\mathrm{F356W}-\mathrm{F410M}$ & 367 & $-0.216\,[-0.238,-0.211]$ & $0.012\,[0.011,0.014]$ & $0.056\,[0.051,0.085]$ \\
\bottomrule
\end{tabular*}

\vspace{0.8em}

\begin{tabular*}{0.8\textwidth}{@{\extracolsep{\fill}}lrrrr@{}}
\toprule
Colour & $A$ & $p$ & $k_1$ & $k_2$ \\
\midrule
$\mathrm{F150W}-\mathrm{F480M}$ & $0.022\,[0.009,0.099]$ & $2.559\,[2.016,2.939]$ & $18.613\,[17.639,19.060]$ & $21.326\,[21.000,22.871]$ \\
$\mathrm{F150W}-\mathrm{F460M}$ & $0.089\,[0.010,0.260]$ & $2.098\,[1.642,2.971]$ & $18.282\,[17.302,18.860]$ & $23.242\,[21.000,24.279]$ \\
$\mathrm{F150W}-\mathrm{F444W}$ & $0.063\,[0.007,0.253]$ & $2.234\,[1.647,3.000]$ & $18.390\,[18.063,19.266]$ & $23.589\,[21.407,24.747]$ \\
$\mathrm{F150W}-\mathrm{F410M}$ & $0.013\,[0.006,0.127]$ & $2.761\,[1.887,3.000]$ & $18.564\,[18.414,19.453]$ & $22.295\,[21.247,24.239]$ \\
$\mathrm{F200W}-\mathrm{F480M}$ & $0.140\,[0.041,1.163]$ & $1.878\,[1.020,2.362]$ & $20.708\,[17.896,20.629]$ & $22.708\,[22.034,26.395]$ \\
$\mathrm{F200W}-\mathrm{F460M}$ & $0.258\,[0.026,1.111]$ & $1.661\,[1.020,2.515]$ & $17.115\,[16.764,18.184]$ & $24.220\,[21.994,26.123]$ \\
$\mathrm{F200W}-\mathrm{F444W}$ & $0.738\,[0.163,1.176]$ & $1.205\,[1.020,1.893]$ & $19.636\,[18.620,21.000]$ & $26.046\,[24.516,26.888]$ \\
$\mathrm{F200W}-\mathrm{F410M}$ & $0.901\,[0.088,1.009]$ & $1.079\,[1.020,2.020]$ & $13.000\,[13.000,21.000]$ & $27.201\,[24.332,27.164]$ \\
$\mathrm{F277W}-\mathrm{F480M}$ & $0.482\,[0.091,0.589]$ & $1.079\,[1.020,1.715]$ & $16.770\,[14.586,20.548]$ & $24.302\,[22.429,24.651]$ \\
$\mathrm{F277W}-\mathrm{F460M}$ & $0.592\,[0.193,0.705]$ & $1.020\,[1.020,1.501]$ & $13.000\,[13.000,20.984]$ & $24.414\,[22.711,24.467]$ \\
$\mathrm{F277W}-\mathrm{F444W}$ & $0.239\,[0.051,0.485]$ & $1.293\,[1.020,1.902]$ & $16.863\,[14.388,17.279]$ & $24.659\,[22.949,25.713]$ \\
$\mathrm{F277W}-\mathrm{F410M}$ & $0.097\,[0.032,0.376]$ & $1.537\,[1.020,1.920]$ & $16.480\,[14.321,17.083]$ & $23.917\,[22.461,26.294]$ \\
$\mathrm{F356W}-\mathrm{F480M}$ & $0.102\,[0.053,0.303]$ & $1.428\,[1.020,1.616]$ & $16.348\,[15.328,20.632]$ & $21.370\,[21.000,24.957]$ \\
$\mathrm{F356W}-\mathrm{F460M}$ & $0.246\,[0.101,0.344]$ & $1.079\,[1.020,1.406]$ & $19.807\,[15.477,20.062]$ & $23.772\,[21.862,24.111]$ \\
$\mathrm{F356W}-\mathrm{F444W}$ & $0.070\,[0.033,0.180]$ & $1.389\,[1.020,1.846]$ & $15.631\,[15.057,18.192]$ & $21.000\,[21.000,25.261]$ \\
$\mathrm{F356W}-\mathrm{F410M}$ & $0.120\,[0.085,0.127]$ & $1.020\,[1.020,1.038]$ & $15.277\,[14.913,16.393]$ & $21.511\,[21.000,22.617]$ \\
\bottomrule
\end{tabular*}

\end{table*}

\begin{table}
\centering
\setlength{\tabcolsep}{2pt}
\renewcommand{\arraystretch}{0.95}
\caption{Diagnostic colours from T6 to Y4 for the reported colour--spectral-type relations in AB magnitude. We also give colour differences to convert the reported AB magnitude to Vega magnitude (Vega = AB + $\Delta$).}
\label{tab:T6_Y4_diagnostic_colours_mnras}
\begin{tabular}{@{}llccc@{}}
\toprule
Colour & SpT & Median & 68.27\% range & 95.45\% range \\
       &     & (mag)  & (mag)        & (mag) \\
\midrule
$\mathrm{F150W}-\mathrm{F480M}$ & T6 & $0.337$ & $[-0.185,0.859]$ & $[-0.706,1.381]$ \\
$\Delta=2.207$ & T7 & $0.929$ & $[0.358,1.501]$ & $[-0.214,2.073]$ \\
& T8 & $1.750$ & $[1.122,2.378]$ & $[0.494,3.005]$ \\
& T9 & $2.820$ & $[2.130,3.511]$ & $[1.440,4.201]$ \\
& Y0 & $4.161$ & $[3.400,4.922]$ & $[2.640,5.682]$ \\
& Y1 & $5.791$ & $[4.952,6.630]$ & $[4.113,7.469]$ \\
& Y2 & $7.729$ & $[6.890,8.569]$ & $[6.050,9.408]$ \\
& Y3 & $9.992$ & $[9.153,10.832]$ & $[8.313,11.671]$ \\
& Y4 & $12.597$ & $[11.757,13.436]$ & $[10.918,14.276]$ \\
\addlinespace[1pt]
\hline
\addlinespace[1pt]
$\mathrm{F150W}-\mathrm{F460M}$ & T6 & $0.132$ & $[-0.406,0.671]$ & $[-0.945,1.209]$ \\
$\Delta=2.134$ & T7 & $0.712$ & $[0.123,1.300]$ & $[-0.466,1.889]$ \\
& T8 & $1.526$ & $[0.881,2.171]$ & $[0.236,2.816]$ \\
& T9 & $2.579$ & $[1.871,3.286]$ & $[1.164,3.993]$ \\
& Y0 & $3.874$ & $[3.097,4.651]$ & $[2.320,5.428]$ \\
& Y1 & $5.416$ & $[4.561,6.270]$ & $[3.707,7.125]$ \\
& Y2 & $7.207$ & $[6.352,8.062]$ & $[5.497,8.917]$ \\
& Y3 & $9.251$ & $[8.395,10.106]$ & $[7.540,10.961]$ \\
& Y4 & $11.550$ & $[10.694,12.405]$ & $[9.839,13.260]$ \\
\addlinespace[1pt]
\hline
\addlinespace[1pt]
$\mathrm{F150W}-\mathrm{F444W}$ & T6 & $0.745$ & $[0.352,1.139]$ & $[-0.041,1.532]$ \\
$\Delta=2.008$ & T7 & $1.224$ & $[0.797,1.651]$ & $[0.370,2.078]$ \\
& T8 & $1.936$ & $[1.471,2.401]$ & $[1.006,2.866]$ \\
& T9 & $2.895$ & $[2.387,3.402]$ & $[1.880,3.910]$ \\
& Y0 & $4.111$ & $[3.556,4.667]$ & $[3.000,5.223]$ \\
& Y1 & $5.596$ & $[4.986,6.205]$ & $[4.377,6.815]$ \\
& Y2 & $7.358$ & $[6.748,7.968]$ & $[6.138,8.578]$ \\
& Y3 & $9.405$ & $[8.795,10.015]$ & $[8.185,10.625]$ \\
& Y4 & $11.745$ & $[11.135,12.355]$ & $[10.525,12.965]$ \\
\addlinespace[1pt]
\hline
\addlinespace[1pt]
$\mathrm{F150W}-\mathrm{F410M}$ & T6 & $1.075$ & $[0.732,1.418]$ & $[0.389,1.761]$ \\
$\Delta=1.874$ & T7 & $1.476$ & $[1.105,1.847]$ & $[0.734,2.218]$ \\
& T8 & $2.086$ & $[1.683,2.488]$ & $[1.281,2.891]$ \\
& T9 & $2.936$ & $[2.498,3.375]$ & $[2.059,3.814]$ \\
& Y0 & $4.059$ & $[3.580,4.539]$ & $[3.100,5.019]$ \\
& Y1 & $5.485$ & $[4.959,6.012]$ & $[4.433,6.538]$ \\
& Y2 & $7.244$ & $[6.717,7.770]$ & $[6.190,8.297]$ \\
& Y3 & $9.363$ & $[8.836,9.889]$ & $[8.309,10.416]$ \\
& Y4 & $11.871$ & $[11.344,12.398]$ & $[10.817,12.924]$ \\
\addlinespace[1pt]
\hline
\addlinespace[1pt]
$\mathrm{F200W}-\mathrm{F480M}$ & T6 & $0.595$ & $[0.035,1.155]$ & $[-0.525,1.715]$ \\
$\Delta=1.735$ & T7 & $1.305$ & $[0.642,1.969]$ & $[-0.022,2.632]$ \\
& T8 & $2.210$ & $[1.418,3.003]$ & $[0.625,3.796]$ \\
& T9 & $3.305$ & $[2.349,4.260]$ & $[1.393,5.216]$ \\
& Y0 & $4.584$ & $[3.421,5.746]$ & $[2.259,6.908]$ \\
& Y1 & $6.044$ & $[4.620,7.468]$ & $[3.196,8.892]$ \\
& Y2 & $7.683$ & $[6.257,9.109]$ & $[4.831,10.535]$ \\
& Y3 & $9.499$ & $[8.073,10.925]$ & $[6.647,12.351]$ \\
& Y4 & $11.488$ & $[10.062,12.914]$ & $[8.636,14.340]$ \\
\addlinespace[1pt]
\hline
\addlinespace[1pt]
$\mathrm{F200W}-\mathrm{F460M}$ & T6 & $0.376$ & $[-0.211,0.962]$ & $[-0.798,1.549]$ \\
$\Delta=1.662$ & T7 & $1.090$ & $[0.404,1.776]$ & $[-0.282,2.462]$ \\
& T8 & $2.018$ & $[1.211,2.825]$ & $[0.403,3.633]$ \\
& T9 & $3.131$ & $[2.173,4.088]$ & $[1.216,5.046]$ \\
& Y0 & $4.412$ & $[3.268,5.555]$ & $[2.124,6.699]$ \\
& Y1 & $5.849$ & $[4.474,7.223]$ & $[3.100,8.598]$ \\
& Y2 & $7.434$ & $[6.058,8.810]$ & $[4.681,10.187]$ \\
& Y3 & $9.160$ & $[7.784,10.537]$ & $[6.407,11.913]$ \\
& Y4 & $11.022$ & $[9.646,12.398]$ & $[8.269,13.775]$ \\
\addlinespace[1pt]
\hline
\addlinespace[1pt]
$\mathrm{F200W}-\mathrm{F444W}$ & T6 & $0.912$ & $[0.480,1.344]$ & $[0.048,1.776]$ \\
$\Delta=1.536$ & T7 & $1.504$ & $[1.003,2.005]$ & $[0.502,2.505]$ \\
& T8 & $2.411$ & $[1.826,2.997]$ & $[1.241,3.582]$ \\
& T9 & $3.494$ & $[2.805,4.183]$ & $[2.116,4.872]$ \\
& Y0 & $4.673$ & $[3.856,5.490]$ & $[3.039,6.307]$ \\
& Y1 & $5.918$ & $[4.943,6.894]$ & $[3.967,7.869]$ \\
& Y2 & $7.216$ & $[6.239,8.192]$ & $[5.262,9.169]$ \\
& Y3 & $8.558$ & $[7.581,9.534]$ & $[6.604,10.511]$ \\
& Y4 & $9.939$ & $[8.962,10.916]$ & $[7.985,11.892]$ \\
\bottomrule
\end{tabular}
\end{table}

\addtocounter{table}{-1}
\begin{table}
\centering
\setlength{\tabcolsep}{2pt}
\renewcommand{\arraystretch}{0.95}
\caption[]{Continued.}
\begin{tabular}{@{}llccc@{}}
\toprule
Colour & SpT & Median & 68.27\% range & 95.45\% range \\
       &     & (mag)  & (mag)        & (mag) \\
\midrule
$\mathrm{F200W}-\mathrm{F410M}$ & T6 & $1.492$ & $[1.051,1.933]$ & $[0.610,2.374]$ \\
$\Delta=1.402$ & T7 & $1.829$ & $[1.323,2.334]$ & $[0.818,2.839]$ \\
& T8 & $2.510$ & $[1.928,3.091]$ & $[1.346,3.673]$ \\
& T9 & $3.491$ & $[2.817,4.164]$ & $[2.143,4.838]$ \\
& Y0 & $4.608$ & $[3.824,5.392]$ & $[3.040,6.176]$ \\
& Y1 & $5.780$ & $[4.863,6.696]$ & $[3.947,7.612]$ \\
& Y2 & $6.980$ & $[6.063,7.897]$ & $[5.145,8.814]$ \\
& Y3 & $8.200$ & $[7.282,9.117]$ & $[6.365,10.034]$ \\
& Y4 & $9.435$ & $[8.518,10.352]$ & $[7.601,11.269]$ \\
\addlinespace[1pt]
\hline
\addlinespace[1pt]
$\mathrm{F277W}-\mathrm{F480M}$ & T6 & $1.400$ & $[1.072,1.728]$ & $[0.744,2.056]$ \\
$\Delta=1.110$ & T7 & $2.003$ & $[1.604,2.402]$ & $[1.204,2.802]$ \\
& T8 & $2.638$ & $[2.146,3.130]$ & $[1.654,3.622]$ \\
& T9 & $3.288$ & $[2.674,3.903]$ & $[2.060,4.517]$ \\
& Y0 & $3.950$ & $[3.174,4.726]$ & $[2.398,5.502]$ \\
& Y1 & $4.620$ & $[3.629,5.611]$ & $[2.638,6.602]$ \\
& Y2 & $5.297$ & $[4.304,6.290]$ & $[3.312,7.282]$ \\
& Y3 & $5.981$ & $[4.988,6.973]$ & $[3.995,7.966]$ \\
& Y4 & $6.670$ & $[5.677,7.663]$ & $[4.685,8.655]$ \\
\addlinespace[1pt]
\hline
\addlinespace[1pt]
$\mathrm{F277W}-\mathrm{F460M}$ & T6 & $1.155$ & $[0.808,1.502]$ & $[0.461,1.849]$ \\
$\Delta=1.037$ & T7 & $1.776$ & $[1.365,2.188]$ & $[0.953,2.600]$ \\
& T8 & $2.423$ & $[1.929,2.916]$ & $[1.436,3.410]$ \\
& T9 & $3.077$ & $[2.479,3.676]$ & $[1.880,4.274]$ \\
& Y0 & $3.735$ & $[3.002,4.469]$ & $[2.268,5.203]$ \\
& Y1 & $4.396$ & $[3.488,5.304]$ & $[2.580,6.212]$ \\
& Y2 & $5.058$ & $[4.149,5.968]$ & $[3.240,6.877]$ \\
& Y3 & $5.723$ & $[4.813,6.632]$ & $[3.904,7.541]$ \\
& Y4 & $6.388$ & $[5.479,7.298]$ & $[4.570,8.207]$ \\
\addlinespace[1pt]
\hline
\addlinespace[1pt]
$\mathrm{F277W}-\mathrm{F444W}$ & T6 & $1.775$ & $[1.475,2.075]$ & $[1.176,2.375]$ \\
$\Delta=0.911$ & T7 & $2.247$ & $[1.891,2.603]$ & $[1.535,2.958]$ \\
& T8 & $2.783$ & $[2.357,3.209]$ & $[1.930,3.636]$ \\
& T9 & $3.362$ & $[2.847,3.878]$ & $[2.331,4.394]$ \\
& Y0 & $3.975$ & $[3.346,4.605]$ & $[2.717,5.234]$ \\
& Y1 & $4.616$ & $[3.842,5.390]$ & $[3.068,6.164]$ \\
& Y2 & $5.282$ & $[4.506,6.057]$ & $[3.731,6.832]$ \\
& Y3 & $5.970$ & $[5.194,6.745]$ & $[4.419,7.520]$ \\
& Y4 & $6.678$ & $[5.902,7.453]$ & $[5.127,8.228]$ \\
\addlinespace[1pt]
\hline
\addlinespace[1pt]
$\mathrm{F277W}-\mathrm{F410M}$ & T6 & $2.102$ & $[1.777,2.428]$ & $[1.451,2.753]$ \\
$\Delta=0.777$ & T7 & $2.474$ & $[2.106,2.842]$ & $[1.739,3.210]$ \\
& T8 & $2.897$ & $[2.480,3.314]$ & $[2.063,3.732]$ \\
& T9 & $3.363$ & $[2.888,3.838]$ & $[2.412,4.314]$ \\
& Y0 & $3.867$ & $[3.324,4.411]$ & $[2.780,4.954]$ \\
& Y1 & $4.406$ & $[3.783,5.030]$ & $[3.159,5.654]$ \\
& Y2 & $4.978$ & $[4.354,5.602]$ & $[3.729,6.226]$ \\
& Y3 & $5.580$ & $[4.956,6.204]$ & $[4.332,6.829]$ \\
& Y4 & $6.211$ & $[5.587,6.836]$ & $[4.963,7.460]$ \\
\addlinespace[1pt]
\hline
\addlinespace[1pt]
$\mathrm{F356W}-\mathrm{F480M}$ & T6 & $0.417$ & $[0.054,0.780]$ & $[-0.309,1.143]$ \\
$\Delta=0.621$ & T7 & $0.733$ & $[0.305,1.160]$ & $[-0.122,1.587]$ \\
& T8 & $1.072$ & $[0.565,1.578]$ & $[0.058,2.085]$ \\
& T9 & $1.432$ & $[0.826,2.037]$ & $[0.221,2.643]$ \\
& Y0 & $1.811$ & $[1.082,2.541]$ & $[0.353,3.270]$ \\
& Y1 & $2.209$ & $[1.325,3.093]$ & $[0.441,3.977]$ \\
& Y2 & $2.624$ & $[1.739,3.509]$ & $[0.854,4.394]$ \\
& Y3 & $3.055$ & $[2.170,3.940]$ & $[1.285,4.825]$ \\
& Y4 & $3.501$ & $[2.616,4.387]$ & $[1.731,5.272]$ \\
\addlinespace[1pt]
\hline
\addlinespace[1pt]
$\mathrm{F356W}-\mathrm{F460M}$ & T6 & $0.263$ & $[-0.129,0.656]$ & $[-0.522,1.049]$ \\
$\Delta=0.548$ & T7 & $0.617$ & $[0.170,1.063]$ & $[-0.276,1.509]$ \\
& T8 & $0.980$ & $[0.471,1.490]$ & $[-0.039,2.000]$ \\
& T9 & $1.351$ & $[0.765,1.936]$ & $[0.180,2.522]$ \\
& Y0 & $1.726$ & $[1.050,2.401]$ & $[0.374,3.077]$ \\
& Y1 & $2.105$ & $[1.321,2.888]$ & $[0.538,3.671]$ \\
& Y2 & $2.487$ & $[1.703,3.271]$ & $[0.919,4.055]$ \\
& Y3 & $2.872$ & $[2.088,3.656]$ & $[1.304,4.440]$ \\
& Y4 & $3.260$ & $[2.476,4.044]$ & $[1.692,4.828]$ \\
\bottomrule
\end{tabular}
\end{table}

\addtocounter{table}{-1}
\begin{table}
\centering
\setlength{\tabcolsep}{2pt}
\renewcommand{\arraystretch}{0.95}
\caption[]{Continued.}
\begin{tabular}{@{}llccc@{}}
\toprule
Colour & SpT & Median & 68.27\% range & 95.45\% range \\
       &     & (mag)  & (mag)        & (mag) \\
\midrule
$\mathrm{F356W}-\mathrm{F444W}$ & T6 & $0.852$ & $[0.685,1.019]$ & $[0.518,1.185]$ \\
$\Delta=0.422$ & T7 & $1.090$ & $[0.894,1.285]$ & $[0.698,1.481]$ \\
& T8 & $1.340$ & $[1.109,1.572]$ & $[0.878,1.803]$ \\
& T9 & $1.602$ & $[1.326,1.878]$ & $[1.051,2.154]$ \\
& Y0 & $1.875$ & $[1.543,2.206]$ & $[1.212,2.538]$ \\
& Y1 & $2.157$ & $[1.756,2.558]$ & $[1.355,2.960]$ \\
& Y2 & $2.449$ & $[2.047,2.851]$ & $[1.646,3.252]$ \\
& Y3 & $2.749$ & $[2.348,3.151]$ & $[1.946,3.553]$ \\
& Y4 & $3.058$ & $[2.656,3.460]$ & $[2.255,3.861]$ \\
\addlinespace[1pt]
\hline
\addlinespace[1pt]
$\mathrm{F356W}-\mathrm{F410M}$ & T6 & $1.126$ & $[1.013,1.239]$ & $[0.900,1.352]$ \\
$\Delta=0.288$ & T7 & $1.309$ & $[1.184,1.433]$ & $[1.060,1.557]$ \\
& T8 & $1.492$ & $[1.355,1.629]$ & $[1.218,1.766]$ \\
& T9 & $1.676$ & $[1.524,1.827]$ & $[1.373,1.978]$ \\
& Y0 & $1.860$ & $[1.692,2.028]$ & $[1.524,2.195]$ \\
& Y1 & $2.044$ & $[1.857,2.231]$ & $[1.671,2.417]$ \\
& Y2 & $2.229$ & $[2.042,2.415]$ & $[1.855,2.602]$ \\
& Y3 & $2.413$ & $[2.227,2.600]$ & $[2.040,2.787]$ \\
& Y4 & $2.598$ & $[2.412,2.785]$ & $[2.225,2.972]$ \\
\bottomrule
\end{tabular}
\end{table}

\section{Forward modelling results}\label{app:forward_modelling_results}

Here we present the results of SED fitting with forward models. The fitting methods are described in Section~\ref{subsec:free_grid}. Table~\ref{tab:gsbd5_freegrid_prism_miri} presents the results from fitting the NIRSpec/PRISM spectrum and MIRI photometry. Table~\ref{tab:gsbd5_freegrid_nircam_all_miri} summarises the results from fitting the NIRCam and MIRI photometry, while Table~\ref{tab:gsbd5_evolution_guided_nircam_all_miri} presents fits to the same photometry with additional priors derived from evolutionary models.

\clearpage
\onecolumn
\begin{landscape}
\thispagestyle{plain}

\vspace*{\fill}

\noindent\makebox[\linewidth][c]{%
\begin{minipage}{0.9\GSBDLandscapeTableWidth}
\setlength{\tabcolsep}{3pt}
\renewcommand{\arraystretch}{0.92}

\GSBDtablecaption{table}{Single-component free atmosphere-grid fits to the calibrated NIRSpec/PRISM spectrum and MIRI photometry of GSBD5. The listed atmospheric parameters and $\chi^2_\nu$ are the maximum-likelihood values for the best-$\chi^2$ branch of each model family, whereas $\ln Z$ is the Bayesian evidence within the family. Here $d_{1R_{\rm J}}$ and $\varpi_{1R_{\rm J}}$ are based on assuming $R=1R_{\rm J}$. $N/k$ gives the number of fitted data points and the number of parameters.}
\label{tab:gsbd5_freegrid_prism_miri}
{\centering

\resizebox{\linewidth}{!}{%
\begin{tabular}{llccccccccccccc}
\toprule
Model & Best branch/config. & $N/k$ &
$T_{\rm eff}$ & $\log g$ & $[\mathrm{M/H}]$ & C/O &
$(\mathrm{C/O})/(\mathrm{C/O})_\odot$ & $\log K_{zz}$ & $f_{\rm sed}$ &
$\log_{10}(R/d)^2$ & $d_{1R_{\rm J}}$ & $\varpi_{1R_{\rm J}}$ &
$\chi^2_\nu$ & $\ln Z_{\rm family}$ \\
 & & & (K) & (cgs) & (dex) & & &
($\mathrm{cm^2\,s^{-1}}$) & & & (pc) & (mas) & & \\
\midrule
ATMO2020
 & weak DEQ & 192/3 & 366.0 & 3.64 & -- & -- & -- & -- & --
 & -22.1699 & 281.7 & 3.549 & 23.142 & $6910.445\pm0.111$ \\
ATMO2020++
 & -- & 192/4 & 375.2 & 4.00 & 0.058 & -- & -- & -- & --
 & -21.6577 & 156.2 & 6.40 & 12.416 & $7922.118\pm0.143$ \\
Sonora Bobcat
 & rainout EQ & 192/4 & 533.0 & 5.25 & $-0.500^{*}$ & -- & -- & -- & --
 & -23.5015 & 1305.1 & 0.766 & 25.430 & $6710.927\pm0.096$ \\
Sonora Elf Owl$^{\dagger}$
 & -- & $174/6^{\dagger}$ & 405.6 & $5.50^{*}$ & -0.500 & -- & 2.077 & 8.334 & --
 & -22.7358 & 540.5 & 1.850 & 24.427 & $6400.448\pm0.144$ \\
Sonora Flame Skimmer
 & DEQ & 192/6 & 380.7 & 2.03 & -0.592 & -- & $0.500^{*}$ & 6.750 & --
 & -21.8458 & 194.0 & 5.155 & 14.467 & $7736.933\pm0.161$ \\
Exo-REM cloudless
 & -- & 192/5 & 450.0 & 4.57 & $-0.49485^{*}$ & $0.800^{*}$ & -- & -- & --
 & -22.8930 & 647.8 & 1.544 & 23.847 & $6860.989\pm0.134$ \\
Exo-REM simple cloud
 & -- & 192/5 & $500.0^{*}$ & 4.22 & $-0.49485^{*}$ & 0.796 & -- & -- & --
 & -23.0896 & 812.3 & 1.231 & 23.734 & $6871.952\pm0.134$ \\
Exo-REM varying-$f_{\rm sed}$
 & -- & 192/6 & $500.0^{*}$ & 4.23 & $-0.49485^{*}$ & $0.800^{*}$ & -- & -- & $1.000^{*}$
 & -23.0260 & 754.9 & 1.325 & 23.207 & $6925.268\pm0.156$ \\
Lacy \& Burrows (2023)
 & thin-cloud DEQ & 192/4 & 394.1 & 3.76 & $-0.500^{*}$ & -- & -- & -- & --
 & -22.2781 & 319.1 & 3.134 & 20.685 & $7148.074\pm0.134$ \\
\bottomrule
\end{tabular}%
}}

\vspace{1pt}
\begin{minipage}{\linewidth}
\textit{$^*$} Maximum-likelihood value lying on a fitted-grid boundary.
$^{\dagger}$The Sonora Elf Owl model covers only 0.79~$\mu$m to 15~$\mu$m and thus uses less data than other models.
\end{minipage}

\vspace{8pt}

\GSBDtablecaption{table}{Single-component free atmosphere-grid fits to the measured NIRCam all-band and MIRI photometry of GSBD5. Columns are the same as in Table~\ref{tab:gsbd5_freegrid_prism_miri}.}
\label{tab:gsbd5_freegrid_nircam_all_miri}

\resizebox{\linewidth}{!}{%
\begin{tabular}{llccccccccccccc}
\toprule
Model & Best branch/config. & $N/k$ &
$T_{\rm eff}$ & $\log g$ & $[\mathrm{M/H}]$ & C/O &
$(\mathrm{C/O})/(\mathrm{C/O})_\odot$ & $\log K_{zz}$ & $f_{\rm sed}$ &
$\log_{10}(R/d)^2$ & $d_{1R_{\rm J}}$ & $\varpi_{1R_{\rm J}}$ &
$\chi^2_\nu$ & $\ln Z_{\rm family}$ \\
 & & & (K) & (cgs) & (dex) & & &
($\mathrm{cm^2\,s^{-1}}$) & & & (pc) & (mas) & & \\
\midrule
ATMO2020
 & weak DEQ & 23/3 & 298.2 & 2.61 & -- & -- & -- & -- & --
 & -20.8113 & 59.0 & 16.961 & 21.929 & $-293.857\pm0.114$ \\
ATMO2020++
 & -- & 23/4 & 315.2 & $2.50^{*}$ & -0.721 & -- & -- & -- & --
 & -21.1425 & 86.3 & 11.583 & 10.961 & $-183.873\pm0.129$ \\
Sonora Bobcat
 & rainout EQ & 23/4 & 363.7 & $3.25^{*}$ & $-0.500^{*}$ & -- & -- & -- & --
 & -21.5013 & 130.5 & 7.664 & 55.102 & $-601.316\pm0.122$ \\
Sonora Elf Owl$^{\dagger}$
 & -- & $19/6^{\dagger}$ & 300.0 & $3.00^{*}$ & -0.158 & -- & 1.297 & 8.943 & --
 & -20.9399 & 68.4 & 14.627 & 25.496 & $-222.703\pm0.148$ \\
Sonora Flame Skimmer
 & DEQ & 23/6 & 310.9 & $2.00^{*}$ & -0.254 & -- & 0.909 & $9.000^{*}$ & --
 & -20.8711 & 63.2 & 15.832 & 5.984 & $-136.350\pm0.142$ \\
Exo-REM cloudless
 & -- & 23/5 & 264.2 & $3.00^{*}$ & -0.414 & $0.800^{*}$ & -- & -- & --
 & -20.6226 & 47.4 & 21.076 & 22.043 & $-280.590\pm0.136$ \\
Exo-REM simple cloud
 & -- & 23/5 & 264.6 & $3.00^{*}$ & -0.417 & $0.800^{*}$ & -- & -- & --
 & -20.6269 & 47.7 & 20.972 & 21.923 & $-280.037\pm0.137$ \\
Exo-REM varying-$f_{\rm sed}$
 & -- & 23/6 & 301.8 & $3.00^{*}$ & -0.284 & $0.800^{*}$ & -- & -- & $1.000^{*}$
 & -20.8632 & 62.6 & 15.978 & 5.467 & $-136.301\pm0.157$ \\
Lacy \& Burrows (2023)
 & thick-cloud DEQ & 23/4 & 276.1 & 3.75 & 0.000 & -- & -- & -- & --
 & -20.6604 & 49.6 & 20.179 & 23.041 & $-302.584\pm0.135$ \\
\bottomrule
\end{tabular}%
}

\vspace{1pt}
\begin{minipage}{\linewidth}
${^*}^{\dagger}$Same as in Table~\ref{tab:gsbd5_freegrid_prism_miri}.
\end{minipage}

\end{minipage}%
}

\vspace*{\fill}

\end{landscape}
\clearpage
\twocolumn

\clearpage
\onecolumn
\begin{landscape}
\thispagestyle{plain}

\vspace*{\fill}

\noindent\makebox[\linewidth][c]{%
\begin{minipage}{0.9\GSBDLandscapeTableWidth}
\setlength{\tabcolsep}{3pt}
\renewcommand{\arraystretch}{0.92}

\GSBDtablecaption{table}{Single-component, evolution-prior-guided atmosphere-grid fits to the measured NIRCam all-band and MIRI photometry of GSBD5. The joint evolutionary posterior in $T_{\rm eff}$ and $\log g$ was used as an informative prior but was allowed to be updated by the atmosphere likelihood; the scale $q=(R/d)^2$ was fitted independently with the same proper log-uniform prior for every model.}
\label{tab:gsbd5_evolution_guided_nircam_all_miri}
{\centering

\resizebox{\linewidth}{!}{%
\begin{tabular}{llccccccccccccc}
\toprule
Model & Best branch/config. & $N/k$ &
$T_{\rm eff}$ & $\log g$ & $[\mathrm{M/H}]$ & C/O &
$(\mathrm{C/O})/(\mathrm{C/O})_\odot$ & $\log K_{zz}$ & $f_{\rm sed}$ &
$\log_{10}(R/d)^2$ & $d_{1R_{\rm J}}$ & $\varpi_{1R_{\rm J}}$ &
$\chi^2_\nu$ & $\ln Z_{\rm family}$ \\
 & & & (K) & (cgs) & (dex) & & &
($\mathrm{cm^2\,s^{-1}}$) & & & (pc) & (mas) & & \\
\midrule
ATMO2020
 & strong DEQ & 23/3 & 284.8 & 3.417 & -- & -- & -- & -- & --
 & -20.8091 & 58.8 & 17.003 & 35.784 & $-444.621\pm0.128$ \\
ATMO2020++
 & -- & 23/4 & 350.0 & 4.048 & $0.285^{*}$ & -- & -- & -- & --
 & -21.1546 & 87.5 & 11.424 & 13.672 & $-211.580\pm0.150$ \\
Sonora Bobcat & rainout EQ & 23/4 & 335.3 & 3.768 & 0.004 & -- & -- & -- & -- & -21.3120 & 104.9 & 9.530 & 67.088 & $-720.048\pm0.135$ \\
Sonora Elf Owl$^{\dagger}$ & -- & $19/6^{\dagger}$ & $277.9^{*}$ & 3.427 & 0.143 & -- & 0.934 & $8.983^{*}$ & -- & -20.7186 & 53.0 & 18.871 & 42.062 & $-336.335\pm0.189$ \\
Sonora Flame Skimmer & DEQ & 23/6 & 288.6 & 3.420 & 0.223 & -- & 0.817 & 8.302 & -- & -20.7533 & 55.2 & 18.131 & 31.599 & $-362.292\pm0.189$ \\
Exo-REM cloudless
 & -- & 23/5 & 281.0 & 3.415 & $-0.454^{*}$ & $0.789^{*}$ & -- & -- & --
 & -20.8645 & 62.7 & 15.953 & 34.280 & $-399.259\pm0.179$ \\
Exo-REM simple cloud
 & -- & 23/5 & 281.0 & 3.415 & $-0.457^{*}$ & $0.797^{*}$ & -- & -- & --
 & -20.8682 & 63.0 & 15.884 & 34.318 & $-400.167\pm0.180$ \\
Exo-REM varying-$f_{\rm sed}$
 & -- & 23/6 & 281.1 & 3.414 & $-0.403^{*}$ & $0.800^{*}$ & -- & -- & $1.001^{*}$
 & -20.8453 & 61.3 & 16.309 & 32.744 & $-379.052\pm0.210$ \\
Lacy \& Burrows (2023)
 & thick-cloud DEQ & 23/4 & 276.2 & 3.750 & 0.000 & -- & -- & -- & --
 & -20.6645 & 49.8 & 20.083 & 23.046 & $-304.556\pm0.160$ \\
\bottomrule
\end{tabular}%
}}

\vspace{1pt}
\begin{minipage}{\linewidth}
${^*}^{\dagger}$Same as in Table~\ref{tab:gsbd5_freegrid_prism_miri} and \ref{tab:gsbd5_freegrid_nircam_all_miri}.
\end{minipage}

\end{minipage}%
}

\vspace*{\fill}

\end{landscape}
\clearpage
\twocolumn

\section{Data used for the colour-magnitude diagrams}
\label{app:cmd_ref}

Here we detail the data and references used to make the four colour-magnitude diagram panels in Figure~\ref{fig:cmd}.

\textit{Upper left:} 2MASS $J-K_{\rm s}$ colour vs. 2MASS $J$ absolute magnitude. Comparison samples comprise M, L and T dwarfs from UltracoolSheet v2.1.0 \citep{Best2021} and the 20-pc census \citep{Kirkpatrick2024} with parallaxes $\mathrm{S/N>3}$; direct-imaging M, L, T and untyped companions from the same catalogues; SUDA SPHEREx synthetic photometry for M--T dwarfs \citep{Tu2026}; \textit{JWST}/NIRSpec late-T/Y dwarfs from \citet{Beiler2024}; MKO late-T/Y objects \citep{Leggett2025} converted to 2MASS following \citet{Stephens2004}; and the coldest dwarfs WISE 0855 \citep{Luhman2024}, WISE 1828 \citep{Kirkpatrick2021} and WISE 1534 \citep{Faherty2025}. 

\textit{Upper right:} \textit{JWST}/NIRCam $F150W-F444W$ vs. $F444W$ absolute magnitude. Data include SUDA L/T dwarfs; NIRSpec synthetic late-T/Y dwarfs \citep{Beiler2024}; JADES T/Y candidates \citep{hainline2026b}; NIRSpec L--Y dwarfs from RUBIES \citep{Morrissey2026} and UNCOVER \citep{Burgasser2024}; brown dwarfs MoM-239450 \citep{Hainline2026c}, WISE 2220 \citep{Albert2025}, WISE 1828, and WISE 1534; Y-dwarf Capotauro \citep{Liu2026}; and NIRSpec synthetic L-type planetary-mass companions TWA 27B \citep{Luhman2023} and VHS 1256 b \citep{Miles2023}. 

\textit{Bottom left:} \textit{JWST}/NIRCam $F150W-F480M$ vs. $F480M$ absolute magnitude. Data include SUDA M--T dwarfs; NIRSpec late-T/Y dwarfs from \citet{Beiler2024}; NIRCam/NIRISS late-T/Y dwarfs from \citet{Albert2025}; an NIRSpec L--Y compilation from GTO~1189, GTO~1275, GO~3486, GO~3670, GO~4539, GO~5118 and GO~6084; metal-poor ultra-cool dwarfs from GO~4668; M--T dwarfs with SED-fitted distances from \citet{Tu2025b}; direct-F480M Y-dwarf candidates from \citet{hainline2026b}; time-variable L/early-T objects WISE~1049AB \citep{Biller2024, X_Chen2025} and SIMP-0136 \citep{Nasedkin2025}; and two young M dwarfs from \citet{Manjavacas2024}. 

\textit{Bottom right:} \textit{JWST}/NIRCam $F200W-F444M$ vs. $F444M$ absolute magnitude. Data include SUDA, \citet{Beiler2024}, \citet{Tu2025b}, \citet{hainline2026b}, the same NIRSpec program compilation, synthetic planetary-mass companions TWA~27B, VHS~1256b \citep{Miles2023}, WD~0806~b \citep{Lew2026} and COCONUTS-2~b \citep{Kiman2026a}; direct-imaging companions 14 Her c \citep{BardalezGagliuffi2025}, TWA 7 b \citep{Lagrange2025, Crotts2025}, AF~Lep~b \citep{Franson2024}, 51~Eri~b \citep[synthesized from its SPHERE spectrum;][]{BrownSevilla2023, Madurowicz2025} and RX~J0534~b \citep{FerrerChavez2026}; NIRSpec brown dwarfs in the Bullet cluster field \citep{bradac2026}; and late-T planetary-mass companion Ross 458 C \citep{Meynardie2025}.

\section{Photometry of other Y-dwarf candidates in JADES}
\label{app:y_photo}

\begin{table*}
\centering
\caption{Photometry of all other nine Y-dwarf candidates in the JADES fields as selected by \citet{Hainline2024b}, now updated with PSF photometry and more \textit{JWST} NIRCam and MIRI bands. The spectral types (SpTs) are given by fitting the photometry to spectral templates as described in Section~\ref{subsec:frontier}. Flux densities and $1\sigma$ uncertainties are given in nJy.}
\label{tab:photometry_summary}
\footnotesize
\setlength{\tabcolsep}{2.2pt}
\renewcommand{\arraystretch}{1.15}

\begin{tabular}{llccccccccc}
\hline
Source ID & SpT & F070W & F090W & F115W & F150W & F162M & F182M & F187N & F200W & F210M \\
\hline
JADES-GN-BD-5$^{\dagger}$ & Y0 & -- & $-0.52\pm0.43$ & $0.41\pm0.34$ & $-0.80\pm0.48$ & -- & $0.44\pm0.90$ & $-5.13\pm8.84$ & $-0.41\pm1.66$ & $1.48\pm1.58$ \\
JADES-GN-BD-8 & Y1(0.5)$^{\ddagger}$ & $0.73\pm0.87$ & $-0.15\pm0.51$ & $1.11\pm0.41$ & $0.53\pm0.52$ & $0.35\pm0.35$ & -- & -- & $0.04\pm0.44$ & -- \\
JADES-GN-BD-9$^{\dagger}$ & Y1(0.5)$^{\ddagger}$ & -- & $-0.02\pm0.35$ & $0.92\pm0.51$ & $-0.22\pm0.68$ & -- & -- & -- & $0.05\pm0.44$ & -- \\
JADES-GN-BD-11 & Y0 & -- & $-0.03\pm0.28$ & $0.34\pm0.62$ & $-0.41\pm0.62$ & -- & -- & -- & $-0.06\pm0.48$ & -- \\
JADES-GN-BD-16 & Y1(0.5)$^{\ddagger}$ & $1.89\pm1.04$ & $-0.17\pm0.62$ & $1.15\pm0.66$ & $-0.69\pm0.63$ & -- & $-1.08\pm1.33$ & $1.03\pm8.28$ & $1.86\pm0.79$ & -- \\
JADES-GS-BD-4 & Y0 & $-0.27\pm0.44$ & $0.13\pm0.39$ & $1.27\pm0.36$ & $1.29\pm0.25$ & $1.47\pm0.52$ & $0.50\pm0.45$ & -- & $0.34\pm0.46$ & $0.70\pm0.53$ \\
JADES-GS-BD-8 & Y1 & -- & $0.37\pm0.27$ & $3.77\pm0.23$ & $1.40\pm0.25$ & -- & $0.31\pm0.54$ & -- & $0.64\pm0.31$ & $1.63\pm0.52$ \\
JADES-GS-BD-21 & Y1(0.5)$^{\ddagger}$ & $-0.30\pm0.90$ & $0.15\pm0.39$ & $-0.44\pm0.34$ & $-0.01\pm0.39$ & -- & -- & -- & $0.50\pm0.33$ & -- \\
PANO-GS-BD-18 & Y0 & -- & -- & $14.16\pm1.86$ & $9.37\pm1.51$ & -- & -- & -- & $4.04\pm1.83$ & -- \\
\hline
\end{tabular}

\vspace{0.8em}

\begin{tabular}{lcccccccccc}
\hline
Source ID & F250M & F277W & F300M & F335M & F356W & F410M & F430M & F444W & F460M & F480M \\
\hline
GNBD5$^{\dagger}$ & -- & $0.49\pm0.66$ & -- & $1.75\pm1.12$ & $7.12\pm0.61$ & $37.56\pm1.72$ & -- & $35.22\pm1.60$ & -- & -- \\
GNBD8 & -- & $2.08\pm0.53$ & -- & $0.32\pm1.65$ & $12.28\pm0.81$ & $83.47\pm3.07$ & -- & $95.01\pm3.34$ & -- & -- \\
GNBD9$^{\dagger}$ & -- & $0.75\pm0.81$ & -- & $-0.28\pm2.11$ & $3.45\pm0.93$ & $24.90\pm2.06$ & -- & $25.74\pm1.34$ & -- & -- \\
GNBD11 & -- & $-0.12\pm0.78$ & -- & $0.88\pm1.29$ & $2.02\pm0.81$ & $11.45\pm1.61$ & -- & $8.15\pm0.87$ & -- & -- \\
GNBD16 & -- & $0.80\pm0.76$ & -- & $1.41\pm1.86$ & $1.74\pm0.88$ & $13.83\pm1.31$ & -- & $17.95\pm1.50$ & -- & -- \\
GSBD4 & $-0.14\pm0.78$ & $0.07\pm0.52$ & $1.09\pm0.76$ & $-0.96\pm0.77$ & $4.98\pm1.75$ & $24.84\pm1.19$ & -- & $25.18\pm1.08$ & -- & $22.57\pm2.84$ \\
GSBD8 & -- & $0.98\pm0.28$ & -- & $-0.76\pm0.61$ & $7.74\pm0.55$ & $50.94\pm1.68$ & $67.82\pm3.18$ & $49.93\pm1.63$ & $38.29\pm3.79$ & $38.43\pm2.59$ \\
GSBD21 & $-2.16\pm3.21$ & $0.30\pm0.45$ & -- & $1.44\pm1.42$ & $1.37\pm0.56$ & $11.70\pm1.35$ & -- & $15.78\pm1.13$ & -- & -- \\
GSBD18 & -- & $6.53\pm1.50$ & -- & -- & $38.98\pm2.14$ & -- & -- & $236.63\pm7.48$ & -- & -- \\
\hline
\end{tabular}
\begin{minipage}{0.8\textwidth}
\vspace{0.05in}
$^{\dagger}$ Expected to have future JWST/NIRSpec observations as part of Cycle 5 GO~10264 (PI Finkelstein).\\
\vspace{0.05in}
$^{\ddagger}$ The best spectral type match is to the Y0.5 templates in \citet{Beiler2024} when adding them to the fit.
\end{minipage}

\end{table*}

\bsp	
\label{lastpage}

\end{document}